\documentclass[12pt,a4paper]{report}

\usepackage[english]{babel}
\usepackage{amsmath}
\usepackage{feynmf}
\usepackage{latexsym}
\usepackage[dvips]{graphicx}
\usepackage{amssymb}
\usepackage{textcomp}
\usepackage{hyperref}
\usepackage{changepage}

\begin{document}

 \def\be{\begin{equation}}
 \def\ee{\end{equation}}
 \def\l{\lambda}
 \def\a{\alpha}
 \def\b{\beta}
 \def\g{\gamma}
 \def\d{\delta}
 \def\e{\epsilon}
 \def\m{\mu}
 \def\n{\nu}
 \def\t{\tau}
 \def\p{\partial}
 \def\s{\sigma}
 \def\r{\rho}
 \def\sl{\ds}
 \def\ds#1{#1\kern-1ex\hbox{/}}
 \def\sla{\raise.15ex\hbox{$/$}\kern-.57em}
 \def\nn{\nonumber}
 \def\bea{\begin{eqnarray}}
 \def\eea{\end{eqnarray}}
 \newcommand{\bth}{{\bf 3}}
 \newcommand{\btw}{{\bf 2}}
 \newcommand{\bon}{{\bf 1}}
 \def\QQ{{Q_Q}}
 \def\QU{{Q_{U^c}}}
 \def\QD{{Q_{D^c}}}
 \def\QL{{Q_L}}
 \def\QE{{Q_{E^c}}}
 \def\QHu{{Q_{H_u}}}
 \def\QHd{{Q_{H_d}}}
 \def\cA{\mathcal{A}}
 \def\Tr{\textnormal{Tr}}
 \def\th{\theta}
 \def\pd{\partial}
 \def\thth{\theta^2 \bar\theta^2}
 \def\sb{\bar{\s}}
 \def\psib{{\bar{\psi}}}
 \def\f{\phi}
 \def\Eps{\epsilon^{\mu\nu\rho\sigma}}
 \def\({\left(}
 \def\){\right)}
 \def\[{\left[}
 \def\]{\right]}
 \def\lb{{\bar\l}}
 \def\half{\frac{1}{2}}
 \def\la{\langle}
 \def\ra{\rangle}
 \def\numeq{n^{eq}}
 \def\D{\Delta}

\title{}{\center{\Large{Universit\`a di Roma Tor Vergata}\\
         \tiny{\rule{13.7cm}{.3pt}}\\
         \Large{Facolt\`a di Scienze Matematiche, Fisiche e Naturali\\}
                 \Large{Corso di Dottorato in Fisica\\}
                 \quad\\
         \quad\\
         \quad\\
         \tiny\quad\\
         \Huge\textbf{Anomalous $U(1)$, Dark Matter and Asymmetry}
         \quad\\
         \large\quad\\
         \quad\\}
\large{\quad\\
%            \qquad \quad Candidato: \qquad \qquad \qquad \qquad \qquad \quad Relatore:\\}
%            \qquad \ \,\,Candidato: \qquad \qquad \qquad \qquad \ \ \,Relatore: \emph{Chiar. Prof.}\\}
             \qquad \qquad Student: \qquad \qquad \qquad \qquad \quad \quad Advisor: \emph{Prof.}\\}
         \Large\textbf{Andrea Mammarella \qquad \quad \   Francesco
         Fucito\\}
         \quad\\
         \quad\\
         \quad \quad \quad \large{Coordinator: \emph{Prof.}\\}
%        \quad\\
         \quad \quad \quad  \Large\textbf{Piergiorgio Picozza}\\
         \quad\\
         \quad\\
         \large\quad\\
         \quad\\
         \quad\\
         \center{
         \tiny\quad\\
         \quad\\
         \tiny{\rule{13.7cm}{.3pt}}\\
         \large{Academic year 2010-2011}\\
         \thispagestyle{empty}

         \newpage
%         \center\normalsize{\quad\\}
         \thispagestyle{empty}
         \newpage

        \thispagestyle{empty}
        \newpage
\quad           }
   }

\newpage
\qquad \qquad \qquad \quad \Huge{\textbf{Ringraziamenti}} \normalsize \\
\quad \\
Un dottorato \`{e} un traguardo importante, uno di quelli di cui si rimane fieri per 
una vita intera. Per questo inizio questa dedica da me stesso, conscio di quello che ho fatto per arrivare
a questo risultato.\\ Voglio poi dedicarlo alla mia famiglia, che non mi ha mai fatto mancare affetto
e serenit\`{a}. Ai miei genitori, per avermi sostenuto ed 
incoraggiato ogni giorno. A Silvia, che \`{e} cresciuta insieme a me e mi fa ricordare che talvolta i piedi
devono stare per terra. A Gian Luigi, perfetto compagno di svago, che condivide molte delle mie passioni e
con cui ho trascorso infinite ore di sfida e di divertimento.\\
Restando in ambito familiare, voglio dedicare questa tesi ai miei zii, Margherita e Paolo, ed alle mie cugine,
Federica e Giulia. Spesso essere parenti non significa nulla di particolare, ma nel loro caso non \`{e} stato cos\`{\i}.
Il vostro affetto \`{e} stato ed \`{e} importante.\\
Una dedica va anche a Francesco Fucito, che oltre ad accompagnarmi in questo percorso, mi ha insegnato quasi tutto
quello che so sul lavoro del fisico. Se quello che era un sogno diventer\`{a} il mio futuro, sar\`{a} anche grazie a lui.\\
Ringrazio anche Andrea Lionetto, Antonio Racioppi e Daniel Ricci Pacifici, che a pi\`{u} riprese hanno lavorato e 
condiviso questo percorso con me.\\
Voglio dedicare questo risultato a Matteo, amico e compagno di squadra da anni, con cui ho condiviso momenti
importanti, con e senza un pallone a spicchi tra le mani.\\
Da ultimo, ma non per importanza, dedico questa tesi ad Ilaria, per la sua presenza, il suo sorriso, la sua
fiducia incrollabile in me e soprattutto per l'amore profondo che mi dimostra ogni giorno.       
         \thispagestyle{empty}
         \newpage
\quad
\newpage
 \tableofcontents

\chapter{Introduction}
Astrophysical observations show the existence of gravitational
effect of an unknown type of matter that does not emitt 
electromagnetic radiation, thus being invisible to every 
direct relevation with the actual experimental instruments.
For this reason this kind of matter has been called dark matter.\\
There are many evidences of its existence: the rotational curves
of the galaxies differ from the theoretical predictions made
taking only count of the visible matter, but they are explained if
we add a dark matter contribution to the calculation.\\
An estimate of the dark matter abundance in the Universe
can be made studying the CMB (Cosmic Microwave Background).
This is one of the goals of WMAP (Wilkinson Microwave Anisotropy Probe)
experiment, that has given very important constraints on
the Universe composition. Analyzing these data it has been
calculated that the abundance of baryons and of matter in the 
actual universe are, respectively, 
 $\Omega_b h^2=0.024 \pm
0.001$ and $\Omega_M h^2=0.14 \pm 0.02$, where $\Omega_M h^2$ 
is defined as  $\Omega_M h^2=\frac{\r}{\r_{crit}}$. The sum of 
these contributions is about $20\%$ of the Universe mass,
so the existence of Dark Matter is necessary to justify these data.\\
Many hypothesis have been made to try to explain the nature
of dark matter: primordial black holes, massive neutrinos, boson
stars, brown dwarfs, non-luminous matter (black holes, non luminous gases, ...). 
However, one of the most natural explanation is given by
the WIMPs and XWIMPs, particles appearing in some supersymmetric
models. \\
A WIMP  (weakly interacting massive particle) is a massive particle that have
only weak or gravitational interactions; an XWIMP (extra weak interacting massive
particle) is a massive particle whose interactions are weaker than those 
of a WIMP, so are not of SM origin. \\
Both types are not predicted by the SM. They appear if we extend 
the SM with supersymmetry. We will show that supersymmetry
implies the existence of massive fermions called neutralinos, that are
superpartner of W, Z, or Higgs bosons. We will show that a new 
symmetry, the R-parity, grants the stability of the lightest of these
particles (the so-called LSP, lightest supersymmetric particles),
that in this way becomes a candidate to explain the dark matter. \\
We will start describing the Minimal Supersymmetric Standard Model
(MSSM), focusing on certain aspects that will be useful for our
purposes, as the neutralinos sector, that will be central in this
thesis. Then we recap the results for the cross-sections 
of the annihilation $\chi_0 \chi_0 \rightarrow f\bar{f}$
and how these affect the LSP relic density calculation.\\
Then we will examinate a different model, in which the MSSM is
extended by an additional $U(1)$ gauge group. We will describe
the changes brought by this extension, especially in the neutralinos
sector, then we will calculate the new interaction vertices and the
related cross sections that affect the LSP relic density calculation.
We will focus on the case in which the LSP is mainly a combination
of particles introduced by the extension, while the LSP is mainly
a MSSM neutralinos combination.
Then we will modify the DarkSUSY package to perform calculations in this
extended framework and we will discuss the results in the most
general case.
In the second part of the thesis we will study another phenomenological
signature of our extended model, the asymmetry. This is one of the main
observables that will be studied at the LHC because can be used to impose constrains
on the new physics beyond the SM and to distinguish among different models 
predictions. \\ We will start describing 
the asymmetry at the LHC, where the simmetry of the initial state ($pp$)
imposes cuts on the parameter space that lead to different asymmetry definitions.
We will explain these definitions. \\
We will calculate the cross section of the process $pp \rightarrow e^+ e^-$
and we will use these results
to calculate the asymmetry at the LHC for all the definitions previously
described. We will optimize this computation calculating the significance related to each
definition and then choosing the cut on the parameter space in order to maximize its 
value.\\
In this way there will remain only the dependence of the various asymmetries
from the free charges of our extended model, so we will study its behaviour
with respect to couple of them, showing bidimensional plot of the results.
Finally we will study the general case, in which we will have dependence
from three free charges.

\chapter{MSSM dark matter}
\section{MSSM}
\subsection{Model setup}
In the SM the radiative corrections to the Higgs mass are proportional to
the square of a
renormalization cutoff, that we will call $\Lambda^2_{UV}$. This is an 
ultraviolet cutoff, needed to deal with the loop corrections of the theory.\\
However, this implies that its value is large, much larger than the order of 
magnitude expected for the Higgs mass, that is around $(150~GeV)$. This implies
that the corrections are larger than the tree level value, making the Higgs
mass larger and larger. Because the fermions and bosons masses in the SM are
related to the Higgs mass, this problem will propagate also on their masses, 
affecting all the mass spectrum of the model.\\
 This problem can be solved extending
the SM i.e. imaging that it is not valid at every energy scale but it is 
only the low energy approximation of a more general theory.
One of the most promising extensions is supersymmetry. \\
It is not a symmetry similar to those of the SM because it connects
boson to fermion and viceversa. Thus the first request for a theory
to be supersymmetric is that the number of fermions equals the
number of bosons.\\
The MSSM is the minimal possible supersymmetrization of the SM,
i.e. using the minimal number of particle to include the SM and to
have a well defined theory.\\
The first idea is to connect each boson to a fermion of the
SM. Unfortunately the simple counting of the number of bosons 
and fermions in the SM shows that the first ones are 28:

\begin{table}[h]
\centering
\begin{tabular}[h]{|c|c|c|c|c|}
\hline  & $G_{\m}^a(SU(3))$ & $A_{\m}^a(SU(2))$ & $B_{\m}(U(1))$ & Higgs \\
\hline degrees of freedom & $8 \times 2$ & $3 \times 2$ & $1 \times 2$ & $2 \times 2$ \\
\hline
\end{tabular}
\end{table}

\newpage   %messo per l'impaginazione

while the fermions are 45:

\begin{table}[h]
\centering
\begin{tabular}[h]{|c|c|c|c|c|c|c|c|}
\hline  & $e_L$ & $u_L^a$ & $d_L^a$ & $e_R$ & $u_R^a$ & $d_R^a$ & $\n_L$ \\
\hline degrees of freedom & $1 \times 3$ & $3 \times 3$ & $3 \times 3$ & $1 \times 3$ & $3 \times 3$ & $3 \times 3$ & $1 \times 3$ \\
\hline
\end{tabular}
\end{table} Because supersymmetry associates a fermionic field to each bosonic 
field it is obvious that we cannot require this symmetry without changing
the number of particles of the theory. Thus the MSSM is built imposing
that each SM field is a component of a chiral superfields if it is
a fermion or a bosonic superfield if it is a boson. In this way each particle
has naturally its own superpartner.\\
This method does not apply on the Higgs field: there is a general theorem
that states that in a supersymmetric theory terms with  $\varphi^{\dagger}$
are forbidden in the superpotential. So it was introduced a second higgs
with the quantum numbers of the complex conjugate of the first one.
In this way we can combine the two higgses to obtain terms in 
the lagrangian that are invariant under the usual gauge symmetries.\\
For completeness we give the higgs-fermion couplings after the introduction
of the second Higgs field:

\be W=\l_{ij} U_i^a H_a \bar{U}_j + \l'_{ij} U_i^a \tilde{H}^b
\e_{ab} \bar{D}_i + \l''_{ij} L_i^a \tilde{H}^b \e_{ab} \bar{E}_j
+\m H_a \tilde{H}^a \label{higgs} \ee where $U^a, U, D, L^a, E$
are the chiral superfields of quarks left, quarks up right, quarks
down right, leptons left, electron (or mu or tau) right, $H_a, H^b$
are the two higgses.

\subsection{R-symmetry and R-parity} The R-symmetry is a global
$U(1)$ symmetry. Its action is defined in the sequent way:

\be R\phi(x, \theta)=e^{2in\a}\phi(e^{-i\a}\theta, x) \ee in which
$\phi(x, \theta)$ is a chiral supermultiplet, that we know contains
a fermion, its bosonic superpartner and an auxiliary field. The
action of the symmetry on the components is:

\bea &&A \rightarrow e^{2in\a}A \nn \\
&&\psi \rightarrow e^{2i(n-\frac{1}{2})\a}\psi \\
&&F \rightarrow e^{2i(n-1)\a}F \nn \eea where $A$is the scalar component, 
$\psi$ is the spinorial component and $F$ is the auxiliary field.\\
The discrete version of this symmetry \hypertarget{Martin}{\cite{Martin}} is called R-parity
and assigns a quantum number to each particle
in the sequent way:

\be P_R=(-1)^{3(B-L)+2s}\ee where $B$ is the baryonic
number of the particle, $L$ is the lepton number, $s$ is its spin.
We can verify that for each SM particle and for the higgses 
$P_R=1$, while for their superpartners  $P_R=-1$. If this quantum
number is conserved, as happens in almost all supersymmetric models, there
are interesting consequences in which we are interested
and that will be one of the subjects of this thesis. 
The point is the sequent: the LSP, i.e. the lightest particle to have $P_R=-1$,
is absolutely stable. This is obvious: because  $P_R$ is conserved
each particle with $P_R=-1$ cannot decay in one or more particles 
with  $P_R=1$. So the lightest sparticle must be stable.\\
Furthermore, if the LSP is electrically neutral, it interacts
only by weak interactions with ordinary matter,
becoming a perfect candidate to describe dark matter.

\subsection{Soft terms}
The soft terms are added by hands to the Lagrangian of the
MSSM to break the supersymmetry without losing the propriety of
absence of quadratic divergences and respecting gauge invariance.
It was shown \hypertarget{Girardello:1981wz}{\cite{Girardello:1981wz}} that the soft terms
permitted in the MSSM are:

\be \mathcal{L}_{soft}=-(\frac{1}{2}M_a \l^a \l^a+ \frac{1}{6}
a^{ijk} \phi_i \phi_j \phi_k +\frac{1}{2} b^{ij} \phi_i \phi_j +
t^i \phi_i +h.c.) +(m^2)^i_j \phi^{j*} \phi_i \label{soft} \ee where 
$\l^a$ are the gauginos and $\phi_i$are the scalar fields. Note that
the soft terms introduce masses for the gauginos, then there is
a mass for the neutralinos, in which we are interested.

\section{Neutralinos}
Neutralinos are the superparters or the neutral bosons.
Thus in the MSSM there are four neutralinos: two higgsinos
and two gauginos. In the basis $(\tilde{B}, \tilde{W}^0, \tilde{H}_a, \tilde{H}^b)$ 
the neutralinos mass matrix is \hypertarget{Martin}{\cite{Martin}}:

\be \ M_{\tilde{N}}=\left(\begin{array}{cccc}  \ M_1 & 0 & -\frac{g_1 v_d}{2} & \frac{g_1 v_u}{2} \\
 \dots & M_2 & \frac{g_2
v_d}{2} & -\frac{g_2 v_u}{2} \\  \dots & \dots & 0 & -\m \\
 \dots & \dots & \dots & 0 \end{array} \right) \label{neumass}\ee $g_1$ e $g_2$ are the coupling costants of the groups
$U(1)$ and $SU(2)$ of the SM. $v_u$ and $v_d$ are the expectation
values of the two higgses. $M_1, M_2$ come from the gaugino mass soft 
terms in (\ref{soft}), $\m$ comes from the higgsinos mass term in  (\ref{higgs}), 
the terms proportional to  $g_1, g_2$ comes from
the Higgs-higgsino-gaugino couplings.\\
The values of the masses and the related eigenstates are
obtained diagonalizing the mass matrix. So the LSP, 
 $\chi_0$, will be a linear combination of the four gauge 
eigenstates:

\be \chi_0=a\tilde{B}+b\tilde{W}_3^0+c\tilde{H}_a+d\tilde{H}^b \ee
The presence of these four components make possible many 
annihilation possibilities for a couple of LSPs. Considering only the tree
level there are the following channels:

\bea &&\chi_0 \chi_0 \rightarrow hh \quad \chi_0 \chi_0
\rightarrow
HH \quad \chi_0 \chi_0 \rightarrow hH \nn \\
&&\chi_0 \chi_0 \rightarrow AA \quad \chi_0 \chi_0 \rightarrow hA
\quad \chi_0 \chi_0 \rightarrow HA \nn \\
&&\chi_0 \chi_0 \rightarrow H^+ H^- \quad \chi_0 \chi_0
\rightarrow
W^{\pm}H^{\mp} \quad \chi_0 \chi_0 \rightarrow ZH  \\
&&\chi_0 \chi_0 \rightarrow Zh \quad \chi_0 \chi_0 \rightarrow ZA
\quad \chi_0 \chi_0 \rightarrow W^{+}W^{-} \nn \\
&&\chi_0 \chi_0 \rightarrow ZZ \quad \chi_0 \chi_0 \rightarrow
f\bar{f} \nn \eea $h, H, A, H^+, H^-$ are the five higgs components
that remain after giving mass to the gauge bosons (initially
the higgses, each composed by two complex fields, consisted
in eight degrees of freedom). $H^+, H^-$
are the charged higgs, the others are neutral.
$A$ is parity odd, the others are parity even.\\
In \hypertarget{Rosk}{\cite{Rosk}}we can find a complete calculation of the amplitudes
for all the annihilation processes of the LSP at tree level.

\chapter{Minimal anomalous U(1) extension of MSSM \label{miaussm}}
String theory suggests \cite{Pradisi:1988xd}-\cite{Kiritsis:2007zz} the existence of extra U(1) symmetries,
so we want to extend the MSSM. We are going to introduce an extra $U(1)$
symmetry. We will have an extra abelian vectorial multiplet and a
St\"{u}ckelberg multiplet, because in this model we use the St\"{u}ckelberg
mechanism \cite{Zhang:2008xq} to break the extra $U(1)$ symmetry
and because the St\"{u}ckelberg
particle is needed to cancel the anomalies via the Generalized Chern-Simons mechanism. 
So this extension is related to the fact
that many string or GUT theories have extra symmetries with respect to
the SM. If these symmetries appear in the high energy theory
they can influence the low energy theory. Now we want to describe
the features of the model that we will study in the rest of this thesis. 
A detailed treatment can be found in \hypertarget{LioRac}{\cite{LioRac}}.

\section{Model setup}
In this section, we discuss how to extend the Minimal Supersymmetric Standard Model (MSSM) to accommodate an
additional abelian vector multiplet $V^{(0)}$. We
assume that all the MSSM fields are charged under the additional vector multiplet $V^{(0)}$, with charges that are
given in Table~\ref{QTable}, where $Q_i, L_i$ are the left handed quarks and leptons respectively while $U^c_i,
D^c_i, E^c_i$ are the right handed up and down quarks and the electrically charged leptons. The superscript $c$ stands
for charge conjugation. The index $i=1,2,3$ denotes the three different families. $H_{u,d}$ are the two Higgs
scalars.
  \begin{table}[h]
  \centering
  \begin{tabular}[h]{|c|c|c|c|c|}
   \hline & SU(3)$_c$ & SU(2)$_L$  & U(1)$_Y$ & ~U(1)$^{\prime}~$\\
   \hline $Q_i$   & $\bth$       &  $\btw$       &  $1/6$   & $Q_{Q}$ \\
   \hline $U^c_i$   & $\bar \bth$  &  $\bon$       &  $-2/3$  & $Q_{U^c}$ \\
   \hline $D^c_i$   & $\bar \bth$  &  $\bon$       &  $1/3$   & $Q_{D^c}$ \\
   \hline $L_i$   & $\bon$       &  $\btw$       &  $-1/2$  & $Q_{L}$ \\
   \hline $E^c_i$   & $\bon$       &  $\bon$       &  $1$     & $Q_{E^c}$\\
   \hline $H_u$ & $\bon$       &  $\btw$       &  $1/2$   & $Q_{H_u}$\\
   \hline $H_d$ & $\bon$       &  $\btw$       &  $-1/2$  & $Q_{H_d}$ \\
   \hline
  \end{tabular}
  \caption{Charge assignment.}\label{QTable}
  \end{table}

Since our model is an extension of the MSSM, the gauge invariance
of the superpotential, that contains the Yukawa couplings and a
$\m$-term, put constraints on the above charges
  \bea
   \QU &=& - \QQ - \QHu  \nn\\
   \QD &=& - \QQ + \QHu  \nn\\
   \QE &=& -\QL  + \QHu  \nn\\
   \QHd  &=& - \QHu \label{Qconstraints}
  \eea
Thus, $\QQ$, $\QL$ and $\QHu$ are free parameters of the model.

\subsection{Anomalies} \label{anomalies}

As it is well known, the MSSM is anomaly free. All the anomalies
that involve only the $SU(3)$, $SU(2)$ and $U(1)_Y$ factors vanish
identically. However, triangle diagrams with $U(1)'$ current in the external legs
in general are potentially anomalous. These anomalies are\footnote{We are working in an effective field theory framework and we ignore troughout the paper all the gravitational effects.  In particular, we do not consider the gravitational anomalies which, however, could be canceled by the Green-Schwarz mechanism.}
\bea
   U(1)'-U(1)'-U(1)'~~~:     &&\ \cA^{(0)} = \sum_f Q_f^3                        \label{Triangles1}\\
   U(1)'-U(1)_Y - U(1)_Y~~~: &&\ \cA^{(1)} = \sum_f Q_f Y_f^2                     \label{Triangles2}\\
   U(1)'-SU(2)-SU(2)~~~:     &&\ \cA^{(2)} = \sum_f Q_f \Tr[T_{k_2}^{(2)} T_{k_2}^{(2)}] \label{Triangles3}\\
   U(1)'-SU(3)-SU(3)~~~:     &&\ \cA^{(3)} = \sum_f Q_f \Tr[T_{k_3}^{(3)} T_{k_3}^{(3)}] \label{Triangles4}\\
   U(1)'-U(1)'-U(1)_Y~~~:    &&\ \cA^{(4)} = \sum_f Q_f^2  Y_f
   \label{Triangles5}
\eea
where $f$ runs over the fermions in Table \ref{QTable}, $Q_f$ is
the corresponding $U(1)'$ charge, $Y_f$ is the hypercharge and
$T_{k_a}^{(a)}$, $a=2,3;\,\, k_a=1,\ldots,{\rm dim G}^{(a)}$ are
the generators of the $G^{(2)}=SU(2)$ and $G^{(3)}=SU(3)$ algebras
respectively. In our notation $\Tr[T_j^{(a)} T_k^{(a)}] = {\frac{1}{2}}\d_{jk}$. 
All the remaining anomalies that involve $U(1)'$s
vanish identically due to group theoretical arguments
(see Chapter 22 of \hypertarget{Weinberg2}{\cite{Weinberg2}}). Using the charge constraints
(\ref{Qconstraints}) we get
  \bea
   \cA^{(0)} &=& 3\ \Big\{ Q_{H_u}^3 + 3 \QHu Q_L^2 + Q_L^3 - 3 Q_{H_u}^2\ \( \QL + 6 \QQ \) \Big\} \label{A0}\\
   \cA^{(1)} &=& -{\frac{3}{2}} \(3\QQ + \QL  \) \label{A1}\\
   \cA^{(2)} &=&  {\frac{3}{2}} \(3\QQ  + \QL \) \label{A2}\\
   \cA^{(3)} &=& 0 \label{A3}\\
   \cA^{(4)} &=& -6 \QHu \(3\QQ + \QL  \)
 \label{A4} \eea
Notice that the mixed anomaly between the anomalous $U(1)$ and the $SU(3)$ nonabelian factors $\cA^{(3)}$ vanishes
identically.

\subsubsection{Anomalous U(1)'s and the St\"uckelberg mechanism}
We assume that the $U(1)'$ is anomalous, i.e. (\ref{A0})-(\ref{A4}) do not vanish. Consistency of the model is achieved
by the contribution of a St\"uckelberg field $S$ and its appropriate couplings to the anomalous $U(1)'$.
The St\"uckelberg lagrangian reads \hypertarget{Klein:1999im}{\cite{Klein:1999im}}
\bea
   \mathcal{L}_{axion} &=& {\frac{1}{4}} \left. \( S + S^\dagger + 4 b_3 V^{(0)} \)^2 \right|_{\thth} \\
                &&- {\frac{1}{4}} \left\{ \[\sum_{a=0}^2 b^{(a)}_2 S ~\Tr\( W^{(a)} W^{(a)} \) + b^{(4)}_2 S ~W^{(1)} ~W^{(0)} \]_{\th^2} +h.c. \right\}~
   \label{Laxion} \nn \eea  

The St\"uckelberg multiplet is chiral:

\be
    S =  s+ i\sqrt2 \th \psi_S + \th^2 F_S - i \th \s^\m \bar\th \pd_\m s +
                \frac{\sqrt2}{2}  \th^2 \bar\th \bar\s^\m \pd_\m \psi_S - \frac{1}{4} \thth \Box s \label{Smult}
 \ee We decouple the scalar component $s$ in real and imaginary parts:
$s=\a + i \phi$. We assume that the real part assume an expectation 
value from an high energy potential \newline The $S$ sector of the Lagrangian is:

\bea
    \mathcal{L}_{axion} &=& \frac{1}{2} ( \pd_\m \f +2 b_3 V^{(0)}_\m )^2
                   +\frac{i}{4} \psi_S \s^\m \pd_\m \psib_S +{\frac{i}{4}} \psib_S \sb^\m \pd_\m \psi_S \label{axion}\\
                 && +\frac{1}{2} F_S \bar F_S + 2 b_3 \langle\a\rangle D^{(0)}-\sqrt2 b_3(\psi_S \l^{(0)}+h.c.)\nn\\
                 &&- \frac{1}{4} \f \, \Eps \sum_{a=0}^2 b^{(a)}_2 \Tr (  F_{\m \n}^{(a)} F_{\r \s}^{(a)} )- \frac{1}{4} b^{(4)}_2 \Eps \f F_{\m \n}^{(1)} F_{\r \s}^{(0)}\nn\\
                 &&+\frac{1}{2} b^{(4)}_2 \langle\a\rangle F_{\m \n}^{(1)} F_{\m \n}^{(0)}- b^{(4)}_2  \langle\a\rangle D^{(1)} D^{(0)} \nn\\
                 &&-\frac{1}{2} \left\{\sum_{a=0}^2b^{(a)}_2 \[- 2\f \Tr \( \l^{(a)} \s^\m D_\m \lb^{(a)} \) +
                       \frac{i}{\sqrt2} \Tr \( \l^{(a)} \s^\m \sb^\n F_{\m \n}^{(a)} \) \psi_S \right. \right. \nn\\
                 &&\left. - F_S \Tr \(\l^{(a)} \l^{(a)}\) - \sqrt2 \psi_S \Tr \(\l^{(a)} D^{(a)}\)\]\nn\\
                 &&+ b^{(4)}_2 \bigg[ \(-\f \l^{(1)} \s^\m \pd_\m \lb^{(0)}
                 +i\langle\a\rangle  \l^{(1)} \s^\m \pd_\m \lb^{(0)}  -\frac{1}{2}F_S \l^{(1)} \l^{(0)}\right.\nn\\
                 && \left. \left.- \frac{1}{sqrt2} \psi_S  \l^{(1)} D^{(0)}
                 +{\frac{i}{2}\sqrt2} \l^{(1)} \s^\m \sb^\n F_{\m \n}^{(0)} \psi_S\)
                 + (0 \leftrightarrow 1) \ \bigg] +h.c.
                 \right\}\nn
   \eea $b_3$ is a theory free parameter, $b_2^{(a)}$ are parameters that have to be fixed in
order to remove the anomaly. Note that $b_2^{(3)}=0$ because there isn't anomaly related to
$SU(3)$.

Without details, we give the condition on the $b_2^{(a)}$ to eliminate all the anomalies:

\bea b^{(1)}_2 b_3 = - \frac{\cA^{(1)}}{128 \pi^2} \qquad
 b^{(0)}_2 b_3 =-\frac{\cA^{(0)}}{384\pi^2} \label{anomalia}\\
 b^{(2)}_2 b_3 = -\frac{\cA^{(2)}}{64 \pi^2} \qquad
 b^{(4)}_2 b_3 = -\frac{\cA^{(4)}}{128 \pi^2} \nn
\eea

\subsection{Soft terms}
The soft terms of the model are:

\be
\mathcal{L}_{soft}=\mathcal{L}_{soft}^{MSSM}+\mathcal{L}_{soft}^{new}
\ee The expression of the new introduced soft terms is:

\bea \mathcal{L}_{soft}^{new}=- \frac{1}{2}  (M_0 \l^{(0)} \l^{(0)} + h.c.
) - \frac{1}{2}  (\frac{M_S}{2} \psi_S \psi_S  + h.c. )
   \eea

\subsection{Kinetic mixing of the U(1)s}
In the equation \ref{axion} we can see the presence of mixing
between the 2 superfields of the $U(1)s$. We have:

\be
  \left. ( \frac{1}{4}  W^{(0)} W^{(0)} + \frac{1}{4} W^{(1)} W^{(1)}   +\frac{\d}{2} W^{(1)} W^{(0)} ) \right|_{\th^2}
\ee
with $\d = - 4 b^{(4)}_2 g_0 g_1  \langle \a\rangle $. To diagonalize
the kinetic terms we use the matrix:

\be
      \( \begin{array}{c} V^{(0)}\\
                          V^{(1)} \end{array} \) = \( \begin{array}{cc} C_\d &  0 \\
                         -S_\d  &  1 \end{array} \)
            \( \begin{array}{c} V_C\\
                                V_B \end{array} \)
 \label{cdelta} \ee
  where:
  \bea &&C_\d = 1/\sqrt{1-\d^2} \label{mixing} \\ &&S_\d =  \d C_\d \nn \eea

\subsection{D-terms}
The D-terms of the model come from the kinetic terms of the chiral multiplets and from 
(\ref{axion}). Their Lagrangian is:

\bea
    \mathcal{L}_D&=&\frac{1}{2}\sum_{a=0}^3 D^{(a)}_{k_a} D^{(a)}_{k_a} + \sum_{a=0}^3 g_a D^{(a)}_{k_a} z_i^\dag (T^{(a)}_{k_a})_j^i z^j
    + 4 g_0 b_3 \langle\a\rangle D^{(0)} + \d D^{(1)} D^{(0)}+\nn\\
                 &&+2 \[\sum_{a=0}^2 g_a^2 \ b^{(a)}_2  \sqrt2 \psi_S \Tr \(  \l^{(a)} D^{(a)} \)
                + g_0 g_1 \frac{b^{(4)}_2}{\sqrt2} \psi_S \( \l^{(1)} D^{(0)} + \l^{(0)} D^{(1)} \) +h.c. \]\nn\\
\eea $a=0,1,2,3$ runs over the gauge groups, $z_i$ are the lower components
of the $i$-th chiral multiplet (except that of the St\"{u}ckelberg field) and $T^{(a)}_{k_a}$, $k_a=1,\ldots,{\rm dim
G}^{(a)}$ are the generators of the gauge group
${\rm G}^{(a)}$. \newline Using the equation of motion we obtain the expressions:

\bea
    \mathcal{L}_{D_C} ~&=&- {\frac{1}{2}} \left\{   \[C_\d g_0 \sum_f Q_f |z_f|^2 - S_\d g_1 \sum_f Y_f |z_f|^2 \] \right.
                  + C_\d 4 g_0 b_3 \langle\a\rangle \nn\\
                  && ~~~~~~ +2\sqrt2 b^{(0)}_2 g_0^2 \[ \psi_S \( C_\d^2 \l_C\)+h.c. \]
                  +2\sqrt2 b^{(1)}_2 g_1^2 \[ \psi_S \( S_\d^2\l_C -S_\d \l_B \)+h.c. \]\nn\\
                  &&~~~~~~+\sqrt2 b^{(4)}_2 g_0 g_1 \[ \psi_S \( C_\d \l_B - 2 C_\d S_\d \l_C\)+h.c. \]\Bigg\}^2\label{dcterm}\\
\mathcal{L}_{D_B} ~&=&- {\frac{1}{2}} \Bigg\{ g_1 \sum_f Y_f |z_f|^2
                +2\sqrt2 b^{(1)}_2 g_1^2 \[ \psi_S \( \l_B -S_\d \l_C  \)+h.c. \]+\nn\\
               &&~~~~~~+\sqrt2 b^{(4)}_2 g_0 g_1 \[  \psi_S  C_\d  \l_C +h.c. \]\Bigg\}^2\label{dbterm}\\
\mathcal{L}_{D^{(2)}} &=& -\frac{1}{2} \sum_k\left\{g_2 z_i^\dag
(T^{(2)}_k)_j^i z^j
          + b^{(2)}_2 g_2^2 \[  \sqrt2 \psi_S  \l^{(2)}_k  + h.c. \]\right\}^2\label{d2term}\\
      \mathcal{L}_{D^{(3)}} &=& -\frac{1}{2} \sum_k \left\{ g_3 z_i^\dag (T^{(3)}_k)_j^i z^j
          \right\}^2\label{d3term}
     \eea

\subsection{Higgs sector}

It is worth noting that in our model there is no st\"{u}ckelber-higgs mixing.
This is due to the fact that with our field content, $N=1$ supersymmetry and
the gauge invariance with respect of the extra $U(1)$ there can not be a term
that couple the st\"{u}ckelberg field with the higgses
(on the contrary to \hypertarget{Coriano':2005js}{\cite{Coriano':2005js}}).
In fact the only permitted term that contains the Higgses and the St\"{u}ckelberg superfields
is:
\be
W\propto e^{-k S}H_u H_d 
\ee
Calculating the terms of the scalar potential generated by this superpotential, we
can found that there appears only the real part of $S$, that in our model is fixed, while
the imaginary part, $\phi$, that is the propagating field, is not coupled with the higgses.
After the electroweak symmetry breaking we have four gauge
generators that are broken, so we have four longitudinal degrees of freedom. One of them
is the axion, while the other three are the usual NG bosons coming
from the Higgs sector.

The Higgs scalar fields consist of two complex $SU(2)_L$-doublets, or eight real,
scalar degrees of freedom. When the electroweak symmetry is
broken, three of them are the would-be NG bosons
$G^0$, $G^\pm$. The remaining five Higgs scalar mass eigenstates
consist of two CP-even neutral scalars $h^0$ and $H^0$, one CP-odd
neutral scalar $A^0$ and a charge $+1$ scalar $H^+$ as well as its
charge conjugate $H^-$ with charge $-1$.\footnote{
We define $G^{-} = G^{+*}$ and $H^- = H^{+*}$. Also, by convention, $h^0$ is lighter
than $H^0$.} The gauge-eigenstate fields can be expressed in terms
of the mass eigenstate fields as
    \bea
     \( \begin{array}{c} h_u^0 \\
                         h_d^0 \end{array} \) &=& {\frac{1}{\sqrt{2}}} \(\begin{array}{c} v_u \\
                                                                           v_d \end{array}\) +
                                          {1\over \sqrt{2}} R_\alpha \(\begin{array}{c} h^0 \\
                                                                                      H^0 \end{array}\) +
                                          \frac{i}{\sqrt{2}} R_{\beta_0}\(\begin{array}{c} G^0 \\
                                                                                        A^0 \end{array}\)\label{HiggsMin}\\
     \( \begin{array}{c} h_u^+ \\
                         h_d^{-*} \end{array} \) &=&  R_{\beta_\pm}\(\begin{array}{c} G^+ \\
                                                                                        H^+ \end{array}\)
    \eea
where the orthogonal rotation matrices $R_\alpha, R_{\beta_0}, R_{\beta_\pm}$ are the same as in \hypertarget{Martin}{\cite{Martin}}
Acting with these matrices on the gauge eigenstate fields we obtain the diagonal mass terms.
Expanding around the minimum (\ref{HiggsMin}) one finds that
$\beta_0 = \beta_\pm = \beta$, and replacing the tilde parameters
 we obtain the masses
    \bea
     m_{A^0}^2 &=& 2|\m|^2 + m^2_{h_u} + m^2_{h_d}\\
     m^2_{h^0, H^0} &=& \frac{1}{2} \Bigg\{ m^2_{A^0} +
     \(\( g_0 X_\d \)^2 +\frac{1}{4}(g_1^2+g_2^2) \)v^2 \nn\\
              &&\mp \[\(m_{A^0}^2
              - \(\( g_0 X_\d \)^2 +\frac{1}{4}(g_1^2+g_2^2) \)v^2\)^2 \right. \nn\\
     &&\left. + 4 \(\( g_0 X_\d \)^2 +\frac{1}{4}(g_1^2+g_2^2) \)v^2
     m_{A^0}^2  \sin^2 (2\beta)\]^\half \Bigg\}\>\>\>\>\>{} \label{eq:m2hH}\\
     m^2_{H^\pm} &=& m^2_{A^0} + m_W^2 =  m^2_{A^0} +  g_2^2 \frac{v^2}{4} \label{eq:m2Hpm}
    \eea
and the mixing angles
    \bea
    \frac{\sin 2\alpha}{\sin 2 \beta} &=& -\frac{m_{H^0}^2 + m_{h^0}^2}{m_{H^0}^2 - m_{h^0}^2}  \nn\\
     \quad \frac{\tan 2\a}{\tan 2 \b} &=&
           \frac{m_{A^0}^2+
           \(\( g_0 X_\d \)^2 +\frac{1}{4}(g_1^2+g_2^2) \)v^2}{m_{A^0}^2-
     \(\( g_0 X_\d \)^2 +\frac{1}{4}(g_1^2+g_2^2) \)v^2}
    \eea
Notice that only the $h^0$ and $H^0$ masses get modified with
respect to the MSSM, due to the additional anomalous $U(1)'$.

\subsection{Neutral Vectors}   \label{vectmasssol}

There are two mass-sources for the gauge bosons: (i) the
St\"uckelberg mechanism and (ii) the Higgs mechanism. In this extension of
the MSSM, the mass terms for the gauge fields are given by
     \be
      \L_M = \frac{1}{2} \(C_\m \ B_\m \ V^{(2)}_{3\m} \) M^2
                          \( \begin{array}{c} C^\m\\ B^\m\\ V^{(2)\m}_{3} \end{array} \)
     \ee
$C_\mu ,\, B_\mu$ are the lowest components of the vector multiplets $V_C,\,V_B$.
The gauge boson mass matrix is
     \be
      M^2= \( \begin{array}{ccc} M^2_C & ~~~g_0 g_1 \frac{v^2}{2} X_\d& ~~~-g_0 g_2 \frac{v^2}{2} X_\d  \\
                 ... & g_1^2 \frac{v^2}{4} & -g_1 g_2 \frac{v^2}{4}   \\
                 ... & ...  & g_2^2 \frac{v^2}{4}  \\\end{array} \)
  \label{BosonMasses} \ee
where $M^2_C =16 g_0^2 b_3^2 C_\d^2 + g_0^2 (v^2) X_\d^2$ and the lower dots denote the obvious terms under
symmetrization. After diagonalization, we obtain the
eigenstates
 \bea
   A_\m &=&\frac{g_2 B_\m + g_1 V^{(2)}_{3\m}}{\sqrt{g_1^2+g_2^2}}  \label{photon}\\
    Z_{0\m} &=& \frac{g_2 V^{(2)}_{3\m} - g_1 B_\m}{\sqrt{g_1^2+g_2^2}}+g_0 \QHu\frac{\sqrt{g_1^2+g_2^2}  v^2}{2 M_{V^{(0)}}^2} C_\m+{\cal O}[g_0^3,M_{V^{(0)}}^{-3}]  \label{Z0}\\
Z'_\m  &=& C_\m +\frac{g_0 \QHu v^2}{2 M_{V^{(0)}}^2}\(g_1 B_\m- g_2 V^{(2)}_{3\m}\) +{\cal O}[g_0^3,M_{V^{(0)}}^{-3}]
\label{Zprime}
    \eea
and the corresponding masses
   \bea
   M^2_{\g}&=&0\\
    M^2_{Z_0} &=&\frac{1}{4} \(g_1^2+g_2^2\) v^2
                 -(\QHu)^2\frac{\(g_1^2+g_2^2\) g_0^2  v^4}{4 M_{V^{(0)}}^2}+{\cal O}[g_0^3,M_{V^{(0)}}^{-3}]
\label{Z0mass}\\
M^2_{Z'}  &=&M_{V^{(0)}}^2+g_0^2 \[(\QHu)^2 \(1+\frac{g_1^2 v^2+g_2^2
v^2}{4M_{V^{(0)}}^2}\)-\frac{\langle\a\rangle g_1^3 \cA^{(4)}}{64 \pi ^2M_{V^{(0)}}}\] v^2+{\cal
O}[g_0^3,M_{V^{(0)}}^{-3}]~~~~~~~~~
  \label{Zpmass} \eea
where $M_{V^{(0)}}=4 b_3 g_0$ is the mass parameter for the anomalous $U(1)$ and it is assumed to be in the TeV
range. Due to their complicated form, the eigenstates and eigenvalues of $M^2$ (\ref{BosonMasses}) are expressed
as power expansions in  $g_0$ and $1/M_{V^{(0)}}$ keeping only the leading terms. Higher terms are denoted
by ${\cal O}[g_0^3,M_{V^{(0)}}^{-3}]$.

The first eigenstate (\ref{photon}) corresponds to the photon and it is exact to all orders. It slightly
differs from the usual MSSM expression due to the kinetic mixing between $V^{(0)}$ and $V^{(1)}$.

For the rest of the thesis, we neglect the kinetic mixing contribution since they are higher loop
effects which go beyond our scope.
Then the rotation matrix from the hypercharge to the photon
basis, up to ${\cal O}[g_0^3,M_{V^{(0)}}^{-3}]$  is
    \bea
     \( \begin{array}{c} Z'_\m\\
                         Z_{0 \m}\\
                         A_\m \end{array} \)
       &=&O_{ij}
     \( \begin{array}{c} V^{(0)}_\m\\
                         V^{(1)}_\m\\
                         V^{(2)}_{3\m} \end{array} \)\label{Oij}\\
      &=&\( \begin{array}{ccc}      1& g_1 \frac{g_0 \QHu v^2}{2 M_{V^{(0)}}^2}  ~~& -g_2 \frac{g_0 \QHu v^2}{2 M_{V^{(0)}}^2}  \\
                                g_0 \QHu \frac{\sqrt{g_1^2+g_2^2}  v^2}{2M_{V^{(0)}}^2} ~~~& - \frac{g_1}{{\sqrt{g_1^2+g_2^2}}}
& \frac{g_2}{{\sqrt{g_1^2+g_2^2}}}  \\
                                0 & \frac{g_2}{{\sqrt{g_1^2+g_2^2}}} & \frac{g_1}{{\sqrt{g_1^2+g_2^2}}}  \\ \end{array} \)
     \( \begin{array}{c} V^{(0)}_\m\\
                         V^{(1)}_\m\\
                         V^{(2)}_{3\m} \end{array} \) \nn \eea
where $i,j=0,1,2$.

\subsection{Sfermions}

In general, the contributions to the sfermion masses are coming
from (i) the D and F terms in the superpotential and (ii) the
soft-terms. However, in our case, the contribution comes only
from the $D_C$ terms
    \bea
     V^{D_C}_\text{mass} = \bigg\{ \( C_\d g_0 \QHu + \half S_\d g_1 \) \( \frac{v_u^2-v_d^2}{2}\) + 4 C_\d g_0 b_3 \langle\a\rangle  \bigg\}
                 \bigg\{ \sum_f  \( C_\d g_0  Q_f - S_\d g_1 Y_f \) |y_f|^2 \bigg\} \nn\\
    \eea
where the $y_f$ stand for all possible sfermions.

\section{Neutralinos sector}
We have already showed that, with respect to the MSSM, we have two new fields in the neutralinos
sector: $\psi_S$ and
$\l^{(0)}$. Subsequently we will have:

    \be
    \mathcal{L}_{\mbox{neutralino mass}} = -\frac{1}{2} (\psi^{0})^T {\bf M}_{\tilde N} \psi^0 + h. c.
    \ee in which we have chosen the basis:

   \be(\psi^{0})= (\psi_S, \ \l_C,\ \l_B,\ \l^{(2)},\ \tilde h_d^0,\ \tilde h_u^0) \label{neutrbase}
    \ee $\bf {M}_{\tilde N}$ receives contributions from many terms of the Lagrangian:\\
1) MSSM terms \\
2) soft terms coming from $\mathcal{L}^{new}_{soft}$ \\
3) $h-\tilde{h}-\l'$ terms\\
4) terms coming from $\mathcal{L}_{axion}$\\
5) D-terms \newline So we have the symmetric mass matrix:

\be
    {\bf M}_{\tilde N}
     =   \(\begin{array}{cccccc}
           \frac{M_S}{2} & m_{SC} & m_{SB} & \frac{2}{\sqrt{2}} g_2^3 b_2^{(2)} \, \Delta v^2 & 0 & 0 \\
           \dots & M_0 C_\d^2+M_1 S_\d^2 &  -M_1 S_\d  & 0 & - g_0 v_d X_\d  & g_0 v_u X_\d   \\
           \dots & \dots & M_1 & 0 & -\frac{g_1 v_d}{2} & \frac{g_1 v_u}{2} \\
           \dots & \dots& \dots & M_2 & \frac{g_2 v_d}{2} & -\frac{g_2 v_u}{2} \\
           \dots & \dots & \dots & \dots & 0 & -\m  \\
          \dots & \dots & \dots & \dots & \dots & 0
         \end{array}\) \label{massmatrix} ~~~~
\ee The $4 \times 4$ matrix in the low-right corner is the mass matrix
of the neutralinos of the MSSM. The diagonal terms proportional to
$M_S$ and $M'_0$ come from the soft terms arisen with the anomalous extension.
The terms with 
$C_{\d}$ and $S_{\d}$ comes from the $U(1)$'s kinetic mixing.
The off-diagonal terms, except $m_{SC}$ and $m_{SB}$
come from the couplings $h-\tilde{h}-\l'$. The $X_{\d}$ definition is:

\be g_0 X_{\d}=C_{\d} g_0 Q_{H_u}-\frac{1}{2} g_1 S_{\d} \ee $g_0$
is the coupling constant of the anomalous $U(1)$, $g_1$
is the coupling constant of the standard $U(1)$. \newline The terms $m_{SC}$ e $m_{SB}$
receives many contributions from different terms of the Lagrangian. Their
complete expressions are:

\bea
    m_{SC}&=&\sqrt2\Bigg\{2\(C_\d^2 g_0^2 b_2^{(0)}+S_\d^2 g_1^2 b_2^{(1)}- C_\d S_\d g_0 g_1 b_2^{(4)} \)
                   \( g_0 X_\d \, \Delta v^2 +  C_\d M_{V^{(0)}} \a\)\nn\\
         &&+\half \(-2S_\d g_1^2 b_2^{(1)}+C_\d g_0 g_1 b_2^{(4)} \)g_1 \, \Delta v^2+  \frac{C_\d}{2}M_{V^{(0)}}\Bigg\}\label{msc}\\
     m_{SB} &=& \sqrt2\left\{ \( C_\d g_0 g_1 b_2^{(4)}-2 S_\d g_1^2 b_2^{(1)} \)
                \(g_0 X_\d \, \Delta v^2  + C_\d M_{V^{(0)}} \a\)
               + b_2^{(1)} g_1^3 \, \Delta v^2\right\}\nn
\eea These expressions contain terms of order higher than 1 in the coupling constant.
So they are negligible with respect to those in ${\bf M}_{\tilde N}$ that come 
from tree level. They can be neglected at
the leading order.

\chapter{LSP decay}
\section{Pure st\"{u}ckelino LSP annihilation to photons \label{stuanni}}

We want to investigate the changes brought in the model by the
anomalous extension. So we require that the LSP comes from the anomalous
neutralino sector introduced adding the new $U(1)$. We start requesting
for simplicity
that the  $2\times 2$ sector related to the extension is  decoupled from the 
MSSM sector.

\subsection{Lagrangian and Feynman rules} To get a pure st\"{u}ckelino LSP
we need some approximations. So we rewrite
the neutralinos mass matrix at tree level in the anomalous $U(1)$
extension:

\be \ M_{\tilde{N}}=\left(\begin{array}{cccccc} \frac{M_S}{2} &
m'_{SC} & m_{SB} &
0 & 0 & 0 \\
\dots & M_0 C_\d^2+M_1 S_\d^2 & -m_1 S_\d & 0 & -g_0 v_d X_\d &
g_0 v_u X_\d \\ \dots & \dots & M_1 & 0 & -\frac{g_1 v_d}{2} &
\frac{g_1 v_u}{2} \\ \dots & \dots & \dots & M_2 & \frac{g_2
v_d}{2} & -\frac{g_2 v_u}{2} \\ \dots & \dots & \dots & \dots & 0 & -\m \\
\dots & \dots & \dots & \dots & \dots & 0 \end{array} \right) \ee
We want to find the limit in which the LSP is a pure st\"{u}ckelino.
We start searching the limit of decoupling between the anomalous
and the MSSM sectors. Remembering formula (\ref{mixing}) the first
observation that we made is that, at tree-level:

\be \d\sim0 \Rightarrow C_\d \sim 1 , S_\d \sim 0 , X_\d \sim
Q_{H_u} \label{Q} \ee Substituting in (\ref{msc}) and eliminating
higher order terms with respect to the tree-level, we obtain:

\bea &m_{SC}=\sqrt{2}[2g_0b_3]
\\
&m_{SB}=0 \eea To complete the decoupling remember (\ref{Q}).
All the terms off the blocks  $2 \times 2$ and $4 \times 4$ different
from zero are proportional to $Q_{H_u}$. \newline So we impose
$Q_{H_u}=0$. Now the mass matrix has the form:

\be \ M_{\tilde{N}}=\left(\begin{array}{cccccc} \frac{M_S}{2} &
2\sqrt{2}g_0b_3 & 0 & 0 & 0 & 0 \\
\dots & M_0 & 0 & 0 & 0 & 0 \\ \dots
& \dots & M_1 & 0 & -\frac{g_1 v_d}{2} & \frac{g_1 v_u}{2} \\
\dots & \dots & \dots & M_2 & \frac{g_2
v_d}{2} & -\frac{g_2 v_u}{2} \\ \dots & \dots & \dots & \dots & 0 & -\m \\
\dots & \dots & \dots & \dots & \dots & 0 \end{array} \right) \ee

Furthermore we impose the LSP to be a pure st\"{u}ckelino. 
Then $M_S<<M_0$, that implies: $M_{LSP}\sim M_{S}$.
Now we are in the desired situation, so we can study the terms in the Lagrangian  that 
contribute to the decay. The part of the Lagrangian that contains these terms is:

\be
  \mathcal{L}=-\frac{1}{2}b_2^{(a)}[\frac{i}{\sqrt{2}}tr(\l^{(a)}\s^\m \bar
{\s}^\n F^{(a)}_{\m \n})\Psi_S + h.c.] \label{l1}\ee We want to
express (\ref{l1}) in Dirac notation, so we introduce a
convention for the gamma matrices:

\be
 \g^\m= \left(\begin{array}{cc} 0 & \s^\m \\
                           \bar{\s}^\m & 0 \end{array} \right)
\ee Using this rule, calculating the traces and remembering
the formula $\g^\m \g^\n F_{\m \n}=[\g^\m,\g^\n]\p_n B_\n$ ($ B_\n $ is the gauge field
related to $F_{\m \n}$) we obtain:

 \be
 \ i\mathcal{L}=\frac{b_2^{(1)}}{\sqrt{2}} g_1^2 \(\l^y\)^T \g_5 [\g^\m,\g^\n](\p_\m B_\n)
 \Psi_S + \frac{b_2^{(2)}}{\sqrt{2}} g_2^2 \(\l^3\)^T [\g^\m,\g^\n](\p_\m
 W_\n^3) \Psi_S
 \label{nnblag2}\ee The interaction vertex associated to this lagrangian is:

\be
 \ C \g_5 [\g^\m,\g^\n]ik_\m \label{lpg}
 \ee $C$ contains all the constants:

\be \ C=\frac{1}{\sqrt{2}}g_{(a)}^2 b_2^{(a)}wf \quad a=1,2
\label{vertice}\ee Among the others, $C$ contains $b_2^{(a)}$, 
parameter coming from anomalous $U(1)$ that implies $C<<g_{(a)}$ \cite{LioRac}.
The notation $g_{(a)}$ and $b_2^{(a)}$ means that the index can be both
that of the hypercharge and of the group $SU(2)$.
$w$ is a factor that, based on which group we are considering, will be:

\be \ w=\{ \begin{array}{ll} cos\theta_w \\ sin\theta_w
\end{array} \ee $f$ is the gaugino coefficient in the mass eigenbasis.
For a generical gaugino $\l^{(a)}$ it will be:

\be \l^{(a)}=c_1^{(a)}\chi_1 +c_2^{(a)}\chi_2
+c_3^{(a)}\chi_3+c_4^{(a)}\chi_4 \ee $\chi_1, \chi_2, \chi_3,
\chi_4$ are the MSSM eigenstates of neutralino mass matrix. $f$ is
the weight related to the gaugino that we are considering.
\newline The propagator, remembering that both st\"{u}ckelino and 
gauginos are Majorana spinors, are:

\be
 \ \frac{i(\g^\m p_\m - m)}{p^2-m^2+i\e}
 \ee Finally, the spin projectors related to Majorana spinors are:

\be \ \Lambda^\pm=\frac{\g^\m p_\m \pm m}{2m} \ee

\newpage

\subsection{Cross-section calculation}
The annihilation is represented in figure 
\ref{figura1}:
\begin{figure}[h]
\centering
      \includegraphics[scale=0.8]{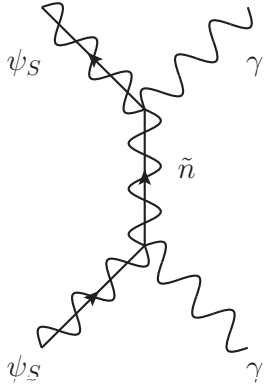}
      \caption{Annichilazione LSP-LSP}
      \label{figura1}
\end{figure} The St\"{u}ckelino mass is $M_S$, while $m$ is the mass of
the generic exchanged gaugino. Applying Feynman rules to this diagram
we have:

\be
 \ \bar{u}(p_2)C \g_5 \s^{\d\r}k_{2\d} \e_\r(k_2) \frac{i(\sl p_1 -\sl
 k_1-m)}{(p_1-k_1)^2-m^2} \e_\n C \g_5 \s^{\m \n} k_{1 \m} u(p_1)
 \ee
We define: $f_1^2=(p_1-k_1)^2$. Now, for generality, we calculate
the product of two diagrams in which two different neutralinos are exchanged,
obtaining:

\bea \ &&\mathcal{MM'}=\frac{2^4 C^2 C'^2}{(f_1^2-m^2)(f_1^2-m'^2)} \times \label{1} \\ \nn 
&&tr[(\sl p_2-M_S)\g_5 \s^{\d \r} k_{2 \d} i
   (\sl f_1 -m) \g_5 \s^{\m \n}
k_{1 \m} (\sl p_1 +M_S) k_1^\e \s_{\e \n} \g_5 i (\sl f_1 -m') k_2^\t
\s_{\t \d} \g_5]  \eea Now we concentrate only on the trace evaluation.
Using the identities:

\bea \ &&\s^{\d \r}k_{1 \d}=\frac{i}{2}(\sl k_1 \g^\r -k_1^\r) \\
&&\g_5 \g_{\m}=-\g_{\m} \g_5 \eea
in formula (\ref{1}) we obtain (we do not write the $(1/2)^4$ coming from the
first identity, assuming it semplifies the factor $2^4$ in (\ref{1})):

\bea \ tr[(-\sl p_2 -M_S)(\sl k_2 \g^\r-k_2^\r)(\sl f_1 -m)(\sl
k_1 \g^\n -k_1^\n) \nn \\ \times (-\sl p_1 +M_S)(\sl k_1 \g_\n -
k_{1 \n})(\sl f_1 -m')(\sl k_2 \g_\r -k_{2 \r})] \eea Before going
on with this calculation we want to list the formulas that we are
going to use:

\bea \ &&\g^\m \sl A \g_\m=-2 \sl A \label{traccia3}\\
&&\g^\m \sl A \sl B \g_\m=4(AB) \label{traccia4}\\
&&\g^\m \sl A \sl B \sl C \g_\m=-2(\sl C \sl B \sl A) \label{traccia5} \\
&&tr(\sl A \sl B)=4(AB) \label{traccia1}\\
&&tr(\sl A \sl B \sl C \sl D)=4[(AB)(CD)-(AC)(BD)+(AD)BC)]
\label{traccia2}\eea Using these relations we obtain:

\be \ (\sl k_1 \g^\n-k_1^\n)(-\sl p_1 +M_S)(\sl k_1 \g^\n - k_{1
\n})=-4\sl k_1 (p_1 k_1) \ee Substituting this result:

\bea \ -4(p_1 k_1)tr[(-\sl p_2 +M_S)(\sl k_2 \g^\r-k_2^\r)(\sl f_1 -
m)(\sl k_1)(\sl f_1 -m')(\sl k_2 \g_\r - k_{2 \r})] \\ \nn =4(p_1
k_1)tr[(\sl p_2 \sl k_2 \g^{\r} \sl p_1 \sl k_1 \sl p_1 \sl k_2 \g_{\r}) +4
mm'(k_2 k_1)(\sl p_2 \sl k_2)] \eea We have used the relations:

\bea \ \sl f_1=\sl p_1 - \sl k_2 \\
\sl k_2 \sl k_2=0 \\
\sl k_1 \sl k_1=0 \eea Let's evaluate the term with 8 gamma matrices:
\bea \ &&tr[(\sl p_2 \sl k_2 \g^{\r} \sl p_1 \sl k_1 \sl p_1 \sl k_2 \g_{\r}]= \\ \nn 
&&=tr[\sl p_2 \sl k_2 (2 \sl k_2 \sl p_1 \sl k_1 \sl p_1-2 \sl p_1 \sl k_1 \sl p_1 \sl k_2)]= \\ \nn
&&=tr[-2 \sl p_2 \sl k_2 \sl p_1 \sl k_1 \sl p_1 \sl k_2]= \\ \nn
&&=32(p_1 k_1)(p_2 k_2)(p_1 k_2)-16(k_1 k_2)(k_2 p_2) M_S^2 \eea We now demonstrate that $(p_2
k_1)=(p_1 k_2)$:

\bea \ p_1-k_1=p_2-k_2 \Rightarrow p_1+k_2=p_2+k_1 \Rightarrow \\
\Rightarrow p_1^2+2(p_1k_2)+k_2^2=p_2^2+2(p_2k_1)+k_1^2
\Rightarrow (p_1k_2)=(p_2k_1) \eea Calculating also the term with only 2 gamma matrices
we obtain:

\bea \ \mathcal{MM'}=\frac{64 C^2 C'^2 (p_1
k_1)}{(M_S^2-m^2-2(p_2 k_2))(M_S^2-m'^2-2(p_2))} \times \\
\nn [mm'(k_2 k_1)(p_2 k_2)+2(p_1 k_1)(p_2 k_2)(p_1 k_2)-M_S^2(p_2
k_2)(k_2 k_1)] \label{noan}\eea To obtain the cross-section we have 
to multiply it with a suitable kinematical factor
that relates $\mathcal{MM'}$ to the cross section. According to \hypertarget{HalzMart}{\cite{HalzMart}}
the general relation between the amplitude and the cross-section is:

\bea &&d \s= \frac{|\mathcal{M}|^2}{F} dQ \nn \\
&&F=4(|\overrightarrow{p_A}| E_A+|\overrightarrow{p_B}|E_B) \label{prefQ} \\ 
&&dQ=(2 \pi)^4 \d^{(4)}(p_C+p_D-p_A-p_B)\frac{d^3 p_C d^3 p_D}{(2 \pi)^3 2 E_C (2 \pi)^3 2 E_D} \nn \eea Indices $A$,$B$
refer to initial particles, indices $C$,$D$ refer to to final particles. Now we concentrate on the center of mass frame. So
we have these additional relations:

\bea 
\overrightarrow{p_A}=-\overrightarrow{p_B} \nn \\
\overrightarrow{p_C}=-\overrightarrow{p_D} \label{CoM}\\
\sqrt{s}=E_C+E_D=E_A+E_B \nn \eea The last equation comes from the second and
the fact that, in our case $m_C=m_D$. Using these relations we specify the (\ref{prefQ})
to our case:

\be F=4 |\overrightarrow{p_A}|(E_A+E_B)= |\overrightarrow{p_A}|\sqrt{s}\ee Now, we define:

\be W \equiv  \sqrt{s}= E_C+E_D=(m_C^2+|\overrightarrow{p_C}|^2)^{1/2}
+(m_D^2+|\overrightarrow{p_D}|^2)^{1/2} \ee Using the (\ref{CoM}) we can calculate:

\be \frac{dW}{d|\overrightarrow{p_C}|}=
|\overrightarrow{p_C}|\left( \frac{1}{E_C}+ \frac{1}{E_D} \right) \Rightarrow 
d|\overrightarrow{p_C}|=\frac{dW}{|\overrightarrow{p_C}|}\frac{E_C E_D}{E_C + E_D}\ee Now, before
re-writing $dQ$, we remember the formula of spatial differential in polar coordinates:

\be d^3 p_C= |\overrightarrow{p_C}|^2 d|\overrightarrow{p_C}| d \Omega \ee Substituting these relations and 
using the spatial part of the $\d^{(4)}$, we obtain:

\be dQ= \frac{1}{4 \pi^2} \frac{1}{4} \frac{|\overrightarrow{p_C}|}{E_C E_D} \frac{\overrightarrow{E_C E_D}}{E_C + E_D}
dW d\Omega \d (W-E_C-E_D)= \frac{1}{16 \pi^2} \frac{|\overrightarrow{p_C}|}{\sqrt{s}} d\Omega \ee So
the complete prefactor is:

\be \frac{1}{64 \pi^2} \frac{|\overrightarrow{p_C}|}{|\overrightarrow{p_A}| s} \label{prefattore}\ee

If we call
$\omega_1$ and $\omega_2$ the energies of final photons, $E_1$
and $E_2$ the energies of the initial st\"{u}ckelinos, we have in the center 
of mass:

\be \ \vec{p}_1=-\vec{p}_2 \ee So the differential cross section is:

\bea \ \frac{d\s}{d\Omega}=\frac{\omega_1}{16 \pi^2
(\omega_1+\omega_2)^2(\sqrt{M_S^2-E_2^2})}\sum_{i,j}\mathcal{M}_i \mathcal{M}_j \\
2(k_1 k_2)=(\omega_1 +\omega_2)^2 \\
(p_2 k_2)=E_2 \omega_2+(\sqrt{M_S^2-E_2^2})\omega_2 \cos \theta \\
(p_2 k_1)=E_2 \omega_1-(\sqrt{M_S^2-E_2^2})\omega_1\cos \theta
\label{sezax} \eea The indices $i,j$  vary from 1 to 4, because we have gauginos 
and each of them generate 2 vertices, with slight different constant
factors. These are the expression in each case:

\bea \ C_1=\sqrt{2}g^{(1)} b^{(1)}_2 c_w a^{(1)} \quad m_1=m_{\chi_1} \\
C_2=\sqrt{2}g^{(1)} b^{(1)}_2 c_w b^{(1)} \quad m_2=m_{\chi_2} \\
C_3=\sqrt{2}g^{(2)}b_2^{(2)}s_w a^{(2)} \quad m_3=m_{\chi_3} \\
C_4=\sqrt{2}g^{(2)}b_2^{(2)}s_w b^{(2)} \quad m_4=m_{\chi_4} \eea
The parameter $a^{(1)},a^{(2)},b^{(1)},b^{(2)}$ are the coefficient of the
neutralinos expressed with respect to the mass eigenstates.

\section{Pure St\"{u}ckelino LSP and NLSP coannihilation}
As we can see from formula (\ref{noan}), the cross section of the
st\"{u}ckelinos annihilation is proportional to
$C^2C'^2$. These are the typical couplings of the extra $U(1)$ that
are at least one order smaller than the electro-weak couplings of the
MSSM. So we can expect that the cross section will be 3-4 orders of 
magnitude smaller than the electro-weak counterparts.
We will show that this is unacceptable if we want to obtain a relic density
estimate in agreement with the experimental results. So we want to
study the situation in which we have a NLSP coming from the MSSM with a mass
comparable to that of the st\"{u}ckelino. As we will show in a following
chapter in which we will study the Boltzmann Equation, this situation can 
lead to a relic density that satisfies the WMAP constraints.

\subsection{Lagrangian and Feynman rules} The possible coannihilating
processes are:

\bea \Psi_S \l\rightarrow f \bar{f} \\
\Psi_S \l\rightarrow W^+ W^-  \\ 
\Psi_S \l\rightarrow h^0 h^0  \\ 
\Psi_S \l\rightarrow H^0 H^0
\Psi_S \l\rightarrow H^0 h^0
\Psi_S \l\rightarrow A^0 A^0
\Psi_S \l\rightarrow H^+ H^- \eea $f$ and $\bar{f}$ can be
any fermion anti-fermion pair of the SM, $W^+$ and $W^-$ are the electro-weak charged boson,
 $h^0$ and $H^0$ are respectively the lighter and the heavier parity-even Higgs, $H^+$ and $H^-$ are the
charged Higgses and $A^0$ is the
parity odd Higgs. In this section we want to obtain an estimate for the cross section, so
we consider only the process $\Psi_S \l\rightarrow f \bar{f}$
because it is by far the main contribution to the total cross section. We will study the other processes
when we will deal with the general case.\\
The $\Psi \l \g$ and $\Psi \l Z^0$ vertices are the same ones of
the st\"{u}ckelino annihilation, (\ref{lpg}).
Because for this simplified calculation we consider the exchange of $\g$ and $Z^0$ we call $C_{Z}$
the costant related to the $Z^0$ exchange and $C_{\g}$ that of the photon exchange.
From the SM we already know the vertices in which appear the pair $f\bar{f}$.
The vertex $f \bar{f} \g$ is:

\be ieQ\g^\m \ee In this expression $e$ is the unit of electric charge, 
$Q$ is the electric quantum number of the fermion anti-fermion pair considered.\\
The vertex $f \bar{f} Z_0$ is:

\be -\frac{ie}{2\sin{\theta_W}\cos{\theta_W}}\g^{\m}(g_V-g_A\g_5)
\ee $e$ is, as in the previous formula, the unit of electric charge,
$\theta_W$ is the Weinberg angle, $g_V$ and $g_A$ are the vector
and axial couplings, which depend on which fermion we are considering:

\newpage

\begin{table}[h]
\centering
\begin{tabular}[h]{|c|c|c|}
\hline  & $g_V$ & $g_A$ \\
\hline $\n_e$, $\n_\m$, $\n_\t$ & $1/2$ & $1/2$ \\
\hline $e$, $\m$, $\t$ & $-1/2 +2sin^2{\theta_W}$ & $-1/2$ \\
\hline $u$, $c$, $t$ & $1/2 -4/3sin^2{\theta_W}$ & $1/2$ \\
\hline $d$, $s$, $b$ & $-1/2+2/3sin^2{\theta_W}$ & $-1/2$ \\
\hline
\end{tabular} 
\end{table} Now we write the propagator. The photon one is:

\be \frac{-ig^{\m \n}}{k^2} \ee $k^2$ is the momentum of the virtual photon.\\
The $Z^0$ propagator instead, after fixing the gauge, is:

\be \frac{-ig^{\m \n}}{k^2-m_Z^2} \ee where $k^2$ represents the virtual $Z^0$ momentum.
\subsection{Cross-section calculation}
he two possible diagrams that we consider in this estimation of the $\Psi_S-\lambda$ cross-section of the process
$\Psi_S \l\rightarrow f \bar{f}$ are:

\begin{figure}[h]
\centering
\includegraphics[scale=0.8]{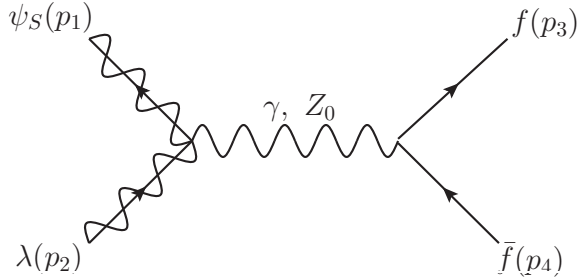}
      \caption{LSP-NLSP coannihilation}
      \label{figuracoann}
\end{figure}

In this section we perform the calculation of the cross-section.\\
The amplitude is:

\bea && \mathcal{M}=-i k^{\m} \bar{v}_1 \g_5 [\g_{\m},\g_{\n}] u_2 
[ e q_f C_{\g} \frac{\eta^{\n \r}}{k^2} \bar{u}_3 \g_{\r} v_4 \\
&&+ \frac{g_{Z_0}}{2} C_{Z_0} \frac{\eta^{\n \r}}{k^2-M_{Z_0^2}} \bar{u}_3 \g_{\r}
(v_f^{Z_0}-a_f^{Z_0} \g_5) v_4 ] \nn \label{coanniampli} \eea The square modulus is:

\be
|\mathcal{M}|^2=\left[ T_a \left( \frac{g_{Z_0}}{2} C_{Z_0} \frac{a_f^{Z_0}}{k^2-M_{Z_0^2}} \right)^2+
T_v \left(\frac{e q_f C_{\g}}{k^2} + \frac{v_f^{Z_0}\frac{g_{Z_0}}{2} C_{Z_0}}{k^2-M_{Z_0^2}} \right)^2 \right] 
\label{moduloquadro}\ee $T_v$ and $T_a$ are:

\be T_v=B_{\n \a} A_1^{\n \a} \qquad T_a=B_{\n \a} A_2^{\n \a} \ee where
$B_{\n \a}$ is the result of the trace over the initial particles, $A_1^{\n \a}$
and $A_2^{\n \a}$ are respectively the results of the vector and axial part of the trace 
over final particles. Now we are going to calculate separately the traces. We
start from $B_{\n \a}$:

\be B^{\n \a}=tr[(-\sl p_1-M_S) (\sl k \g_{\n}- \g_{\n}\sl k) (\sl p_2+m_{\l})(\sl k \g_{\a}- \g_{\a}\sl k)] \label{bnualfa}\ee to
obtain this formula we have used the $\g_5$ property:

\bea \g_5 \g_{\m}=-\g_{\m} \g_5 \\
(\g_5)^2=1 \eea Now we develop the product in the trace:

\bea &&B_{\n \a}= -tr[\sl p_1 \sl k \g_{\n} \sl p_2 \sl k \g_{\a} - 
\sl p_1 \g_{\n} \sl k \sl p_2 \sl k \g_{\a}-
\sl p_1 \sl k \g_{\n} \sl p_2 \g_{\a} \sl k+
\sl p_1 \g_{\n} \sl k \sl p_2 \g_{\a} \sl k]- \nn \\
&& m_{\l} M_S tr[ \sl k \g_{\n} \sl k \g{\a} -
\sl k \g_{\n} \g_{\a} \sl k + \g_{\n} \sl k \g{\a} \sl k-
\g_{\n} \sl k \sl k \g_{\a} ]  \eea The second term we want to 
calculate is $A_1^{\n \a}$:

\be A_1^{\n \a}=tr[(\sl p_3+m_f) \g^{\n} (\sl p_4 - m_f) \g^{\a} ] \ee Again we
perform the product to obtain:

\be A_1^{\n \a}=tr[\sl p_3 \g^{\n} \sl p_4 \g^{\a}]- m_f^2 tr[ \g^{\n}\g^{\a} ] \ee The third term, $A_2^{\n \a}$ is:

\be A_2^{\n \a}=-tr[(\sl p_3+m_f) \g^{\n} \g_5 (\sl p_4 - m_f)\g_5 \g^{\a} ] \ee After developing
the products it becomes:

\be A_2^{\n \a}=tr[\sl p_3 \g^{\n} \sl p_4 \g^{\a}]+ m_f^2 tr[ \g^{\n}\g^{\a} ] \ee Calculating 
the trace using the formulas (\ref{traccia1}), (\ref{traccia2}) and performing the product,
we obtain:

\bea
 T_v &=&m_f^4 (p_{\l_1} p_S)
     +M_1 M_S \Big[2 m_f^4+3 (p_f p_{\bar f}) m_f^2+(p_f p_{\bar f})^2\Big]+\nn\\
&&-(p_f p_{\bar f})
   \Big[(p_{\l_1} p_f) (p_f p_S)+(p_{\l_1} p_{\bar f}) (p_{\bar f} p_S)\Big] +m_f^2 \Big[(p_{\l_1} p_S) (p_f p_{\bar f})+\nn\\
&&-2 (p_{\l_1} p_f) (p_f p_S)-(p_{\l_1} p_{\bar f})
   (p_f p_S)-(p_{\l_1} p_f) (p_{\bar f} p_S)-2 (p_{\l_1} p_{\bar f}) (p_{\bar f} p_S)\Big] \nn\\
 T_a &=& \Big[(p_{\l_1} p_{\bar f}) (p_f p_S)+(p_{\l_1} p_f) (p_{\bar f} p_S)\Big] m_f^2-M_1 M_S
   \[m_f^4-(p_f p_{\bar f})^2\]+\nn\\
&&-(p_f p_{\bar f}) \Big[(p_{\l_1} p_f) (p_f p_S)+(p_{\l_1}
p_{\bar f})(p_{\bar f} p_S) \Big] \eea

Substituting these results in eq. (\ref{moduloquadro}), we have the complete expression of the amplitude.
Now we multiply the amplitude for the prefactor (\ref{prefattore}). In the center of mass we obtain:

\be \frac{d\s}{d \Omega}=\sum_{f} c_f
\frac{\sqrt{(E_3-m_{f})^2}}{64 \pi^2 (E_1+E_2)^2
\sqrt{(E_1^2-M_{S}^2)}}(\mathcal{M}^2)
\label{sezcoan}\ee The sum is over all SM fermions.
$c_f$ is the color factor of each fermion and its value is 3
for the quarks, 1 for the leptons.

\section{Mixed LSP and NLSP cohannihilation \label{mixinganoformulas}}
In this section we want to study the general case, in which the LSP is not a pure
st\"{u}ckelino, but it is a mix of all neutralinos. We also want to deal with the mixing 
among neutral vectors and thus we will consider interactions that appear at the tree-level but
we have not considered in the previous sections because they
are numerically less important than those we already
mentioned.
\subsection{Boson mixing and currents}
Inverting equations (\ref{photon}),(\ref{Z0}) and (\ref{Zprime}), we obtain:
\be
      \( \begin{array}{c} B^{\m}\\
                          W^{3 \m} \\ C^{\m} \end{array} \) = M
            \( \begin{array}{c} A^{\m}\\
                                Z_0^{\m}\\ Z'^{\m} \end{array} \)
\label{neumix}  \ee Defining $an \equiv g_0 ~ Q_{H_u} \frac{2 v^2}{2 M^2_{Z'}}$ we have:
\be  M= \( \begin{array}{ccc} \frac{cos(\theta_W)+an^2~ g_2 \sqrt{g_1^2+g_2^2}}{1+an^2 (g_1^2+g_2^2)} & -\frac{sin(\theta_W)}{1+an^2 (g_1^2+g_2^2)} & \frac{an ~ sin(\theta_W) \sqrt{g_1^2 + g_2^2}}{1+an^2 (g_1^2+g_2^2)} \\
\frac{sin(\theta_W)+an^2~ g_1 \sqrt{g_1^2+g_2^2})}{1+an^2 (g_1^2+g_2^2)} & \frac{cos(\theta_W)}{1+an^2 (g_1^2+g_2^2)}  & \frac{an ~ cos(\theta_W) \sqrt{g_1^2 + g_2^2}}{1+an^2 (g_1^2+g_2^2)} \\
  0 & \frac{an ~(cos(\theta_W) g_2+ sin(\theta_W) ~g_1)}{1+an^2 (g_1^2+g_2^2)} & \frac{1}{1+an^2 (g_1^2+g_2^2)}    \end{array} \)
\label{neumass1} \ee This leads to slight changes in the currents that we want to calculate.\\
Now we use this mixing matrix to express the interaction lagrangian that involves bosons and their currents with respect to
the mass eigenstates:

\bea &&\mathcal{L}= -ig_1j_{\m}^Y ~B^{\m}-ig_2j_{\m}^3 ~W^{3\m}-ig_0j_{\m}^0 ~Z'^{\m}= \nn \\
&&-ig_1j_{\m}^Y (M_{11}A^{\m}+M_{12}Z_0^{\m}+M_{13}Z'^{\m})\\ &&-ig_2j_{\m}^3(M_{21}A^{\m}+M_{22}Z_0^{\m}+M_{23}Z'^{\m}) \nn \\
&&-ig_0j_{\m}^0 (M_{31}A^{\m}+M_{32}Z_0^{\m}+M_{33}Z'^{\m}) \nn
\eea Considering only the terms proportional to $A^{\m}$ we have:
\bea -i(g_2 M_{21}j^3_{\m}+g_1M_{11}j^Y_{\m})A^{\m}= \nn \\ 
=-ie\(\frac{g_2}{e}M_{21}j^3_{\m}+\frac{g_1}{e}M_{11}j^Y_{\m} \)A^{\m} \label{photcurr}\eea Experimental constraints
on the neutral mixing imply that the photon current must be equal to that of the SM. This can be done
easily, because the same constraints \hypertarget{langacker}{\cite{langacker}} permit us to not consider terms of powers greater
than 1 in $an$. Later we are going to show that this is all we need to keep the photon current
equal to that of SM. For simplicity we perform the calculation using the electron,
but it is identical for each SM fermion. Using (\ref{photcurr}), we have:
\be j^{em}_{\m}=\frac{g_2}{e}M_{21}\(-\frac{1}{2}\)\bar{e_L}\g_{\m}e_L+
\frac{g_1}{e}M_{11}\(-\frac{1}{2}\bar{e_L}\g_{\m}e_L-\bar{e_R}\g_{\m}e_R \) \label{photcurr1}\ee where we
have used the SM quantum numbers of $e_L$ and $e_R$. Now we want to express left and right components
with respect to the Dirac spinor, using the standard definitions:
\be \Psi_{L/R}=\frac{1}{2}(1\mp\g_5)\Psi \Rightarrow 
\Psi_{L/R}\g_{\m}\Psi_{L/R}=\Psi\g_{\m}\(\frac{1\mp\g_5}{2}\)\Psi\ee Substituting these
definitions in (\ref{photcurr1}), we obtain:
\bea j^{em}_{\m}= \frac{g_2}{e}M_{21}\(-\frac{1}{2}\)\bar{e}\g_{\m}\( \frac{1-\g_5}{2} \)e \nn \\
+\frac{g_1}{e}M_{11}\(-\frac{1}{2} \bar{e}\g_{\m}\( \frac{1-\g_5}{2} \)e- \bar{e}\g_{\m}\( \frac{1+\g_5}{2} \)e\)\eea Now 
we can read the analogous of the vector and axial couplings of the SM and we can impose the desired equivalence:
\bea &&vec=\frac{1}{2}\(-\frac{1}{2}\frac{g_2}{e}M_{21}-\frac{1}{2}\frac{g_1}{e}M_{11}-\frac{1}{2}\frac{g_1}{e}M_{11} \)=-1 \\
&&ax= \frac{1}{2}\(-\frac{1}{2}\frac{g_2}{e}M_{21}-\frac{1}{2}\frac{g_1}{e}M_{11}+\frac{1}{2}\frac{g_1}{e}M_{11} \)=0 \eea This
impositions can be realized if and only if $g_2 M_{21}=g_1 M_{11}=e$. As promised, we now show  
that neglecting $an$ powers higher than 1 is sufficient to achieve our goal:
\bea &&0=g_2 M_{21}-g_1 M_{11}= \nn \\ &&g_2\( \frac{sin(\theta_W)+an^2~ g_1 \sqrt{g_1^2+g_2^2})}{1+an^2 (g_1^2+g_2^2)}\)
-g_1\(\frac{cos(\theta_W)+an^2~ g_2 \sqrt{g_11^2+g_2^2}}{1+an^2 (g_1^2+g_2^2)}\) \nn \\
 &&\cong g_2 sin(\theta_W)-g_1 cos(\theta_W) \Rightarrow g_2 sin(\theta_W)=g_1 cos(\theta_W) \eea The latter is identical
to the SM relation and also implies (substituting in (\ref{photcurr})):
\be j^{em}_{\m}=j^{3}_{\m}+j^{Y}_{\m} \ee In the anomalous extension the photon current is identical to that 
of the SM up to terms of second order in $an$, which are experimentally negligible.\\
Thus the related Feynman rule will be the same of the SM:

\begin{figure}[h!]
\centering
\includegraphics[scale=0.8]{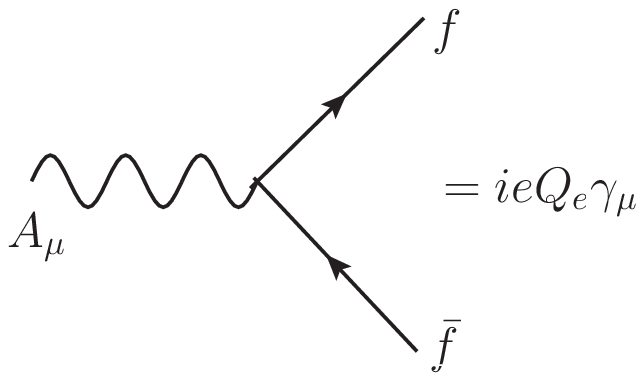} \be \label{gammaff} \ee
\end{figure} 

Now we consider the terms proportional to $Z_0^{\m}$:
\bea &&-i(g_2 M_{22}j^3_{\m}+g_1M_{12}j^Y_{\m}+g_0M_{32}j^0_{\m})Z_0^{\m}=  \\ 
&&-i\frac{g_2}{cos(\theta_W)}\(cos(\theta_W)M_{22}j^3_{\m}+\frac{g_1cos(\theta_W)}{g_2}M_{12}j^Y_{\m}+
\frac{g_0cos(\theta_W)}{g_2}M_{32}j^0_{\m} \)Z_0^{\m} \label{z0curr} \nn \eea Again we try to reconstruct the
SM structure of the current, so we have evidenced $\frac{g_2}{cos(\theta_W)}$· Now we use the same calculations
of the SM case to express the neutral current with respect to $j^3_{\m}$ and $j^{em}_{\m}$(instead
of $j_{\m}^Y$), plus $j^0_{\m}$ introduced by our extension.
\bea &&j_{\m}^{NC}=cos(\theta_W)M_{22}j_{\m}^3+sin(\theta_W)M_{12}j_{\m}^Y+\frac{g_0}{g_2}cos(\theta_W)M_{32}j_{\m}^0= \nn \\
&&cos(\theta_W)M_{22}j_{\m}^3+sin(\theta_W)M_{12}(j_{\m}^{em}-j_{\m}^{3})+\frac{g_0}{g_2}cos(\theta_W)M_{32}j_{\m}^0= \nn \\ 
&&(cos(\theta_W)M_{22}-sin(\theta_W)M_{12})j_{\m}^3+sin(\theta_W)M_{12}j_{\m}^{em}+\frac{g_0}{g_2}cos(\theta_W)M_{32}j_{\m}^0 
\label{z0curr1}  \eea Now we want to eliminate the terms with powers higher than 2 in $an$. Using their definition
in (\ref{neumass1}) we note that there are simplifications in $M_{12}$ and $M_{22}$, while the $M_{32}$ term remains
unchanged:
\bea &&cos(\theta_W)M_{22}-sin(\theta_W)M_{12}=1 \\ &&sin(\theta_W)M_{12}=-sin^2(\theta_W)\eea Remembering that
$(j_{\m}^{NC})^{SM}=j_{\m}^3-sin^2(\theta_W)j_{\m}^{em}$ this relations implies:
\be j_{\m}^{NC}=(j_{\m}^{NC})^{SM}+\frac{g_0}{g_2}cos(\theta_W)M_{32}j_{\m}^0\ee where the latter term is the
anomalous contribution. Now we have to calculate how this term changes the axial and vector couplings of SM. We only
compute the variation, because obviously the analogous term in the SM will give the known contributions.
Defining $Q_L$, $Q_R$ as the anomalous $U(1)$ quantum numbers of the left and right fermions respectively:

\bea &&\frac{g_0}{g_2} cos(\theta_W) M_{32} \( Q_L \bar{e_L}\g_{\m}e_L+Q_R \bar{e_R}\g_{\m}e_R  \)= \nn \\
&&\frac{g_0}{g_2} cos(\theta_W) M_{32} \( Q_L \bar{e} \g_{\m} \( \frac{1-\g_5}{2} \)e+Q_R \bar{e} \g_{\m} \( \frac{1+\g_5}{2} \)e \)= \nn \\
&&\frac{g_0}{g_2} cos(\theta_W) M_{32}\(\( \frac{Q_L+Q_R}{2} \)\bar{e}\g_{\m}e-\( \frac{Q_L-Q_R}{2} \)\bar{e}\g_{\m}\g_5e\)
\eea So we have:
\bea \Bigg\{ \begin{array}{c}
         g_V=g_V^{SM}+\frac{g_0}{g_2} cos(\theta_W) M_{32} Q_V \\
         g_A=g_A^{SM}+\frac{g_0}{g_2} cos(\theta_W) M_{32} Q_A
        \end{array}
 \label{gvgaano}\eea Where $Q_V$, $Q_A$ are defined\footnote{Note that in (\ref{QTable}) are listed the right-handed
charges of the anti-particles, so we have to change their signs to use them in these definitions}:
\bea Q_V=\frac{Q_L+Q_R}{2} \\ Q_A=\frac{Q_L-Q_R}{2} \label{qvqaano}\eea So the Feynman rule is the usual:

\begin{figure}[h!]
\centering
\includegraphics[scale=0.8]{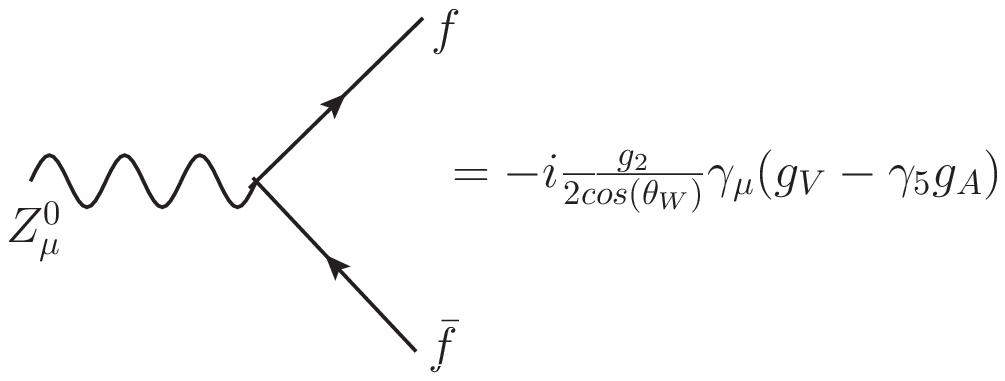} \be \label{z0ff} \ee
\end{figure}

It remains only to calculate the current of the $Z'$. The lagrangian term is:
\bea &&-i(g_2 M_{23}j^3_{\m}+g_1M_{13}j^Y_{\m}+g_0M_{33}j^0_{\m})Z'^{\m}=  \\ 
&&-ig_0\(\frac{g_2}{g_0} M_{23}j^3_{\m}+\frac{g_1}{g_0}M_{13}j^Y_{\m}+
M_{33}j^0_{\m} \)Z'^{\m} \label{zpcurr} \nn \eea \\ Again, neglecting the terms in (\ref{neumass1}) of order higher than 2 in $an$,
we obtain that $M_{13}=an\sqrt{g_1^2+g_2^2}sin(\theta_W)$,$M_{23}=-an\sqrt{g_1^2+g_2^2}cos(\theta_W)$ and $M_{33}=1$.
As in the previous case we substitute $j_{\m}^Y$ with $j_{\m}^{em}-j_{\m}^{3}$, obtaining:
\be j_{\m}^{Z'}=j_{\m}^0+\( \frac{g_2}{g_0}M_{23}-\frac{g_1}{g_0}M_{13} \)j_{\m}^3+\frac{g_1}{g_0}M_{13}j_{\m}^{em} \ee For
the quantum numbers this implies (omitting the calculation that is analogous to the previous case):
\bea \Bigg\{ \begin{array}{c}
              Q_V^{mix}=Q_V+T^3\( \frac{g_2}{g_0}M_{23}-\frac{g_1}{g_0}M_{13} \)+q_{el}\frac{g_1}{g_0}M_{13} \\
              Q_A^{mix}=Q_A+T^3\( \frac{g_2}{g_0}M_{23}-\frac{g_1}{g_0}M_{13} \)
             \end{array}
 \eea The related Feynman rule is:

\begin{figure}[h!]
\centering
\includegraphics[scale=0.8]{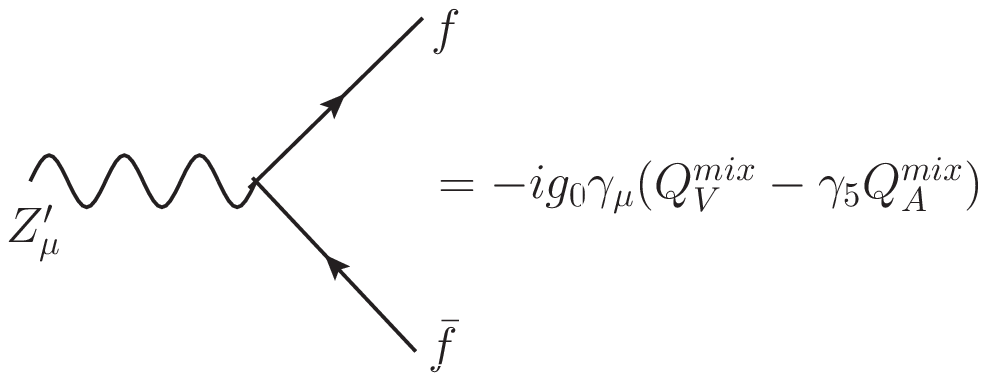} \be \label{z'ff} \ee
\end{figure}

\subsection{Neutralino-neutralino-boson anomalous vertices}
We have already studied the situation where the LSP is a pure st\"{u}ckelino. However in the general case we can have 
mass eigenstates that are an undefined mix of gauge eigenstates. In this situation we have to account for this mixing
in the interaction vertices. To achieve this goal we write the relevant terms of (\ref{axion}):

\bea &&i \mathcal{L}=\frac{b_2^{(0)}}{\sqrt{2}} g_0^2 (\l^{(0)})^T \g_5 [\g^{\m},\g^{\n}] \pd_{\m} C_{\n} \psi_S+
\frac{b_2^{(1)}}{\sqrt{2}} g_1^2 (\l^{(1)})^T \g_5 [\g^{\m},\g^{\n}] \pd_{\m} B_{\n} \psi_S+ \nn \\ 
&&\frac{b_2^{(2)}}{\sqrt{2}} g_2^2 (\l^{(2)})^T \g_5 [\g^{\m},\g^{\n}] \pd_{\m} W^3_{\n} \psi_S  +
\frac{b_2^{(4)}}{2\sqrt{2}} g_1 g_0 [(\l^{(1)})^T \g_5 [\g^{\m},\g^{\n}] \pd_{\m} C_{\n}+\\
&&(\l^{(0)})^T \g_5 [\g^{\m},\g^{\n}] \pd_{\m} B_{\n}] \psi_S \nn \label{nnblag} \eea This is the lagrangian in the gauge basis.
Now we need the rotation matrices that maps bosons and neutralinos to the mass basis. For the bosons we have written the
solution in the previous section, in formula (\ref{neumass1}). For the neutralinos we cannot write an explicit matrix,
because it is model-dependent. So we call the matrix $N$ and, ordering the basis as $\l^{(1)},\l^{(2)},\l^{h_1},\l^{h_2},\psi_S,
\l^{(0)}$, we have:
\bea &&\l^{(1)}=\sum_i N^+_{1i} n_i=\sum_i N^*_{i1} n_i \nn \\
&&\l^{(2)}=\sum_i N^*_{i2} n_i \qquad \qquad \l^{(0)}=\sum_i N^*_{i6} n_i  \\ &&\psi_S=\sum_i N^*_{i5} n_i 
\qquad \qquad i=1,\dots,6 \nn \eea where 
$n_i$ are the mass eigenstates. Substituting this expression in (\ref{nnblag}) we obtain:
\bea &&i \mathcal{L}=\frac{b_2^{(0)}}{\sqrt{2}} g_0^2 \sum_i n_i^T N^*_{6i}\g_5 [\g^{\m},\g^{\n}]\pd_{\m}(M_{32}Z^0_{\n}+M_{33}Z'_{\n})\sum_j N^*_{j5}n_j+ \nn \\
&&\frac{b_2^{(1)}}{\sqrt{2}} g_1^2 \sum_i n_i^T N^*_{1i}\g_5 [\g^{\m},\g^{\n}]\pd_{\m}(M_{11}A_{\n}+M_{12}Z^0_{\n}+M_{13}Z'_{\n})\sum_j N^*_{j5}n_j+ \nn \\
&&\frac{b_2^{(2)}}{\sqrt{2}} g_2^2 \sum_i n_i^T N^*_{2i}\g_5 [\g^{\m},\g^{\n}]\pd_{\m}(M_{21}A_{\n}+M_{22}Z^0_{\n}+M_{23}Z'_{\n})\sum_j N^*_{j5}n_j+ \nn \\
&&\frac{b_2^{(4)}}{2\sqrt{2}} g_0 g_1 \sum_i n_i^T N^*_{1i}\g_5 [\g^{\m},\g^{\n}]\pd_{\m}(M_{32}Z^0_{\n}+M_{33}Z'_{\n})\sum_j N^*_{j5}n_j+ \\
&&\frac{b_2^{(4)}}{2\sqrt{2}} g_0 g_1 \sum_i n_i^T N^*_{6i}\g_5 [\g^{\m},\g^{\n}]\pd_{\m}(M_{11}A_{\n}+M_{12}Z^0_{\n}+M_{13}Z'_{\n})\sum_j N^*_{j5}n_j \nn\eea This
expression is very long, but if we collect the three different interactions it can be written in this more
compact way:
\bea &&i \mathcal{L}=\sum_{i,j}C_A(i,j) n_i^T \g_5 [\g^{\m},\g^{\n}]\pd_{\m} A_{\n} n_j+ \\
&&\sum_{i,j}C_Z(i,j) n_i^T \g_5 [\g^{\m},\g^{\n}]\pd_{\m} Z^0_{\n} n_j+
\sum_{i,j}C_P(i,j) n_i^T \g_5 [\g^{\m},\g^{\n}]\pd_{\m} Z'_{\n} n_j \nn \label{nnblag1} \eea where:
\bea  &&C_A(i,j)= \frac{1}{\sqrt{2}} (b_2^{(1)} g_1^2 N^*_{1i} N^*_{j5} M_{11}+b_2^{(2)} g_2^2 N^*_{2i} N^*_{j5} M_{21}+
 \nn \\&&\frac{b_2^{(4)}}{2} g_0 g_1 N^*_{6i} N^*_{j5} M_{11}+(i\leftrightarrow j) )  \\
&&C_Z(i,j)= \frac{1}{\sqrt{2}} (b_2^{(0)} g_0^2 N^*_{6i} N^*_{j5} M_{32}+b_2^{(1)} g_1^2 N^*_{1i} N^*_{j5} M_{12}+
b_2^{(2)} g_2^2 N^*_{2i} N^*_{j5} M_{22}+ \nn \\ &&\frac{b_2^{(4)}}{2} g_0 g_1 N^*_{1i} N^*_{j5} M_{32}+
\frac{b_2^{(4)}}{2} g_0 g_1 N^*_{6i} N^*_{j5} M_{12}+(i\leftrightarrow j) )  \\  
&&C_P(i,j)= \frac{1}{\sqrt{2}} (b_2^{(0)} g_0^2 N^*_{6i} N^*_{j5} M_{33}+b_2^{(1)} g_1^2 N^*_{1i} N^*_{j5} M_{13}+
b_2^{(2)} g_2^2 N^*_{2i} N^*_{j5} M_{23}+\nn \\ &&\frac{b_2^{(4)}}{2} g_0 g_1 N^*_{1i} N^*_{j5} M_{33}+
\frac{b_2^{(4)}}{2} g_0 g_1 N^*_{6i} N^*_{j5} M_{13}+(i\leftrightarrow j) )  \label{caczcp} \eea As
we can see comparing (\ref{nnblag1}) with (\ref{nnblag2}) we have the same structure of the lagrangian: the only changes are
in the couplings just written. So the vertices are all equal except that we have to use the general couplings.
We'll have three possible interactions:

\begin{figure}[h!]
\includegraphics[scale=0.8]{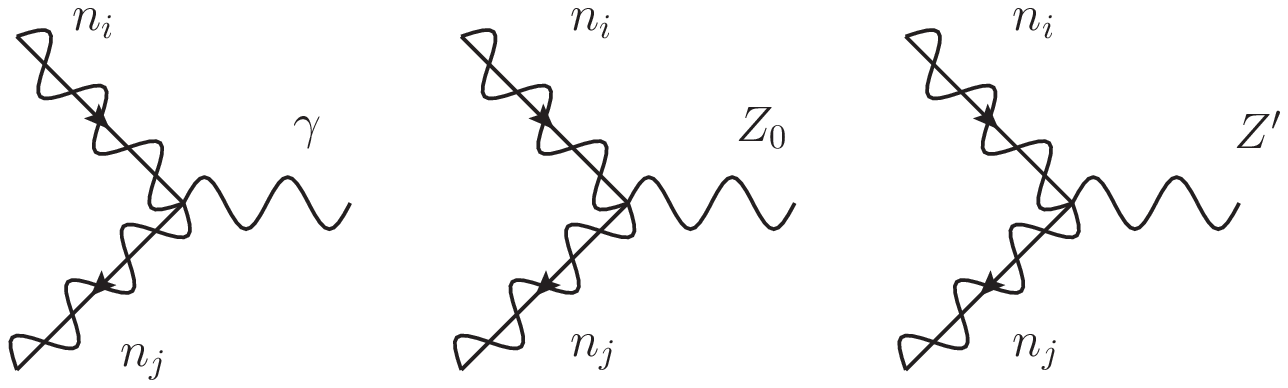} \be \label{nnb} \ee
\end{figure} with respective Feynman rules:

\be C_A(i,j) \g_5 [\g^\m,\g^\n]ik_\m \qquad C_Z(i,j) \g_5 [\g^\m,\g^\n]ik_\m \qquad 
C_P(i,j) \g_5 [\g^\m,\g^\n]ik_\m \label{nnbrules}\ee in 
which $k_{\m}$ is the momentum of the vector boson.

\subsection{Anomalous changes to higgs-higgs-boson and boson-boson-boson vertices}
We want to calculate the changes in the interactions of higgses and bosons caused by the 
anomalous extension. We start calculating the higgs-higgs-boson vertices. The lagrangian can be written
in the same way of the fermion-fermion-boson one:
\be j_{\m}^{H_u}\equiv Q^{gen} H_u^*\overleftrightarrow{\pd}H_u \Rightarrow \mathcal{L}=-ig_1 B_{\m} Y_{H_u} j^{\m}_{H_u}
-ig_2 W^3_{\m} T^3_{H_u} j^{\m}_{H_u}-ig_0 C_{\m} Q_{H_u} j^{\m}_{H_u}  \ee We have also an analogous term for $H_d$.
Because the lagrangian is equal to that already studied, the currents will change in the known way when we use the mass 
eigenstates:
\bea &&j_{\m}^{em}=j_{\m}^{3}+j_{\m}^{Y} \nn \\ &&j_{\m}^{NC}=(j_{\m}^{NC})^{SM}+\frac{g_0}{g_2} cos(\theta_W) M_{32}~
j_{\m}^{0} \\ &&j_{\m}^{Z'}=j_{\m}^{0}+\( \frac{g_2}{g_0} M_{23}-\frac{g_1}{g_0} M_{13} \)j_{\m}^{3} +
\frac{g_1}{g_0} M_{13}~ j_{\m}^{em} \nn \label{currents}\eea In this expression we have the physical vector bosons,
but the higgses are the gauge eigenstates. Using the formulas of \hypertarget{Martin}{\cite{Martin}}, we obtain:
\bea H_u=\( \begin{array}{c}
             H_u^{+} \\ H_u^0
            \end{array}
  \)=\( \begin{array}{c}
         cos \b~ H^+ \\\frac{1}{\sqrt{2}}(cos\a~ h^0+sin\a ~H^0)+\frac{i}{\sqrt{2}}cos \b ~A^0
        \end{array}
  \) \\ H_d=\( \begin{array}{c}
             H_d^{0} \\ H_d^-
            \end{array}
  \)=\( \begin{array}{c}
         \frac{1}{\sqrt{2}}(-sin\a~ h^0+cos\a ~H^0)+\frac{i}{\sqrt{2}}sin \b ~A^0 \\ sin~ \b~ H^-
        \end{array}
  \)  \eea Remembering that $ \phi\overleftrightarrow{\pd}_{\m}\chi=\phi \pd \chi-\chi \pd \phi $ we know that if 
we have the same fields this derivative is $0$. Using this fact we calculate:
\bea &&H_u^*\overleftrightarrow{\pd}_{\m}H_u= (H_u^+)^*\overleftrightarrow{\pd}_{\m}H_u^+ + (H_u^0)^*\overleftrightarrow{\pd}_{\m}H_u^0=
 \\ &&cos^2\b~H^-\overleftrightarrow{\pd}_{\m}H^+ +i~cos~\a~cos~\b~h^0\overleftrightarrow{\pd}_{\m}A^0+
+i~sen~\a~cos~\b~H^0\overleftrightarrow{\pd}_{\m}A^0 \nn \\ &&H_d^*\overleftrightarrow{\pd}_{\m}H_d=
 (H_d^-)^*\overleftrightarrow{\pd}_{\m}H_d^- + (H_d^0)^*\overleftrightarrow{\pd}_{\m}H_d^0=
 \\ &&sin^2\b~H^+\overleftrightarrow{\pd}_{\m}H^- -i~sin~\a~sin~\b~h^0\overleftrightarrow{\pd}_{\m}A^0+
+i~cos~\a~sin~\b~H^0\overleftrightarrow{\pd}_{\m}A^0 \nn \eea Using formulas (\ref{currents}) and remembering that
$Q_{H_d}=-Q_{H_u}$ the total interaction lagrangian becomes:
\bea &&\mathcal{L}^{hhv}=-i~e~A_{\m}H^-\overleftrightarrow{\pd}_{\m}H^+ \nn \\
&&-i\frac{g_2}{cos(\theta_W)} Z_{\m}^0
\Big[ \( \frac{1}{2}-sin^2(\theta_W)+\frac{g_0}{g_2}cos(\theta_W) M_{32}Q_{H_u} \)H^-\overleftrightarrow{\pd}_{\m}H^+ + \nn \\
&&i\(\frac{1}{2}-\frac{g_0}{g_2}cos(\theta_W) M_{32}Q_{H_u} \)
(-sin~\a~sin\b-cos~\a~cos~\b)h^0\overleftrightarrow{\pd}_{\m}A^0+ \nn \\ 
&&i\(\frac{1}{2}-\frac{g_0}{g_2}cos(\theta_W) M_{32}Q_{H_u} \)
(-cos~\a~sin\b-sin~\a~cos~\b)H^0\overleftrightarrow{\pd}_{\m}A^0\Big]+ \\
&&-i~g_0Z'_{\m}\Big[\(Q_{H_u}+\frac{1}{2}\(\frac{g_2}{g_0}M_{23}-\frac{g_1}{g_0}M_{13}\)+
\frac{g_1}{g_0}M_{13}\)H^-\overleftrightarrow{\pd}_{\m}H^+ \nn \\
&&+i\(Q_{H_u}-\frac{1}{2}\(\frac{g_2}{g_0}M_{23}-\frac{g_1}{g_0}M_{13}\) \)
(cos~\a~cos~\b+sin~\a~sin~\b)h^0\overleftrightarrow{\pd}_{\m}A^0+ \nn \\
&&+i\(Q_{H_u}-\frac{1}{2}\(\frac{g_2}{g_0}M_{23}-\frac{g_1}{g_0}M_{13}\) \)
(sin~\a~cos~\b-cos~\a~sin~\b)H^0\overleftrightarrow{\pd}_{\m}A^0 \Big]\nn \eea These
are all the interaction terms between a vector boson and two higgses. The constant in each term
is the constant of the respective vertex, whose Feynman diagram and rule are (from \hypertarget{mandl}{\cite{mandl}}):

\begin{figure}[h!]
\centering
\includegraphics[scale=0.8]{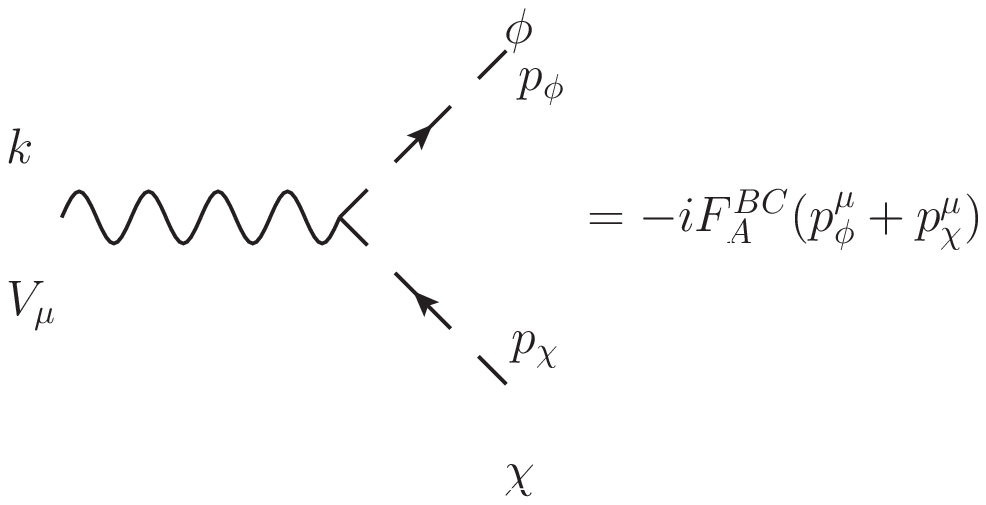} 
\be \label{vphichi} \ee
\end{figure} 
 Now we are going to list all the couplings of these
vertices. We call them $F_A^{BC}$, where $A$ is the vector boson, $BC$ is the couple of higgs entering the vertex:

\bea  &&F_{\g}^{H^-H^+}=e \nn \\ &&F_{Z_0}^{H^-H^+}=\frac{g_2}{cos(\theta_W)}\Big(\frac{1}{2}-
\frac{g_0}{g_2}M_{32}Q_{H_u}cos(\theta_W)-sen^2(\theta_W)  \Big) \nn \\ &&F_{Z'}^{H^-H^+}=g_0
\Big( Q_{H_u}+\frac{1}{2}\Big( \frac{g_2}{g_0}M_{23}-\frac{g_1}{g_0}M_{13} \Big)+\frac{g_1}{g_0}M_{13} \Big)
\nn \\ &&F_{\g}^{h^0A^0}=0 \\ &&F_{Z_0}^{h^0A^0}=\frac{g_2}{cos(\theta_W)}\Big(\frac{1}{2}-
\frac{g_0}{g_2}M_{32}Q_{H_u}cos(\theta_W) \Big)(-sin(\a)sin(\b)-cos(\a)cos(\b)) \nn \\
&&F_{Z'}^{h^0A^0}=g_0 \Big( Q_{H_u}-\frac{1}{2}\Big( \frac{g_2}{g_0}M_{23}-
\frac{g_1}{g_0}M_{13} \Big) \Big)(cos(\a)cos(\b))+sin(\a)sin(\b) \nn \\ 
&&F_{\g}^{H^0A^0}=0 \nn \\ &&F_{Z_0}^{H^0A^0}=\frac{g_2}{cos(\theta_W)}\Big(\frac{1}{2}-
\frac{g_0}{g_2}M_{32}Q_{H_u}cos(\theta_W) \Big)(cos(\a)sin(\b)-sin(\a)cos(\b)) \nn \\
&&F_{Z'}^{H^0A^0}=g_0 \Big( Q_{H_u}-\frac{1}{2}\Big( \frac{g_2}{g_0}M_{23}-
\frac{g_1}{g_0}M_{13} \Big) \Big)(sin(\a)cos(\b))-cos(\a)sin(\b) \nn \label{hhbcoup}\eea Now we want 
to calculate the boson-boson-boson vertices. The lagrangian in the gauge eigenstates basis is the 
same of the SM:

\bea &&\mathcal{L}=ig_2[(W_{\a}^{\dag}W_{\b}-W_{\b}^{\dag}W_{\a})\pd^{\a}W^{3\b}+(\pd_{\a}W_{\b}-
\pd_{\b}W_{\a})W^{\dag \b}W^{3\a}-\nn \\ &&(\pd_{\a}W_{\b}^{\dag}-\pd_{\b}W_{\a}^{\dag})W^{\b}W^{3\a}]  \eea As usual,
to go to the mass eigenstates basis we have to use the formulas (\ref{neumix}) and (\ref{neumass}), neglecting terms of
order higher or equal to 2 in $an$. In this way the lagrangian becomes:

\bea &&\mathcal{L}=ig_2cos(\theta_W)[(W_{\a}^{\dag}W_{\b}-W_{\b}^{\dag}W_{\a})\pd^{\a}Z_0^{\b}+(\pd_{\a}W_{\b}-
\pd_{\b}W_{\a})W^{\dag \b}Z_0^{\a}- \nn \\ &&(\pd_{\a}W_{\b}^{\dag}-\pd_{\b}W_{\a}^{\dag})W^{\b}Z_0^{\a}]+ \nn \\
&&ie[(W_{\a}^{\dag}W_{\b}-W_{\b}^{\dag}W_{\a})\pd^{\a}A^{\b}+(\pd_{\a}W_{\b}-
\pd_{\b}W_{\a})W^{\dag \b}A^{\a}- \nn \\ &&(\pd_{\a}W_{\b}^{\dag}-\pd_{\b}W_{\a}^{\dag})W^{\b}A^{\a}]+ \\
&&ig_2M_{23}[(W_{\a}^{\dag}W_{\b}-W_{\b}^{\dag}W_{\a})\pd^{\a}Z'^{\b}+(\pd_{\a}W_{\b}-
\pd_{\b}W_{\a})W^{\dag \b}Z'^{\a}- \nn \\ &&(\pd_{\a}W_{\b}^{\dag}-\pd_{\b}W_{\a}^{\dag})W^{\b}Z'^{\a}] \nn  \eea The first
two terms give the vertices $\g-W^+-W^-$ and $Z_0-W^+-W^-$ and are exactly the same of the SM, so we obtain the well known
Feynman rules:

\begin{figure}[h!]
\includegraphics[scale=0.8]{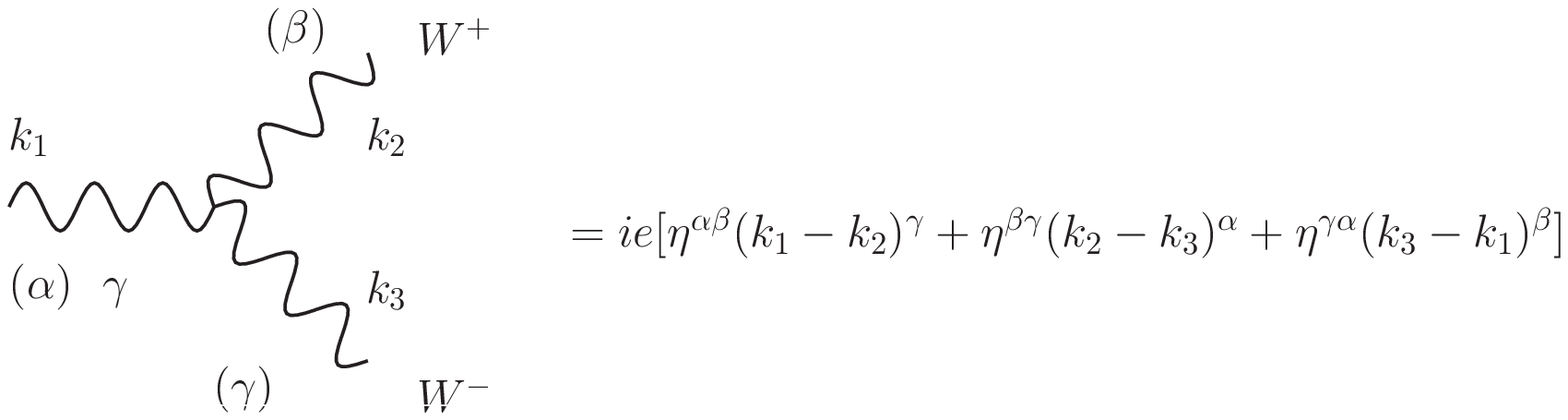} \be \label{gammaWW}\ee
\end{figure}

\begin{figure}[h!]
\includegraphics[scale=0.8]{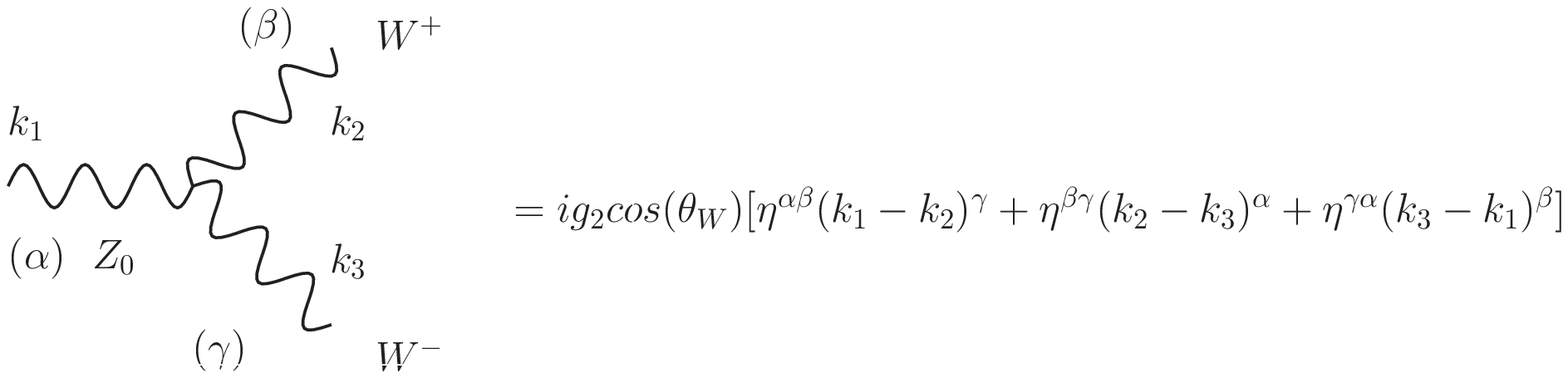} \be \label{Z0WW} \ee
\end{figure} The third term, related to the $Z'$, is due to the anomalous extension. However its
structure is identical to that of the first two terms so its Feynman rule will be:

\begin{figure}[h!]
\includegraphics[scale=0.8]{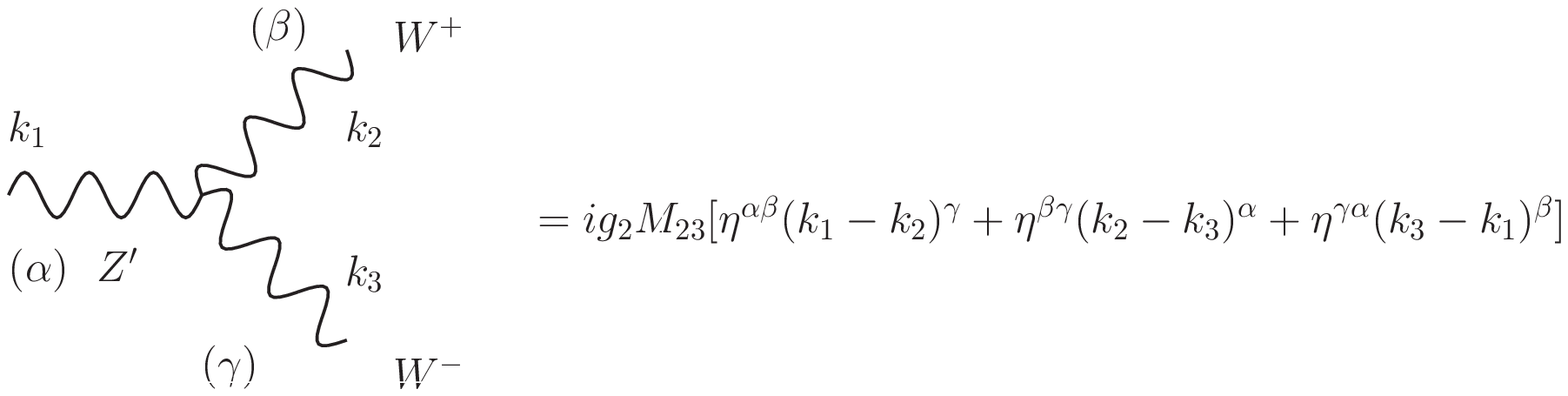} \be \label{Z'WW} \ee
\end{figure}

\subsection{Calculation of coannihilations in fermions}
In this section we want to calculate all the extra $U(1)$-related contributions to the coannihilation of two
neutralinos in a couple fermion-antifermion. The rules (\ref{nnb}),(\ref{gammaff}),(\ref{z0ff}) and (\ref{z'ff}) imply
that there are three s-channel coannihilation processes:

\begin{figure}[h!]
\centering
\includegraphics[scale=0.8]{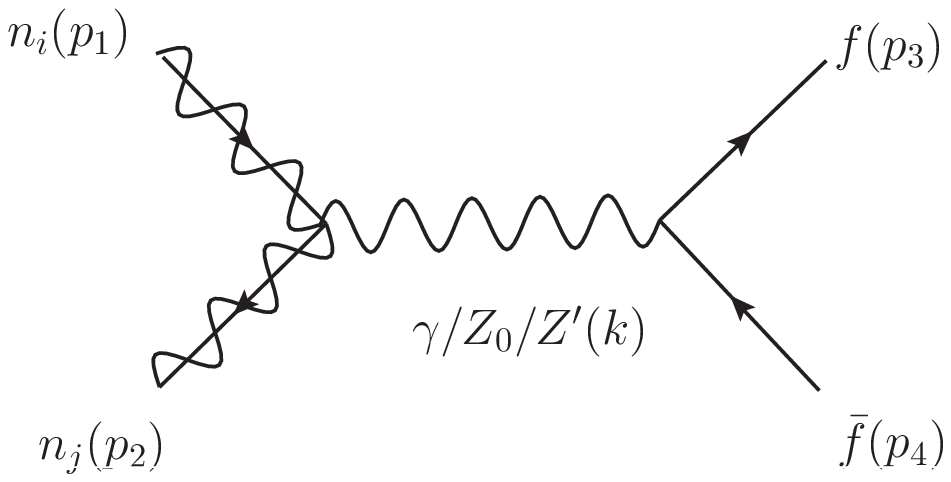}
\end{figure} Summing this three contributions we have:

\bea && \mathcal{M}=-i k^{\m} \bar{v}_1 \g_5 [\g_{\m},\g_{\n}] u_2 
[ e q_f C_{A}(i,j) \frac{\eta^{\n \r}}{k^2} \bar{u}_3 \g_{\r} v_4 
+  \\ && \frac{g_{2}}{2cos(\theta_W)} C_{Z}(i,j) \frac{\eta^{\n \r}}{k^2-M_{Z_0^2}} \bar{u}_3 \g_{\r}
(g_V-g_A \g_5) v_4+ g_0C_P(i,j) \frac{\eta^{\n \r}}{k^2-M_{Z'^2}} \bar{u}_3 \g_{\r}
(Q_V-Q_A \g_5) v_4 ] \nn \eea The tensorial structure is the same of (\ref{coanniampli}) so we have the same $T_v$
and $T_a$. The difference with this latter expression is in the couplings and in the propagators, that gives us the amplitude
square modulus:

\bea &&|\mathcal{M}|^2=\big[ T_a \big( \frac{g_2}{2cos(\theta_W)} C_{Z}(i,j) \frac{g_A}{k^2-M_{Z_0^2}}+
 g_0C_{P}(i,j) \frac{Q_A}{k^2-M_{Z'^2}} \big)^2+ \nn \\
&&T_v \big(\frac{e q_f C_{A}(i,j)}{k^2} + \frac{g_{2}}{2 cos(\theta_W)} \frac{g_VC_{Z}(i,j)}{k^2-M_{Z_0^2}}
+g_0 \frac{Q_VC_{P}(i,j)}{k^2-M_{Z'^2}} \big)^2 \big] \label{coanniff} \eea Multiplying this result for the prefactor 
(\ref{prefattore}) we obtain the differential cross section in the center of mass frame:
\be \frac{d\s}{d\Omega}= \frac{1}{64 \pi^2} \frac{|\overrightarrow{p_3}|}{|\overrightarrow{p_1}| s}
 |\mathcal{M}|^2 \ee

\subsection{Calculation of coannihilation in vector bosons}
There are three diagrams for the cohannihilation in $W^+-W^-$. They are analogous to that of the cohannihilation in
$f-\bar{f}$. The only difference is in the final states, which are a couple of vector instead of a couple of fermions:

\begin{figure}[h!]
\centering
\includegraphics[scale=0.8]{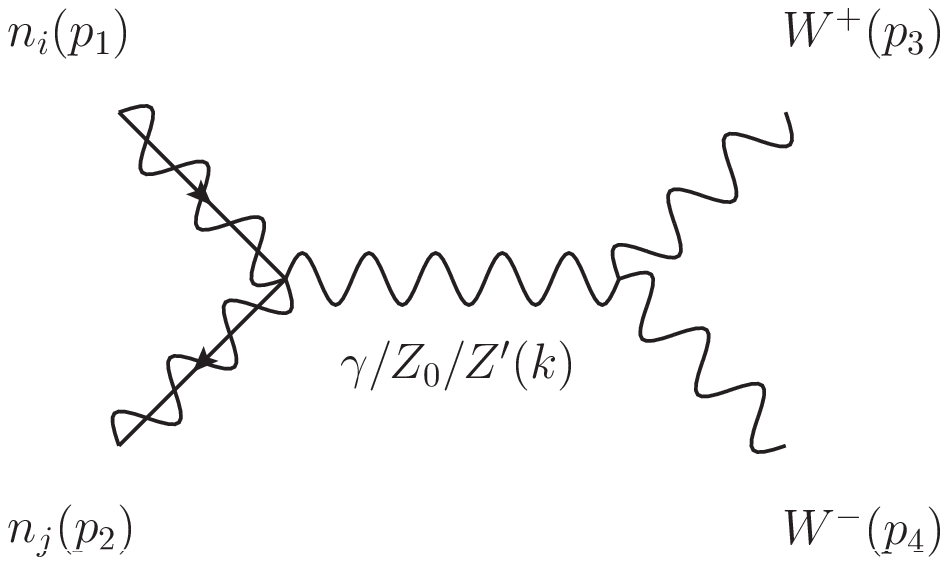}
\end{figure} Using the rules (\ref{nnb}),(\ref{gammaWW}),(\ref{Z0WW}) and (\ref{Z'WW}) 
and summing these three contributions we have:

\bea && \mathcal{M}=-i k^{\m} \bar{v}_1 \g_5 [\g_{\m},\g_{\n}] u_2 
\big( \frac{eC_A(i,j)\eta^{\n\r}}{k^2}+\frac{g_2cos(\theta_W)C_Z(i,j)\eta^{\n\r}}{k^2-M_{Z_0}^2}+ \\
&&\frac{g_2M_{23}C_P(i,j)\eta^{\n\r}}{k^2-M_{Z'}^2} \big)[\eta_{\r\a}(k-p_3)_{\b}+ 
\eta_{\a\b}(p_3-p_4)_{\r}+\eta_{\b\r}(p_4-p_k)_{\a}]\epsilon^{\a}_+ \epsilon^{\b}_- \nn  \eea The square 
of this amplitude gives:

\be |\mathcal{M}|^2=A^{\m\a}B_{\m\a}\big[ \frac{eC_A(i,j)}{k^2}+\frac{g_2cos(\theta_W)C_Z(i,j)}{k^2-M_{Z_0}^2}+
\frac{g_2M_{23}C_P(i,j)}{k^2-M_{Z'}^2} \big]^2  \label{coannivv}\ee As for the fermionic final states, we have the product of two 
tensor terms. $B_{\m\a}$ belongs to the incoming neutralinos, and it is equal to (\ref{bnualfa}).
$A^{\m\a}$ instead is one of the vectors in the final states. To calculate its expression we have to 
remember that, summing over the polarization \hypertarget{mandl}{\cite{mandl}}, there are the identities:
\be \epsilon^{+}_{\a}\epsilon^{+}_{\b}=\eta_{\a\b} \ee Using this formula, and making the square modulus
of the vector terms, we obtain:

\bea &&A^{\a \m}=[\eta^{\a}_{\n}(k-p_3)_{\b}+\eta_{\n\b}(p_3-p_4)^{\a}+\eta^{\a}_{\b}(p_4-k)_{\n}] \nn \\
 &&[\eta^{\m\n}(k-p_3)^{\b}+\eta^{\n\b}(p_3-p_4)^{\m}+\eta^{\a\b}(p_4-k)^{\n}]= \nn \\
&&\eta^{\a \m}(k-p_3)^2+(k-p_3)^{\a}(p_3-p_4)^{\m}+(p_4-k)^{\a}(k-p_3)^{\m}+  \\
&&(p_3-p_4)^{\a}(k-p_3)^{\m}+4(p_3-p_4)^{\a}(p_3-p_4)^{\m}+(p_3-p_4)^{\a}(p_4-k)^{\m}+ \nn \\
&&(k-p_3)^{\a}(p_4-k)^{\m}+(p_4-k)^{\a}(p_3-p_4)^{\m}+\eta^{\a \m}(p_4-k)^2 \nn \eea Computing
the product $A^{\a \m}B_{\a \m}$ we obtain:

\bea  &&|\mathcal{M}|^2=\big[ \frac{eC_A}{k^2}+\frac{g_2cos(\theta_W)C_Z}{k^2-M_{Z_0}^2}+
\frac{g_2M_{23}C_P}{k^2-M_{Z'}^2} \big]^2 \big[16 \big(4 k^4 \left(2 m_{\lambda } m_{\psi
   _S}+(p_{1}p_2)\right) \nn \\ &&-k^2 \big(2 k^2 \left(m_{\lambda } m_{\psi
   _S}+(p_{1}p_2)\right)+7 (kp_3) m_{\lambda } m_{\psi _S}+7
   (kp_4) m_{\lambda } m_{\psi _S}+3 (kp_3) (p_1p_2)+ \nn \\ &&3
   (kp_4) (p_1p_2)+(kp_2)
   \left((p_1p_3)+(p_1p_4)\right)+(kp_1) \left(8
   (kp_2)+(p_2p_3)+(p_2p_4)\right)+\nn \\ &&6 (p_3p_4) \left(m_{\lambda }
   m_{\psi _S}+(p_1p_2)\right)-8 (p_1p_2) m_W^4-12 m_W^4 m_{\lambda
   } m_{\psi _S}+4 (p_1p_3) (p_2p_3)-\nn \\ &&6 (p_1p_4) (p_2p_3)-6 (p_2p_3)
   (p_2p_4)+4 (p_1p_4) (p_2p_4)\big)+(kp_1) \big((p_2p_3)
   \left(k^2+4 (kp_3)-6 (kp_4)\right)+\nn \\ &&(p_2p_4)
   \left(k^2-6 (kp_3)+4 (kp_4)\right)+8 (kp_2)
   \left((kp_3)+(kp_4)-2 m_W^4\right)+\\ &&12 (kp_2)
   (p_3p_4)\big)+k^2 (kp_3) m_{\lambda } m_{\psi _S}+k^2
   (kp_4) m_{\lambda } m_{\psi _S}+k^2 (kp_3)
   (p_1p_2)+k^2 (kp_4) p_{12}+\nn \\ &&k^2 (kp_2)
   (p_1p_3)+(kp_2) (p_1p_4) \left(k^2-6 (kp_3)+4
   (kp_4)\right)-2 k^2 (p_1p_2) m_W^4-\nn \\ &&2 m_W^4 m_{\lambda
   } m_{\psi _S}k^2-2 (kp_3)^2 m_{\lambda } m_{\psi _S}-2
   (kp_4)^2 m_{\lambda } m_{\psi _S}+6 (kp_3)
   (kp_4) m_{\lambda } m_{\psi _S}-\nn \\ &&2 (kp_3)^2 (p_1p_2)-2
   (kp_4)^2 p_{12}+6 (kp_3) (kp_4) (p_1p_2)+4
   (kp_2) (kp_3) (p_1p_3)-\nn \\ &&6 (kp_2) (kp_4)
   (p_1p_3)\big)\big]  \nn \eea As usual the differential cross-section in the CM frame is:

\be \frac{d\s}{d\Omega}= \frac{1}{64 \pi^2} \frac{|\overrightarrow{p_3}|}{|\overrightarrow{p_1}| s}
 |\mathcal{M}|^2 \ee

\subsection{Calculation of coannihilation with final Higgs particles}
Considering the coannihilation in a couple of Higgs fields there are different possibilities, with related diagrams:

\begin{figure}[h!]
\centering
\includegraphics[scale=0.8]{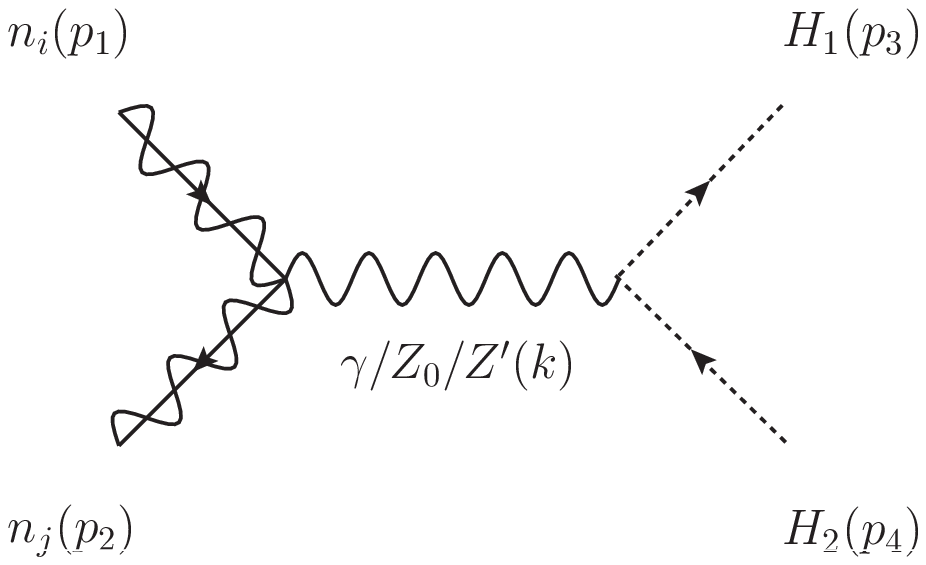} 
\end{figure} Where $H_1$ and $H_2$ can be the pair: $H^+~H^-$,$h^0~A^0$ or $H^0~A^0$.
Using the Feynman rules (\ref{nnbrules}) and (\ref{vphichi}), we compute the amplitude:

\bea &&\mathcal{M}^{H_1H_2}(i,j)=-ik^{\m}\bar{v}_1\g_5[\g_{\m},\g_{\n}]u_2\Bigg[\frac{C_A(i,j)F_A^{H_1H_2}}{k^2}
+ \nn \\ &&\frac{C_Z(i,j)F_{Z_0}^{H_1H_2}}{k^2-M_{Z_0}^2}+\frac{C_P(i,j)F_{Z'}^{H_1H_2}}{k^2-M_{Z'}^2} \Bigg]
\eta^{\n \r}(p_3+p_4)_{\r}  \eea Now we want to calculate the tensor term, remembering the kinematic
relation $k^{\m}=p_3^{\m}+p_4^{\m}$:

\bea &&k^{\m}\bar{v}_1\g_5[\g_{\m},\g_{\n}]u_2 \eta^{\n \r}(p_3+p_4)_{\r}= \nn \\
&&\bar{v}_1[\sl k, \sl k]u_1=0   \eea So the cross-section is $0$ and there is not an higgs
contribution to neutralino coannihilations.

\chapter{Calculation of LSP relic density}
We want to give an estimate of the LSP relic density and then to
compare it with the observed relic density of Dark Matter. Now we want to 
describe the tools needed for the relic density calculations.
We'll use a $(+,-,-,-)$ metric.
\section{Calculation method} \label{calcmet}
To calculate the relic density of a certain particle we have to 
consider the number of particles per unit of volume, $n$. We will 
also imagine to have coannihilation, that is the most general situation.
This means that N particles are near in mass to
the particle we are interested in.
The evolution of the $n$ in such a system is 
governed by the Boltzmann equation \hypertarget{Gondolo}{\cite{Gondolo}}:

\be \frac{dn}{dt}=-3Hn-\sum_{ij}\langle \s_{ij} v_{ij} \rangle
(n_i n_j - n^{eq}_i n^{eq}_j) \label{boltz3} \ee $\s_{ij}$ are the annihilation cross section
of the processes:

\be X_i X_j \rightarrow f\bar{f} \ee $v_{ij}$ are the absolute values
of relative velocity between i-th and j-th particle, defined by:
$v_{ij}=|v_i-v_j|$ , $n_i$ is the number of particle per unit of
volume of the i-th specie (index 1 refers to LSP),
 $n_i^{eq}$ are the number of particles per unit of
volume of the i-th specie in thermal equilibrium. $n=\sum_i n_i$,
$H$ is the Hubble constant and $v$ is the relative velocity of the 
initial particles. \newline The equation \ref{boltz3} is usually 
approximated considering $n_i/n=n_i^{eq}/n^{eq}$.
We also define the thermal average of effective cross section:

\be \langle \s_{eff} v \rangle \equiv \sum_{ij}\langle \s_{ij}
v_{ij} \rangle \frac{n_i^{eq}}{n^{eq}} \frac{n_j^{eq}}{n^{eq}} \ee
Substituting we obtain:

\be \frac{dn}{dt}=-3Hn-\langle \s_{eff} v \rangle (n^2 -
(n^{eq})^2) \label{Boltz} \ee $\langle \s_{eff} v_{ij} \rangle$
can be re-written in the sequent way \hypertarget{Ullio}{\cite{Ullio}}:

\be \langle \s_{eff} v \rangle=\frac{\sum_{ij}\langle \s_{ij}
v_{ij} \rangle n_i^{eq} n_j^{eq}}{n_{eq}^2}=\frac{A}{n^2_{eq}}
\label{sdef} \ee $A$ is the annihilation rate per unit of volume
at temperature $T$. We want to calculate it, using the classical gas
approximation, that is the Maxwell-Boltzmann distribution:

\be A=\sum_{ij}\langle \s_{ij} v_{ij} \rangle n_i^{eq}
n_j^{eq}=\sum_{ij} \frac{g_i g_j}{(2 \pi)^6} \int d^3\vec{p_i} d^3
\vec{p_j} e^{-E_i/T} e^{-E_j/T} \s_{ij} v_{ij} \ee where $g_i$ are
the degrees of freedom of the i-th particle. Now we re-write $A$:

\be A=\sum_{ij} \int W_{ij} \frac{g_i e^{-E_i/T} d^3\vec{p_i} }{(2
\pi)^3 2 E_i} \frac{g_j e^{-E_j/T} d^3 \vec{p_j}}{(2 \pi)^3 2 E_j}
\label{A}\ee In this expression we have introduced:
 \be W_{ij}=4E_i E_j \s_{ij} v_{ij} \label{Wij}\ee Now we define a
function associated to the relative velocity, that is going to be
useful in the calculation:

\be p_{ij}=\frac{v_{ij} E_i E_j}{\sqrt{s}} \Rightarrow W_{ij}=4
p_{ij} \sqrt{s} \s_{ij} \label{pij}\ee $s$ is the Mandelstam
variable that represents the energy in the CM frame, and it is defined as 
$s=(p_i+p_j)^2$. Remembering that the relative velocity between two 
relativistic particles is:

\be v_{ij}=\frac{\sqrt{(p_i p_j)^2-m_i^2 m_j^2}}{E_i E_j} \ee
Substituting in (\ref{pij}) and re-writing with respect to $s$ we obtain:

\bea &&p_{ij}=\frac{\sqrt{(p_i p_j)^2-m_i^2
m_j^2}}{\sqrt{s}}=\frac{\sqrt{(s-m_i^2-m_j^2)^2-m_i^2 m_j^2}}{2
\sqrt{s}}= \nn
\\ &&=\frac{[s-(m_i+m_j)^2]^{1/2} [s-(m_i-m_j)^2]^{1/2}}{2
\sqrt{s}} \label{pij1}\eea From this relation and from (\ref{pij})
it is clear that $W_{ij}$ is a function of $s$ only.\\

Now we want to re-write the integral (\ref{A}) and change integration
variables to obtain a one-dimensional integral with respect to 
the variable $s$. We will use spherical variables for  $p_i$, modified 
spherical coordinates in which the z axis is generated by $\vec{p_i}$, 
so the $\theta$ angle is between  $\vec{p_j}$ and $\vec{p_i}$ 
and the $\phi_j$ angle is on the plane orthogonal to $\vec{p_i}$.
Re-writing the differential in this variables we obtain:

\be d^3 \vec{p_i} d^3 \vec{p_j}= |\vec{p_i}|^2 d|\vec{p_i}| d\cos
\theta_i d \phi_i |\vec{p_j}|^2 d|\vec{p_j}| d \phi_j d \cos\theta
\ee We can integrate all angular variables, except $\theta$, because
the integrand in the equation (\ref{A}) is independent from them.
So we have:

\be d^3 \vec{p_i} d^3 \vec{p_j}= 4 \pi |\vec{p_i}|^2 d|\vec{p_i}|
4 \pi |\vec{p_j}|^2 d|\vec{p_j}| \frac{1}{2} d \cos\theta \ee
Now we use the equation $|\vec{p_i}|=\sqrt{E_i^2-m_i^2}$ and
its analogous for $|\vec{p_j}|$ to change the variable in the differential
from momentum to energy:

\be d|\vec{p_i}|=\frac{1}{|\vec{p_i}|}2E_i dE_i \label{vecpi}\ee
Using this result we obtain:

\be  d^3 \vec{p_i} d^3 \vec{p_j}= 4 \pi |\vec{p_i}| E_i dE_i 4 \pi
|\vec{p_j}| E_j dE_j \frac{1}{2} d \cos\theta \label{diff}\ee
Now we want to change variables again. The target is to have 
a set of variables that includes $s$. So we define:

\bea \left\{ \begin{array}{c} E_+=E_i + E_j \\
E_-=E_i - E_j  \\
s=m_i^2+m_j^2+2E_i E_j -2 |\vec{p_i}||\vec{p_j}|cos\theta \end
{array} \right. \label{def} \eea To express the differential in 
these new variables we have to calculate the Jacobian determinant: 

\be \ J=\left(\begin{array}{ccc} \frac{\pd E_i}{\pd E_+} & \frac{\pd E_i}{\pd E_-} & \frac{\pd E_i}{\pd s} \\
\frac{\pd E_j}{\pd E_+} & \frac{\pd E_j}{\pd E_-} & \frac{\pd E_j}{\pd s}  \\
\frac{\pd \cos \theta}{\pd E_+} & \frac{\pd \cos \theta}{\pd E_-}
& \frac{\pd \cos \theta}{\pd s} \end{array} \right) \ee To calculate
the determinant we have to invert the system
(\ref{def}). For the first two line the inversion is trivial:

\bea \left\{ \begin{array}{c} E_i=\frac{E_+ + E_-}{2} \\
E_j=\frac{E_+ - E_-}{2}  \\
 \end
{array} \right.  \eea We clearly see that $E_i$ ed $E_j$ depend
only on $E_+$ ed $E_-$. Given that
$|\vec{p_i}|=\sqrt{E_i^2-m_i^2}$, $|\vec{p_i}|$, that
$|\vec{p_j}|$  dependz only on $E_+$ ed $E_-$, if we invert the
third equation of the system
(\ref{def}) the only dependence from $s$ is the explicit one:

\be \cos \theta= \frac{-s +m_i^2 +m_j^2 +
1/2(E_+^2-E_-^2)}{2|\vec{p_i}||\vec{p_j}|} \ee So we have
$\frac{\pd \cos \theta }{\pd
s}=\frac{-1}{2|\vec{p_i}||\vec{p_j}|}$. Now we can write down
the known terms of the Jacobian matrix:

\be \ J=\left(\begin{array}{ccc} \frac{1}{2} & \frac{1}{2} & 0 \\
\frac{1}{2} & -\frac{1}{2} & 0  \\
\frac{\pd \cos \theta}{\pd E_+} & \frac{\pd \cos \theta}{\pd E_-}
& \frac{-1}{2|\vec{p_i}||\vec{p_j}|} \end{array} \right) \ee
Using the third row we can calculate the determinant without
knowing the remaining derivatives:

\be
det(J)=\frac{-1}{2|\vec{p_i}||\vec{p_j}|}(-\frac{1}{4}+-\frac{1}{4})=\frac{1}{4|\vec{p_i}||\vec{p_j}|}
\ee Using this result the equation (\ref{diff}) reads:

\be  d^3 \vec{p_i} d^3 \vec{p_j}= 2 \pi^2 E_i   E_j dE_+ dE_- d s
\ee The volume element is:

\be \frac{d^3\vec{p_i} }{(2 \pi)^3 2 E_i} \frac{d^3 \vec{p_j}}{(2
\pi)^3 2 E_j}=\frac{dE_+ dE_- ds}{8(2 \pi)^4} \ee Now we have to
find the integration limits. From the definition of spherical 
coordinates we know that $-1 \le \cos \theta \le 1$. Now we want
to find the conditions on $E_i$ (the conditions on $E_j$ will be analogous)
starting from those on $|\vec{p_i}|$:

\be |\vec{p_i}| \ge 0 \Rightarrow \sqrt{E_i^2-m_i^2}\ge 0
\Rightarrow E_i\ge m_i \ee Now, using $E_i\ge m_i$, $E_j\ge m_j$
e $|\cos \theta| \le 1$ we have to find the conditions on 
$E_+$, $E_-$ and $s$. \newline
%%%%%%%%%%%%%%%%%%%%%%%%%%%%%%
The integration limits become:

\bea s\ge(m_i+m_j)^2 \\
     E_+\ge\sqrt{s} \\
     |E_- - E_+ \frac{m_j^2-m_i^2}{s}|\le2p_{ij}
     \sqrt{\frac{E_+^2-s}{s}} \label{limiti}\eea
Now we can calculate $A$. The equation (\ref{A}), in the new variables, reads:

\be A=\sum_{ij} \int g_j g_i W_{ij} \frac{e^{-E_+/T} dE_+ dE_-
ds}{8 (2 \pi)^4} \ee We have already demonstrated that $W_{ij}$ is a function of
$s$ only, the exponential is a function of $E_+$ only, so we
can easily integrate with respect to $E_-$. It's sufficient using (\ref{limiti}) to derive:

\be \int dE_- =4p_{ij}
     \sqrt{\frac{E_+^2-s}{s}} \ee The integral becomes:

\be A=\sum_{ij} \int g_j g_i W_{ij} 4p_{ij}
     \sqrt{\frac{E_+^2-s}{s}} \frac{e^{-E_+/T} dE_+ ds}{8 (2
\pi)^4} \label{AA} \ee These are the integral representation
of the modified Bessel functions of the second type, that are
crucial for the integration with respect to $E_+$:

\be K_{\n}(az)=\frac{z}{\n! a^{\n}} \int_a^{\infty} dt e^{-zt}
t^{\n-1} \sqrt{t^2-a^2} \ee We are interested in:

\bea   &&\int^{\infty}_{\sqrt{s}} dE_+ e^{-E_+/T} \sqrt{E_+^2-s}=
T \sqrt{s} K_1\(\frac{\sqrt{s}}{T}\) \\
&&\int^{\infty}_{m_i} dE_i e^{-E_i/T} E_i \sqrt{E_i^2-m_i^2}=
\frac{T m_i^2}{2} K_2\(\frac{m_i}{T}\) \label{seconda}\eea
Using the first formula we perform the integration with respect to
$E_+$in (\ref{AA}), obtaining:

\be A=\frac{T}{32 \pi^4} \sum_{ij} \int_{(m_i+m_j)^2}^{\infty} ds
g_i g_j p_{ij} W_{ij} K_1\(\frac{\sqrt{s}}{T}\) \ee At this point we
define $W_{eff}$:

\be \sum_{ij}g_i g_j p_{ij} W_{ij}=g_1^2 p_{eff} W_{eff} \ee where:

\be p_{eff}=\frac{1}{2}\sqrt{s-4M_S^2} \label{peff}\ee Remember that 
$M_S$ is the LSP mass. In other words, using (\ref{pij1}):

\be W_{eff}=\sum_{ij}
\sqrt{\frac{[s-(m_i-m_j)^2][s-(m_i+m_j)^2]}{s(s-4M_S^2)}}\frac{g_i
g_j}{g_1^2}W_{ij} \label{Weff}\ee From equation (\ref{peff})
follows $ds=8p_{eff} dp_{eff}$, so:

\be A=\frac{g_1^2 T}{4 \pi^4} \int dp_{eff} p^2_{eff} W_{eff}
K_1\(\frac{\sqrt{s}}{T}\) \label{annirate}\ee Now we have to
calculate the denominator of the equation (\ref{sdef}), using again
the classical gas approximation and the relative Maxwell-Boltzmann
statistic:

\be n_{eq}=\sum_i n_i^{eq}=\sum_i \frac{g_i}{(2 \pi)^3} \int
d^3\vec{p_i} e^{-E_i/T} \ee Using the (\ref{seconda}) and (\ref{vecpi}),
we go to spherical coordinates, obtaining:

\bea &&n_{eq}=\sum_i \frac{g_i}{(2 \pi)^3} \int 4 \pi
|\vec{p_i}|^2 d \vec{p_i} e^{-E_i/T}=\sum_i \frac{g_i}{\pi^2}
\int_{m_i}^{\infty} dE_i E_i |\vec{p_i}| e^{-E_i/T}= \nn \\
&&=\sum_i \frac{g_i}{\pi^2} \int_{m_i}^{\infty} dE_i E_i
\sqrt{E_i^2-m_i^2} e^{-E_i/T}=\sum_i \frac{g_i m_i^2 T}{2 \pi^2}
K_2\(\frac{m_i}{T} \) \label{neq} \eea Substituting
(\ref{neq}) and (\ref{annirate}) equations in (\ref{sdef}), we obtain:

\be \langle \s_{eff}v \rangle (T) =\frac{\int dp_{eff} p^2_{eff}
W_{eff} K_1(\frac{\sqrt{s}}{T})}{M_S^4 T [\sum_i \frac{g_i}{g_1}
\frac{m_i^2}{M_S^2} K_2(\frac{m_i}{T})]^2} \label{media} \ee For example,
if we have two particles coannihilating, the formula (\ref{Weff}) reads:

\be W_{eff}=\sum_{i,j=1}^2
\sqrt{\frac{[s-(m_i-m_j)^2][s-(m_i+m_j)^2]}{s(s-4M_S^2)}}\frac{g_i
g_j}{g_1^2}W_{ij} \label{weff1}\ee Indices $1$ and $2$ refer to LSP and NLSP.\\

Now we have formulas for $\langle \s_{eff} v \rangle$. We have to use
this result to calculate the relic density. This is the normalization
to 1 of the comoving number of particle per units of volume:

\be \Omega_{\psi_S} h^2=\frac{\r_{\psi_S}}{\r_{crit}}
\label{omega}\ee $\r_{crit}$ is the density of comoving particles
in universe. The expression for $\Omega_{\psi_S} h^2$
is calculated solving the equation (\ref{Boltz}) with appropriate
approximations. The first step is to define $Y=n/s$,
where $s=S/R^3$ is the entropy density. Assuming that the universe expansion
is adiabatic, we have: $sR^3=cost$, where $R$ is the scale factor of the
universe. We will also use the Hubble law:
$H=\frac{\dot{R}}{R}$. The dot means time derivative.\\
We calculate:

\be \dot{Y}=\frac{\dot{n}}{s}-\frac{n}{s^2}\dot{s} \label{Y} \ee
Furthermore:

\be \dot{\(\frac{1}{s}\)}=\frac{3R^2}{cost}\dot{R}=3\frac{H}{s}
\label{1S}\ee Substituting this result in (\ref{Y}) we obtain:

\be \dot{Y} s=\dot{n}+3Hn \ee So $\dot{Y} s$ is equal to the sum
of the two terms in equation (\ref{Boltz}), that now, expressed 
in terms of $Y$, reads:

\be \dot{Y}=-s \langle \s_{eff} v \rangle (Y^2-Y^2_{eq}) \ee
We want to switch from the time derivative to the temperature one:

\be \frac{dY}{dt}=\frac{dY}{dT} \frac{dT}{dt}=\frac{dY}{dT}
\frac{dS}{dt} \frac{dT}{dS}=\frac{dY}{dT} (-3HS) \frac{dT}{dS} \ee
In the last step we have used (\ref{1S}). The Boltzmann Equation
now reads:

\be \frac{dY}{dT}=\frac{1}{3H} \frac{dS}{dT} \langle \s_{eff} v
\rangle (Y^2-Y^2_{eq}) \label{2}\ee At this point we want to use adimensional
variables, so we define:

\be x=\frac{M_S}{T} \Rightarrow \frac{dx}{dT}=-\frac{M_S}{T^2}
\Rightarrow \frac{dY}{dT}=-\frac{dY}{dx}\frac{M_S}{T^2} \ee
Substituting in equation (\ref{2}) we obtain:

\be \frac{dY}{dx}=-\frac{1}{3H}\frac{m}{x^2} \frac{dS}{dT} \langle \s_{eff} v
\rangle (Y^2-Y^2_{eq}) \ee We have to discuss the term 
$\frac{1}{3H} \frac{dS}{dT}$. For this purpose we will use the Friedmann
equation for $H$ and some parametrization for the functions $S$ and $\r$:

\bea &&H^2=\frac{8 \pi G \r}{3} \\
&&\r=g_{eff}(T)\frac{\pi^2}{30}T^4 \\
&&s=h_{eff}(T) \frac{2 \pi^2}{45} T^3 \label{seff}\eea $g_{eff}$ and
$h_{eff}$ are effective degrees of freedon for the density and
the entropy. They parametrize the deviation from the free gas behaviour 
multiplied by each effective degree of freedom.
Using this relation we obtain:

\bea &&\frac{1}{3H} \frac{ds}{dT}=\frac{1}{3}\sqrt{\frac{3}{8 \pi
G }\frac{30}{g_{eff}(T) \pi^2}}\frac{1}{T^2}\frac{2 \pi^2
}{45}\(\frac{dh_{eff}(T)}{dT}T^3+3T^2 h_{eff}(T)\)= \nn \\
&&=\sqrt{\frac{\pi}{45
G}}\frac{h_{eff}(T)}{\sqrt{g_{eff}(T)}}(\frac{T}{3
h_{eff}(T)}\frac{dh_{eff}(T)}{dT}+1) \eea Now we define the effective degrees
of freedom that we are going to use:

\be g_*^{1/2}=\frac{h_{eff}(T)}{\sqrt{g_{eff}(T)}}\(\frac{T}{3
h_{eff}(T)}\frac{dh_{eff}(T)}{dT}+1\) \ee Experiments show that
$g_{eff}(T)$ and $h_{eff}(T)$ are almost constant, so it is constant 
also $g_*^{1/2}$, with a numerical value of $g_*^{1/2} \sim 9$.
However  we will continue to consider those quantities as functions,
dealing with the most general case. (We only omit to write the variables, so
for example $g_*=g_*(T)$).\newline
%%%%%%%%%%%%%%%%%%%%%%%%%%%%%%%%%%%%%%%%%%%%spiegazione sulla "trasformazione" di g* in numero
Using these results the Boltzmann equation (\ref{boltz3}) becomes:

\be \frac{dY}{dx}=-\frac{M_S}{x^2} \sqrt{\frac{\pi g_*}{45 G}} \langle \s_{eff}
v \rangle (Y^2-Y^2_{eq}) \label{Boltz1}\ee

Using this equation we can calculate the freeze-out point.
It is defined as the point where $\Delta
\sim \d Y_{eq}$, where $\d$ is a parameter of order 1 that is fitted
to match experimental data. This fit gives the estimate
 $\d \sim 1.5$. However we will use the symbolic expression for generality.
The definition of $\Delta$ is: $\Delta=Y-Y_{eq}$.
Now we rewrite the Boltzmann equation with respect to
$\Delta$. We have these identities:

\bea \frac{d Y}{dx}=\frac{d \Delta}{dx}+\frac{d Y_{eq}}{dx} \\
Y^2-Y_{eq}^2=\Delta(\Delta +2 Y_{eq}) \eea Substituting in
(\ref{Boltz1}) we obtain:

\be \frac{d \Delta}{dx}+\frac{d Y_{eq}}{dx}=M_S \sqrt{\frac{\pi
g_*}{45 G}} \langle \s_{eff} v \rangle \Delta(\Delta +2 Y_{eq})
\ee Before the freeze-out $Y \sim Y_{eq} \Rightarrow \Delta \sim
0$, so the term with the derivative with respect to $\Delta$ can
be omitted. Substituting the condition
$x_f \Rightarrow \Delta \sim \d Y_{eq}$, we obtain:

\be \frac{1}{Y_{eq}} \frac{d Y_{eq}}{dx}=-\frac{M_S}{x^2} 
\sqrt{\frac{\pi g_*}{45 G}} \langle \s_{eff} v \rangle 
\d (\d +2) Y_{eq} \label{boltz2}\ee We start
calculating the first member of the equation: for this purpose we
write the explicit dependence  from $x$ of  $Y_{eq}$,
using the equations (\ref{neq}) and (\ref{seff}):

\be \frac{1}{Y_{eq}} \frac{dY_{eq}}{dx}=
\frac{1}{Y_{eq}} \frac{d}{dx}\left(\frac{n_{eq}}{s}\right)= 
\frac{1}{Y_{eq}} \sum_i \frac{45 g_i}{4 \pi^4 } 
\left(\frac{m_i}{M_S}\right)^2
\frac{d}{dx} \left( \frac{x^2 K_2\( \frac{m_i}{M_S} x \)}{h_{eff}\( \frac{M_S}{x} \)} \right) 
 \ee
Now we calculate the derivative, using a property of the modified
Bessel functions of the second kind: 

\be \frac{d K_{\n}(z)}{dz}=-K_{\n-1}(z)-\frac{\n}{z}K_{\n}(z) \ee
Using this equation we obtain:

\bea &&\frac{1}{Y_{eq}} \frac{d Y_{eq}}{dx}=-\frac{1}{Y_{eq}} 
\sum_i \frac{45 g_i}{4 \pi^4 h_{eff}} 
\left( \frac{m_i}{M_S}\right)^2  \frac{x^2 K_2\left(\frac{m_i}{M_S} x \right)}{h_{eff}
\left(\frac{M_S}{x} \right)} \times \nn \\ 
&&\left(\frac{m_i}{M_S} \frac{ K_1 \left( \frac{m_i}{M_S} x \right)}{K_2\left(\frac{m_i}{M_S} x \right)} 
+\frac{d}{dx} log\( h_{eff}\(\frac{M_S}{x} \) \) \right)\eea

Substituting this result in equation (\ref{boltz2}) we have:

\bea &&\frac{1}{Y_{eq}} 
\sum_i \frac{45 g_i}{4 \pi^4 h_{eff}} 
\left( \frac{m_i}{M_S}\right)^2  \frac{x^2 K_2\left(\frac{m_i}{M_S} x \right)}{h_{eff}
\left(\frac{M_S}{x} \right)} 
\left(\frac{m_i}{M_S} \frac{ K_1 \left( \frac{m_i}{M_S} x \right)}{K_2\left(\frac{m_i}{M_S} x \right)} 
+\frac{d}{dx} log\( h_{eff}\(\frac{M_S}{x} \) \) \right)=\nn \\
&&M_S \sqrt{\frac{\pi g_*}{45 G}} \langle \s_{eff} v \rangle \d
(\d +2) Y_{eq} \eea In the non-relativistic case, $x_f>>3$ we can use an asymptotical expansion for the 
modified Bessel functions of the second kind:

\be K_{\n}(z) \propto \sqrt{\frac{\pi}{2}}\frac{e^{-z}}{\sqrt{z}}+
... \ee In this limit this equation implies that
$K_1 \(\frac{m_i}{M_S x}\right)$ and $K_2\(\frac{m_i}{M_S x} \right) $ 
are equal, so, $\frac{ K_1 \left( \frac{m_i}{M_S} x \right)}{K_2\left(\frac{m_i}{M_S}x \right)} \simeq 1 $.
If we assume that comoving entropy density is conserved we also have $\frac{d}{dx} 
log \left( h_{eff}\left(\frac{M_S}{x}\right)\right)=0$.
The equation now reads:
\bea &&\frac{1}{Y_{eq}} 
\sum_i \frac{45 g_i}{4 \pi^4 h_{eff}} 
\left( \frac{m_i}{M_S}\right)^3  \frac{x^2 K_2\left(\frac{m_i}{M_S} x \right)}{h_{eff}
\left(\frac{M_S}{x} \right)} 
=\nn \\
&&M_S \sqrt{\frac{\pi g_*}{45 G}} \langle \s_{eff} v \rangle \d
(\d +2) Y_{eq} \eea 
$x_f$ is the solution of
this equation, that is numerically determined. For weak interactions
the result is almost independent from effective cross section, and is 
$x_f \sim 25$.\\

Now we can calculate the relic density of LSP. We are interested in the
situation prior the freezing-out, when we are not near the equilibrium:
this means $Y>>Y_{eq}$. Neglecting $Y_{eq}$ the Boltzmann equation
becomes: 

\be \frac{dY}{dx}=-\frac{M_S}{x^2} \sqrt{\frac{\pi g_*}{45 G}} \langle \s_{eff}
v \rangle (Y^2) \ee We can find the solution integrating from 
$T_f$ to the actual temperature $T_0$:

\be \frac{1}{Y_{0}}=\frac{1}{Y_f} + \sqrt{\frac{\pi}{45
G}} \int_{T_0}^{T_f} g_*^{1/2} \langle \s_{eff} v \rangle dT \ee This is the
complete approximate solution of the Boltzmann equation.\\
To have an estimate we can assume that $g^*$ is constant and we can also
neglect $Y_f$, obtaining:
\be Y_0 \sim \( \frac{45 G}{\pi g^*}\)^{1/2} \( \int_{T_0}^{T_f} \langle \s_{eff} v \rangle dT \)^{-1} \label{Ya}\ee
Substituting this result in 
(\ref{omega}) we can calculate the relic density, remembering that
(for our LSP) the relic density is given by  $\r_{\psi_S}=M_S s_0 Y_0$,
where $s_0$ represents the entropy density at the actual time.
So we have the expression:

\be \Omega_{\psi_S} h^2=\frac{\r_{\psi_S}}{\r_{crit}}=\frac{M_S
s_0 Y_0}{\r_{crit}} \ee Using now
$\r_{crit}=\frac{3H^2}{8 \pi G}$ and $M_{Pl}=\frac{1}{\sqrt{G}}$, together
with (\ref{Ya}), we can estimate the relic density.
There is also a very naive rule to estimate the relic density.
It consists in the approximation of $\langle \s_{eff} v\rangle$ as a constant;
in this way the integration (\ref{Ya}) is trivial. The resulting
approximate expression for the relic density is:

\be \Omega_{\psi_S} h^2 \approx \frac{3 \times 10^{-27} cm^3
s^{-1}}{\langle \s_{eff} v\rangle} \label{approx}\ee

\section{Relic density}
Our starting points are the annihilation and coannihilations cross-
section already calculated. If we consider only the annihilation, 
from equation (\ref{noan}) we infer that
the cross-section is proportional to $C^2C'^2$. Naive estimates from
(\ref{vertice}) lead to a $C$ and $C'$ of order of magnitude 
smaller than the typical electro-weak couplings.
This implies that the annihilation cross-section is roughly of the order
of $10^{-29} cm^3 s^{-1}$.\newline
Substituting this value in (\ref{approx}) we obtain $\Omega_{\psi_S} h^2 \approx 10^2$
which exceeds the WMAP \cite{wmapdata} constraints. We can exclude that
the annihilation of an LSP coming from our extension could match the
experimental data.\newline
The case in which there are coannihilations is more interesting:
the equation (\ref{moduloquadro}) implies that this cross section
is proportional to $C^2e^2$. Thus, remembering that the masses of 
the LSP and NLSP are free parameters, we want to study if the LSP relic density
can be described by this process.

\subsection{Cross-section and Mandelstam variables}
The formula (\ref{media}) implies that the calculation of the thermal
cross-section requires that the total cross-section was expressed in 
Mandelstam variables. Their definition, with reference to the figure
\ref{figuracoann} is:

\bea s=(p_1+p_2)^2=(p_3+p_4)^2=k^2 \nn \\
t=(p_3-p_1)^2=(p_4-p_2)^2 \\
u=(p_4-p_1)^2=(p_3-p_2)^2 \nn \eea Now we give some important
properties of these quantities in the CM frame, defined by:
\be \vec{p_1}=-\vec{p_2}\ee The momentum conservation implies:

\be \vec{p_3}=-\vec{p_4} \label{uscenti} \ee So we have:

\be (p_1 + p_2)=(E_1+E_2, \vec{0}) \Rightarrow
s=(E_1+E_2)^2=E_{CM}^2=(E_3+E_4)^2 \ee The explicit writing of the variables
permits to calculate many other useful relations:                                                                     

\bea s=p_1^2 +p_2^2 +2(p_1 p_2)=M_S^2 + m_{\l}^2+2(p_1
p_2)=2m_{f}^2 +2(p_3 p_4) \nn \\
t=p_1^2 +p_3^2 -2(p_1 p_3)=M_S^2 + m_{f}^2-2(p_1 p_3)=m_{\l}^2 + m_{f}^2-2(p_2 p_4) \\
u=p_1^2 +p_4^2 -2(p_1 p_4)=M_S^2 + m_{f}^2-2(p_1 p_4)=m_{\l}^2 +
m_{f}^2-2(p_2 p_3) \nn \eea These relations express the Mandelstam 
variables as functions of the masses and scalar products of the momentum of
the external particles. In the scalar products there are the dependencies
from the angular variables, that have to be made explicit in order to
write the total cross-section from the differential cross-section:

\bea &&(p_1 p_2)=E_1 E_2 + |p_1|^2 \nn\\
&&(p_3 p_4)=E_3 E_4 + |p_3|^2 \nn \\
&&(p_1 p_3)=E_1 E_3 - |p_1||p_3|\cos{\theta}  \\
&&(p_1 p_4)=E_1 E_4 + |p_1||p_3|\cos{\theta} \nn \\
&&(p_2 p_3)=E_2 E_3 + |p_1||p_3|\cos{\theta} \nn \\
&&(p_2 p_4)=E_2 E_4 - |p_1||p_3|\cos{\theta} \nn \label{scalarproducts}\eea In these formulas 
$\theta$ is defined as the angle between $p_1$ and $p_3$. At this point
we also want relations that express all non-angular variables as functions
of $s$. We want to solve the general case, in which the two initial particles
(or analogously the two final particles) have different masses.
We have to solve the system:

\bea  \left\{ \begin{array}{c} m_1^2 + m_{2}^2 +2 E_1 E_2 +2 |p_1|^2=s \nn \\
m_1^2=E_1^2-|p_1|^2  \\
m_{2}^2=E_2^2-|p_1|^2 \nn  \end {array}  \right. \label{p1quadro} \eea The second
and third equations give the solutions for $E_1$ and $E_2$with respect to
$|p_1|^2$, that remains the only real unknown quantity. Solving 
the system we obtain:

\be
|p_1|^2=\frac{s}{4}-\frac{m_1^2+m_{2}^2}{2}+\frac{(m_1^2-m_{2}^2)^2}{4s}\ee Now we 
want to calculate how this expression reduces if $m_1=m_2$:

\bea &&|p_1|^2=\frac{s}{4}-m_1^2  \\ &&E_1^2=E_2^2=s/4  \eea Using these
relations we can calculate the total cross-section and express it as a function
of $s$.

\subsection{Relic density examples}
\begin{figure}[t]
\centering
      \includegraphics[scale=0.65]{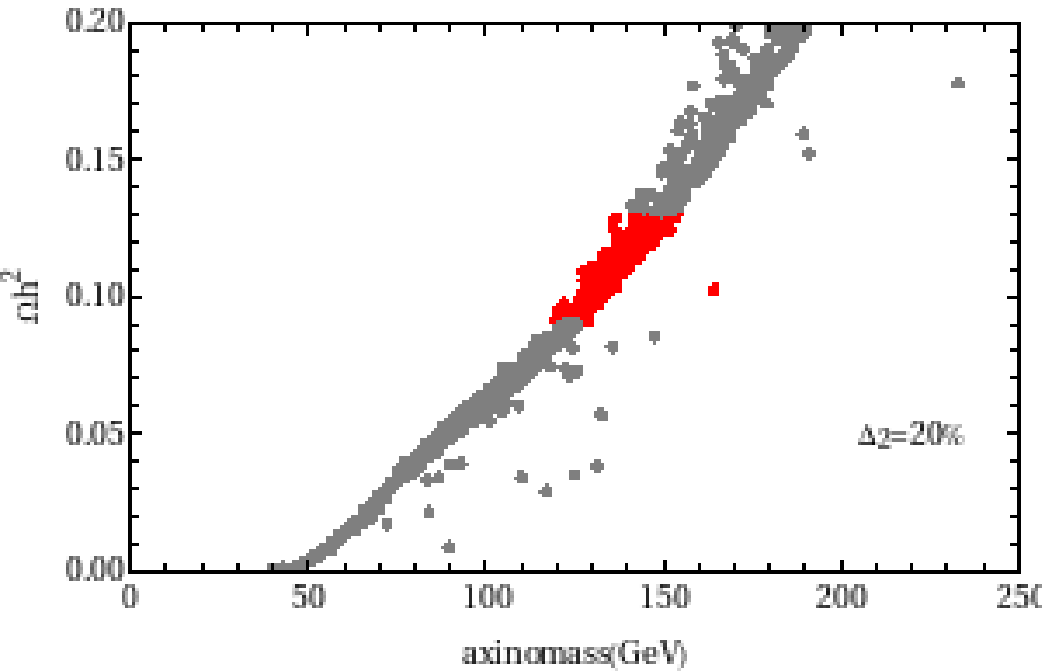}
      \includegraphics[scale=0.65]{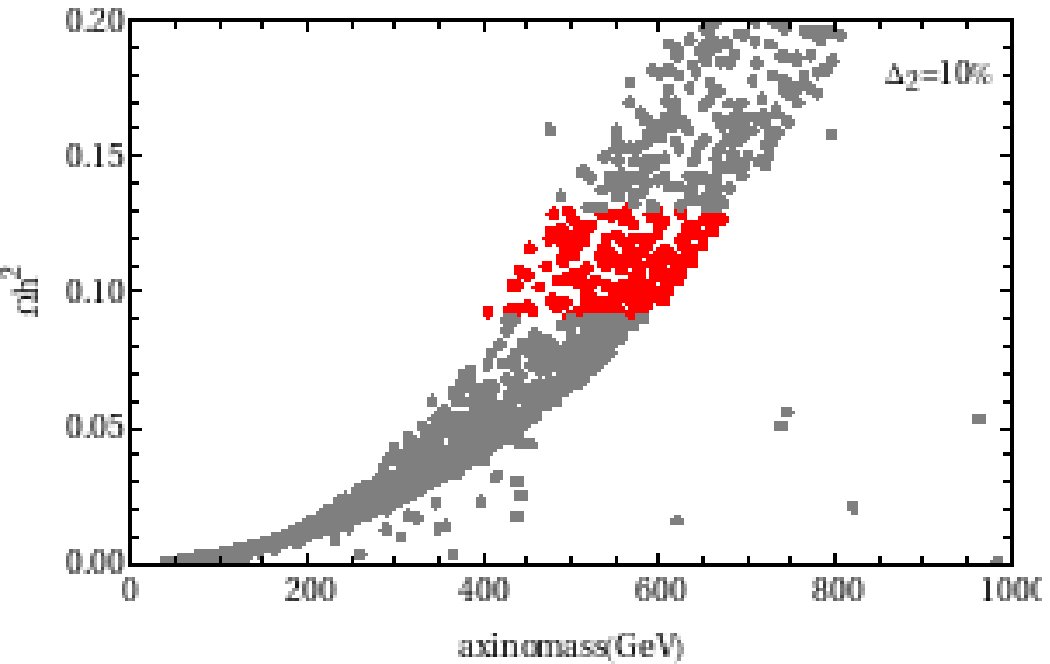}
      \includegraphics[scale=0.65]{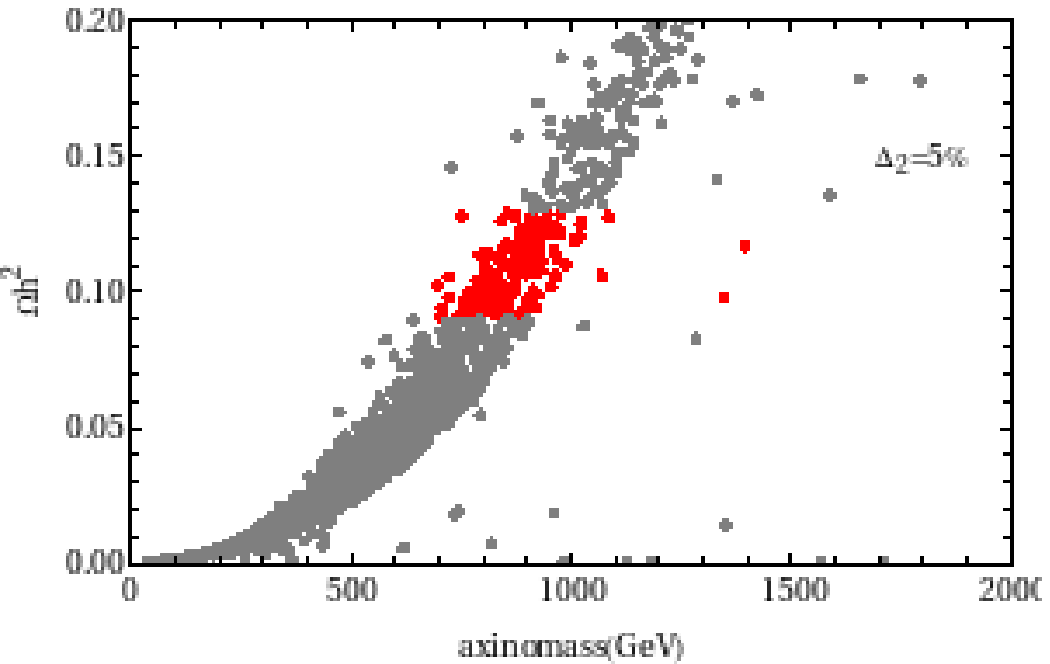}
      \includegraphics[scale=0.65]{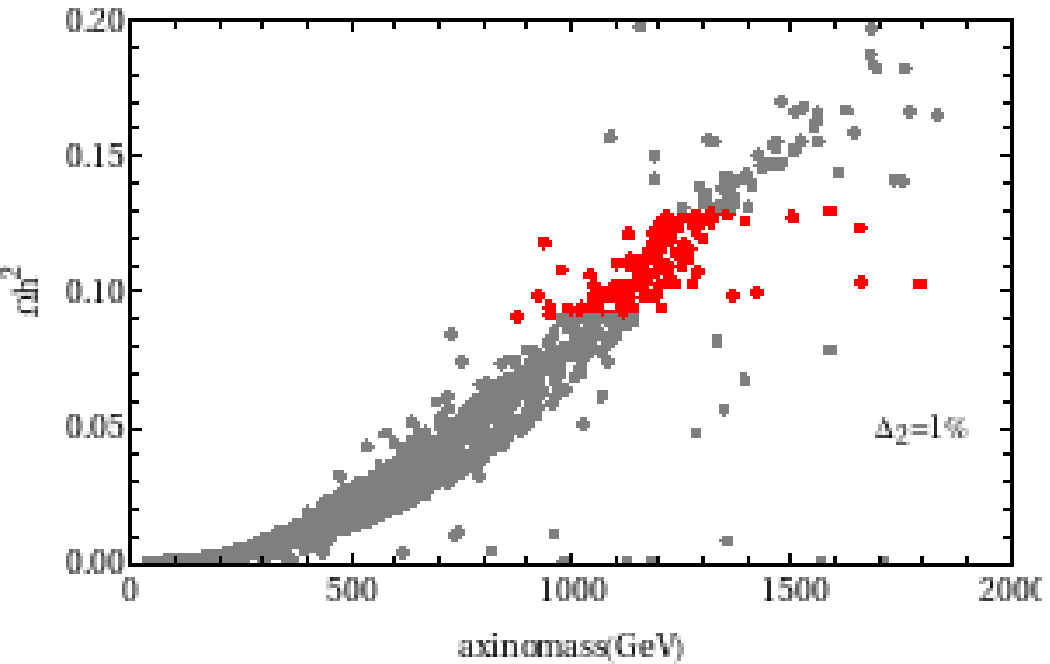}
     \caption{St\"uckelino relic density in the case in which the NLSP is a wino while the NNLSP is the lightest chargino. Red (darker) points denote models which satisfy WMAP data. Upper left panel: $\Delta_2=20\%$. Upper right panel:  $\Delta_2=10\%$. Lower left panel: $\Delta_2=5\%$. Lower right panel: $\Delta_2=1\%$.}
\label{DMwino-chargino}

\end{figure}

Now we want to study the relic density in simple cases of coannihilation. The
simplest possibility is that in which the LSP is a pure st\"{u}ckelino, 
coannihilating with a MSSM bino or a MSSM wino in a couple 
$f-\bar{f}$. In this case $C_{\g}=C^{(1)}cos(\theta_W)$, $C_{Z_0}=-C^{(1)}sin(\theta_W)$, 
$C^{(1)}$ is given by (\ref{vertice}).
We write down the related total cross-section, using the formulas
(\ref{sezcoan}),(\ref{scalarproducts}),(\ref{p1quadro}) and integrating
over the solid angle:

\bea
&&\sigma=c_f \(g_1^2 b_2^{(1)}\)^2 \sqrt{s-4 m_f^2} \times \\
&&\times\frac{ \Big[-2 M_1^4+\(4
   M_S^2+s\) M_1^2-6 M_S s M_1-2 M_S^4+s^2+M_S^2 s \Big]}{12 \pi  \(M_{Z_0}^2-s\)^2 s^{5/2} \sqrt{M_1^4-2
   \(M_S^2+s\) M_1^2+\(M_S^2-s\)^2}} \times\nn\\
&&\times\Bigg[(2 m_f^2+s) \Big(2 \cos\theta_W e q_f (M_{Z_0}^2-s)
  + \sin\theta_W g_{Z_0} v_f s \Big)^2 +  \(\sin\theta_W g_{Z_0} a_f\)^2 s^2 \(s-4 m_f^2\)\Bigg] \nn
\label{seztot}\eea This result can be used for a numerical calculation of the relic density.
From the formula (\ref{sdef}), written in the case $N_{coan}=2$, we obtain:

\bea
 &&\langle \s_{eff}^{(2)} v \rangle = \langle\sigma_{11}v\rangle (\frac{n_1^{eq}}{n^{eq}})^2+
2 \langle\sigma_{12}v\rangle \frac{n_1^{eq}n_2^{eq}}{(n^{eq})^2}+
\langle\sigma_{22}v\rangle (\frac{n_2^{eq}}{n^{eq}})^2= \nn \\
&&\langle\sigma_{22}v\rangle\frac{\langle\sigma_{11}v\rangle/\langle\sigma_{22}v\rangle
+2\langle\sigma_{12}v\rangle/\langle\sigma_{22}v\rangle Q+Q^2}{(1+Q)^2}
\label{sigma2eff}
\eea where $Q=n_2^{eq}/n_1^{eq}$. The first term in the numerator can
be neglected because the St\"uckelino annihilation cross section
is suppressed by a factor $(C^{(a)})^4$ with respect to the MSSM
neutralino annihilations (as already said) and thus
$\langle\sigma_{11}v\rangle\ll \langle\sigma_{22}v\rangle$. The
second term involves the coannihilation cross section and it is the
termal average of (\ref{seztot}). As we have seen in the previous section
the anomalous couplings $C_{\g}$ and $C_{Z_0}$ are both proportional to
$C^{(1)}\sim b_2^{(1)} g_{1}^2$, with $b_2^{(1)}$ given by (\ref{anomalia}):
\be b_2^{(1)}=\frac{3(3Q_Q+Q_L)}{256 \pi^2 b_3}
\ee where $b_3=M_{Z'}/4g_0$. With the assumption $M_{Z'}=1~TeV$ as
in \hypertarget{LioRac}{\cite{LioRac}} we can give the estimate:
\be \frac{C^2_\g}{e^2}\simeq
5.76 \times 10^{-12} (3 g_0 Q_Q + g_0 Q_L)^2 \, {\rm GeV}^{-2} \ee Substituting
this result in (\ref{seztot}), given that the perturbative requirement is:
\be g_0^2\cdot (3 Q_Q +
Q_L)^2<16 \label{chargebound} \ee and given also that the typical weak cross section is
$\langle\sigma_{22}v\rangle\simeq 10^{-9} ~ GeV^{-2}$ we obtain:

\begin{equation}
 \frac{\langle\sigma_{12}v\rangle}{\langle\sigma_{22}v\rangle}\lesssim 10^{-6}
\label{binosigma12neg}
\end{equation}
in the case of a pure bino, while
\begin{equation}
 \frac{\langle\sigma_{12}v\rangle}{\langle\sigma_{22}v\rangle}\lesssim 10^{-5}
\label{winosigma12neg}
\end{equation} in the case of a pure wino. So we can use the formulas (\ref{sigma2eff}) and 
(\ref{approx}) to write:
\be \( \Omega h^2\)^{(2)} \simeq \[
\frac{1+Q}{Q} \]^2 \( \Omega h^2\)^{(1)} \label{2rescaling} \ee in which the 
relic density in the presence of coannihilations is related to that of the MSSM
system without the st\"{u}ckelino. We have calculated the MSSM relic density 
with the DarkSUSY package \hypertarget{Gondolo:2004sc}{\cite{Gondolo:2004sc}} and then we have used it to 
calculate the coannihilating relic density for different mass gaps between LSP
and NLSP. 
\begin{figure}[t]
\centering
      \includegraphics[scale=0.7]{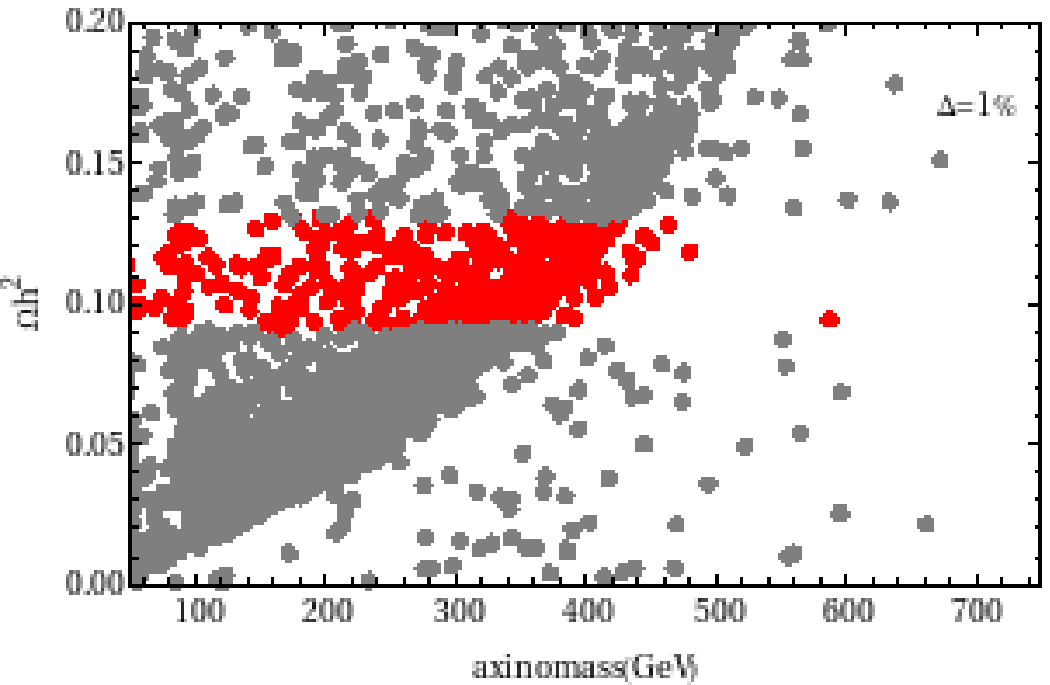}
      \includegraphics[scale=0.7]{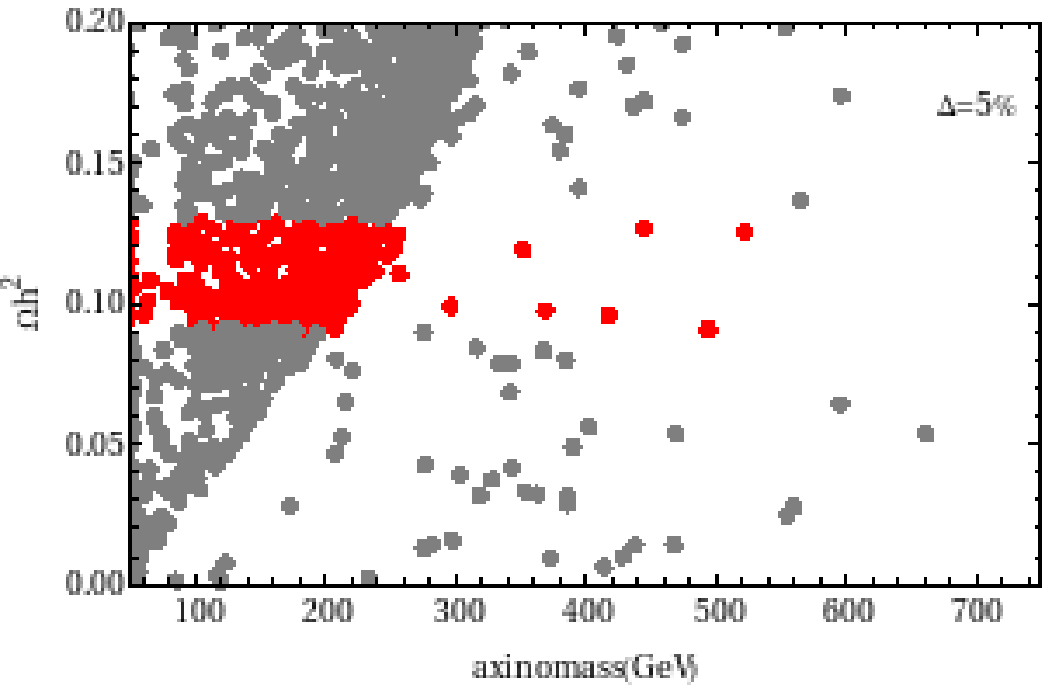}
      \caption{St\"uckelino relic density in the case in which the NLSP is a linear combination bino-higgsino. Red (darker) points denote models which satisfy WMAP data. Left panel: $\Delta_2=1\%$. Right panel:  $\Delta_2=5\%$.}
\label{DMbino-higgsino}
\end{figure} In figure (\ref{DMbino-higgsino}) we have plotted the results.
The red points satisfy the WMAP \cite{wmapdata} constraints:
\be 0.0913\le\Omega h^2\le
0.1285 \ee \\
Now we want to study the case $N=3$. This means that there are two particles with masses 
comparable to the LSP mass. This can be achieved very simply if the NLSP is a nearly 
pure wino: in fact the wino is almost degenerate in mass with the lightest chargino.
Expanding all the terms in the sum~(\ref{sdef}) we get:
\bea
 \la \s_\text{eff}^{(3)} v \ra &=& \la \s_{11} v \ra \g_1^2 + \la \s_{12} v \ra  \g_1 \g_2 + \la \s_{13} v \ra  \g_1 \g_3 +\nn\\
                 &&\la \s_{21} v \ra  \g_2 \g_1 + \la \s_{22} v \ra  \g_2^2 + \la \s_{23} v \ra  \g_2 \g_3 +\nn\\
                 &&\la \s_{31} v \ra  \g_3 \g_1 + \la \s_{32} v \ra  \g_3 \g_2 + \la \s_{33} v \ra  \g_3^2 \nn\\
               &=& \Big[ \la \s_{11} v \ra  (\numeq_1)^2 + \la \s_{12} v \ra  \numeq_1 \numeq_2 + \la \s_{13} v \ra  \numeq_1 \numeq_3 + \nn\\
                 &&\ \la \s_{21} v \ra  \numeq_2 \numeq_1 + \la \s_{22} v \ra  (\numeq_2)^2 + \la \s_{23} v \ra  \numeq_2 \numeq_3 +\nn\\
                 &&\phantom{\Big[}
                  \la \s_{31} v \ra  \numeq_3 \numeq_1 + \la \s_{32} v \ra  \numeq_3 \numeq_2 + \la \s_{33} v \ra  (\numeq_3)^2 \Big]
                 \frac{1}{(\numeq)^2} \nn\\
               &\simeq& %\frac{2 \( \la \s_{12} v \ra  \numeq_1 \numeq_2 + \la \s_{13} v \ra  \numeq_1 \numeq_3 \)}{(\numeq)^2} +\nn\\
%              &&
\frac{ \big[ \la \s_{22} v \ra  (\numeq_2)^2 + 2 \la \s_{23} v \ra  \numeq_2 \numeq_3 +
                 \la \s_{33} v \ra  (\numeq_3)^2 \big] }{(\numeq)^2}
\eea
where in the last line we have neglected the terms $\la \s_{11} v
\ra $, $\la \s_{12} v \ra $ and $\la \s_{13} v \ra $ since these
are the thermal averaged cross sections which involve the
St\"uckelino. In the equilibrium approximation we can write:
\be \numeq_i=g_i(1+\Delta_i)^{3/2}e^{-x_f\Delta_i} \qquad \text{for}~i=2,3  \ee where
$g_i$ are the internal degrees of freedom of the particle species and $\D_i=(m_i-m_1)/m_1$. That
permits to define and approximate the variables:
\be
 Q_i=\frac{\numeq_i}{\numeq_1}= \frac{g_i}{g_1} (1 + \D_i)^{3/2} e^{-x_f \D_i} \qquad \text{for} \ i=2,3
\ee As in the $N=2$ case, we use this variables to rewrite the 
termal cross-section:
\bea
 \la \s_\text{eff}^{(3)} v \ra
                 %&=& \frac{\la \s_{22} v \ra  Q_2^2 + 2 \la \s_{23} v \ra  Q_2 Q_3 + \la \s_{33} v \ra  Q_3^2}{\(1+Q_2+Q_3\)^2}\nn\\
                 %&&\times\( 1 + \frac{2 \( \la \s_{12} v \ra  Q_2 + \la \s_{13} v \ra  Q_3 \)} {\la \s_{22} v \ra  Q_2^2 + 2 \la \s_{23} v \ra  Q_2 Q_3 + \la \s_{33} v \ra  Q_3^2}\) \nn\\
               &\simeq& \frac{\la \s_{22} v \ra  Q_2^2 + 2 \la \s_{23} v \ra  Q_2 Q_3 + \la \s_{33} v \ra  Q_3^2}{\(1+Q_2+Q_3\)^2}
\eea This expression can be further simplified assuming
$(m_3-m_2)/m_1 \ll 1 /x_f$, $Q_3/Q_2 \simeq g_3/g_2$:
\bea
 &&\la \s_\text{eff}^{(3)} v \ra \simeq 
\frac{\Big(\la \s_{22} v \ra  + 2 \frac{g_3}{g_2} \la \s_{23} v \ra  + \(\frac{g_3}{g_2}\)^2 \la \s_{33} v \ra\Big)Q_2^2}{\Big[1+ \Big(1 + \frac{g_3}{g_2}\Big) Q_2\Big]^2}
\nn \\ &&\simeq \frac{Q_2^2}{\Big[1+ \Big(1 + \frac{g_3}{g_2}\Big) Q_2\Big]^2} \la \s_\text{MSSM} v \ra
\label{sigma3eff}
\eea
where
\be
 \la \s_\text{MSSM} v \ra  = \la \s_{22} v \ra  + 2 \frac{g_3}{g_2} \la \s_{23} v \ra  + \(\frac{g_3}{g_2}\)^2 \la \s_{33} v \ra
\ee Now we have to express this quantity with respect to a 2 particles
thermal cross-section, in order to compute a rescaling factor as in the $N=2$
case. The 2 particles cross-section in our case is given by:
\bea
 \la \s_\text{eff}^{(2)} v \ra &=& \frac{ \la \s_{22} v \ra  (\numeq_2)^2 + 2 \la \s_{23} v \ra  \numeq_2 \numeq_3 + \la \s_{33} v \ra  (\numeq_3)^2}{(\numeq)^2} \nn\\
               &=& \frac{ \la \s_{22} v \ra  (\numeq_2)^2 + 2 \la \s_{23} v \ra  \numeq_2 \numeq_3 + \la \s_{33} v \ra  (\numeq_3)^2}{(\numeq_2 + \numeq_3)^2} \nn\\
               &=& \frac{ \la \s_{22} v \ra  (\numeq_2)^2 + 2 \la \s_{23} v \ra  \numeq_2 \numeq_3 + \la \s_{33} v \ra  (\numeq_3)^2}{(\numeq_2)^2 (1 + \numeq_3/\numeq_2)^2}\nn\\
               &=& \frac{ \la \s_{22} v \ra   + 2 \la \s_{23} v \ra  Q_{23} + \la \s_{33} v \ra  Q_{23}^2}{ (1 + Q_{23})^2}
\eea
where
\bea
 Q_{23}&=&\numeq_3/\numeq_2= \frac{g_3}{g_2} \(1 + \frac{m_3-m_2}{m_2}\)^{3/2} e^{-x_f \frac{m_3-m_2}{m_2}}\nn\\
       &\simeq& \frac{g_3}{g_2}
\eea since $(m_3-m_2)/m_1 \ll 1 /x_f$ and $m_2>m_1$ then $(m_3-m_2)/m_2 \ll 1 /x_f$.
Note  that the values of $\numeq_2$, $\numeq_3$ and $\numeq$ are different with respect to those in the former case since now there are only two species in the thermal bath.
We then find
\bea
 \la \s_\text{eff}^{(2)} v \ra &\simeq& \frac{ \la \s_{22} v \ra  + 2 \frac{g_3}{g_2} \la \s_{23} v \ra  + \(\frac{g_3}{g_2}\)^2 \la \s_{33} v \ra  }{ \(1 + \frac{g_3}{g_2}\)^2} \nn\\
                     &\simeq& \frac{\la \s_\text{MSSM} v \ra  }{\(1 + \frac{g_3}{g_2}\)^2}
\eea
and inserting back this relation into~(\ref{sigma3eff}) we obtain
\be
 \la \s_\text{eff}^{(3)} v \ra \simeq \[ \frac{\(1 + \frac{g_3}{g_2}\) Q_2}{1+ \(1 + \frac{g_3}{g_2}\)  Q_2} \]^2 \la \s_\text{eff}^{(2)} v \ra
\ee
The rescaling factor between the three and two particle species relic density is given by the following relation:
\be
 \( \Omega h^2\)^{(3)} \simeq \[ \frac{1+ \(1 + \frac{g_3}{g_2}\)  Q_2}{\(1 + \frac{g_3}{g_2}\) Q_2} \]^2 \( \Omega h^2\)^{(2)}\label{rescrelic2to3}
\ee In the same way of the $N=2$ case we numerically calculate $(\Omega h^2)^{(2)}$
with the DarkSUSY package, rescale the result according to (\ref{rescrelic2to3})
and compare it with the WMAP \cite{wmapdata} constraints for different mass gaps. We give
the results in figures (\ref{DMwino-chargino}).

\chapter{DarkSUSY}
In this chapter we want to describe how the routines of the DarkSUSY package
permit to numerically calculate the relic density of the LSP and thus of 
the supersymmetric candidate to explain the dark matter abundance. Our
aim is to use this package to calculate the relic density in the anomalous
model.\\
First of all we will describe how the routines of the standard package work,
then we will describe how this routines can be modified to perform not only calculations
in the MSSM but in our extended model. Finally, we will list our results obtained
with the extended package.

\section{Relic density with DarkSUSY package}
DarkSUSYis a program package for supersymmetric dark matter calculations.
It contains many set of routines, each with its purpose; the idea of the package
is that the user can choose what set of routines is more useful for his 
calculation and write accordingly his own main program. We are interested
in dark matter relic density and now we want to explain how the related 
routines work.

\subsection{Model defining routines}
\begin{itemize}
 \item Common blocks \\
In DarkSUSY the common blocks, which contain the variables that are used frequently
in the routines, are stored in the folder \textquotedblleft~/include\textquotedblright, that is in the DarkSUSY 
root directory. There are many files, all with the appropriated extension .h,
that contain the common variables related to a certain use.\\
The variables that are used to define the supersymmetric model are in the file
\textquotedblleft dssusy.h\textquotedblright. This file does not contain directly the variables, but calls
four other files, \textquotedblleft dsge.h\textquotedblright,\textquotedblleft dsio.h\textquotedblright,
\textquotedblleft dsdirver.h\textquotedblright and \textquotedblleft dsmssm.h\textquotedblright.
\textquotedblleft dsge.h\textquotedblright contains general numerical values, as $\pi$, \textquotedblleft dsdirver.h\textquotedblright
contains values used by the program during the installation and are not 
used to perform calculations. Even \textquotedblleft dsio.h\textquotedblright is not related directly
to calculations but sets the parameters that fix the unit in which certain
routines will print their results. \textquotedblleft dsmssm.h\textquotedblright is the most important file
from the point of view of the calculations, because it contains the declarations
and definitions of the numerical parameters that identify the particles in the 
routines, the definition of all the physical constants, such as couplings,
Weinberg angles, mixing matrices, masses and so on.\\
The variables that are used in the calculations of the cross sections are contained in 
\textquotedblleft dsandwcom.h\textquotedblright. The variables that set the parameters that will be used in the 
numerical relic density calculations are defined in \textquotedblleft dsrdcom.h\textquotedblright and 
\textquotedblleft dsrncom.h\textquotedblright.

\item Initializing the program \\
The program is initialized by the routine dsinit, that is located in the folder
\textquotedblleft~/src/ini\textquotedblright. This should be the first routine called in any main program, because 
it gives values to all the physical parameters such as particles degrees of freedom, quantum numbers,
SM known masses, constants. Furthermore, in this routine the program sets the 
constants that decide how it will perform certain calculation. To explain this
point, it is important to known that DarkSUSY contains many possibility to perform
the same calculation, some quicker, others with higher 
precision. The user can 
choose the appropriate method for his needs simply changing the related value in
dsinit. 
Furthermore the user can choose if the program has to consider loop effect and other
details. \\
Last, this routine call other subroutines that set constants used for calculations
different from the relic density. We are not interested in those calculations
so we do not deal with these subroutines.

\item Setting model-dependent parameters \\
There are different ways to define a certain model in DarkSUSY.\\
The first one is to introduce each single parameter of the new model. In this way
the user writes directly the value desired for each parameter. This
input is read by DarkSUSY in the \textquotedblleft~/src/su/dsgive\_model.f\textquotedblright routine if we want
an effective model defined at low-energy scale, or by 
\textquotedblleft~/src/rge/\newline dsgive\_model\_isasugra.f\textquotedblright
if we want an mSUGRA model whose low-energy parameters are calculated by RGE evolution.
In the following we will use the first option since we are not interested in the evolution
from a high energy scale.\\
The second way to define a model in DarkSUSY is to generate it randomly.
The related routine is \textquotedblleft random\_model()\textquotedblright and can be found in the file 
\textquotedblleft~/test/dstest.f\textquotedblright. This routine simply generates random values for all the 
model parameters. The range in which these parameters are generated
can be manually changed by the user opening the routine and changing the desired
minimum and maximum values. \\
The last way to define a model is via an SLHA file. In this way the program reads 
the file using the routine \textquotedblleft~/src/slha/dsSLHAread.f\textquotedblright that sets the parameters
according to the given file.

\item Defining all model-dependent quantities from the inputs\\
There are three possibilities: if the inputs were given (manually or randomly)
at high energy and calculated at low energy by RGE you have to use 
\textquotedblleft~/su/rge/dssusy\_isasugra.f\textquotedblright; if the inputs were given at low energy (manually or randomly)
you have to use \textquotedblleft~/src/su/dssusy.f\textquotedblright and if the inputs were given by SLHA file 
you have to use \textquotedblleft~/src/su/dsprep.f\textquotedblright. This routine does not calculate anything
because in the SLHA file already there are all the values of the model parameters, so it
only transfers the values from the file to the program.
From now and on we concentrate only in the 
low energy case, that is the one in which we are interested.

\item Model setup \\
The routine dssusy contains two subroutines: \textquotedblleft~/src/su/dsmodelsetup.f\textquotedblright,
and \textquotedblleft~/src/su/dsprep.f\textquotedblright. We have already explained the content of the latter, that only
transfer the correct values of the parameters in all the variables used in each subroutine.
Dsmodelsetup is more important, because it calculates these values step by step. This is obtained using
many subroutines:
\begin{description}
 \item[Constant calculation]: the related subroutine is \textquotedblleft~/src/su/dssuconst.f\textquotedblright, 
that calculates the values of the couplings,
of the CKM matrix, of the Yukawa parameters, \dots
 \item[Spectrum mass]: the spectrum is calculated calling \textquotedblleft~/src/su/\newline dsspectrum.f\textquotedblright 
that calls many subroutines:
\textquotedblleft~/src/su/dssfesct.f\textquotedblright for the sfermion masses, 
\textquotedblleft~/src/su/dsneusct.f\textquotedblright for the neutralino masses,\newline
 \textquotedblleft~/src/su/dschasct.f\textquotedblright for the chargino masses, 
\textquotedblleft~/src/su/dshigsct.f\textquotedblright`` for the higgs masses. 
These subroutines also check that the resulting spectrum is physically acceptable and stop
the running of DarkSUSY if not. Finally, the routine checks which particle is the LSP.
\item[Vertex]: the vertices are calculated calling \textquotedblleft~/src/su/dsvertx.f\textquotedblright 
that contains two subroutines:
\textquotedblleft~/src/su/dsvertx1.f\textquotedblright that calculates the vertices related to boson vectors, 
neutralinos and charginos 
while \textquotedblleft~/src/su/dsvertx3.f\textquotedblright calculates vertices related to sfermions.
\item[Higgses and sparticles widths]: these widths are calculated calling respectively
\textquotedblleft~/src/su/dshigwid.f\textquotedblright for the higgses and 
\textquotedblleft~/src/su/\newline dsspwid.f\textquotedblright for the sparticles.

\end{description}

\item Experimental bounds check \\
DarkSUSY checks if a certain model satisfies the experimental bounds on the mass spectrum directly
in the routines that calculates it, as already said, but checks the accelerator constraints on particles
interactions in another routine: \textquotedblleft~/src/ac/dsacbnd.f\textquotedblright``. 
This routine compare the results of the program
with the most recent accelerator data and, if a model is rejected, it 
stops the running of the program and prints a message in which
it explains what experimental constraint was not satisfied.

\end{itemize}

\subsection{Relic density calculation routines}
DarkSUSY computes the DM relic density using a specifical function,
\textquotedblleft~/src\newline /rn/dsrdomega.f\textquotedblright. 
The complete definition is: dsrdomega(omtype, fast, xf, ierr, iwar, nfc).
Omtype and fast are inputs that have to be given by the user to set the type of coannihilation 
that the program has to take into account and the numerical accuracy that 
has to be used to perform the calculations. There are many options: 
\begin{itemize}
 \item omtype = 0 - no coann
 \item 1 - includes all relevant coannihilations (charginos, neutralinos and sleptons)
 \item 2 - includes only coannihilations betweeen charginos and neutralinos
 \item 3 - includes only coannihilations between sfermions and the lightest neutralino
\end{itemize}

\begin{itemize}
 \item  fast =   0 - standard accurate calculation (accuracy better than 1\%)
 \item  1 - fast calculation: (recommended unless extreme accuracy is needed)
 \item  2 - faster and less precise method, i.e. expands the annihilation cross section in x
\end{itemize}
The output are dsrdomega, that is $\Omega h^2$, xf that is
$x_f$, ierr and iwar that are the identificator of errors or warnings during the calculation,
and nfc that counts the number of function calls to the effective annihilation cross-section.\\
This function performs all the numerical calculation needed to numerically solve the Boltzmann 
equation, that we have described in detail in section \ref{calcmet}. This calculation is divided 
in many routines, that we are going to describe.\\
Prior to call the routines, dsrdomega calculates the number of coannihilating particles, 
according to omtype. The default ratio between the mass of each particle and that of the LSP 
to be added to the coannihilating particles is $1.5$, but can be modified by hand by the user.
When the routine has ended the scan over the particles and determined all the coannihilating
ones, there is the call of \textquotedblleft~/src/rd/dsrdens.f\textquotedblright`` if fast is equal to 0 or 1, 
\textquotedblleft~src/rd/dsrdquad.f\textquotedblright``
if fast is equal to 2.\\
The latter simply makes an expansion of the effective annihilation rate \ref{weff1}, gaining speed 
but losing precision. Instead, the subroutine dsrdens implements numerical methods to
solve directly the differential equation, with the only approximations given by the numerical precision 
of the method used. It does not directly perform the calculation: first it sets the initial value of the 
variables that are going to be used, as the temperature and the momentum, then calls the subroutine. 
\textquotedblleft~/src/rd/dsrdtab.f\textquotedblright which creates a table in which each value of $x$
 is associated to the related 
value of the effective annihilation rate; after that it calls the subroutine which calculates the limits
of the numerical integration, \textquotedblleft~/src/rd/dsrdthlim.f\textquotedblright``; 
last, it calls the subroutine that perform
the numerical calculation:\textquotedblleft~/src/rd/dsrdeqn.f\textquotedblright.\\
This subroutine numerically solves the Boltzmann equation in the form (\ref{Boltz1})
using a trapezoidal method with adaptive stepsize and termination. Its input is
the effective annihilation cross section, provided by the function \textquotedblleft~/src/rd/dsrdwintp.f\textquotedblright.
These function interpolates the discrete values given by dsrdtab to have all the points needed by the
numerical analysis.\\
It remains to analyze the set of subroutines that calculate the effective annihilation rate, used by dsrdtab.
The values tabulated in this routine are calculated in \textquotedblleft~/src/an/dsanwx.f\textquotedblright, 
that, as usual, does not 
calculate the effective annihilation rate, but calls the subroutines that makes the calculation
and organizes the results. In this case the called subroutine is the first of a set of nested subroutines, each
one performing a certain step of the calculation. The complete list, in order of call, is:
\begin{description}
 \item[dgadap]: its pattern is \textquotedblleft~/src/xcern/dgadap.f\textquotedblright.
It is called by dsanwx to perform the angular integration in the $\theta$ variable. 
It uses an adaptive gaussian method to perform the integration.
\item[dsandwdcosy and dsandwdcoss]: their patterns are: \newline 
\textquotedblleft~/src/an/dsandwcosy.f\textquotedblright 
and \textquotedblleft~/src/an/dsandwdcoss.f\textquotedblright.
Both are used as integrand function in dgadap. They differ only in the fact that the integration variable
in dsandwdcosy is changed from $\theta$ to $y=\frac{1}{m_{LSP}^2+2*p_{cm}^2(1-cos(\theta))}$ to avoid poles
in $cos(\theta)=\pm1$. Both multiply the value of the effective annihilation cross-section by $10^15$, 
a trick useful for the numerical calculations.
\item[dsandwdcos]: its pattern is \textquotedblleft~/src/an/dsandwdcos.f\textquotedblright. 
It is called by dsandwcosy and dsandwdcoss.
It gives the numerical value of the effective annihilation cross-section, summing over all coannihilation
processes cross-sections.
\item[dsandwdcosij]: its pattern is \textquotedblleft~/src/an/dsandwdcos.f\textquotedblright 
(it is a subroutine written in the same file of
dsandwdcos). It is called by dsandwdcos. For each couple of coannihilating particles this function determines
their types (neutralino, sfermion, chargino) and calls the proper subroutine to calculate their cross-section.
\item[dsandwcosnn]: its pattern is \textquotedblleft ~/src/an/dsandwdcosnn.f\textquotedblright.It is called
by dsandwdcosij and calculates the cross-section of neutralino-neutralino annihilation.
\item[dsandwdcoscn]: its pattern is \textquotedblleft ~/src/an/dsandwdcoscn.f\textquotedblright.It is called
by dsandwdcosij and calculates the cross-section of neutralino-chargino annihilation.
\item[dsandwdcoscc]: its pattern is \textquotedblleft ~/src/an/dsandwdcoscc.f\textquotedblright.It is called
by dsandwdcosij and calculates the cross-section of chargino-chargino annihilation.
\item[dsandwdcossfsf]: its pattern is \textquotedblleft ~/src/an/dsandwdcossfsf.f\textquotedblright.It is called
by dsandwdcosij and calculates the cross-section of sfermion-sfermion annihilation.
\item[dsandwdcossfchi]: its pattern is \textquotedblleft ~/src/an/dsandwdcossfchi.f\textquotedblright.It is called
by dsandwdcosij and calculates the cross-section of sfermion-neutralino (or chargino) annihilation.
\item[anstu folder routines]: the pattern is \textquotedblleft~/src/anstu\textquotedblright.
All the routines that calculate the amplitudes of a single physical 
process in a certain channel are stored in this folder. They are called by the various 
routine listed above.
\end{description}

\section{Modification to DarkSUSY}
We want to modify DarkSUSY routines to have the possibility to perform calculations
in the extended model described in chapter \ref{miaussm}. Our scope is to made all the changes
in such a way that permits to perform calculation in the MSSM if we desire it. We will start describing 
how we modify the subroutines, then we will describe some examples of main programs,
using the main programs of the original version of DarkSUSY modified in a convenient
way. During the description we will try to use the same order of the previous section, 
to permit easy comparisons.
\subsection{Model defining routines}
\begin{itemize}
 \item Common blocks \\
As we have seen, in DarkSUSY the common blocks that contain the variables used in the 
model building are defined in \textquotedblleft ~/include/dssusy.h \textquotedblright.
This file contains the variables in four subfiles. For simplicity we added another file,
\textquotedblleft dsanom.h \textquotedblright, called together with the other four.
In this file we have stored all the new common variables. This is the list of all the variables we add
(we write the name used in the program), and their definition.

\begin{enumerate}
 \item Masses \\
mpsiano and m0ano are, respectively, the mass of the st\"{u}ckelino and of the primeino.
ampsiano and am0ano are two support variables that permit to transfer the value of the
related variables from the input to the core of the program. mzp is the mass of the $Z'$.
\item Couplings \\
g0ano is the anomalous coupling $g_0$. gmix(3,3) is the $3 \times 3$ matrix that gives the mixing
among neutral vectors, (\ref{neumass1}). cznn(6,6), cann(6,6) and cpnn(6,6) are the couplings of 
the neutralino-neutralino-boson vertices (\ref{nnbrules}).
\item Anomaly related variables \\
b3ano, b20ano, b21ano, b22ano, b24ano, are respectively the $b_2^{(i)}$ of the (\ref{anomalia});
qhuano, qqano, qlano, quano, qdano, qeano, qhdano are the anomalous charges defined in (\ref{QTable}).
\item Other \\
qfnew(12),afnew(12), vfnew(12),  qvanonew(12), qaanonew(12) are vectors that contains, respectively, 
the electro-magnetic quantum number $q_f$, the vector and axial quantum numbers $g_V$,$g_A$ (\ref{gvgaano})
and the vector and axial quantum numbers for the anomalous $U(1)$, $Q_V$,$Q_A$ (\ref{qvqaano}) 
for all fermions. Dsansumfanof1,dsansfanofvvv1 will contain, respectively, the coannihilation
contributions from (\ref{coanniff}) and (\ref{coannivv}).
\end{enumerate}
\item Initializing the program \\
We slightly modify the routine dsinit: we added two neutralinos, the $Z'$ and their 
related degrees of freedom to the particles considered by the program. Then we fix the values
of some general constant that we have introduced i.e. $g0ano=0.1$, $mzp=1~TeV$, $b3ano=mzp/4 g0ano$.
\item Setting model-dependent parameters \\
In the extended anomalous model we have 5 new free parameter: $M_S~(mpsiano),~M_0~(m0ano),~Q_{H_u}
~(qhuano),~Q_L~(qlano),~Q_Q~(qqano)$. The value of these parameters has to be given for each model
we will generate. So we have modified the input system of DarkSUSY.\\
To give values to the model parameters by hands we have modified the main program to request
an input for each of these 5 new parameter. Furthermore, we have modified dsgive\_model to tranfer 
these inputs to the subroutines. Also we have added in this routine the constraints (\ref{Qconstraints})
and the definitions (\ref{anomalia}).\\
To define the model parameters randomly we have modified the subroutine random\_model(), adding 
the 5 new parameters to those that have to be randomized. Because this subroutine is written in the 
same file of the related main program, we have introduced directly in the main program
the constraints (\ref{Qconstraints}) and the definitions (\ref{anomalia}).\\
We are not interested in using SLHA files to introduce the model parameters, so we do not have
modified that part of the program.
\item Defining all model-dependent quantities from the inputs\\
Being our model an effective model, we skip on the routines that perform RGE evolution
and on the routines that use SLHA files to define the model dependent quantities. Thus 
we concentrate only on dssusy.
\item Model setup \\
We have modified the subroutines called by dsmodelsetup in order to describe the extended
model:
\begin{description}
 \item[Spectrum mass]: first of all we modify the command that checks if the LSP is a neutralino in
such a way that it considers all six neutralinos and not only the four of the MSSM, then we add the definition
of axinofrac as the fraction of anomalous neutralinos in the LSP. We have also modified the subroutines
dsneusct and dshigsct, that calculate the masses of neutralinos and higgses. In dsneusct we have 
rewritten the mass matrix, substituting (\ref{neumass}) with (\ref{massmatrix}) and changing all the cycles
lenghts from 4 to 6 to take into account the two extra neutralinos.\\
Modifying the higgs masses calculation is not straightforward. The reason is that the latest version of
DarkSUSY does not calculate these masses with inner routines but using the external program feynhiggs.
We want to modify tree-level values for higgs masses, implementing (\ref{eq:m2Hpm}). To achieve this goal,
we have to modify a feynhiggs subroutine. DarkSUSY stores the external programs that it utilizes in the folder
\textquotedblleft ~/contrib \textquotedblright. In this folder there are many version of feynhiggs.
The one used by DarkSUSY is 2.6.3. The subroutine in which there is the tree-level higgses mass
calculation is \textquotedblleft ~/contrib/FeynHiggs-2.6.3/src/Main/Para.F \textquotedblright. 
Here we have changed the formulas according to (\ref{eq:m2Hpm}).
\item[Vertex]: we had to change the vertices according to the various equations of section \ref{mixinganoformulas}.
The main routine that implements the vertices in DarkSUSY, as we have already said, is dsvertx, that
contains dsvertx1 and dsvertx3. These are the files that we have to modificate.\\
We start from dsvertx1. First we introduce the definitions of the variables related 
to the new vertices: so we give the definition of gmix according to (\ref{neumass1}),
the definition of vfnew, afnew, qvanonew and qaanonew according respectively to the 
SM definitions \hypertarget{HalzMart}{\cite{HalzMart}}
for the first two, and to (\ref{qvqaano}) for the others two. After that it starts the vertices
definition part of the subroutine. First of all there are the MSSM vertices, already defined in 
DarkSUSY but that have to be modified to describe our model. So we modify the higgs-higgs-vector
boson vertices according to (\ref{hhbcoup}) and the fermion-fermion-vector boson vertices according to
(\ref{z0ff}) and (\ref{z'ff}). Then we extend the length of the cycles in which there is the
definition of the vertices in which appear neutralinos from 4 to 6, to take count of the two extra 
neutralinos of our model.
After that we added the new vertices: $\gamma$-neutralino-neutralino, $Z_0$-neutralino-
neutralino, $Z'$-neutralino-neutralino according to (\ref{caczcp}).\\
In dsvertx3 we have modified the sfermion-sfermion-vector boson vertices according to 
(\ref{z0ff}) and (\ref{z'ff}) and the sfermion-sfermion-higgs vertices according to 
(\ref{gvgaano}).
 
\end{description}

\end{itemize}

\subsection{Relic density calculation routines}
We have not changed the subroutines that perform the numerical integration of the 
Boltzmann equation, as the equation obviously does not change introducing the
anomalous $U(1)$. As we have seen, the changes are the presence of new interactions
and of new particles, as well as some modifications to the MSSM interactions.
DarkSUSY already contains all MSSM interactions, so their modifications, listed in the
previous section, are already taken in account, while we have to implement the new interactions
(\ref{coanniff}) and (\ref{coannivv}).\\
We want to keep separated the routines that we have written to add the contributions to the 
cross section that were not present in the MSSM and consequently in the DarkSUSY package. So 
we have created and added to the makefile the folder 
\textquotedblleft ~/src/anano \textquotedblright, where we have stored the new subroutines
that we are going to describe.\\
\begin{itemize}
 \item dsanscalarproduct \\
This subroutine contains the definitions of the kinematical variables (\ref{scalarproducts}).
It calls dsankinvar1, a modified version of dsankinvar, a DarkSUSY subroutine that contains 
the definitions of the kinematical variables already used by the program.
\item dsankinvar1 \\
It is identical to dsankinvar except for the definition of the mass of the exchanged particles,
that we skipped because we have not common vertices for all the vector boson, as in the
original program, so we have not the need to distinguish among them.
\item dsansumfanof \\
This subroutine resets the value of dsankinvar1 \footnote{It is necessary because
the kinematics changes when we change the studied process}, and it sums the contribution of 
(\ref{coanniff}) over all fermions of the SM.
\item dsansfanofvff \\
This subroutine calculates (\ref{coanniff}) for every couple of neutralinos
and SM fermions.
\item dsansfanofvvv \\
This subroutine calculates (\ref{coannivv}) for every couple of neutralinos.
\end{itemize}
The only change we made in the original DarkSUSY relic densiy routines are in dsandwdcosnn,
where we add the call of dsansumfanof and dsansfanofvvv and a final sum of a $10^{-17}$ that
is a rough estimate of the contribution of the annihilation of an LSP mostly composed by
st\"{u}ckelino. This is needed to avoid the possibility $W_eff=0$ that causes the failure
of the numerical algorithm to solve the Boltzmann equation. Because this contribution is an estimate,
it makes the result of a calculation in which there is only the annihilation of a 
st\"{u}ckelino-type LSP absolutely imprecise, but as we have seen in section \ref{stuanni},
this contribution cannot ever explain WMAP data, while if we have coannihilations this
contribution is negligible, so this approximation does not modify the physical results.

\subsection{Main programs}
We have written two main programs. The first (anomssm3) has the scope to study 
random models, while the second (dsmain) permits
to study a specific model.\\
Because anomssm3 has the purpose to study a complete randomly generated
model we have modified the DarkSUSY subroutine random\_model adding the 
five parameters from the anomalous extensions to be generated along
the seven parameters from MSSM already implemented. The new parameters 
are $Q_Q$, $Q_L$, $Q_{H_u}$, $M_S$ and $M_0$, implemented respectively
as qqano, qlano, qhuano, mpsiano, m0ano. We add an user interface to decide
if we want to perform simulation in the MSSM or in the miaUSSM, if we
want a bino-type or wino-type MSSM LSSP, to introduce
manually the range in which each parameter have to be generated and the number
of models that will be generated. Then we add a command to 
write all the twelve parameters in a file that we had already open in the
main program. In this way we have stored the input
parameters of each model.\\
Because the models will be randomly generated, we have not control on the
mass gap between LSP and NLSP. It will be a result that depends on many
of the model parameters. However we want to have the possibility to study the results
with a \textquotedblleft controlled\textquotedblright mass gap. 
So we add checks after the relic density
calculations that permits to store the results in different files if the mass gap
is in the following intervals: $0-5\%$,$5-10\%$,$10-15\%$, $15-20\%$,$>20\%$
plus a file in which are stored all the results, that can be useful
if we want an ensemble vision.\\
The calculation part of this main program is very simple: first we 
randomly generate the parameters of the model with the subroutine 
random\_model, then we build the model with the subroutines dsmodelsetup
and dssusy, then we check the
accelerator bounds and the constraints on the neutral mixing
(subroutine acbounds),
then we perform the relic density calculation (with dsrdomega) only if the LSP 
satisfies certain conditions (for example we can request that
it is a combination of st\"{u}ckelino and primeino).
After that the relic density results are stored (with other
important quantities calculated, as $x_f$, the spectrum,...)
in the file related to the mass gap. We store separately the
models that satisfy WMAP from those that doesn't.\\
The second program, a modified version of dsmain, a program already present
in the current DarkSUSY distribution, permits to the user to 
enter by hand the values of the twelve parameters that define the 
model. This is made using explicit write and read commands
for each quantity. After that we call the already modified
subroutines dsgive\_model and dssusy, then we call the same 
subroutines of the previous case, acbounds and dsrdomega.
Because this program deals with one user-defined model it
is not necessary to open files for many possible mass gaps,
because we will have only one result.\\
Note that, if in anomssm3 we choose the lower and the higher extremes of 
the parameters range to be equal and if we set the number of models
to 1, the program works in the same way of dsmain. We can use this
property to perform simulations in which we keep constant some parameters
to study the dependence of the model from the others.
\section{Simulations and results}
As for the study of the no mixing case, we perform our simulations
for the different situations of a bino-higgsino NLSP and a wino-higgsino NLSP.
In both cases the LSP will be a random mix of st\"{u}ckelino and primeino.\\
We have started performing an ample
scan (number of model $>10^5$) in a wide area of parameter space, without
any constraint. We have utilized these results to find the regions of parameter 
space in which there is a model with relic density that satisfies or is near to the 
WMAP \cite{wmapdata} results.\\
Tipically, in each region we have found, the mass gap between LSP and NLSP 
of the models that satisfy WMAP data is almost constant, but from region to region it
can change from $1\div2\%$ up to $25\%$.\\
To perform a more accurate study, we have chosen sample region to be explored more carefully 
for each of this mass gap ranges: $0\div5\%$, $5\div10\%$, $10\div15\%$, $15\div20\%$, $>20$. 
Then we have used our modified DarkSUSY to scan these regions keeping all but two parameters constant.\\
In this way we have obtained the dependence of the relic density on each couple of
parameters in regions in which WMAP data are satisfied. Because now we are interested in 
well defined regions of parameters space we have used less ($\sim 10^4$ instead of $\sim 10^5$)
number of models. 

\subsection{General results \label{genres}}
\begin{figure}
 \includegraphics[scale=0.65]{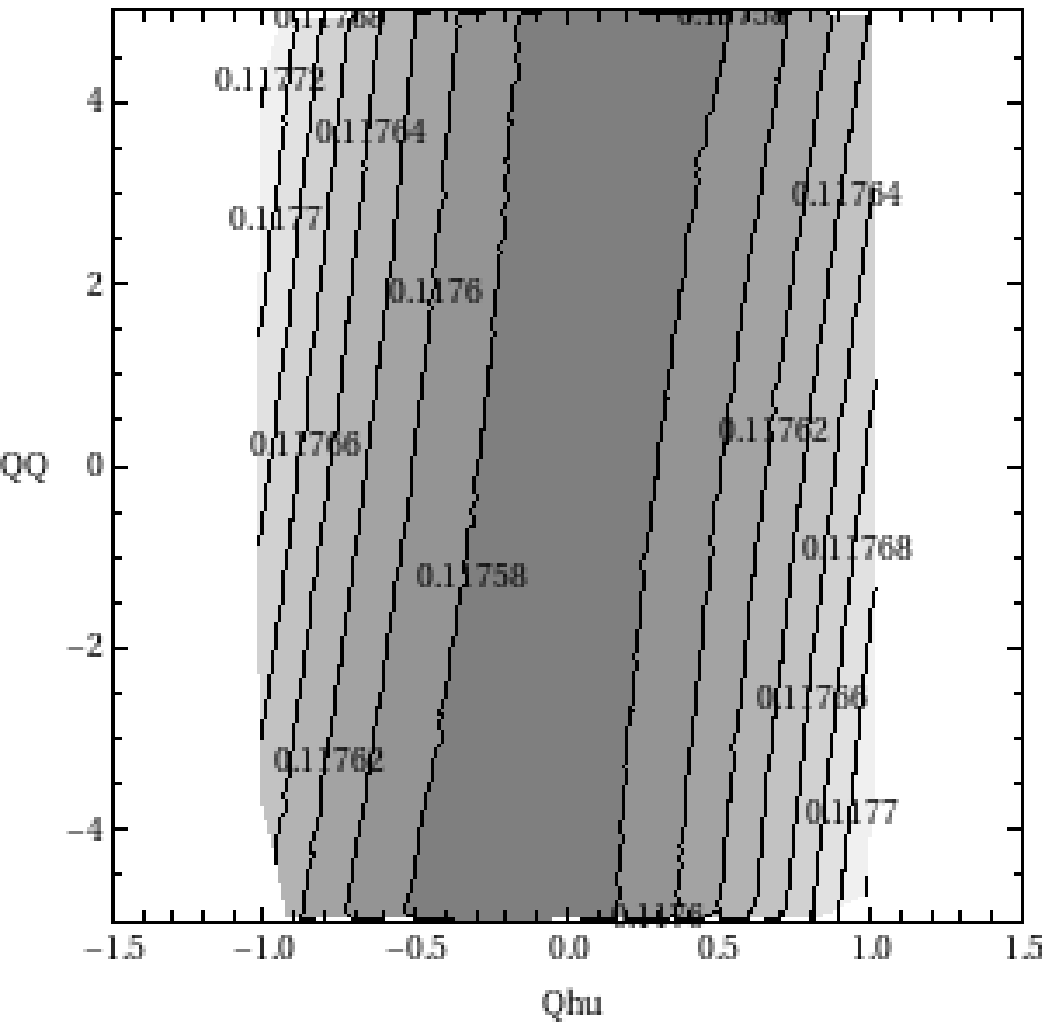}
 \includegraphics[scale=0.65]{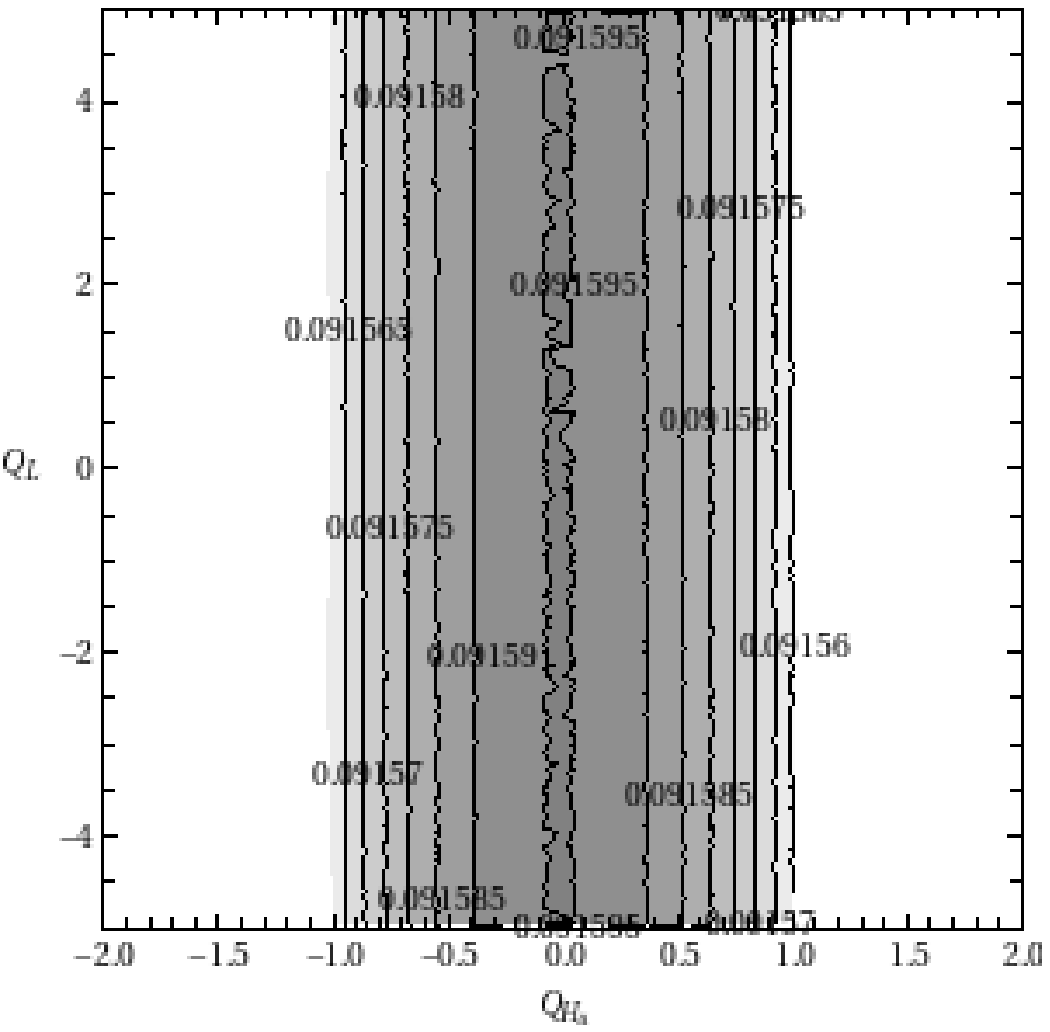}
\caption{Plot of relic density versus $Q_{H_u}$ and $Q_Q$ (left, bino-higgsino NLSP) and versus 
$Q_{H_u}$ and $Q_L$ (right, wino-higgsino NLSP)}
\label{qhuimage}
\end{figure}

In this section we want to list some results that are valid for both types
of NLSP.\\ First of all, given the constraints on the 
neutral mixing described in \cite{Langacker:2008yv}, we want to found
their effects on the model parameters. The results are showed in figure \ref{qhuimage}, for
two sample models.
 These plots show the general result:
\be Constraints~on~the~mixing\Rightarrow -1\lesssim Q_{H_u} \lesssim 1 \ee
As we can see from the figure \ref{qhuimage}, the contribution of the anomalous charges
to the relic density is extremely small.

\subsection{Bino-higgsino NLSP}
\begin{figure}[h!]
\includegraphics[scale=0.65]{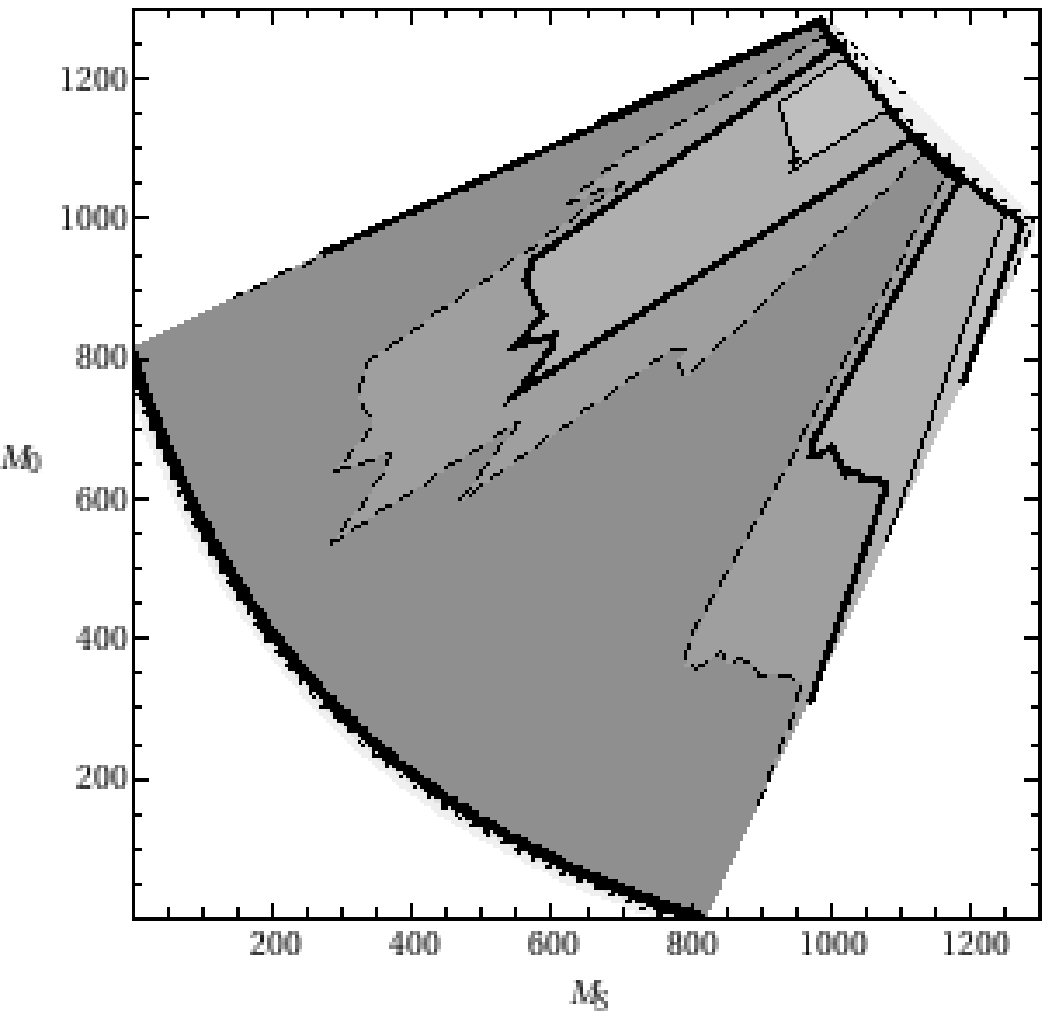}
\includegraphics[scale=0.5]{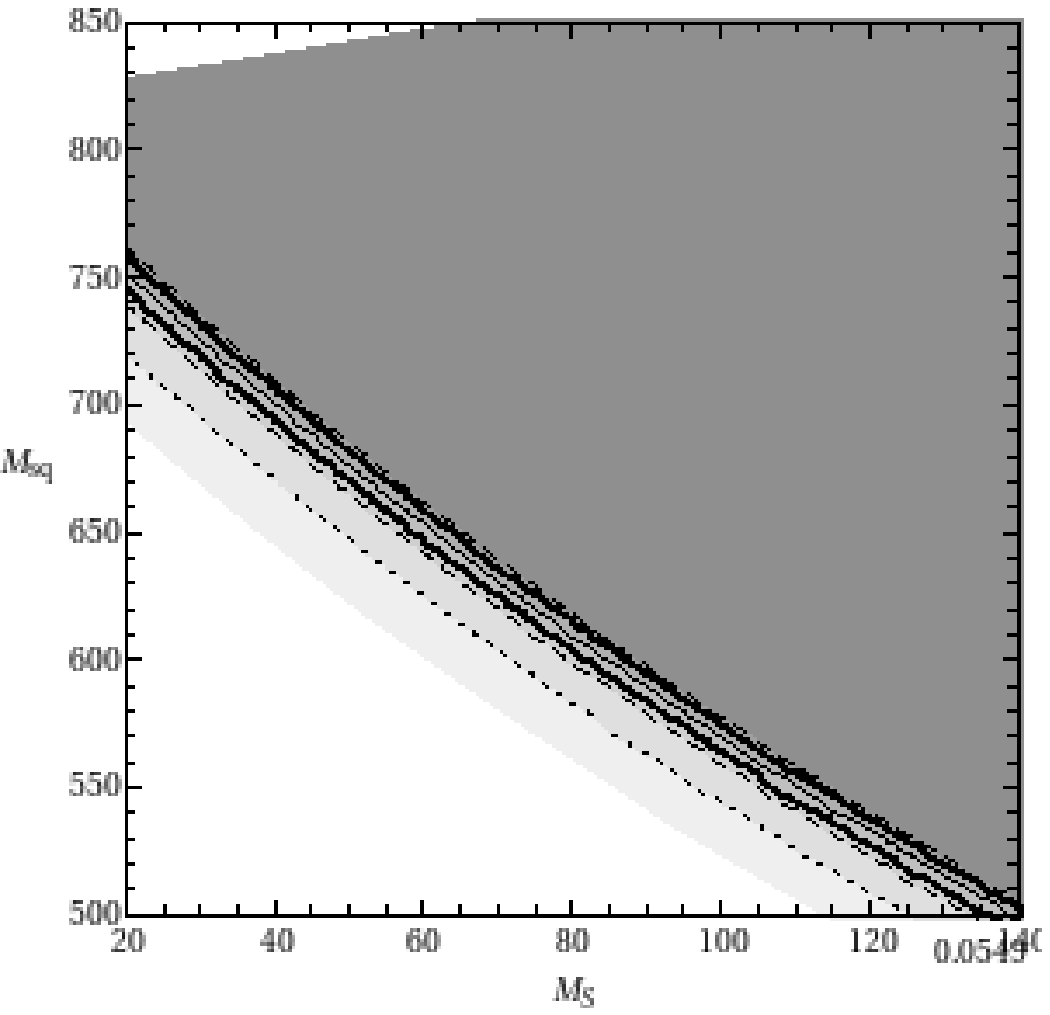}
\includegraphics[scale=0.8]{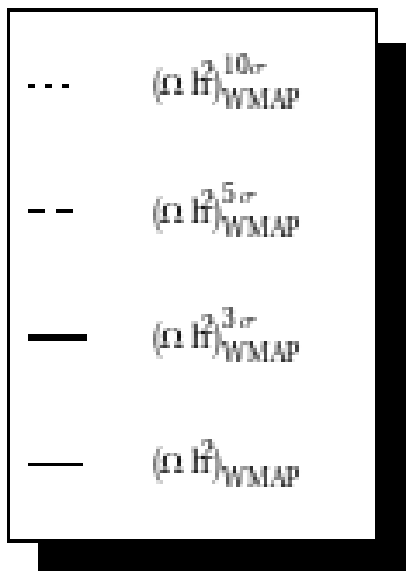}
\includegraphics[scale=0.65]{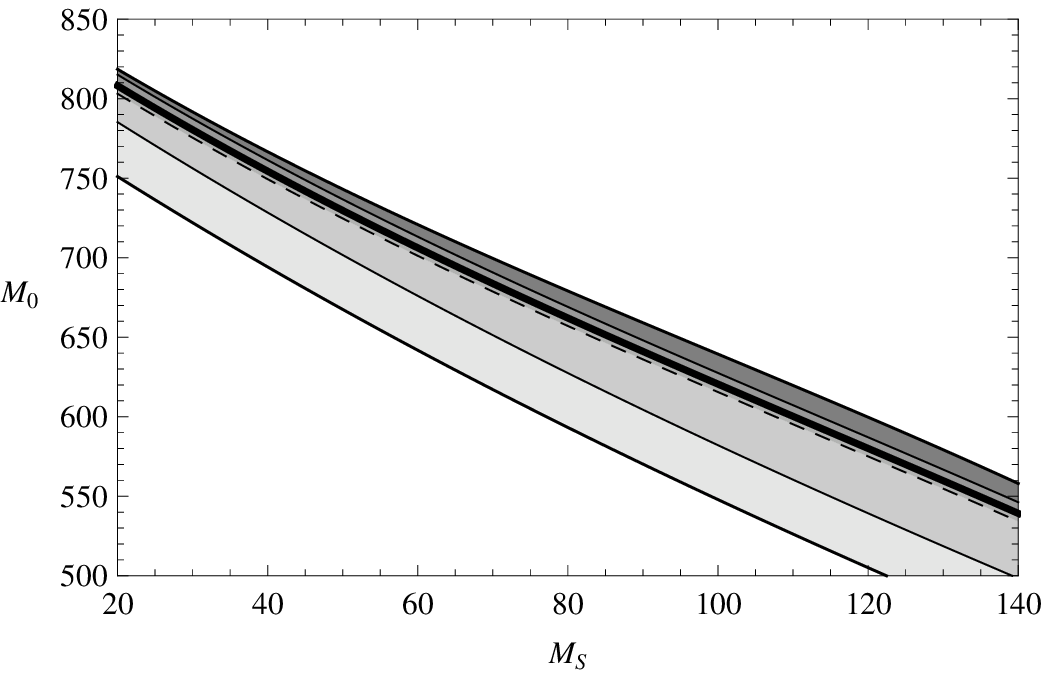}
\caption{Plot of the relic density of the LSP vs the st\"{u}ckelino
and primeino masses in the case of bino NLSP. The first plot shows the case of mass gap 5 $\%$. The second plot
is a zoom of the first in the region between 20-140 GeV for st\"{u}ckelino mass and 500-850 GeV
for primeino mass. The third plot shows the case of mass gap 10 $\%$ in the same region of the second plot.
All the masses are expressed in GeV}
\label{ms-m0}
\end{figure}

\begin{figure}
 \includegraphics[scale=0.65]{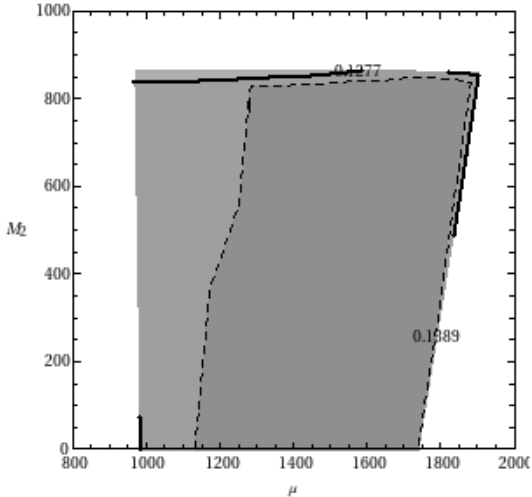}
\caption{Plot of the relic density vs $M_2$ and $\m$, MSSM parameters, for mass gap $10\%$ in
the case of bino NLSP. 
The masses are expressed in GeV}
\label{m2-mu}
\end{figure}

If the NLSP is mostly a bino-higgsino we have a two particle coannihilation.\\
We want to study the relic density with respect to the parameters that 
mostly contribute to the LSP mass ($M_S$ and $M_0$), with respect to the
parameters that mostly contribute to the NLSP mass ($M_2$ and $\m$) and also
with respect to the other MSSM7 parameters. The anomalous charges $Q_{H_u},~Q_L,~Q_Q$,
as we have seen in the previous section, does not contribute in a significant way to the
relic density.\\
We studied the relic density with respect to $M_S$ and $M_0$ for 
different values of the mass gap. The results are shown in fig \ref{ms-m0}.
 As we can see, changing the mass gap from $5\%$ to $10\%$ greatly reduces the
region in which WMAP \cite{wmapdata} data are satisfied. Only in the region showed in the second image the relic
density satisfies WMAP data for both cases.\\
We have also studied the relic density with respect to $M_2$ and $\m$. The results are 
showed in the figure \ref{m2-mu}.
 The image is roughly symmetric with respect to the line $\m=0$, also showing that 
near this line we do not satisfy WMAP data, while in the $TeV$ range for $\m$ we satisfy WMAP data 
in each point that is not forbidden by experimental constraints. Furthermore, we can see
that the region of low $M_2$ is physically unacceptable.\\
The figure obviously refers to a sample model, but the fact that the relic density behaviour is
symmetric with respect to $\m=0$ and that for low $M_2$ there are violation of experimental
constraints is a general result.

\subsection{Wino-higgsino NLSP}
\begin{figure}
 \includegraphics[scale=0.65]{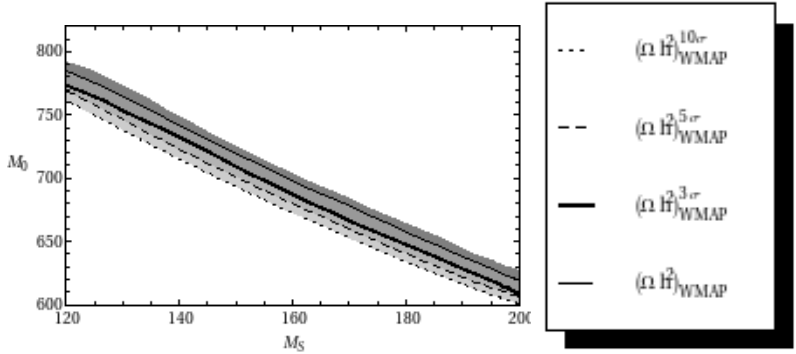}
 \includegraphics[scale=0.8]{legend.eps}
\caption{Plot of the relic density vs $M_S$ and $M_0$ for mass gap $10\%$ in the case of wino
NLSP. 
The masses are expressed in GeV}
\label{ms-m0_w}
\end{figure}

\begin{figure}
 \includegraphics[scale=0.65]{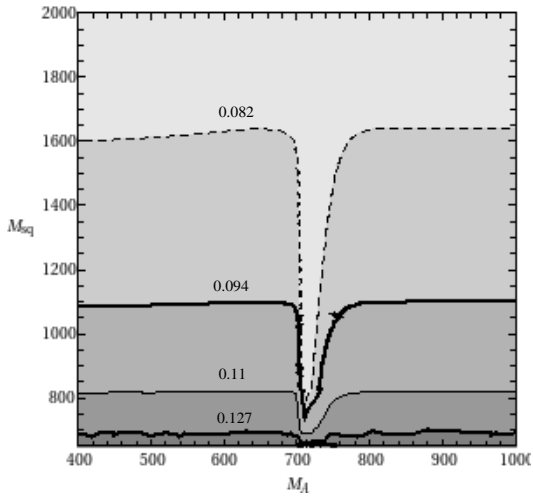}
\caption{Plot of the relic density vs $M_{A_0}$ and $m_0$, MSSM parameters, for mass gap $10\%$
in the case of wino NLSP. 
The masses are expressed in GeV}
\label{mA-m0_w}
\end{figure}

If the NLSP is mostly a wino-higgsino we have a three particle coannihilation, because the wino
is almost degenerate in mass with the lowest chargino. Again in this case the anomalous charges
 $Q_{H_u},~Q_L,~Q_Q$ do not contribute significantly to the the relic density. So we performed 
the same study described in the previous section.\\
The results for the dependence of the relic density on $M_S$ and $M_0$ are very similar to those
of the previous section, so we will only show one of them as an example in fig. \ref{ms-m0_w}.
 Furthermore can be very interesting to show the 
presence of a phenomenon called funnel region. It is not new: it is already predicted in the MSSM.
Substantially it consists in a sort of resonance in the case in which the relation $2 M_{LSP} \sim M_{A_0}$
holds. So there is a peak of the relic density versus the mass of parity odd higgs boson
$A_0$. In the MiAUSSM we have a different LSP, the st\"{u}ckelino-primeino mix, but the NLSP is
the old MSSM LSP, whose interactions contribute greatly to the relic density so we expect the 
funnel region phenomenon to appear in the same way of the MSSM. 
As it is showed in figure \ref{mA-m0_w}, the funnel region appears as expected.

\begin{figure}
 \includegraphics[scale=0.65]{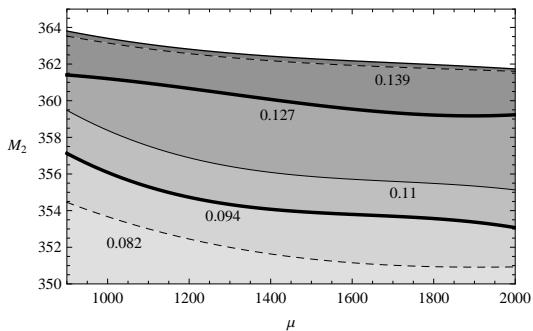}
\caption{Plot of the relic density vs $\m$ and $M_2$, MSSM parameters, for mass gap $10\%$
in the case of wino NLSP. 
The masses are expressed in GeV}
\label{mu-m2_w}
\end{figure}

 We can also show for completeness our result for the dependence
of the relic density on $\m$ and $M_2$ (fig. \ref{mu-m2_w}).

\chapter{Asymmetry} 
\section{Introduction}
\subsection{PDFs}
The study of the asymmetry at the LHC implies dealing with hadronic processes. To calculate cross sections
that involve hadrons we need the Partonic Distribution Functions (PDFs) \hypertarget{Greiner}{\cite{Greiner}}. 
These functions give the distribution of momenta of the elementary constituents of a given hadron in the parton model. 
In this model the elementary constituents of the hadrons, called partons, are point-like and almost free particles.
Because QCD is an asymptotic free theory \hypertarget{Greiner}{\cite{Greiner}} this assumption is accurate for
high momentum transfer (that is the so called Deep Inelastic Scattering Region).
So this model
holds for momentum exchanges much larger than the mass of the particles. We are surely in this
situation at the LHC so we will use this model in the following sections. Parton is a generic
term that indicates the constituents of the hadrons. So a parton can be a quark or a gluon.\\
To introduce the PDFs let's consider a generic process:
\be
A+B \rightarrow c+d+X \label{hadcs}
\ee where $A$ and $B$ are the initial hadrons, $c$ and $d$ are the final particles in which we are
interested and $X$ represents the hadronic remnants. The elementary subprocess associated to the hadronic
process is:
\be
a+b\rightarrow c+d \label{elemcs}
\ee where $a$ and $b$ are elementary constituents of the hadrons. This cross section can be 
calculated with the usual Feynman rules derived from the Lagrangian. We need the PDFs to
obtain the cross section (\ref{hadcs}) from the elementary one. Infact (\ref{hadcs}) is given
by the convolution of (\ref{elemcs}) with the PDFs:
\bea
    \label{eq:convoluzione}
&&d\sigma(A B \to c d X) = \\
&&\sum_{a,b}\int_{0}^{1}dx_1\int_{0}^{1}dx_2[f_{a/A}(x_1)f_{b/B}(x_2)+f_{b/A}(x_1)f_{a/B}(x_2)]
d\hat{\sigma}(a b \to c d) \nn
\eea where:
\begin{itemize}
 \item $\sigma(A B \to c d X)$ is the cross section of the hadronic process
 \item $\hat{\sigma}(a b \to c d)$ is the cross section of the elementary process
 \item $f_{(a/b)/(A,B)}(x_{(1,2)},Q^2)$ is the PDF that gives the probability to have the 
parton $a/b$ in the hadron $A/B$ with momentum fraction $x_{1/2}$ and with momentum transfer 
$Q^2$ (this is sometimes omitted for simplicity). 
\end{itemize} Being $x_i$ a fraction of the total momentum of the hadron that contains the 
related parton, we have necessarily: $0\le x_i\le 1$.\\
The PDFs are obtained using experimental data, because the present numerical calculations
in QCD are not able to give satisfactory results. So there exist many sets of PDF, given
by different research groups worldwide. For the calculations that we are going to perform
in the next sections we will use the PDFs defined in \hypertarget{Alekhin:2006zm}{\cite{Alekhin:2006zm}}
and \cite{Watt:2012tq}.

\subsection{Asymmetry definition}
\begin{figure}
\centering
 \includegraphics[scale=0.5]{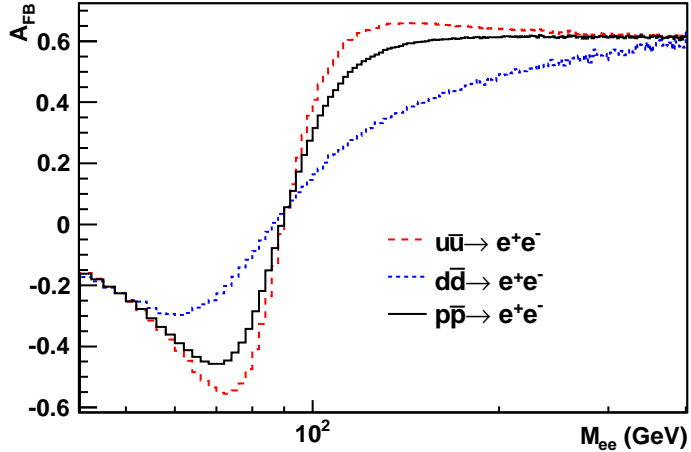}
\caption{The SM LO $A_{FB}$ prediction as a function of the dielectron invariant mass for
$u\bar{u} \rightarrow e^+e^-$, $d\bar{d} \rightarrow e^+e^-$, and
$p\bar{p} \rightarrow e^+e^-$}
\label{SMasym}
\end{figure}

The asymmetry \hypertarget{Langacker:1984dc}{\cite{Langacker:1984dc}} is a quantity that, given a 
certain particle interaction, measures the difference 
between the number of final particles emitted in one direction (\textquotedblleft forward\textquotedblleft) 
and in the opposite direction (\textquotedblleft backward\textquotedblright). The direction is usually 
chosen with respect to that of the initial particles: we define as forward the semisphere identified by 
the beam direction, while backward is identified by the opposite direction. The basic equation is
\be 
A_{FB}=\frac{F-B}{F+B}
\ee where
\bea
&&F= \#~of~particles~going~forward   \\
&&B= \#~of~particles~going~backward
\eea Explicitly:
 \be
 A_{FB}=\frac{\int dx_1dx_2\sum_q f_q(x_1)f_{\bar{q}}(x_2)(F-B)}{\int dx_1dx_2\sum_q f_q(x_1)f_{\bar{q}}(x_2)(F+B)}
\ee with:
 \be
 F=\int_0^1 d\cos\theta \frac{d\sigma(\cos\theta, s)}{d\cos\theta ds}~~~~~~~ 
B=\int_{-1}^0 d\cos\theta \frac{d\sigma(\cos\theta, s)}{d\cos\theta ds}~~~~~~~
\label{FandB}
 \ee In these formulas $x_1$ and $x_2$ are the standard partonic variables, so the integral over them goes
from $0$ to $1$. $f_q(x_i)$ are the partonic distribution functions (PDFs) of the quark $q$ with respect to the variable
$x_i$. $\cos\theta$ is the angle of emission of the electron of the final state in the CM frame.
$ \frac{d\sigma(\cos\theta, s)}{d\cos\theta ds}$ is the differential cross section in the CM frame.\\
The asymmetry has been calculated and measured for the SM with good precision. 
A sample of the results obtained \hypertarget{Abazov:2011ws}{\cite{Abazov:2011ws}} are in figure \ref{SMasym}.
 It has been used for many
phenomenological purposes, such as giving constraints on the mixing angles of SM,
measuring the couplings, setting constraints on the new physics beyond the SM.\\ At the LHC
it can be used in the same ways as well to discover new physics. In the following we aim
to calculate the asymmetry at the LHC to impose constraints on the free charges of the
MiAUSSM and to give a method that, if an asymmetry different from the MSSM predicted one will be
measured, permits to check the consistency of our and other models that extend the MSSM.

 \section{Four Asymmetries at the LHC}

\begin{figure}
\centering
 \includegraphics[scale=0.6]{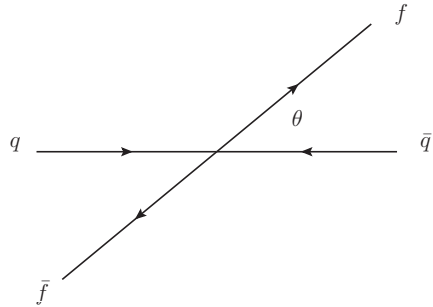}
\caption{Representation of the process $q\bar{q}\rightarrow f\bar{f}$ in the CM frame, with explicit
indication of the angle that we have defined as $\theta$}
\label{FBCM}
\end{figure}

\begin{figure}
\centering
 \includegraphics[scale=0.4]{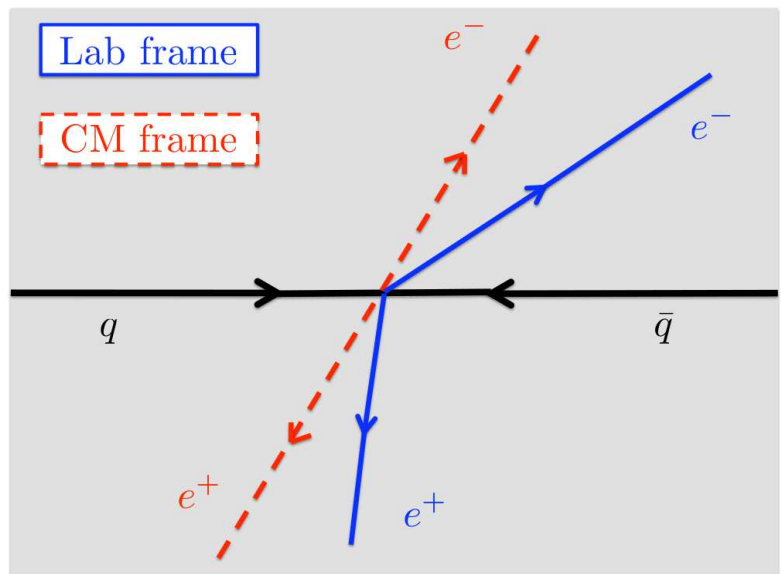}
\caption{Effect of the boost from the CM frame to the lab frame. Note that the final particles
is squeezed in the beam direction while the final antiparticles is squeezed in the transverse direction}
\label{scherm2}
\end{figure}

\begin{figure}
 \centering
 \includegraphics[scale=0.3]{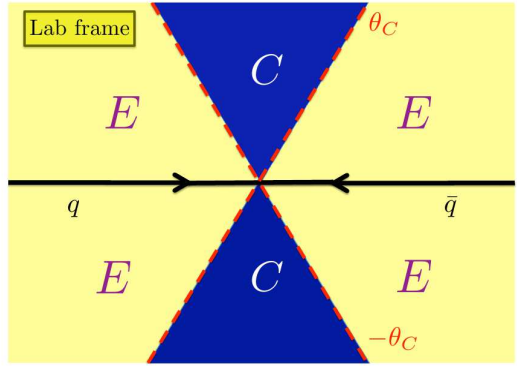}
 \caption{Graphical representation of the Central and Edge regions in which we have to integrate
to obtain the Central and Edge asymmetry, respectively}
\label{C-Eregion}
\end{figure}

In $e^+e^-$ or $p\bar{p}$ colliders, in which the initial state is asymmetric, the asymmetry is naturally different from 0.
However at the LHC the total asymmetry is zero since the initial state, that is proton-proton, is symmetric. 
The elementary subprocess $q\bar{q}\to f\bar{f}$  is instead asymmetric. This property is lost when we
integrate over the full space of partonic and kinematic variables to calculate the asymmetry of the $pp$
process. However if we do not consider the total 
parameter space performing suitable cuts we can obtain a non-zero value for the
asymmetry. Consequently we can have different definitions of asymmetry at the LHC changing the choice
of the cut. 
 In the following we present four definitions of asymmetry viable at LHC, each one associated to a different cut
of the parameter space. 
These can be presented in couples: the Forward-Backward asymmetry with the One-Side asymmetry and the Central asymmetry 
with the Edge asymmetry.\\
 The definition of  the ``Forward-Backward" asymmetry is 
\hypertarget{Langacker:2008yv}{\cite{Langacker:2008yv}}, \hypertarget{Langacker:1984dc}{\cite{Langacker:1984dc}},
\hypertarget{Petriello:2008zr}{\cite{Petriello:2008zr}},\hypertarget{Cvetic:1995zs}{\cite{Cvetic:1995zs}},
\hypertarget{Dittmar:2003ir}{\cite{Dittmar:2003ir}},\hypertarget{Godfrey:2008vf}{\cite{Godfrey:2008vf}}: 
 \be
 A_{RFB}=\frac{\sigma(|Y_f|>|Y_{\bar{f}}|)-\sigma(|Y_f|<|Y_{\bar{f}}|)}{\sigma(|Y_f|>|Y_{\bar{f}}|)+
\sigma(|Y_f|<|Y_{\bar{f}}|)}\Big|_{|Y|>|Y_{cut}|}
  \label{ARFBchinese}
 \ee
 where $\sigma$ is the cross section after having performed all the integrals and $Y_f,~Y_{\bar{f}}$ are the 
pseudorapidies of the outcoming fermion or antifermion, respectively. In the Centre of
Mass (CM) frame their expressions are:
 \be
 Y_{f/\bar{f}}=-\log\left(\tan\left(\frac{\theta_{f/\bar{f}}}{2}\right)\right)
 \ee As already said in the previous section $\theta_{f/\bar{f}}$ is the angle of emission of 
the electron/positron of the final state in the laboratory frame. To perform the calculation we need
to write \ref{ARFBchinese} with respect to $\theta$ that is the angle of the fermion with respect to the 
beam axis and is the simplest angular variable that we can use. This angles are represented
in figures \ref{FBCM} and \ref{scherm2}.\\
 So we have to rewrite the conditions $|Y_f| \gtrless |Y_{\bar{f}}|$ with respect to $\theta$.
In the CM frame $\theta_{\bar{f}}=\theta_f+\pi$, so we have:
\be
\tan\left(\frac{\theta_{\bar{f}}}{2}\right)=\tan\left(\frac{\theta_f+\pi}{2}\right)=
\frac{\sin\left(\frac{\theta_f}{2}+\frac{\pi}{2} \right)}{\cos\left(\frac{\theta_f}{2}+\frac{\pi}{2} \right)}=
\frac{\cos\left( \frac{\theta_f}{2} \right)}{\sin\left( \frac{\theta_f}{2} \right)}=\cot\left(\frac{\theta_f}{2} \right)
\ee Rewriting the related pseudorapidity we obtain:
\bea
&&Y_{\bar{f}}=-\log\left(\tan\left(\frac{\theta_{\bar{f}}}{2}\right)\right)= 
-\log\left(\cot\left(\frac{\theta_{f}}{2}\right)\right)= \nn \\
&&-\log\left(\tan\left(\frac{\theta_{f}}{2}\right)^{-1}\right)=-Y_f
\eea Our conditions $|Y_f| \gtrless |Y_{\bar{f}}|$ become:
\bea
\tan\left(\frac{\theta_f}{2}\right) > \cot\left(\frac{\theta_f}{2}\right) \Rightarrow
-\frac{\pi}{2}<\theta_f<\frac{\pi}{2} \Rightarrow \cos \theta_f>0 \\
\tan\left(\frac{\theta_f}{2}\right) < \cot\left(\frac{\theta_f}{2}\right) \Rightarrow
\frac{\pi}{2}<\theta_f<\frac{3\pi}{2} \Rightarrow \cos \theta_f<0
\eea So we can identify the forward direction as the region with $\cos \theta_f>0$ and
the backward direction as the region with $\cos \theta_f<0$. From now on, we will use
only the angle $\theta_f$ for the calculations in the CM frame, so we will call it
simply $\theta$. Later we will use the expressions $\theta_{f/\bar{f}}$ to indicate
the angles of the outcoming fermions in the lab frame.\\
Having just identified the forward and backward regions, we use the definition 
in the CM frame (\ref{FandB})
 \be
 F=\int_0^1 d\cos\theta \frac{d\sigma(\cos\theta, s)}{d\cos\theta ds}~~~~~~~ 
B=\int_{-1}^0 d\cos\theta \frac{d\sigma(\cos\theta, s)}{d\cos\theta ds}~~~~~~~
 \ee These expressions identify the forward and backward part of the
cross section associated to the fundamental subprocess  $q\bar{q}\to e^+e^-$.
 The relation (\ref{ARFBchinese}) can then be rewritten 
 \be
 A_{RFB}=\frac{\int_{C_{cut}}dx_1dx_2\sum_q f_q(x_1)f_{\bar{q}}(x_2)(F-B)}{\int_{C_{cut}}dx_1dx_2\sum_q f_q(x_1)f_{\bar{q}}(x_2)(F+B)}
\ee
 where $C_{cut}$ is the integration domain and it is dependent from the cut $|Y|>|Y_{cut}|$. As we have already defined 
$f_{q},~f_{\bar{q}}$ are the parton distribution functions (PDFs) 
associated to the quark $q$ or to the antiquark $\bar{q}$; $x_1$ and $x_2$ are the fraction of the proton momentum
carried by the quark or the anti-quark, respectively. $x_1$ and $x_2$ are not the best variables to perform the integration,
because the standard calculation of Feynman amplitudes gives us the differential cross section in terms of kinematic variables,
such as $s$ and $\cos \theta$. So we want to write the integrals in terms of s and the rapidity Y, 
with the definitions:
 \be
 s=S x_1 x_2~~~~~Y=\frac{1}{2}\log\Big(\frac{x_1}{x_2}\Big)
 \label{coord_change}
 \ee
 $S=(14~TeV)^2$ at the LHC. This relation for Y is valid only in the CM system and in the ultra-relativistic limit. In fact 
the general definition for the rapidity is
 \be Y=\frac{1}{2}\log\Big(\frac{E+p_z}{E-p_z}\Big)=\frac{1}{2}\log\Big(\frac{1+v_{rel}}{1-v_{rel}}\Big)\ee
 with $v_{rel}=p_z/E$ being the relative velocity between the two interacting particles.\\
  Using the new variables the formula for the  asymmetry can then be written in the following form
 \be
 A_{RFB}=\frac{\int ds J \Big[ \int_{Y_{cut}}^{+\infty}dY+\int_{-\infty}^{-Y_{cut}}
dY\Big]\sum_q f_q(x_1(s,Y))f_{\bar{q}}(x_2(s,Y))(F-B)}{\int ds J \Big[ \int_{Y_{cut}}^{+\infty}dY+
\int_{-\infty}^{-Y_{cut}}dY\Big]\sum_q f_q(x_1(s,Y))f_{\bar{q}}(x_2(s,Y))(F+B)}
\label{ARFB1} \ee
 $J$ is the Jacobian associated to the change of variables. Using \ref{coord_change} we can obtain
\be
x_1=\sqrt{\frac{s}{S}}e^Y~~~~~~x_2=\sqrt{\frac{s}{S}}e^{-Y}
\ee So the Jacobian is:
\be
J=\left|\left( \begin{array}{cc}
      \pd_s x_1 & \pd_Y x_1 \\
      \pd_s x_2 & \pd_Y x_2 
     \end{array} \right)\right|=\left|\left( \begin{array}{cc}
      \frac{1}{2 \sqrt{s S}}e^Y & \sqrt{\frac{s}{S}}e^Y \\
      \frac{1}{2 \sqrt{s S}}e^{-Y} & -\sqrt{\frac{s}{S}}e^{-Y} 
     \end{array} \right)\right|=\frac{1}{S}
\ee As we can see, in formula \ref{ARFB1} the cut used in $A_{RFB}$
is explicit and consists in excluding the region $-Y_{cut}<Y<Y_{cut}$. We will see in the following that this domain of
integration can be further reduced limiting the range of the variable $s$.
 Now we focus on the integral $\int_{-\infty}^{-Y_{cut}} dY$: if we perform the change of variable $Y\to -Y$, because 
of the relation in (\ref{coord_change}), this corresponds to the exchange $x_1\leftrightarrow x_2$ and consequently to 
the exchange of what is forward with what is backward ($F\leftrightarrow B$). 
  By summarizing, the final effect of $Y\to -Y$ is 
 \be
 \int_{-\infty}^{-Y_{cut}} dY \sum_qf_q(x_1)f_{\bar{q}}(x_2)(F\pm B)=\int_{Y_{cut}}^{\infty} dY \sum_qf_q(x_2)f_{\bar{q}}(x_1)(\pm F+ B)
 \ee
 Finally the FB asymmetry in (\ref{ARFBchinese}) can be rewritten in the useful form
 \be
  A_{RFB}=\frac{\int ds J  \int_{Y_{cut}}^{+\infty}dY \sum_q f^-_{q\bar{q}}(x_1(s,Y),x_2(s,Y))(F-B)} {\int ds J  \int_{Y_{cut}}^{+\infty}dY \sum_q f^+_{q\bar{q}}(x_1(s,Y),x_2(s,Y))(F+B)}   
   \ee
   with
   \be
   f^{\pm}_{q\bar{q}}(x_1,x_2)=f_q(x_1)f_{\bar{q}}(x_2)\pm f_{\bar{q}}(x_1)f_q(x_2)
   \ee
   Another type of asymmetry defined in \hypertarget{Wang:2010tg}{\cite{Wang:2010tg}},
\hypertarget{Wang:2010du}{\cite{Wang:2010du}} is the `` One-side"
 \be
 A_{O}=\frac{\sigma(|Y_f|>|Y_{\bar{f}}|)-\sigma(|Y_f|<|Y_{\bar{f}}|)}{\sigma(|Y_f|>|Y_{\bar{f}}|)+\sigma(|Y_f|<|Y_{\bar{f}}|)}\Big|_{|p_{z,f\bar{f}}|>|p_{z,cut}|}
 \label{ao}
 \ee
 the only difference with the previous one is  in the cut, no longer performed on the rapidity but on the 
longitudinal momentum along the beam axis, $p_z$. We can apply the same steps described before with now  
$\int_{-\infty}^{-p_{z,cut}} dp_z$ instead of $\int_{-\infty}^{-Y_{cut}} dY$ (the exchange is now $p_z\to - p_z$ for 
$x_1\leftrightarrow x_2$). Because $p_z=\frac{\sqrt{S}}{2}(x_1-x_2)$ the change of variable is now
 \be
 x_1=\frac{1}{\sqrt{S}}\Big(p_z+\sqrt{p_z^2+s}\Big)~~~~~x_2=\frac{1}{\sqrt{S}}\Big(-p_z+\sqrt{p_z^2+s}\Big)
 \ee
 with obviously a different definition for J:
\be
J=\left| \left( \begin{array}{cc}
      \pd_s x_1 & \pd_{p_z} x_1 \\
      \pd_s x_2 & \pd_{p_z} x_2 
     \end{array} \right)\right|=\left| \left( \begin{array}{cc}
      \frac{1}{2 \sqrt{S(s+p_z^2)}} & \frac{1}{\sqrt{S}}(1+\frac{p_z}{\sqrt{s+p_z^2}}) \\
      \frac{1}{2 \sqrt{S(s+p_z^2)}} & \frac{1}{\sqrt{S}}(-1+\frac{p_z}{\sqrt{s+p_z^2}})
     \end{array} \right)\right|=\frac{1}{S\sqrt{s+p_z^2}}
\ee This leads to the result:
 \be
  A_{O}=\frac{\int ds  \int_{p_z{_{cut}}}^{+\infty}dp_z J \sum_q f^-_{q\bar{q}}(x_1(s,p_z),x_2(s,p_z))(F-B)} {\int ds \int_{p_{z_{cut}}}^{+\infty}dp_z J \sum_q f^+_{q\bar{q}}(x_1(s,p_z),x_2(s,p_Z))(F+B)}   
   \ee\\ The final two definitions of asymmetry, which explore complementary 
regions in the space of angles, are the ``Central" 
\hypertarget{Ferrario:2009ns}{\cite{Ferrario:2009ns}},\hypertarget{Kuhn:1998jr}{\cite{Kuhn:1998jr}},
\hypertarget{Kuhn:1998kw}{\cite{Kuhn:1998kw}},\hypertarget{Antunano:2007da}{\cite{Antunano:2007da}},
\hypertarget{Ferrario:2008wm}{\cite{Ferrario:2008wm}} and ``Edge" 
\hypertarget{Xiao:2011kp}{\cite{Xiao:2011kp}} asymmetries
 \bea
  \label{ACE}
 A_C(Y_C)=\frac{\sigma_f(|Y_f|<Y_C)-\sigma_{\bar{f}}(|Y_{\bar{f}}|<Y_C)}{\sigma_f(|Y_f|<Y_C)+\sigma_{\bar{f}}(|Y_{\bar{f}}|<Y_C)} \\
 A_E(Y_C)=\frac{\sigma_f(|Y_f|>Y_C)-\sigma_{\bar{f}}(|Y_{\bar{f}}|>Y_C)}{\sigma_f(|Y_f|>Y_C)+\sigma_{\bar{f}}(|Y_{\bar{f}}|>Y_C)}
 \eea
In these two cases the cut is not performed on the rapidity 
nor on the longitudinal momentum but on the angle (therefore on the limits of integration for  F and B). These asymmetries
are defined in the lab frame, because the Lorentz transformation from the CM frame 
\textquotedblleft squezees \textquotedblright the final particles. So in the laboratory (Lab) frame the outgoing $e^-$ and 
$e^+$ have 
no longer the same direction: the angles $\pm\theta$ with respect to the z axis for the $e^{\mp}$ respectively in the 
CM frame are replaced by the different $\theta^{e^-}$ and $\theta^{e^+}$. So the explicit form of the definitions 
(\ref{ACE}) in the lab frame is:
 \bea
  \label{FandBlab}
& ( F/B)_{C}=\int_{-cut}^{+cut}\frac{d\sigma}{d\cos\theta^{e^{(-/+)}} ds}d\cos\theta^{e^{(-/+)}}\\
 & ( F/B)_{E}=\int_{-1}^{-cut}\frac{d\sigma}{d\cos\theta^{e^{(-/+)}} ds}d\cos\theta^{e^{(-/+)}}+\int_{cut}^{+1}\frac{d\sigma}{d\cos\theta^{e^{(-/+)}} ds}d\cos\theta^{e^{(-/+)}}\nonumber
  \eea From this expression we can understand the different names of the two definitions: the ``Central"
asymmetry has the angular integration on the central angular region, the ``Edge" in the complementary region. This
is well shown in figure \ref{C-Eregion}.
 As we have already seen the boost given by the Lorentz transformation from the CM 
frame to the Lab frame ``squeezes" the $e^-$ mainly in the forward  and backward directions, while the $e^+$ are 
left more abundant centrally \hypertarget{Ferrario:2009ns}{\cite{Ferrario:2009ns}}: according to the 
previous definitions $B_C>F_C$ and consequently we expect negative values for the Central asymmetry 
(in all the other three cases we have always positive values).\\
Their expressions in terms of F and B are:
 \be
 A_{C/E}=\frac{\int dx_1dx_2\sum_q f_q(x_1)f_{\bar{q}}(x_2)(F_{C/E}-B_{C/E})}{\int dx_1dx_2\sum_q f_q(x_1)f_{\bar{q}}(x_2)(F_{C/E}+B_{C/E})}
\ee
  Because we have calculated the cross section of the processes in the CM frame we have to calculate how the definitions
(\ref{FandBlab}) appear in the CM frame. In this frame the differential cross section for $e^+$, here denoted with the tilde, 
is obtained from that of the $e^-$ by simply inverting the sign of the cosine of the angle   
  \be
  \frac{d\tilde{\sigma}}{d\cos\theta ds}=\frac{d\sigma}{d\cos\theta ds}\Big|_{\cos\theta \to-\cos\theta}
  \ee
So the previous definitions (\ref{FandBlab}) for F and B in the Lab frame assume the following form in the CM frame:
  \bea 
& F_{C}=\int_{-1}^{f(cut)}\frac{d\sigma}{d\cos\theta ds}d\cos\theta~~~~~~ B_{C}=\int_{-1}^{f(cut)}\frac{d\tilde{\sigma}}{d\cos\theta ds}d\cos\theta \nonumber\\
 &F_{E}=\int_{f(cut)}^{+1}\frac{d\sigma}{d\cos\theta ds}d\cos\theta~~~~~~B_{E}=\int_{f(cut)}^{+1}\frac{d\tilde{\sigma}}{d\cos\theta ds}d\cos\theta
 \eea
 where $f(x)$ is simply the relativistic function which relates the cosines of angles in the Lab frame to those in 
the CM frame, i.e. $\cos\theta^{CM}=f(\cos\theta^{Lab})$: 
 \be
 f(x)=\frac{\gamma^2-(1+\gamma^2)x^2}{-\gamma^2+(-1+\gamma^2)x^2};~~~~~~~\gamma=\frac{1}{\sqrt{1-v_{rel}^2}};~~~~~v_{rel}
=\frac{x_1-x_2}{x_1+x_2} \label{fx}
 \ee
The calculation of this function is a simple application of special relativity. We start from 
the equation that gives the relativistic velocity in the lab frame in function of the relative velocity
with respect to the CM frame and the angle between the initial and the final fermion in the CM frame (this
is the angle that we have used for the first two definition of the asymmetry at the LHC, $\theta$) 
\hypertarget{jackson}{\cite{jackson}}:
\be
\left\{ \begin{array}{c}
   (u_f)_x=\frac{v\sin \theta}{\gamma (1+v^2 \cos \theta)} \\
   (u_f)_z=\frac{v(1+\cos \theta)}{(1+v^2 \cos \theta)} 
  \end{array} \right.
\ee In this formula $z$ is the beam direction, $x$ is the orthogonal direction, $u$ is the velocity of the electron
in the lab frame and $v$ is the relative velocity between the lab and the CM frames. Now we want to calculate
the relation between $\cos \theta_f$ and $\cos \theta$. 
\be
\begin{array}{c}
 \tan \theta_f=\frac{(u_f)_x}{(u_f)_z}=\frac{\sin \theta}{\gamma (1+\cos \theta)}=A(\cos \theta) \\
\tan \theta_f=\frac{\sin \theta_f}{\cos \theta_f}=\frac{\sqrt{1-\cos^2 \theta_f}}{\cos \theta_f}
\end{array} \Rightarrow 1-\cos^2 \theta_f=A^2 \cos^2 \theta_f
\ee Solving the equation on the right we obtain:
\be
\cos \theta_f=\frac{1}{\sqrt{1+\frac{1-\cos^2 \theta}{\gamma^2(1+\cos \theta)^2}}}
\ee If now we invert this relation we have:
\be
\cos \theta=\frac{\gamma^2-(1+\gamma^2)\cos^2 \theta_f}{-\gamma^2+(-1+\gamma^2)\cos^2 \theta_f}
\ee that is the formula (\ref{fx}) that we wanted to demonstrate.\\
 The definition of asymmetry for the Central and Edge cases can therefore be presented as follows:
  \be
 A_{C/E}=\int ds\int_{s/S}^1dx_1 J \sum_q f_{q}(x_1)f_{\bar{q}}\Big(x_2=\frac{s}{S x_1}\Big)\Big(F_{C/E}-B_{C/E}\Big)  
 \ee
 with
\be
J=\left|\left( \begin{array}{cc}
      \pd_s x_1 & \pd_{x_1} x_1 \\
      \pd_s x_2 & \pd_{x_1} x_2 
     \end{array} \right)\right|=\left|
\left( \begin{array}{cc}
      0 & 1 \\
      \frac{1}{2 S x_1} & -\frac{s}{S x_1^2} 
     \end{array} \right)\right|=\frac{1}{2 x_1S} 
\ee
Instead of the angle $cut$ it is convention to write the cut performed on these two definitions of asymmetry 
using the related pseudorapidity 
\be Y_C=-\log\Big(\tan(\theta^{cut}/2)\Big) \ee
where $\theta^{cut}=\arccos(cut)$ This variable (usually denoted with $\eta$ in literature), 
must not be confused with the cut on the rapidity $Y$ previously used for the RFB definition 
of asymmetry.\\ 

\section{Cross-section \label{crsec}}
\begin{figure}[h]
\centering
 \includegraphics[scale=0.65]{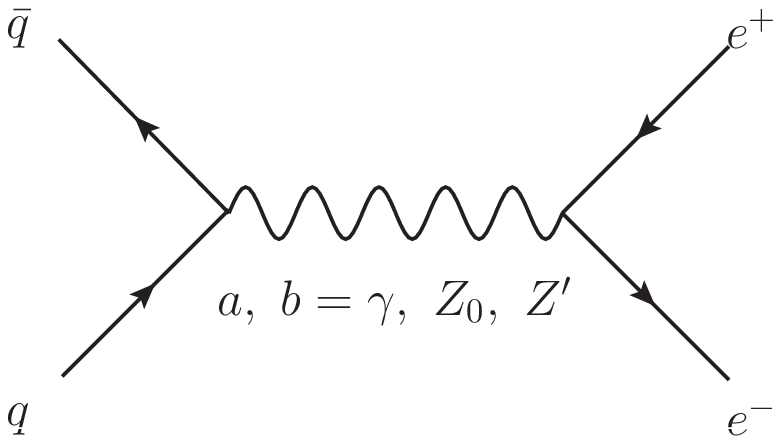}
\label{ffvff} 
\caption{Feynman diagrams for the Drell-Yan process in the MiAUSSM} \end{figure}

As we said earlier we are interested in the cross-section in the CM for the process $pp\rightarrow e^+e^-$.
This is the convolution of the elementary cross-section of the process $q\bar{q}\rightarrow e^+e^-$
with the PDFs of the proton. So we have to calculate the cross-section of 
$q\bar{q}\rightarrow e^+e^-$ for a general
Drell-Yan interaction. This means that we can have the product of diagrams in which can be exchanged the 
$\g$, $Z_0$ and $Z'$, showed in fig. \ref{ffvff}.
  Thus 
we can have 6 possible terms in the amplitude: $\g\g$, $\g Z_0$, $\g Z'$, 
$Z_0 Z_0$, $Z_0 Z'$ and $Z' Z'$. We will sum them all:
\begin{equation}
|M|^2(q)=\frac{1}{3} \frac{1}{4} \sum_{a,b=\gamma,Z,Z'} {g}_{a}^{2} \, {g}_{b}^{2} \, {M}_{ab}
\end{equation} where the fractions $\frac{1}{4}$ and $\frac{1}{3}$ come out from the averages over 
spin and color,  $g_0$ is the coupling associated to the $Z'$. We fix its value
to $0.1$. $M_{ab}$ is the amplitude of each process divided by the couplings.\\
To calculate $M_{ab}$ we have to write down the Feynman amplitudes divided by the
couplings for the processes
in figure \ref{ffvff} for the exchange of any of the three possible vector bosons:
\be
\mathcal{M_a}=i \left[\bar{u}_f \g_{\m}(C^V_{f,a}-\g_5 C^A_{f,a}) v_{\bar{f}} \right] \frac{1}{s-M_a^2}
\left[\bar{v}_{\bar{q}} \g_{\n}(C^V_{q,a}-\g_5 C^A_{q,a}) u_q \right]
\ee $C^V_{f/q,a},~C^V_{f/q,a}$ are the quantum numbers of the quark $q$ or the fermion $f$
with respect to the interation $a$, that goes from $0$ to $2$ and represent the $U(1)'$, $U(1)$ and
$SU(2)$ respectively. Their values are listed (with the electron as the fermion $f$) in table
\ref{anochar}. From now on we use the subfix $e$ instead of $f$ because we are focusing on electrons
in the final state, but the formulae can be used for other fermions simply using the appropriate
quantum numbers. Its square modulus gives:
\bea
&&M_{ab}=\frac{1}{(s-M_a^2)(s-M_b^2)}tr[(\sl p_e+m_e)\g_{\m}(C^V_{e,a}-\g_5 C^A_{e,a})(\sl p_{\bar{f}}-m_{\bar{f}})  \\
&&(C^V_{e,b}+\g_5 C^A_{e,b})\g_{\n}]  tr[(\sl p_q-m_q)\g^{\m}(C^V_{q,a}-\g_5 C^A_{q,a})(\sl p_{\bar{q}}+m_{\bar{q}})
(C^V_{q,b}-C^A_{q,b})\g^{\n}] \nn
\eea Calculating the products, using the properties $\g_5 \g_{\m}=-\g_{\m} \g_5$ and $\g_5^2=1$ and
the fact that $m_q=m_{\bar{q}}$, $m_e=m_{\bar{e}}$ the first trace becomes:
\bea
&&tr[(\sl p_e+m_e)\g_{\m}(C^V_{e,a}-\g_5 C^A_{e,a})(\sl p_{\bar{f}}-m_{\bar{f}})(C^V_{e,b}+\g_5 C^A_{e,b})\g_{\n}]= \nn \\
&&C^V_{e,a} C^V_{e,b}tr[\sl p_{\bar{e}}\g_{\m}p_e \g_{\n}-m_e m_{\bar{e}}\g_{\m} \g_{\n}]+ \nn \\
&&C^V_{e,a} C^A_{e,b}tr[(\sl p_{\bar{e}}\g_{\m}p_e \g_5 \g_{\n}]+ \\
&&C^A_{e,a} C^V_{e,b}tr[\sl p_{\bar{e}}\g_{\m}(-\g_5)p_e \g_{\n}]+ \nn \\
&&C^A_{e,a} C^A_{e,b}tr[\sl p_{\bar{e}}\g_{\m}p_e \g_{\n}+m_e m_{\bar{e}}\g_{\m} \g_{\n}] \nn
\eea In the same way for the second trace we have:
\bea
&&tr[(\sl p_q-m_q)\g^{\m}(C^V_{q,a}-\g_5 C^A_{q,a})(\sl p_{\bar{q}}+m_{\bar{q}})(C^V_{q,b}-C^A_{q,b})\g^{\n}] \nn \\
&&C^V_{q,a} C^V_{q,b}tr[\sl p_{\bar{q}}\g^{\m}p_q \g^{\n}-m_q m_{\bar{q}}\g^{\m} \g^{\n}]+ \nn \\
&&C^V_{q,a} C^A_{q,b}tr[(\sl p_{\bar{q}}\g^{\m}p_q \g_5 \g^{\n}]+ \\
&&C^A_{q,a} C^V_{q,b}tr[\sl p_{\bar{q}}\g^{\m}(-\g_5)p_q \g^{\n}]+ \nn \\
&&C^A_{q,a} C^A_{q,b}tr[\sl p_{\bar{q}}\g^{\m}p_q \g^{\n}+m_q m_{\bar{q}}\g^{\m} \g^{\n}] \nn
\eea Multiplying the two contributions and using the formulae (\ref{traccia3}), (\ref{traccia4}), (\ref{traccia5}), 
(\ref{traccia1}) and (\ref{traccia2}) we obtain the final expression:
 \bea
&&{M}_{ab}=\frac{ 64  \, N_{ab}  \, \Big[(s-m_a^2) \,(s-m_b^2)+(\Gamma_a \,  m_a \,  \Gamma_b  \, m_b)\Big]}{\Big[(s-m_a^2)^2+(\Gamma_a \, m_a)^2\Big ]\Big[(s-m_b^2)^2+(\Gamma_b \, m_b)^2\Big]} \times \nn \\
&& \Bigg \{  2 \,  m_e^2 \,  m_q^2 \big(C^A_{e,a}  \, C^A_{e,b} - C^V_{e,a} \,  C^V_{e,b}\big)\, \big (C^A_{q,a}  \,  C^A_{q,b} - C^V_{q,a} C^V_{q,b}\big)  - 
   \,   m_e^2 \, \big (p_{q}\cdot p_{\bar{q}}\big ) \times \nn \\ 
&&\,\big (C^A_{e,a} \,   C^A_{e,b} - C^V_{e,a}  \,  C^V_{e,b}\big) \,   \big(C^A_{q,a} \,   C^A_{q,b} + C^V_{q,a} \,   C^V_{q,b}\big) + \,\big (p_q \cdot p_{\bar{e}} \big) \, \big(p_{\bar{q}}\cdot p_e \big) \times \nn \\ 
&&\,\big (C^A_{e,b}  \, C^V_{e,a} + C^A_{e,a} \,  C^V_{e,b}\big)  \,\big (C^A_{q,b}  \, C^V_{q,a} + C^A_{q,a} \,  C^V_{q,b}\big) + 
    \, \big(p_q\cdot p_{\bar{e}}\big) \, \big (p_{\bar{q}}\cdot p_e\big)\times\\
  &&  \,  \big(C^A_{e,a}  \, C^A_{e,b} + C^V_{e,a}  \, C^V_{e,b}\big) \,  \big(C^A_{q,a}  \, C^A_{q,b} + C^V_{q,a} \,  C^V_{q,b}\big) -\, \big(p_q\cdot p_e \big) \,\big( p_{\bar{q}}\cdot p_{\bar{e}}\big)  \times \nn \\ 
&&\,  \big(C^A_{e,b} \,  C^V_{e,a} + C^A_{e,a}  \, C^V_{e,b}\big)\, \big(C^A_{q,b} \,  C^V_{q,a} + C^A_{q,a}  \, C^V_{q,b}\big)\,+ \big(p_q\cdot p_e\big) \, \big( p_{\bar{q}}\cdot p_{\bar{e}}\big) \times \nonumber\\
&&\, \big(C^A_{e,a} \,  C^A_{e,b} + C^V_{e,a} \,  C^V_{e,b}\big)  \, (C^A_{q,a}  \, C^A_{q,b} + C^V_{q,a} \,  C^V_{q,b})-\, m_q^2\,\big (p_e\cdot p_{\bar{e}}\big)\, \times \nn \\ 
&&\big (C^A_{e,a}  \, C^A_{e,b} + C^V_{e,a}  \, C^V_{e,b}\big)  \big (C^A_{q,a} \,  C^A_{q,b} - C^V_{q,a} \,  C^V_{q,b}\big) \Bigg\}\nonumber
  \eea In this expression we have substituted the divergent expression of the propagator,
$\frac{1}{s-m_a^2}$, with the regulated one $\frac{1}{s-m_a^2+i \Gamma_a m_a}$ where
$\Gamma_a$ is the width of the boson $a$.
$N_{ab}$ is a multiplicity factor that is equal to $\frac{1}{2}$ if 
the exchanged vector bosons  are identical 
and is equal to $1$ if they are different. The Cs are
simply the vector and axial quantum numbers related to the vector bosons: for the $\g$ and the $Z_0$ they are
the usual SM quantum numbers that can be found in \hypertarget{HalzMart}{\cite{HalzMart}}, while the vector and axial
couplings of the extra $U(1)$ have been previously calculated in (\ref{qvqaano}) and are showed in table \ref{anochar}.
\begin{table}
 \centering
\begin{tabular}[h!]{|c||c|c|}
 \hline & $C^V_{f,Z'}$ & $C^A_{f,Z'}$ \\
 \hline \hline $f=u,c,t$ & $Q_Q+Q_{H_u}/2$ & $-Q_{H_u}/2$ \\
 \hline $f=d,s,b$ & $Q_Q-Q_{H_u}/2$ & $Q_{H_u}/2$ \\
 \hline $f=e,\m,\t$ & $Q_L-Q_{H_u}/2$ & $Q_{H_u}/2$ \\
 \hline 
\end{tabular}
\caption{Vector and axial quantum numbers of the SM fermions with respect to the $Z'$}
\label{anochar}
\end{table} The differential cross section can be found multiplying for the usual
kinematic prefactor and summing this result over the contribution of the 6 possible initial quarks:
\begin{equation}
\frac{\partial^2\sigma}{\partial s\partial \cos\theta}\Big|_{CM}=\sum_q \frac{p_e}{32 \, \pi \,  s \, p_q}|M|^2(q)
\end{equation} To obtain its explicit expression we have to write the
scalar products with respect to $s$ and $\theta$. Using the formulae (\ref{scalarproducts})
we obtain (in the CM frame):
\bea &&(p_q p_{\bar{q}})=E_q E_{\bar{q}} + |p_q|^2 \nn\\
&&(p_e p_{\bar{e}})=E_e E_{\bar{e}} + |p_e|^2 \nn \\
&&(p_q p_e)=E_q E_e - |p_q||p_e|\cos{\theta}  \\
&&(p_q p_{\bar{e}})=E_q E_{\bar{e}} + |p_q||p_e|\cos{\theta} \nn \\
&&(p_{\bar{q}} p_e)=E_{\bar{q}} E_e + |p_q||p_e|\cos{\theta} \nn \\
&&(p_{\bar{q}} p_{\bar{e}})=E_{\bar{q}} E_{\bar{e}} - |p_q||p_e|\cos{\theta} \nn 
\eea Because the couples of incoming and outcoming particles have the same masses
we can use the definition of $s$ to calculate all the quantities in this expression.
In the CM frame we have:
\be
m_e^2=(E_e^2-|p_e|^2)=(E_{\bar{e}}^2-|p_e|^2) \Rightarrow E_e=E_{\bar{e}}
\ee so:
\bea
&&s=(p_e+p_{\bar{e}})^2=(E_e+E_{\bar{e}},\vec{0})^2=(2E_e)^2 \Rightarrow E_e=\frac{\sqrt{s}}{2} \\
&&|p_e|=\sqrt{\frac{s}{4}-m_e^2}
\eea Substituting $e$ with $q$ we obtain the corresponding relations for the quarks.

\section{Asymmetry calculation}

\subsection{Optimized asymmetry}
As we have explained, each definition of the asymmetry at the LHC contains a certain cut on the 
parameter space. The cut acts as a variable of the asymmetry so it has to be fixed in order to obtain
a numerical value from the calculation. The problem is to choose the optimal cut. One can be tempted
to use the value for which the asymmetry is maximum, but this is not the best choice.\\
To find the best cut, we have to define the significance:
\be
 Sig= A \sqrt{{\cal L}~\sigma_{tot}}
\ee with ${\cal L}=100 ~fb^{-1}$ as the integrated luminosity at LHC and $\sigma_{tot}$ the total cross section of the 
process, sum of the forward and backward contributions.
% This quantity measures the possibility that a certain measure is not occurred by chance.
So the optimal cut is the value of $Y_{cut}$,  $p_{z,cut}$ or $Y_C$ for which the significance is maximum.\\
It depends  mainly on the properties of the PDFs and on the mass of the $Z'$. In this work we set $M_{Z'}=1.5~ TeV$. 
Because the best cut is nearly 
independent on the properties of the $Z'$ coupling to fermions 
\hypertarget{Zhou:2011dg}{\cite{Zhou:2011dg}}, we can find it fixing the free charges $Q_i$ associated to the extra $U(1)'$. 
Obviously the values of the asymmetry and the significance are dependent on these 
charges: we will concentrate on them in the next part of our analysis.\\
Furthermore it is useful to restrict the range of $s$ of the studied region: the standard choice
is to use the \textquotedblleft on peak\textquotedblright and the \textquotedblleft off peak\textquotedblright
regions, defined in the sequent way:
\bea
&&\text{\textquotedblleft on peak\textquotedblright} \Leftrightarrow M_{Z'}-3 \Gamma_{Z'}< M_{e^+e^-}<
M_{Z'}+ 3\Gamma_{Z'} \\
&&\text{\textquotedblleft off peak\textquotedblright} \Leftrightarrow \frac{2}{3} M_{Z'}< M_{e^+e^-}<
M_{Z'}- 3\Gamma_{Z'}
\eea We will use only the \textquotedblleft on peak\textquotedblright region for simplicity,
but our calculation can be perfomed in the same way in the \textquotedblleft off peak\textquotedblright
region.\\
As said, we have performed calculations with fixed charges in order to find the best cuts that then
we will use to study the dependence of the asymmetry from the charges. Our choice has been 
$Q_L=1$, $Q_{H_u}=1/2$ and $Q_Q=3/4$.\\
The results for the best cuts, that are obtained from the figures in the following, are:\\
\begin{table}[h]
 \centering
\begin{tabular}[h]{|c||c|c|c|c|}
 \hline & $A_{RFB}$ & $A_{O}$ & $A_C$ & $A_E$  \\
 \hline \hline Best cut & $Y_{f\bar{f}}^{cut}=0.4$ & $p_{z,f\bar{f}}^{cut}=580~ GeV$ & $Y_C=0.8$ & $Y_C=1.4$ \\
 \hline 
\end{tabular}
\caption{Best cuts for
the on-peak $e^+e^-$ asymmetries}
\label{bestcuts}
\end{table} As a check, because the cuts have to be nearly independent from the charges 
\hypertarget{Zhou:2011dg}{\cite{Zhou:2011dg}},
we have calculated the cuts after changing the value of $Q_{H_u}, Q_Q$ and $Q_L$, obtaining the same results
just listed.

\subsubsection{FB asymmetry and significance with fixed charges: the search of the peak for the significance}
\begin{figure}[h!]
\includegraphics[scale=0.65]{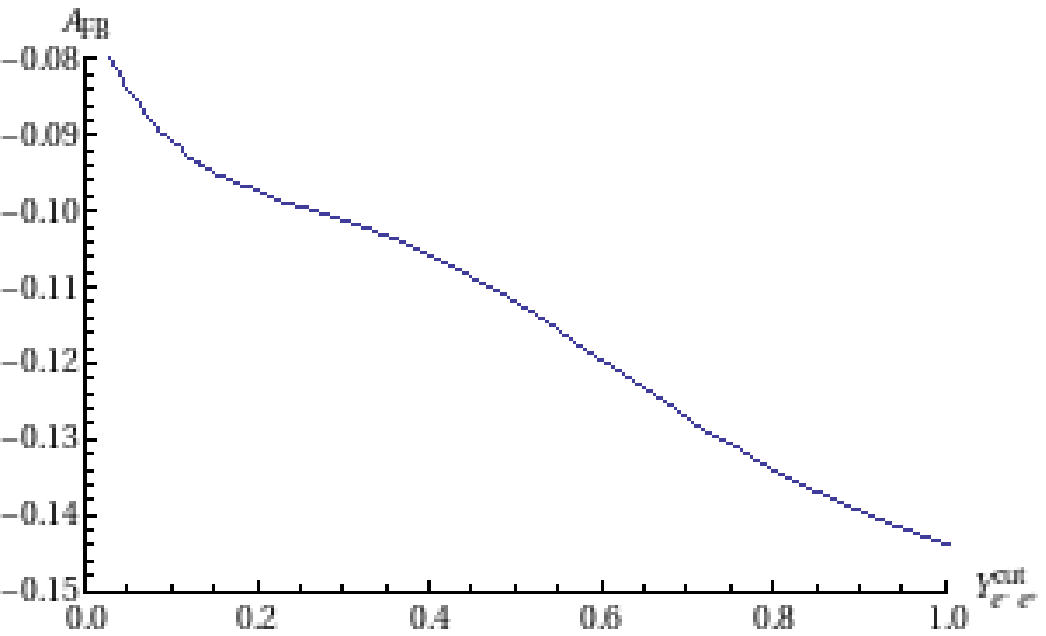}
\includegraphics[scale=0.65]{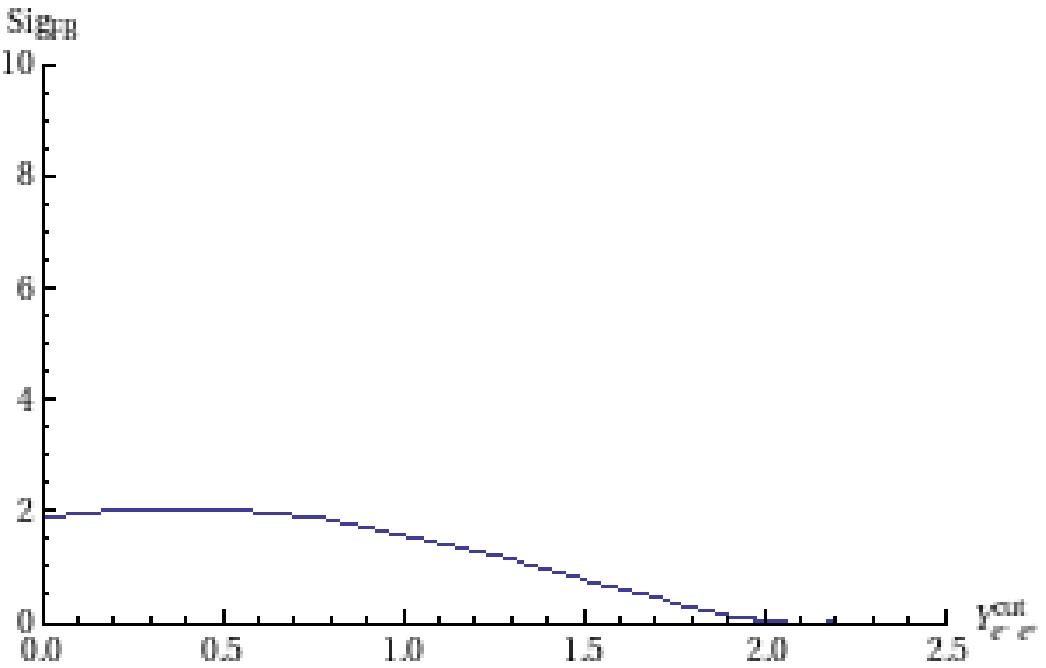}
\caption{On peak ($M_{Z'}-3\Gamma_{Z'}<M_{e^+e^-}<M_{Z'}+3\Gamma_{Z'}$) asymmetry (left) and significance (right) 
versus the cut on  rapidity $Y_{cut}$. The anomalous charges are fixed to the general values $Q_L=1$, $Q_{H_u}=1/2$ 
and $Q_Q=3/4$. The peak for the significance is for $Y_{cut}=0.4$. The position of the peak does not depend on the 
choice of the values for the charges, while the intensity of asymmetry and significance depends on it.}
\label{arfbpeak}
\end{figure}
\begin{figure}[h!]
 \includegraphics[scale=0.65]{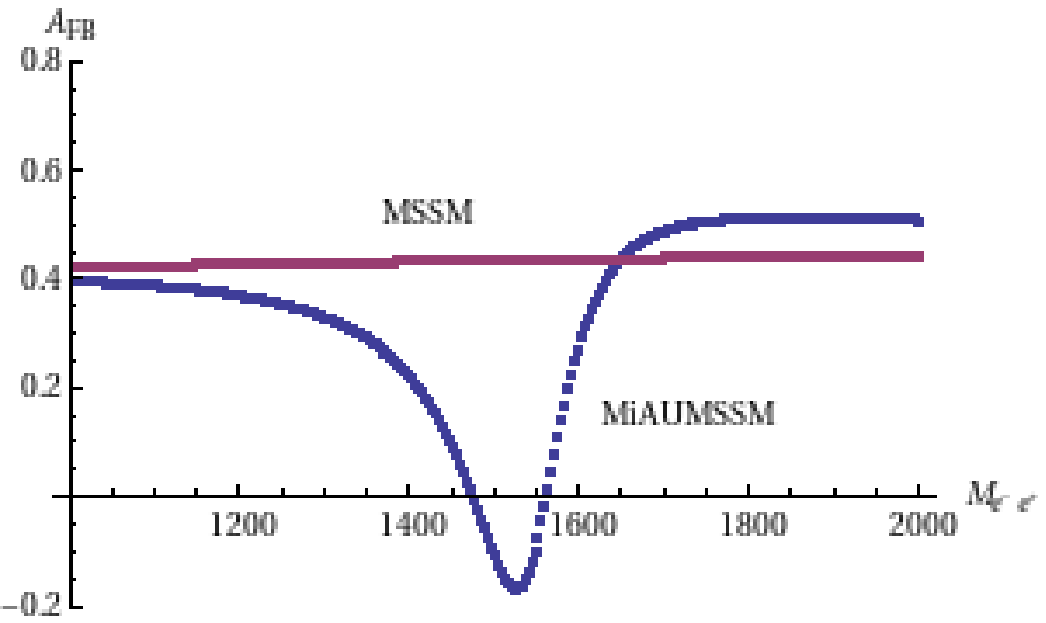}
 \includegraphics[scale=0.65]{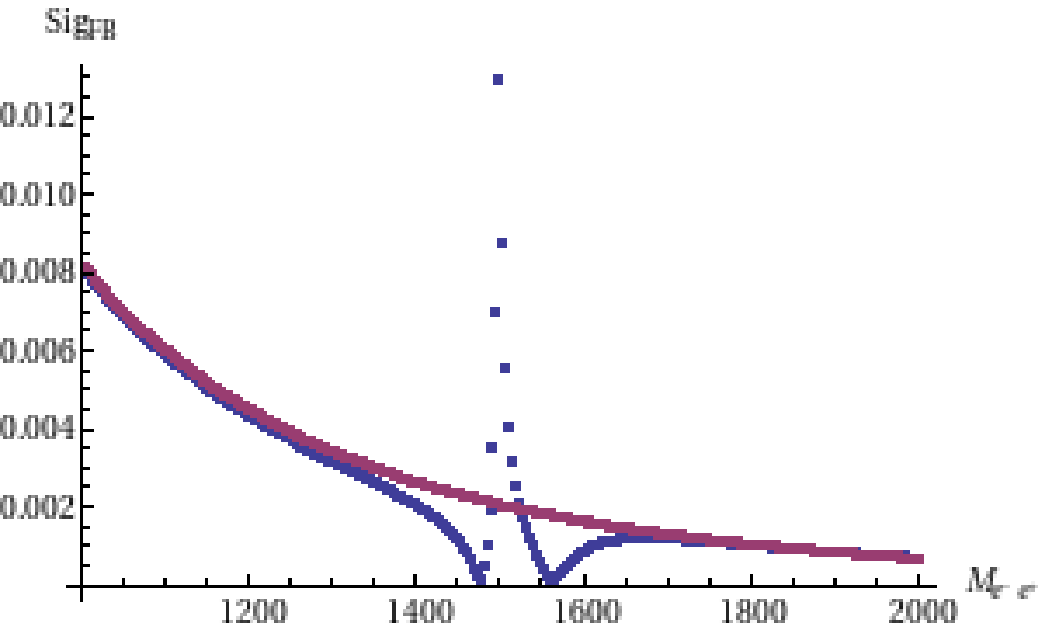}
\caption{Asymmetry (left) and significance (right) in MiAUSSM (blue) and MSSM (purple) around the peak of the Z'
versus the invariant mass $M_{e^+e^-}$. 
$y_{cut}$ is set to 0.4, while the
anomalous charges are fixed to the general values $Q_L=1$, $Q_{H_u}=1/2$ and $Q_Q=3/4$}
\label{confr}
\end{figure} The figure \ref{arfbpeak} shows the dependence of the on peak asymmetry and the significance 
versus the cut $Y_{cut}$. It is clear that the significance has a maximum around $Y_{cut}=0.4$. Using this 
result we have obtained the figure \ref{confr}, in wich we keep fixed the cut and study the dependence of the 
asymmetry and significance on the variable $s$ in the range permitted by the on peak region. In this case 
we also plot the asymmetry and significance of the MSSM to compare the results. It is clear that the minimum
in the asymmetry as function of $s$ is due to the presence of the extra bosonic resonance, $Z'$.

%\newpage

\subsubsection{One side asymmetry and significance with fixed charges: the search of the peak for the significance}
\begin{figure}[h!]
\includegraphics[scale=0.65]{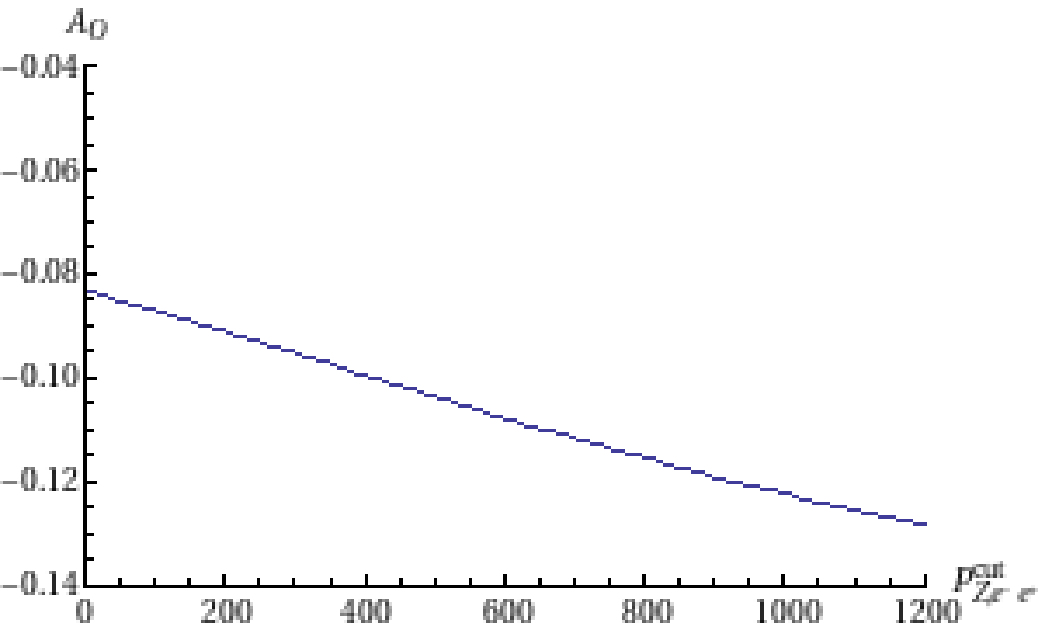}
\includegraphics[scale=0.65]{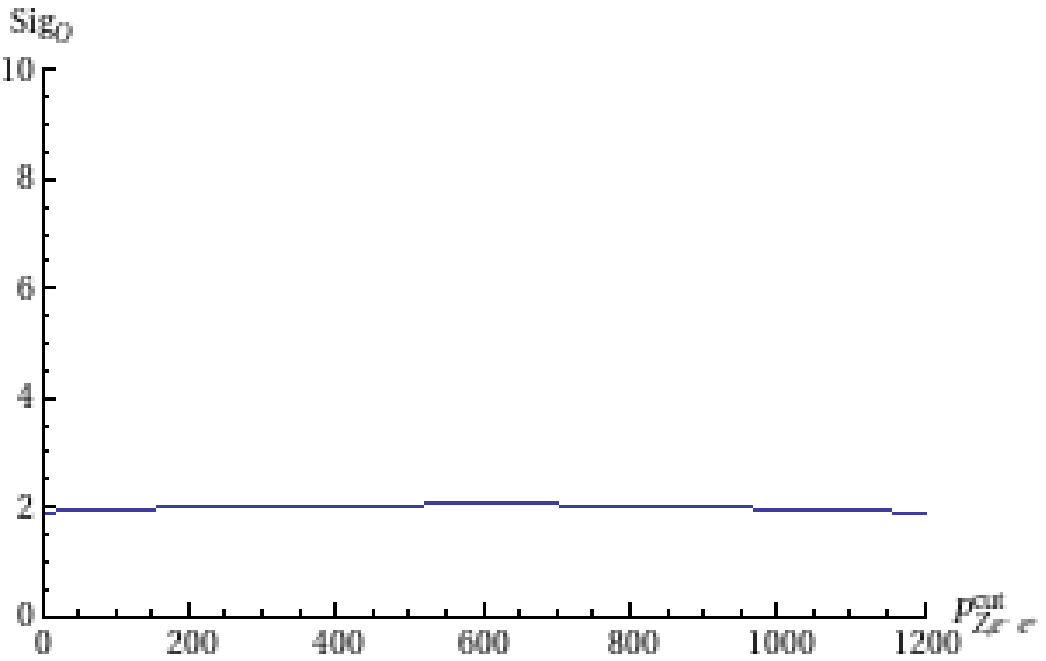}
\caption{On peak ($M_{Z'}-3\Gamma_{Z'}<M_{e^+e^-}<M_{Z'}+3\Gamma_{Z'}$) asymmetry (left) and significance (right) 
versus the cut on  the longitudinal momentum $p_{z,cut}$. The anomalous charges are fixed to the general values 
$Q_L=1$, $Q_{H_u}=1/2$ and $Q_Q=3/4$. The peak for the significance is for $p_{z,cut}=580$ GeV. 
The position of the peak does not depend on the choice of the values for the charges, while the intensity 
of asymmetry and significance depends on it. }
\label{aopeak}
\end{figure}
\begin{figure}[h!]
 \includegraphics[scale=0.65]{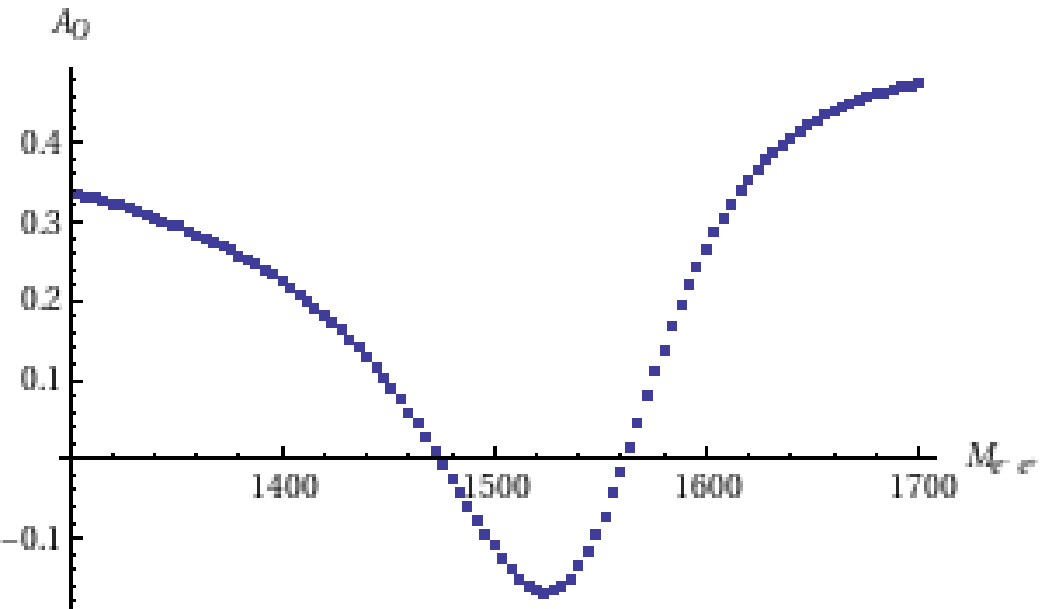}
 \includegraphics[scale=0.65]{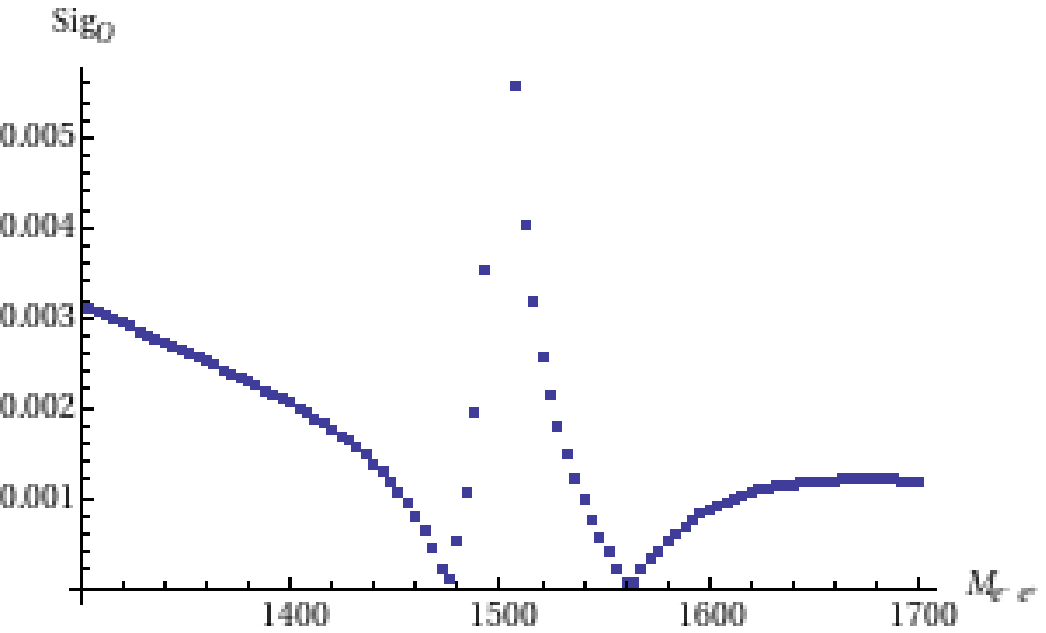}
\caption{Asymmetry (left) and significance (right) in MiAUSSM  around the peak of the Z'. $p_{z,cut}$ is set 
to 580 GeV, while the
anomalous charges are fixed to the general values $Q_L=1$, $Q_{H_u}=1/2$ and $Q_Q=3/4$}
\label{aos}
\end{figure} The figure \ref{aopeak} shows the dependence of the on peak asymmetry and the significance 
versus the cut $Y_{cut}$. It is clear that the significance has a maximum around $p_{z,cut}=580~GeV$. Using this 
result we have obtained the figure \ref{aos}, in which we keep fixed the cut and study the dependence of the 
asymmetry and significance on the variable $s$ in the range permitted by the on peak region.

%\newpage

\subsubsection{Central asymmetry and significance with fixed charges: the search of the peak for the significance}
\begin{figure}[h!]
\includegraphics[scale=0.65]{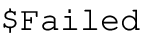}
\includegraphics[scale=0.65]{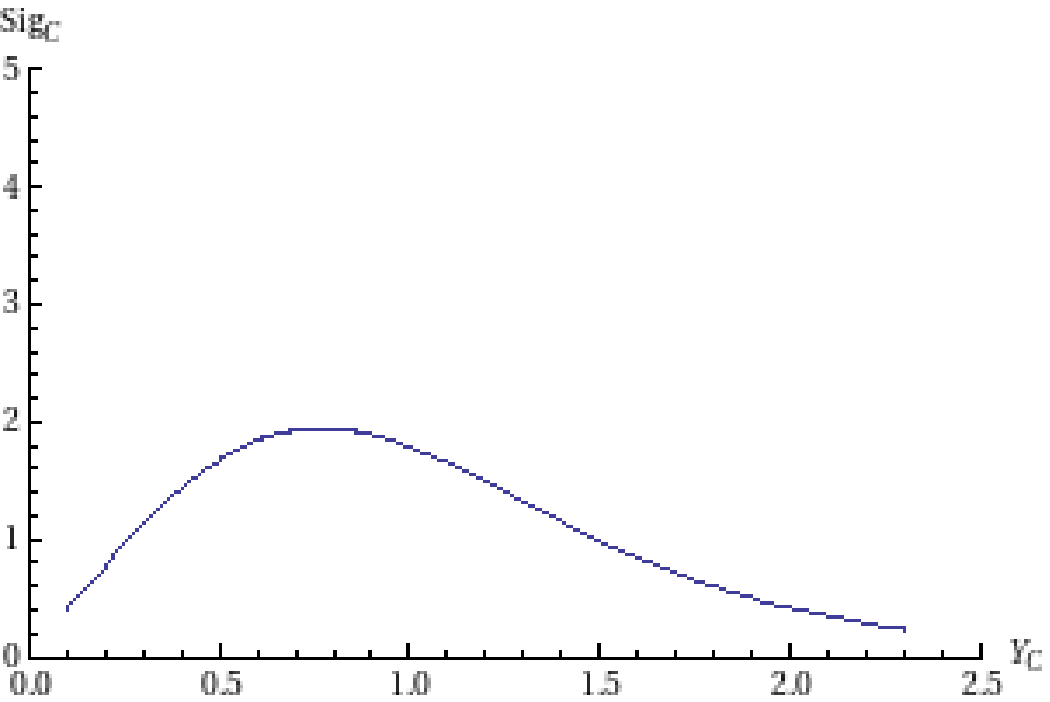}
\caption{On peak ($M_{Z'}-3\Gamma_{Z'}<M_{e^+e^-}<M_{Z'}+3\Gamma_{Z'}$) asymmetry (left) and significance (right) 
versus the cut on  the longitudinal momentum $p_{z,cut}$. The anomalous charges are fixed to the general values 
$Q_L=1$, $Q_{H_u}=1/2$ and $Q_Q=3/4$. The peak for the significance is for $Y_C=0.8$. The position of the peak 
does not depend on the choice of the values for the charges, while the intensity of asymmetry and significance 
depends on it. }
\label{acpeak}
\end{figure} 
\begin{figure}[h!]
\includegraphics[scale=0.65]{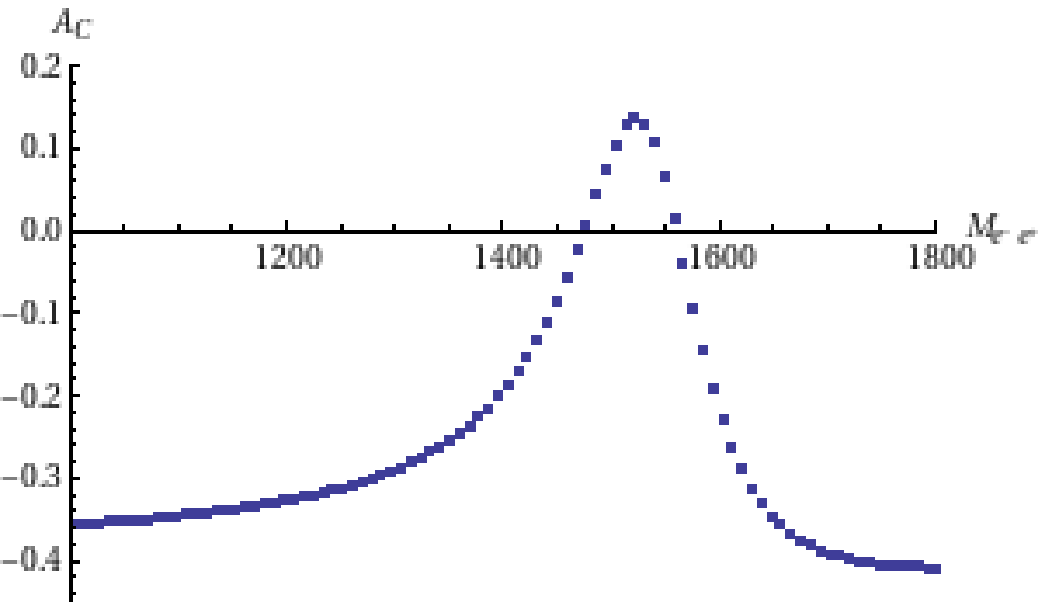}
\includegraphics[scale=0.65]{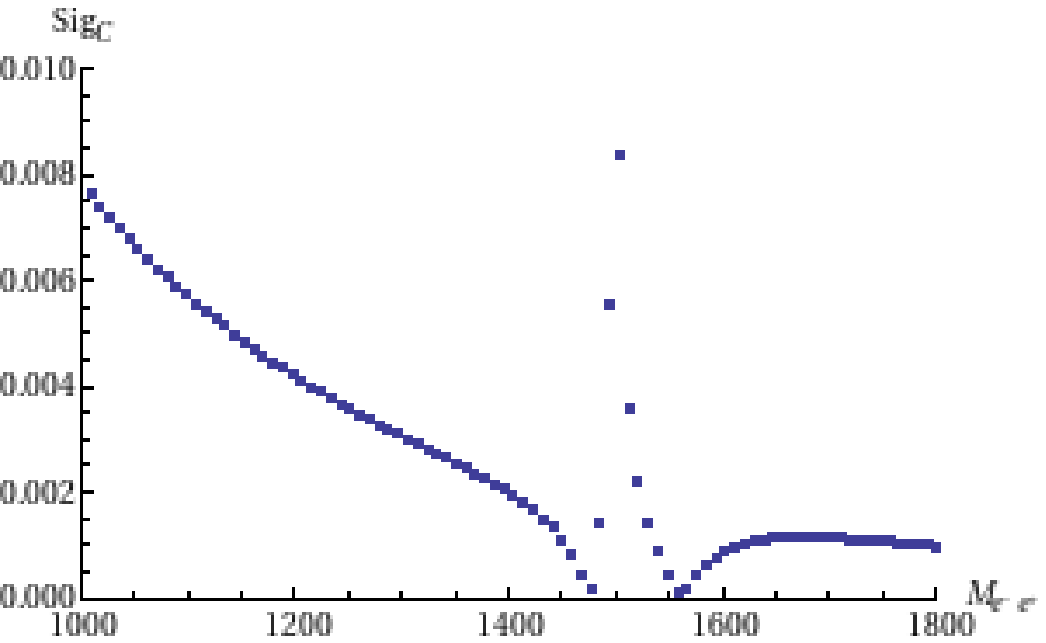}
\caption{Asymmetry (left) and significance (right) in MiAUSSM  around the peak of the Z'. $Y_C$ is set to 0.8, while the
anomalous charges are fixed to the general values $Q_L=1$, $Q_{H_u}=1/2$ and $Q_Q=3/4$}
\label{acs}
\end{figure} The figure \ref{acpeak} shows the dependence of the on peak asymmetry and the significance 
versus the cut $Y_{cut}$. It is clear that the significance has a maximum around $Y_{cut}=0.8$. Using this 
result we have obtained the figure \ref{acs}, in wich we keep fixed the cut and study the dependence of the 
asymmetry and significance on the variable $s$ in the range permitted by the on peak region.

%\newpage

\subsubsection{Edge asymmetry with fixed charges: the search of the peak for the significance}
\begin{figure}[h!]
\includegraphics[scale=0.65]{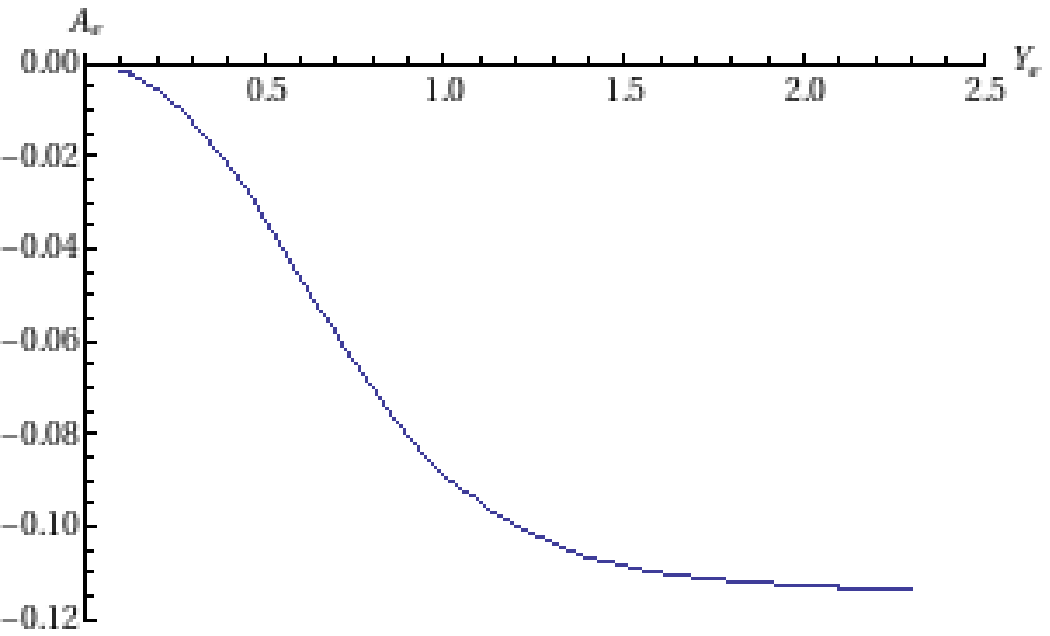}
\includegraphics[scale=0.65]{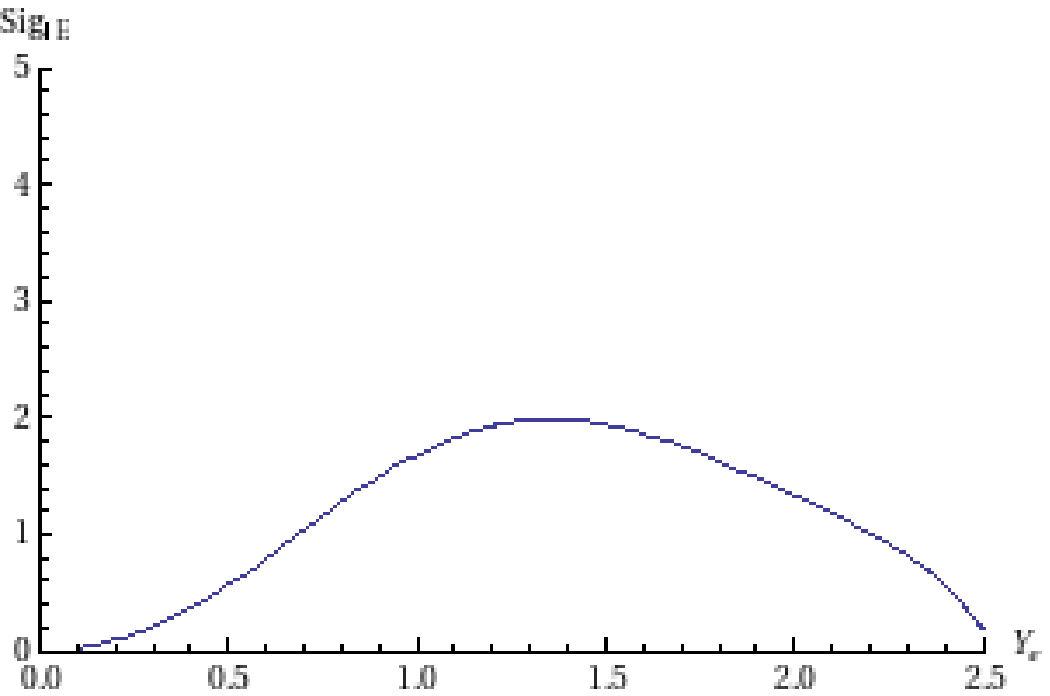}
\caption{On peak ($M_{Z'}-3\Gamma_{Z'}<M_{e^+e^-}<M_{Z'}+3\Gamma_{Z'}$) asymmetry (left) and significance (right) 
versus the cut on  the longitudinal momentum $p_{z,cut}$. The anomalous charges are fixed to the general values $Q_L=1$, $Q_{H_u}=1/2$ and $Q_Q=3/4$. The peak for the significance is for $Y_C=1.4$. The position of the peak does not depend on the choice of the values for the charges, while the intensity of asymmetry and significance depends on it. }
\label{aepeak}
\end{figure}
\begin{figure}[h!]
\includegraphics[scale=0.65]{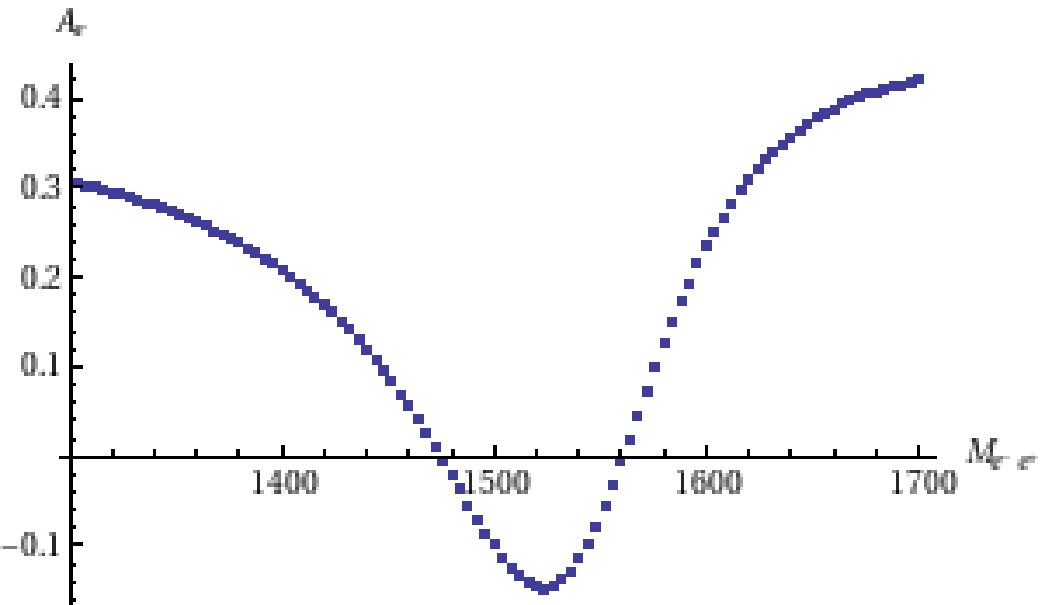}
\includegraphics[scale=0.65]{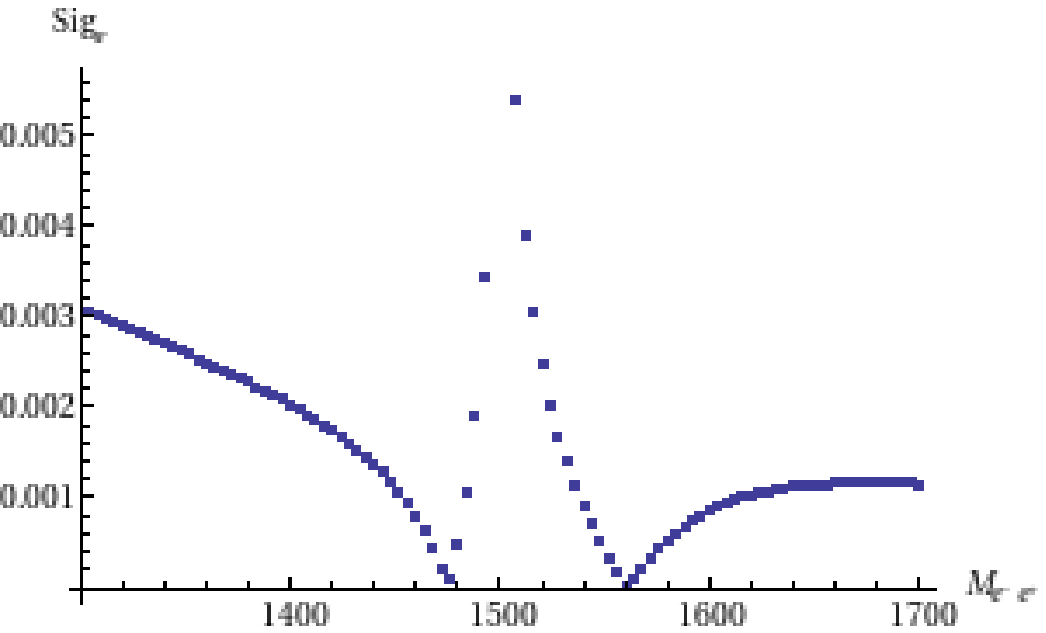}
\caption{Asymmetry (left) and significance (right) in MiAUSSM  around the peak of the Z'. $Y_C$ is set to 1.4, while the
anomalous charges are fixed to the general values $Q_L=1$, $Q_{H_u}=1/2$ and $Q_Q=3/4$}
\label{aes}
\end{figure} The figure \ref{aepeak} shows the dependence of the on peak asymmetry and the significance 
versus the cut $Y_{cut}$. It is clear that the significance has a maximum around $Y_{cut}=1.4$. Using this 
result we have obtained the figure \ref{aes}, in wich we keep fixed the cut and study the dependence of the 
asymmetry and significance on the variable $s$ in the range permitted by the on peak region.

%\newpage

\subsection{Dependence of the asymmetry from couples of the charges}
We have already said that the asymmetry depends on the charges of the model. For the MiAUSSM
we know that the free charges are $Q_{H_u},~Q_Q$ and $Q_L$. In the previous section we have found
the cuts that optimize each definition of asymmetry at the LHC. Now, keeping them fixed to those 
values, we want to study the dependence of the asymmetry on the charges.\\
Because we cannot represent on a single diagram a three variable function we chose to study 
each asymmetry with respect to the three possible couples of variables, i.e. $Q_{H_u},~Q_Q$,
$Q_{H_u},~Q_L$ and $Q_Q,~Q_L$.\\
We have studied the value of the four asymmetries
keeping alternatively one of the charges fixed to $0$ and varying the others two from 
$-1$ to $1$. We limit the value in these ranges because in the SM all the
charges are of this order. Furthermore we have found in sec \ref{genres} that
$-1\lesssim Q_{H_u}\lesssim1$, so for simplicity we have used the same region for the other
two charges.\\
In the following subsections we will show the contourplots of
the results for each definition.

\subsubsection{RFB asymmetry with respect to the charges}

\begin{figure}[h!]
 \includegraphics[scale=0.65]{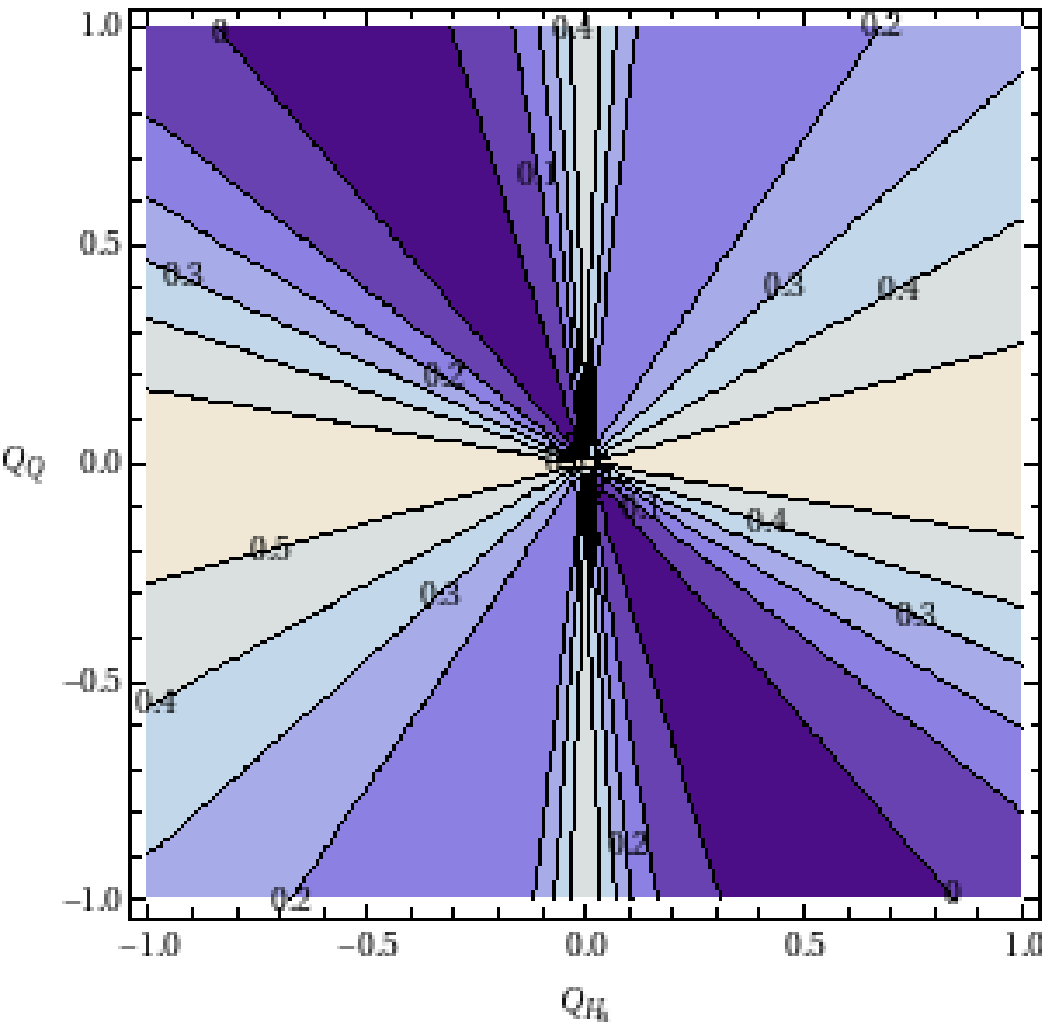}
 \includegraphics[scale=0.65]{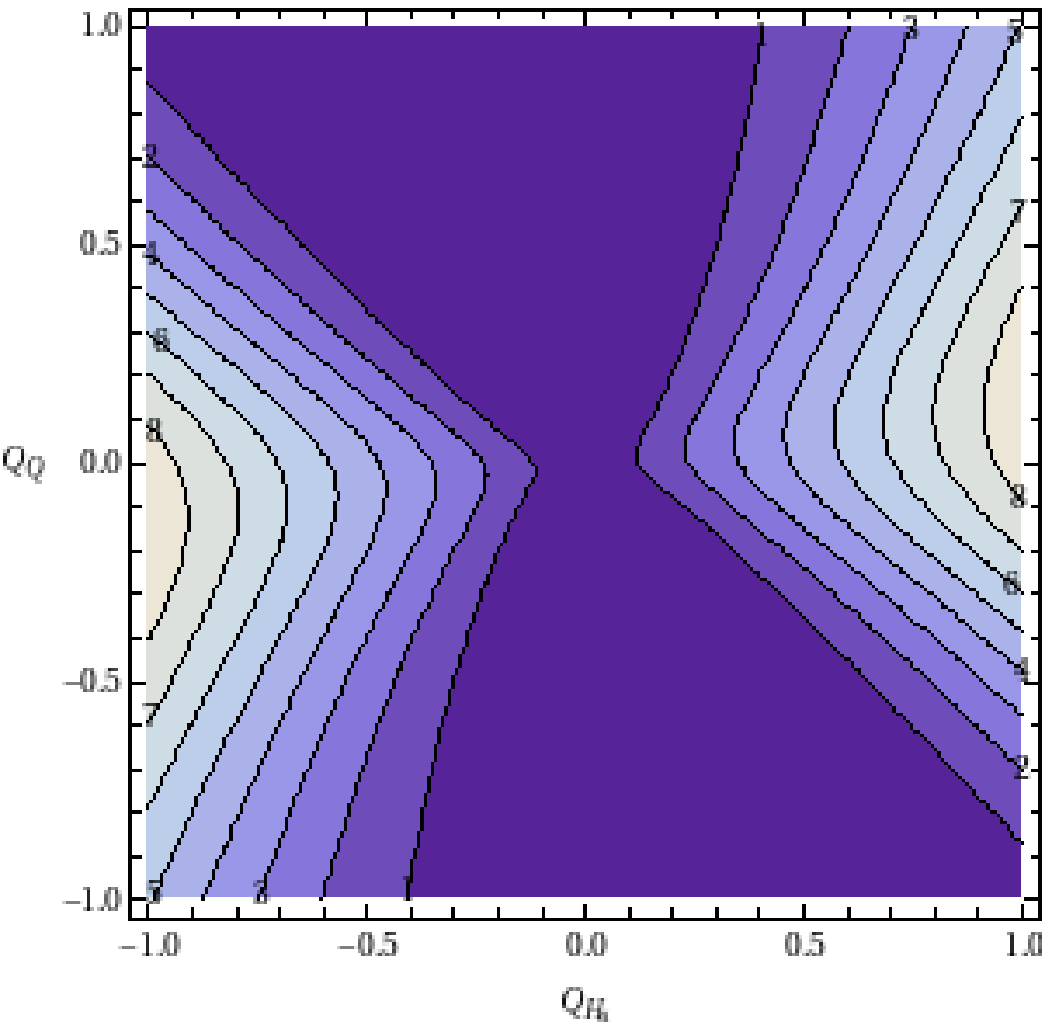}
\caption{FB asymmetry (left) and significance (right) in MiAUSSM versus $Q_{H_u}$ and $Q_Q$ ($Q_L=0$ and $y_{cut}=0.4$)}
\label{arfb1}\end{figure}
\begin{figure}[h!]
 \includegraphics[scale=0.65]{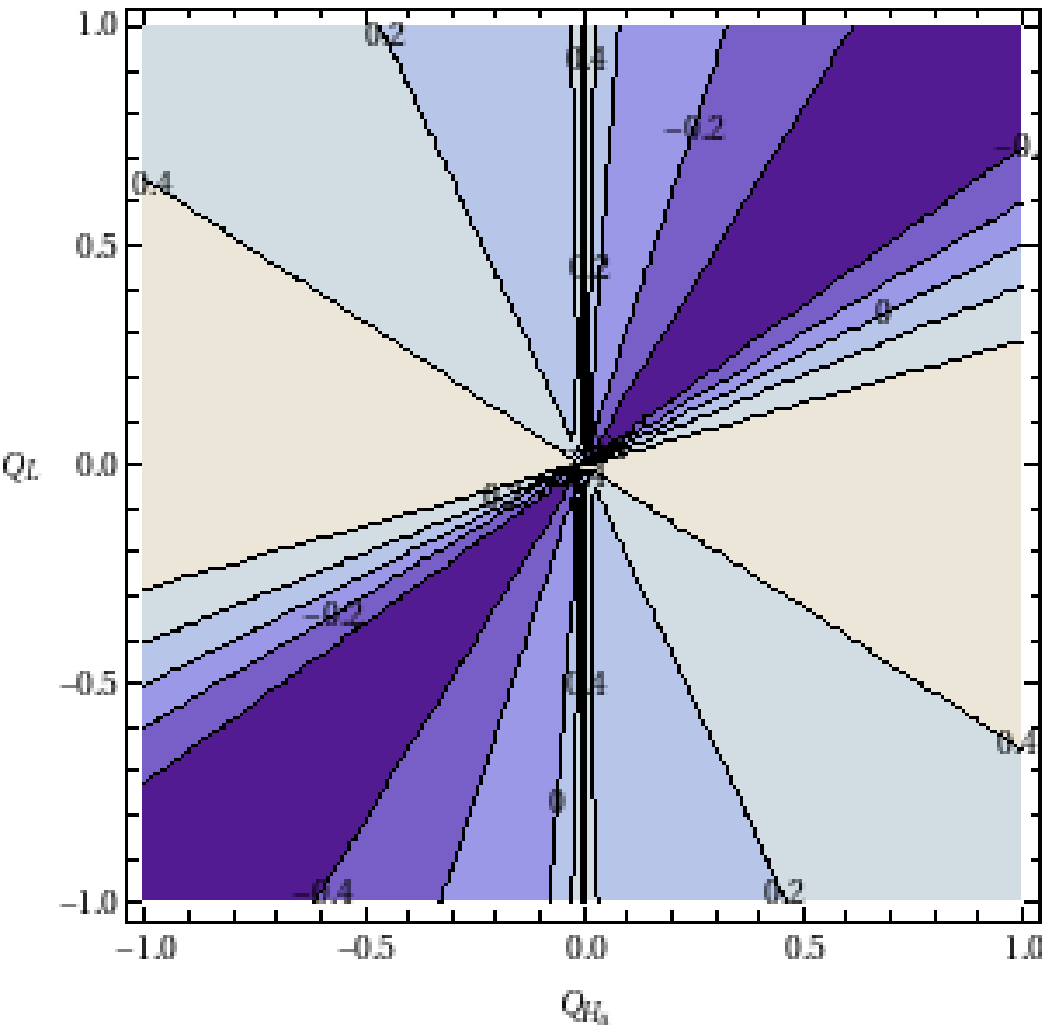}
 \includegraphics[scale=0.65]{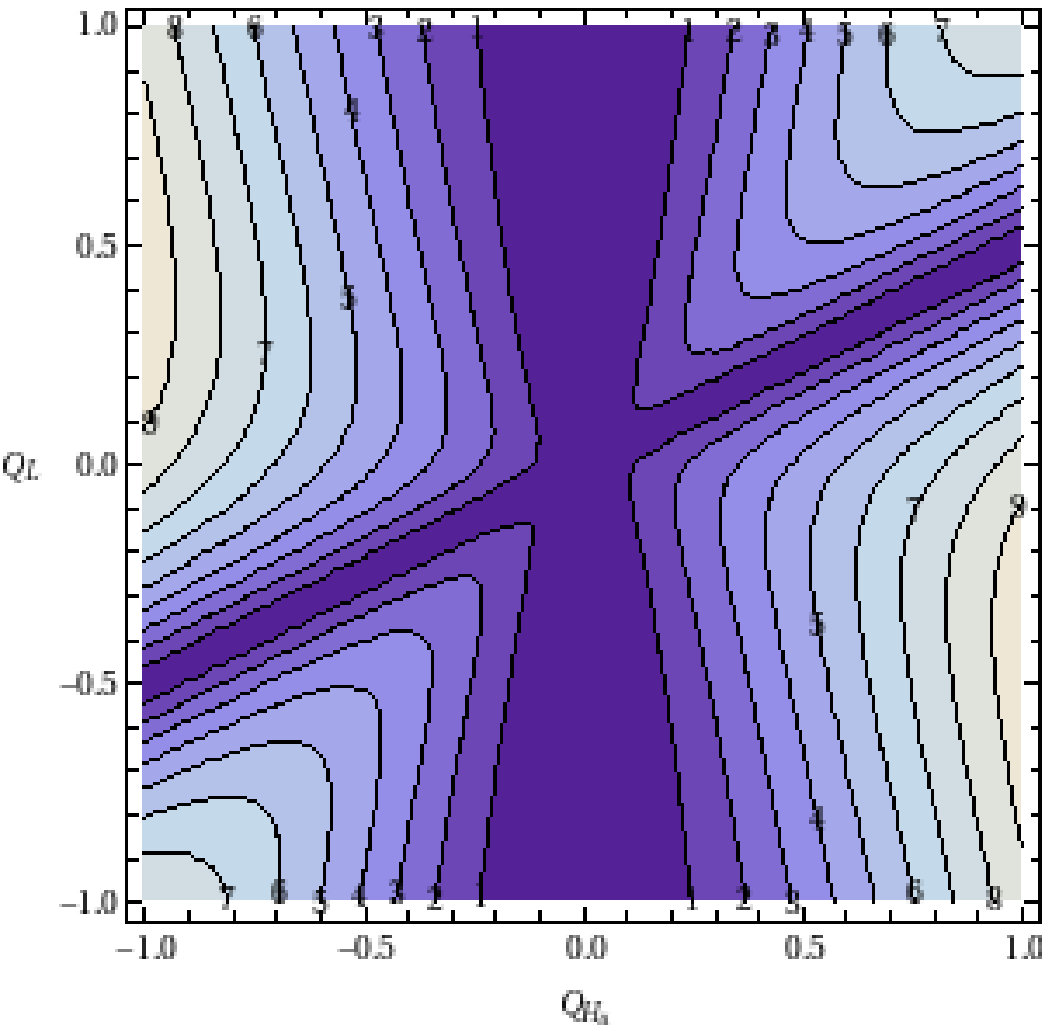}
\caption{FB asymmetry (left) and significance (right) in MiAUSSM versus $Q_{H_u}$ and $Q_L$($Q_Q=0$ and $y_{cut}=0.4$)}
\label{arfb2}\end{figure}

\newpage

\begin{figure}[h!]
 \includegraphics[scale=0.65]{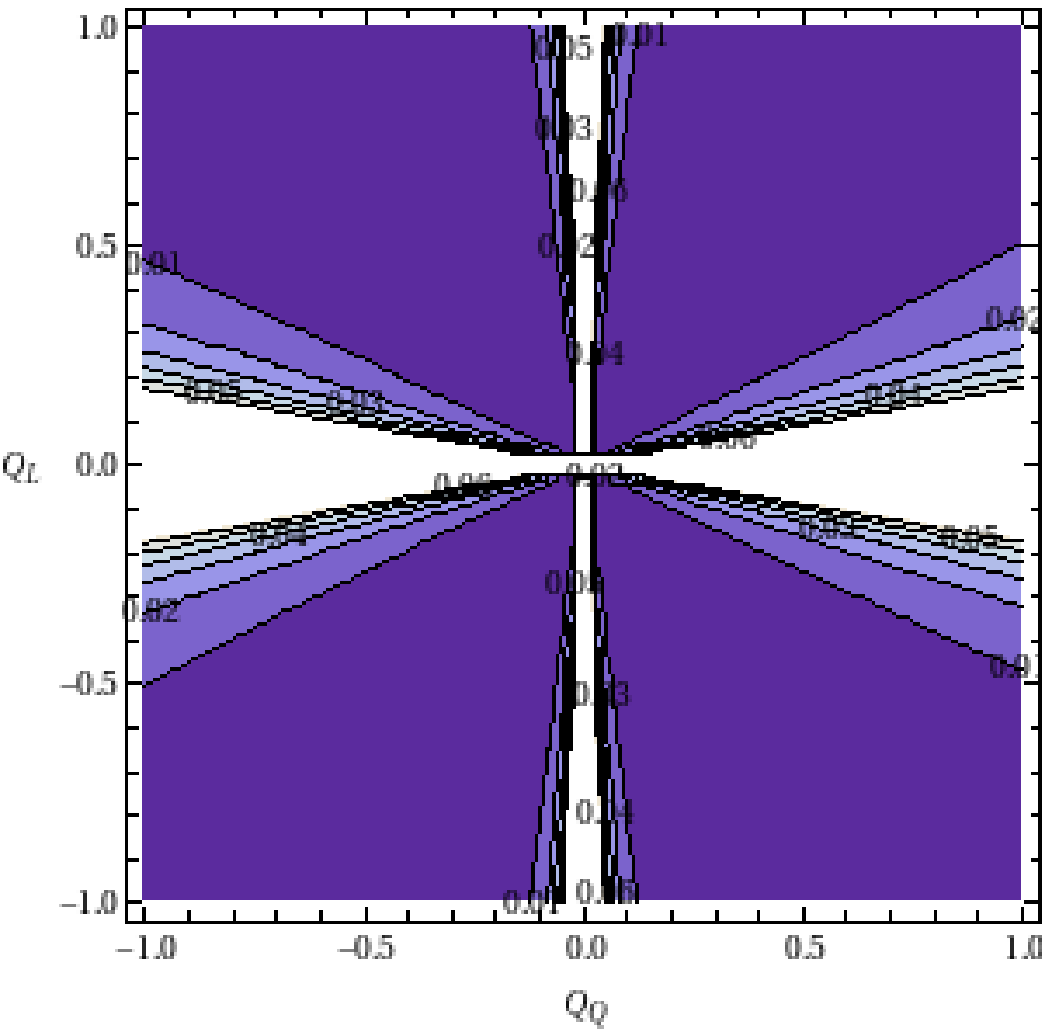}
 \includegraphics[scale=0.65]{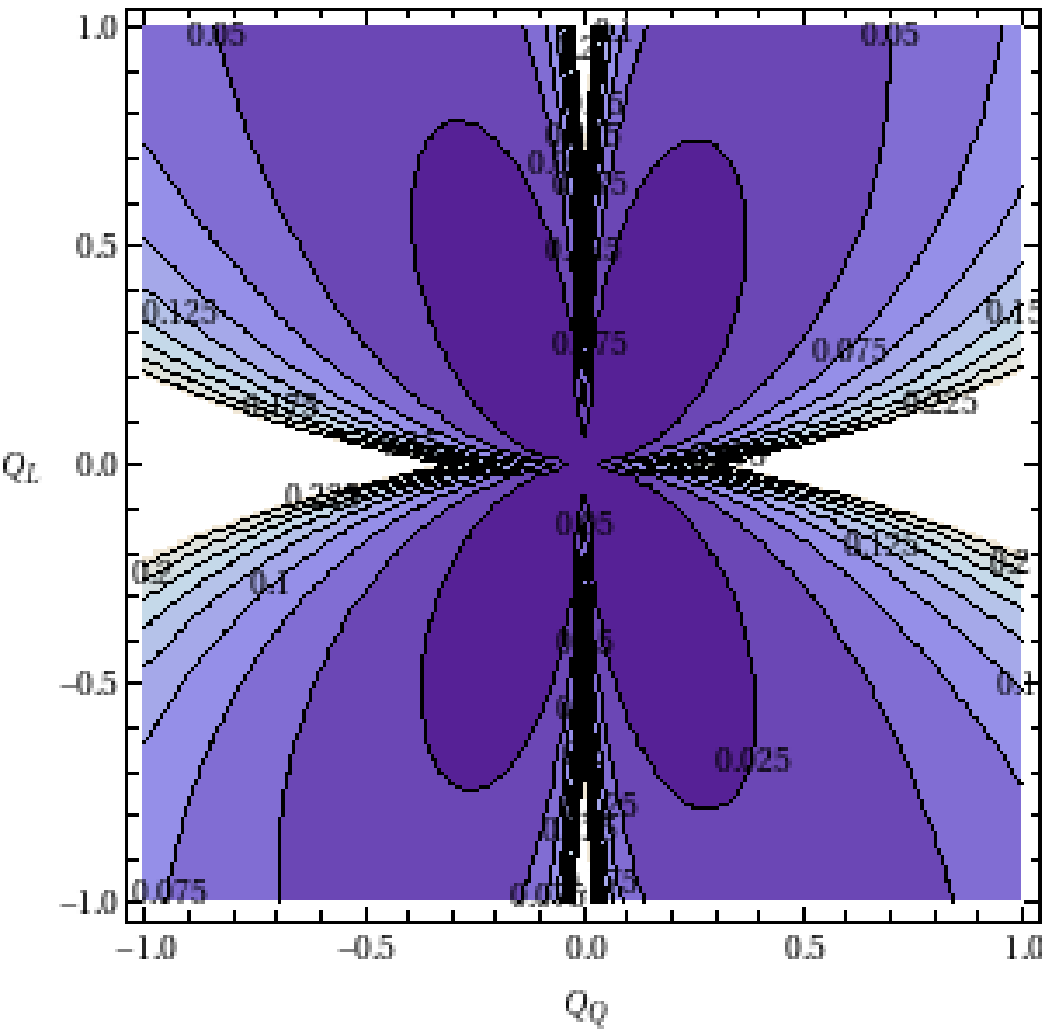}
\caption{FB asymmetry (left) and significance (right) in MiAUSSM versus $Q_Q$ and $Q_L$($Q_{H_u}=0$ and $y_{cut}=0.4$)}
\label{arfb3}\end{figure} In figures \ref{arfb1}, \ref{arfb2} and \ref{arfb3} we can see that $A_{RFB}$ 
is even for the exchanges $(Q_{H_u},~Q_Q,~0)\rightarrow (-Q_{H_u},~-Q_Q,~0)$,
$(Q_{H_u},~0,~Q_L)\rightarrow (-Q_{H_u},~0,~-Q_L)$, $(0,~Q_Q,~Q_L)\rightarrow (0,~-Q_Q,~Q_L)$ and 
$(0,~Q_Q,~Q_L)\rightarrow (0,~Q_Q,~-Q_L)$.

\newpage

\subsubsection{One-side asymmetry with respect to the charges}

\begin{figure}[h!]
 \includegraphics[scale=0.65]{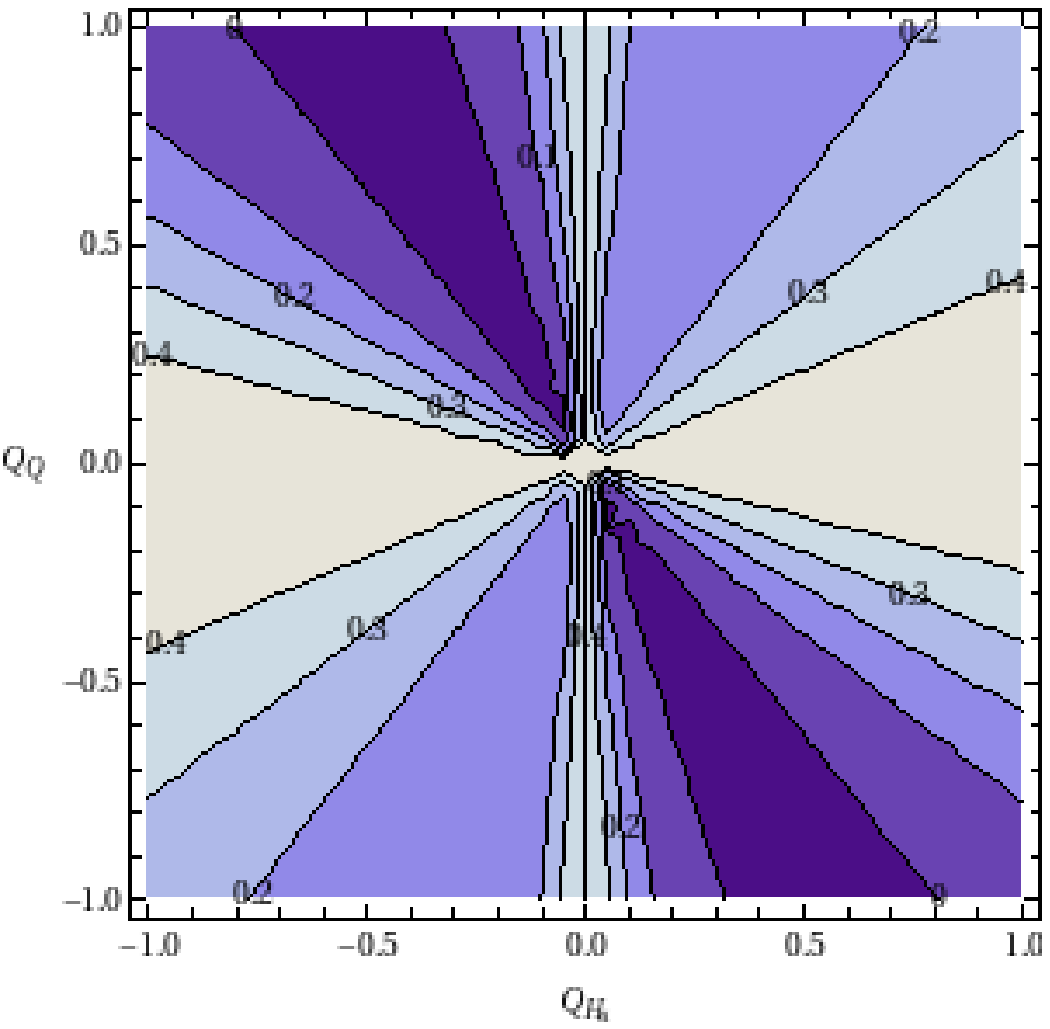}
 \includegraphics[scale=0.65]{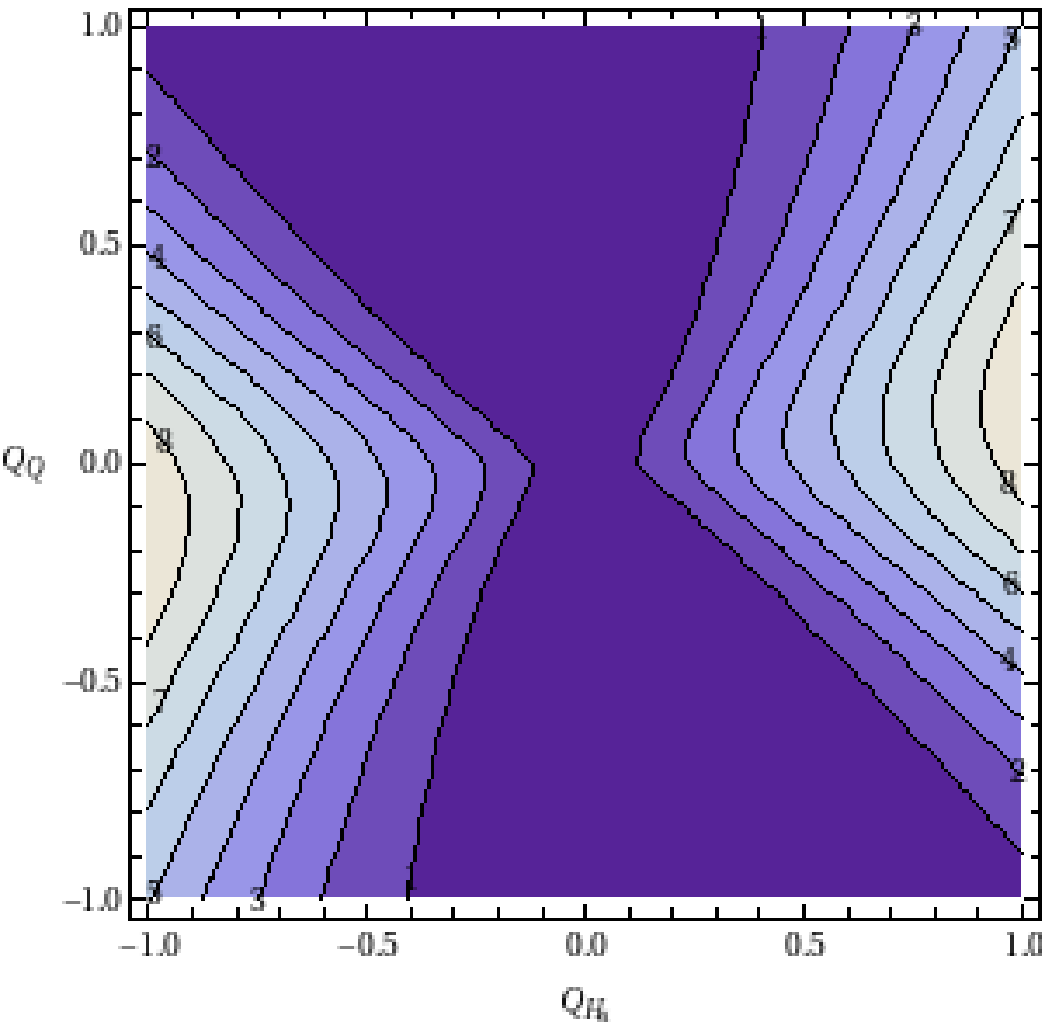}
\caption{One-side asymmetry (left) and significance (right) in MiAUSSM versus $Q_{H_u}$ and $Q_Q$($Q_L=0$ and $p_{z_{cut}}= 580$ GeV)}
\label{ao1}\end{figure}
\begin{figure}[h!]
 \includegraphics[scale=0.65]{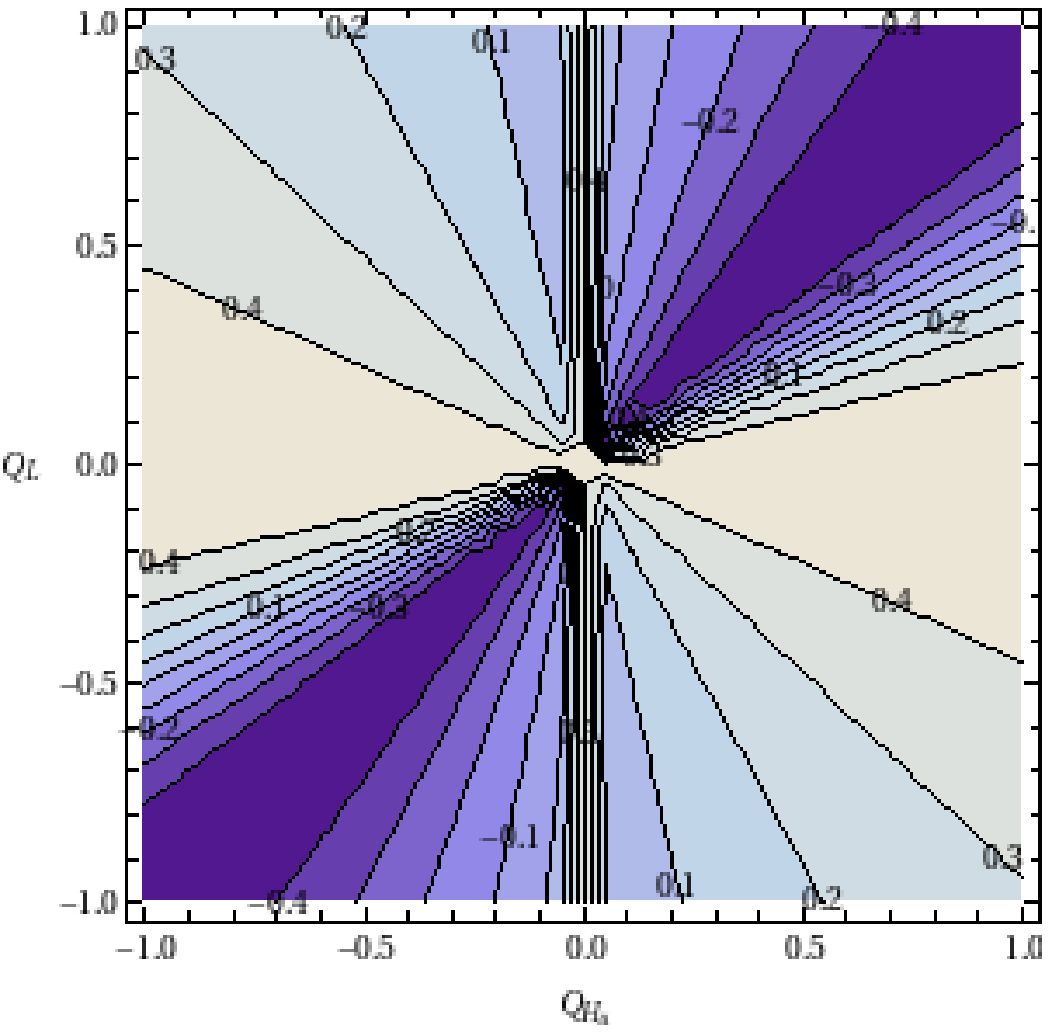}
 \includegraphics[scale=0.65]{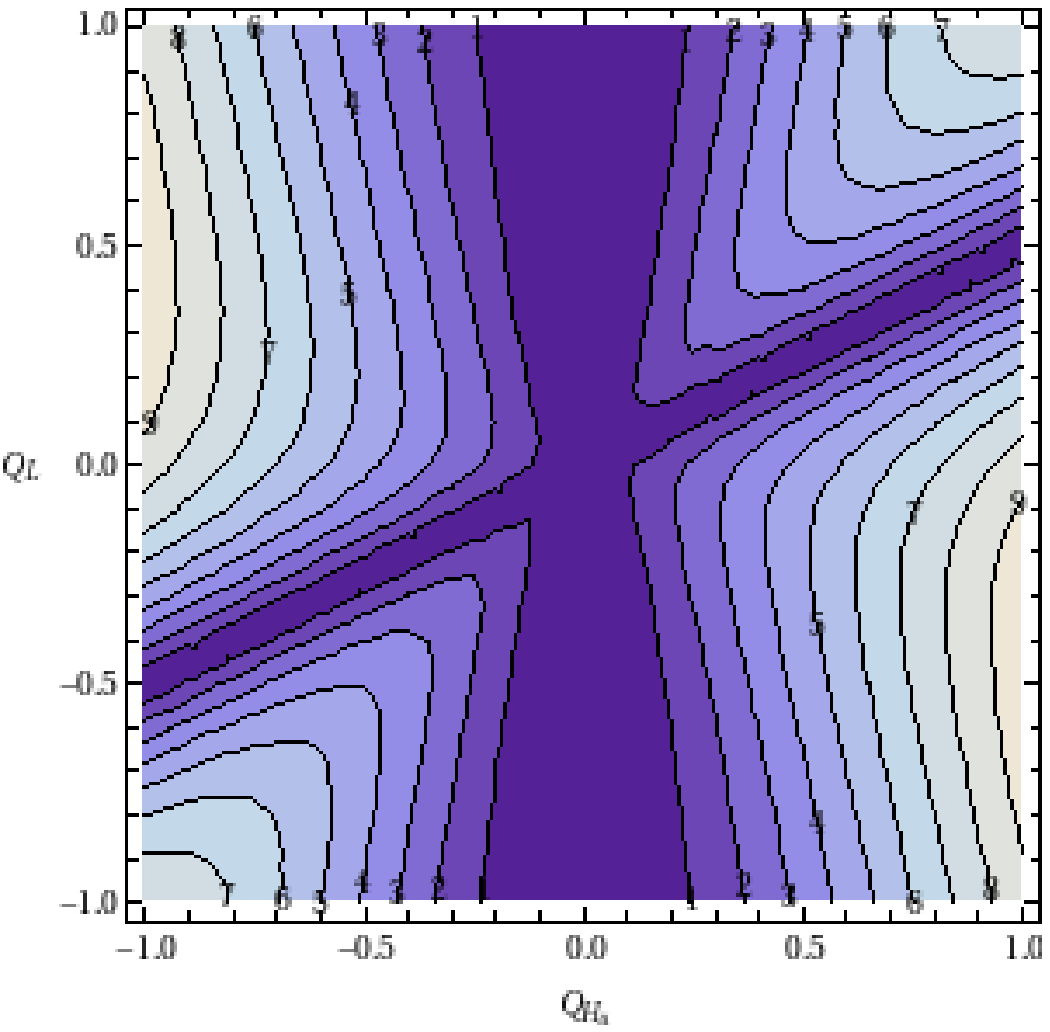}
\caption{One-side asymmetry (left) and significance (right) in MiAUSSM versus $Q_{H_u}$ and $Q_L$($Q_Q=0$ and $p_{z_{cut}}=580$ GeV)}
\label{ao2}\end{figure}
\begin{figure}[h!]
 \includegraphics[scale=0.65]{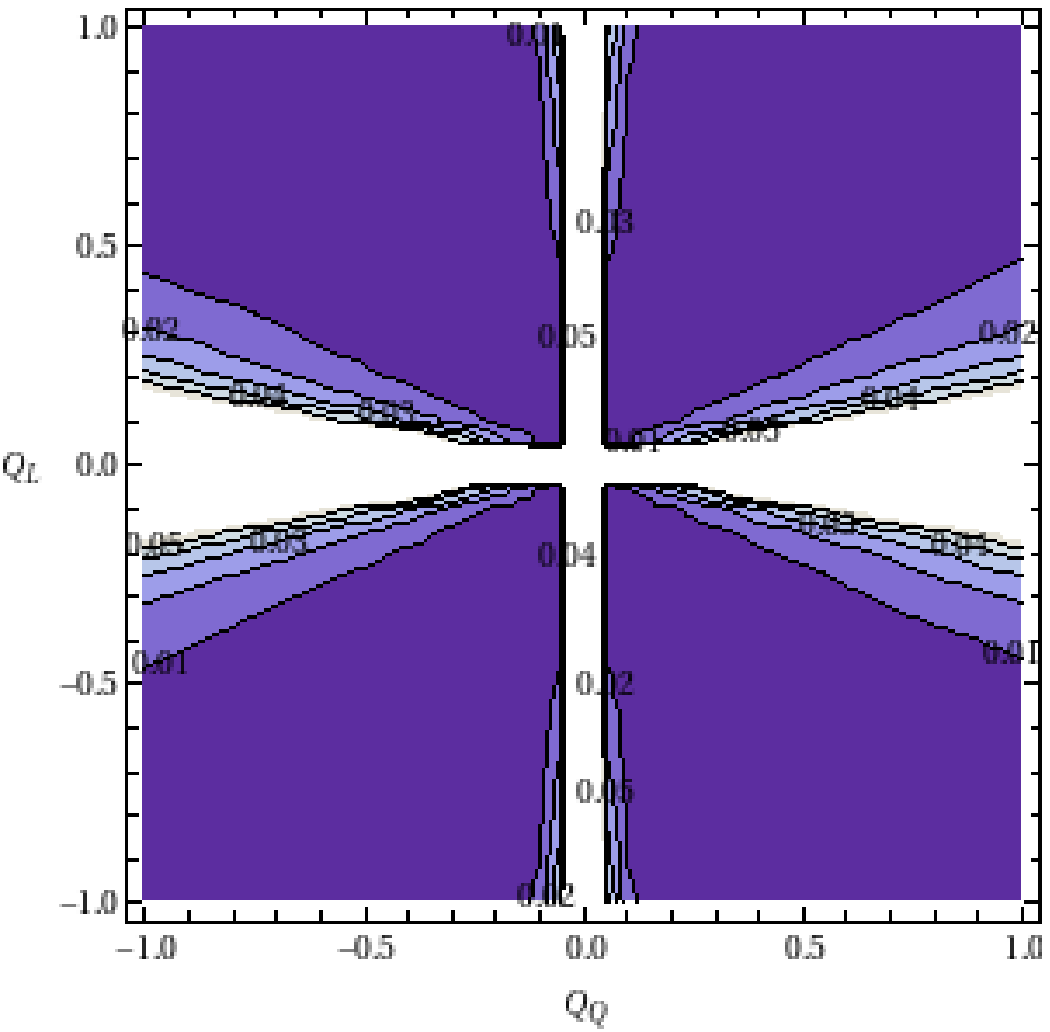}
 \includegraphics[scale=0.65]{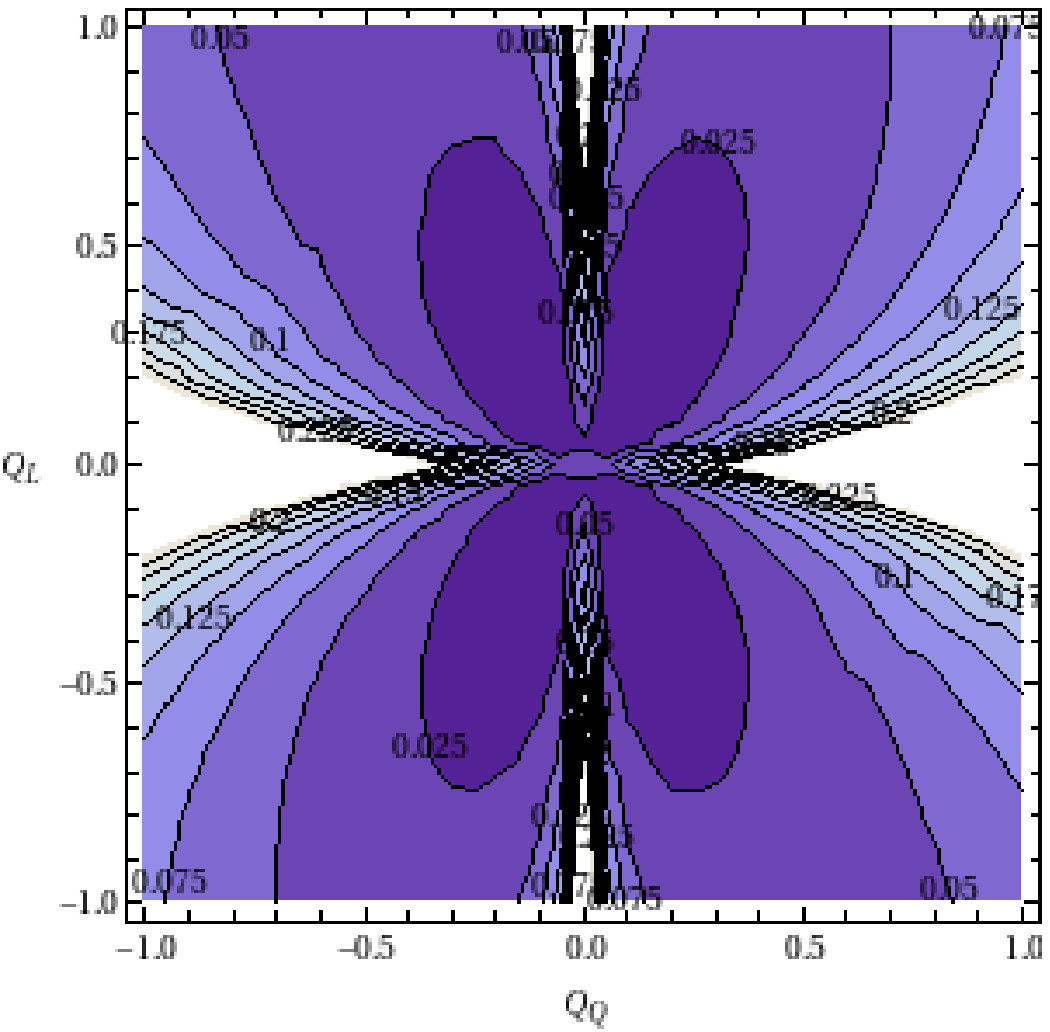}
\caption{One-side asymmetry (left) and significance (right) in MiAUSSM versus $Q_Q$ and $Q_L$($Q_{H_u}=0$ and $p_{z_{cut}}=580$ GeV)}
\label{ao3}\end{figure} In figures \ref{ao1}, \ref{ao2} and \ref{ao3}we can see that $A_{O}$ 
is even for the exchanges $(Q_{H_u},~Q_Q,~0)\rightarrow (-Q_{H_u},~-Q_Q,~0)$,
$(Q_{H_u},~0,~Q_L)\rightarrow (-Q_{H_u},~0,~-Q_L)$, $(0,~Q_Q,~Q_L)\rightarrow (0,~-Q_Q,~Q_L)$ and 
$(0,~Q_Q,~Q_L)\rightarrow (0,~Q_Q,~-Q_L)$.

\newpage

\subsubsection{Central asymmetry with respect to the charges}

\begin{figure}[h!]
 \includegraphics[scale=0.65]{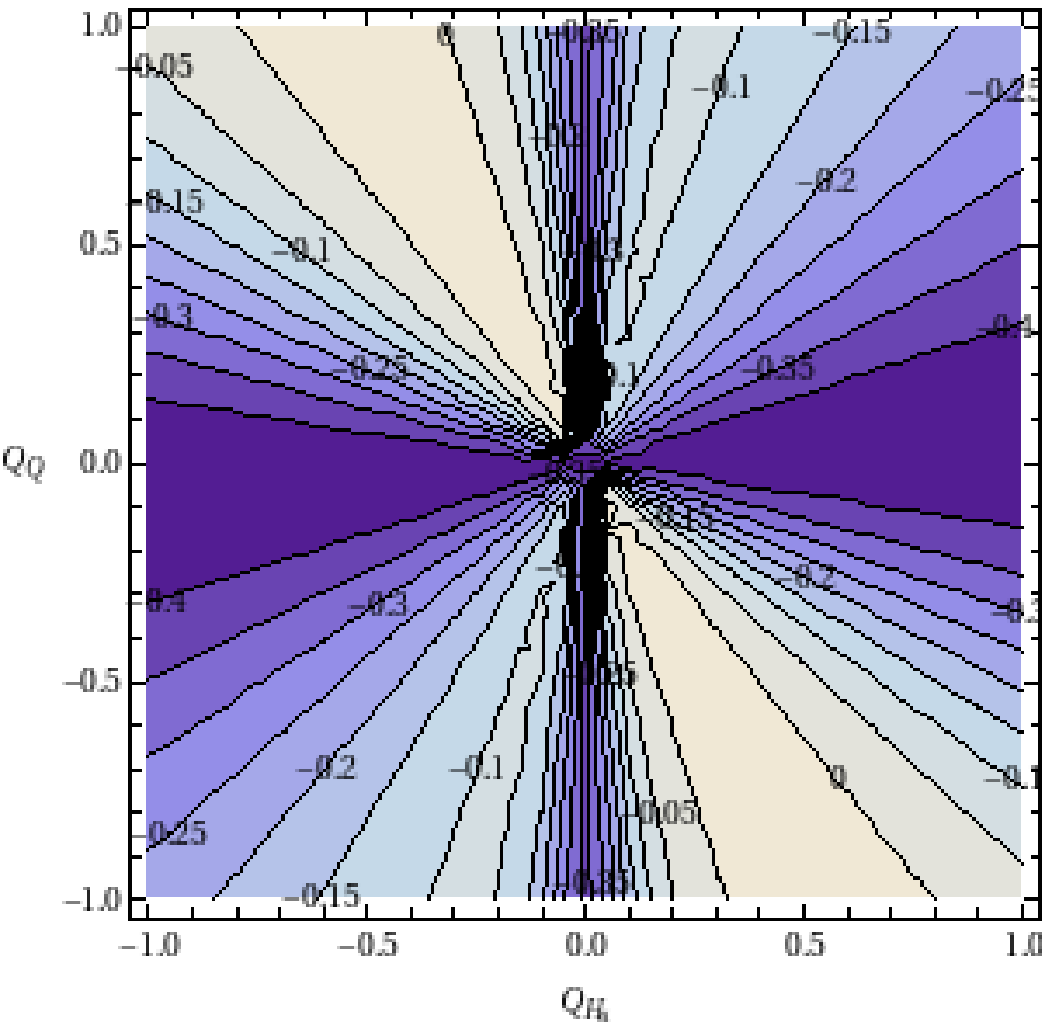}
 \includegraphics[scale=0.65]{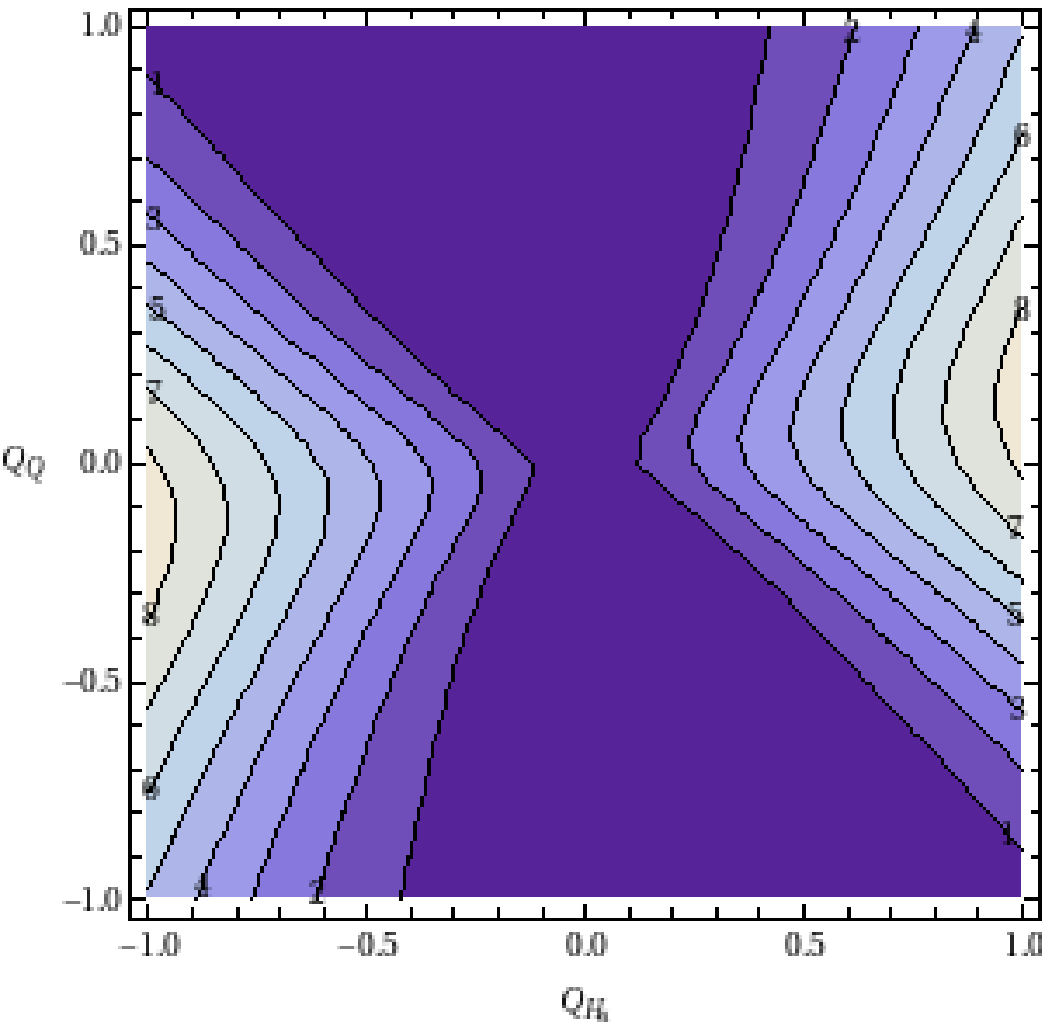}
\caption{Central asymmetry (left) and significance (right) in MiAUSSM versus $Q_{H_u}$ and $Q_Q$($Q_L=0$ and $Y_{C}=0.8$)}
\label{ac1}\end{figure}

\newpage

\begin{figure}[h!]
 \includegraphics[scale=0.65]{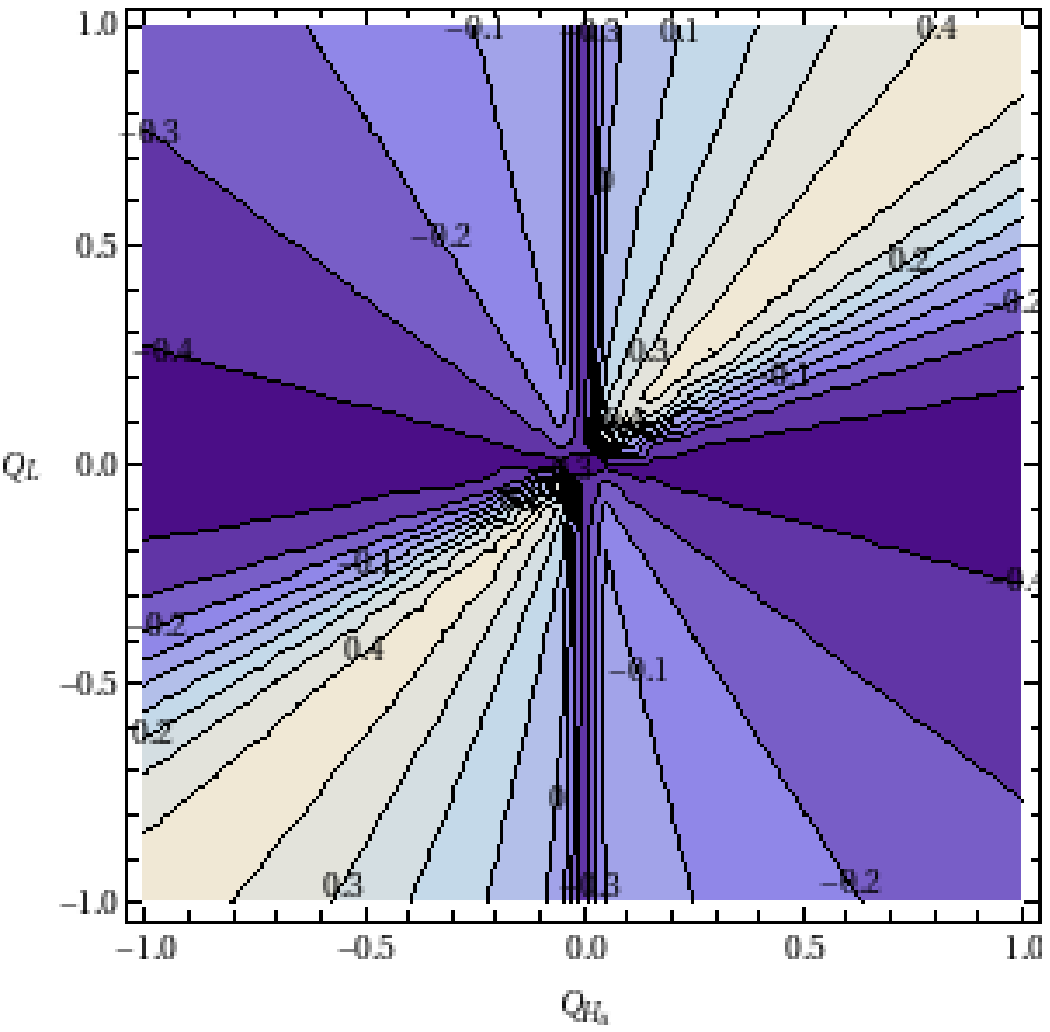}
 \includegraphics[scale=0.65]{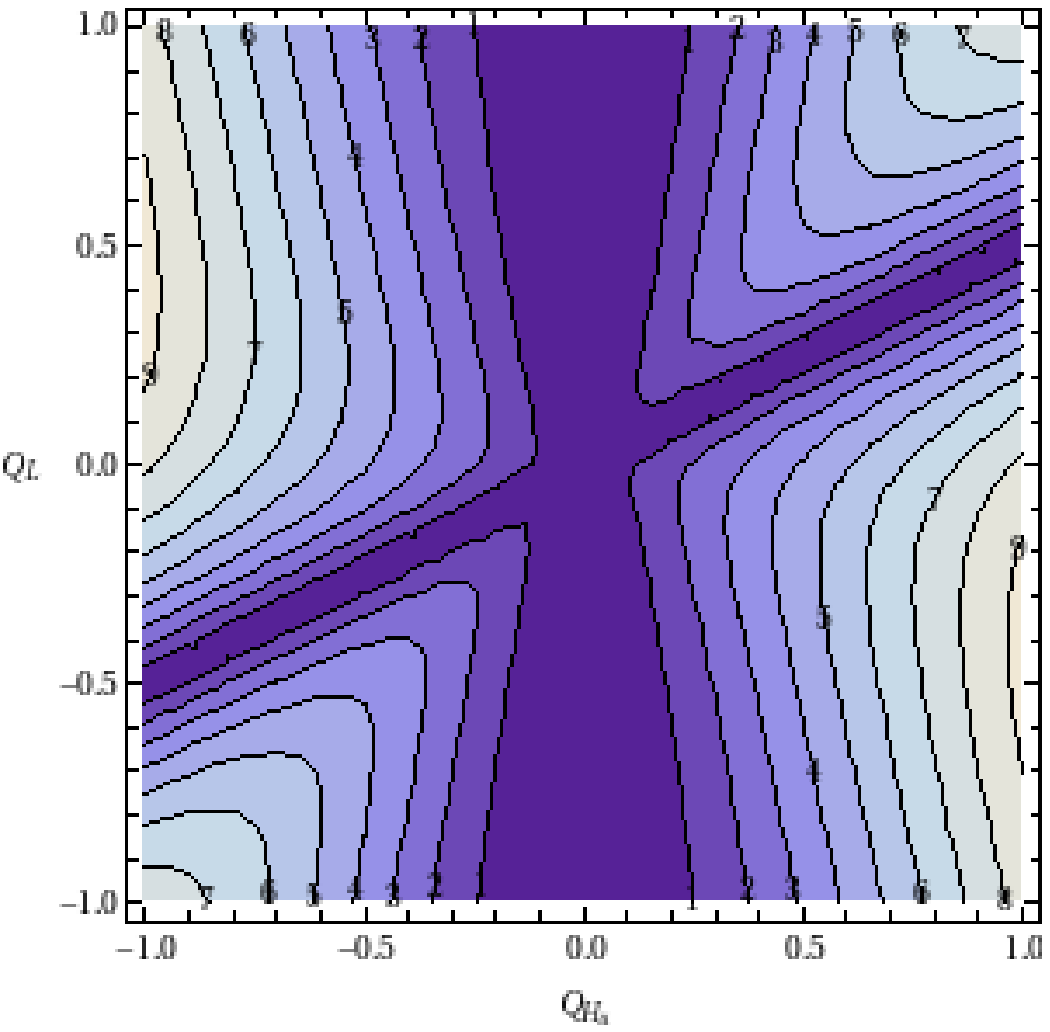}
\caption{Central asymmetry (left) and significance (right) in MiAUSSM versus $Q_{H_u}$ and $Q_L$($Q_Q=0$ and $Y_{C}=0.8$)}
\label{ac2}\end{figure}
\begin{figure}[h!]
 \includegraphics[scale=0.65]{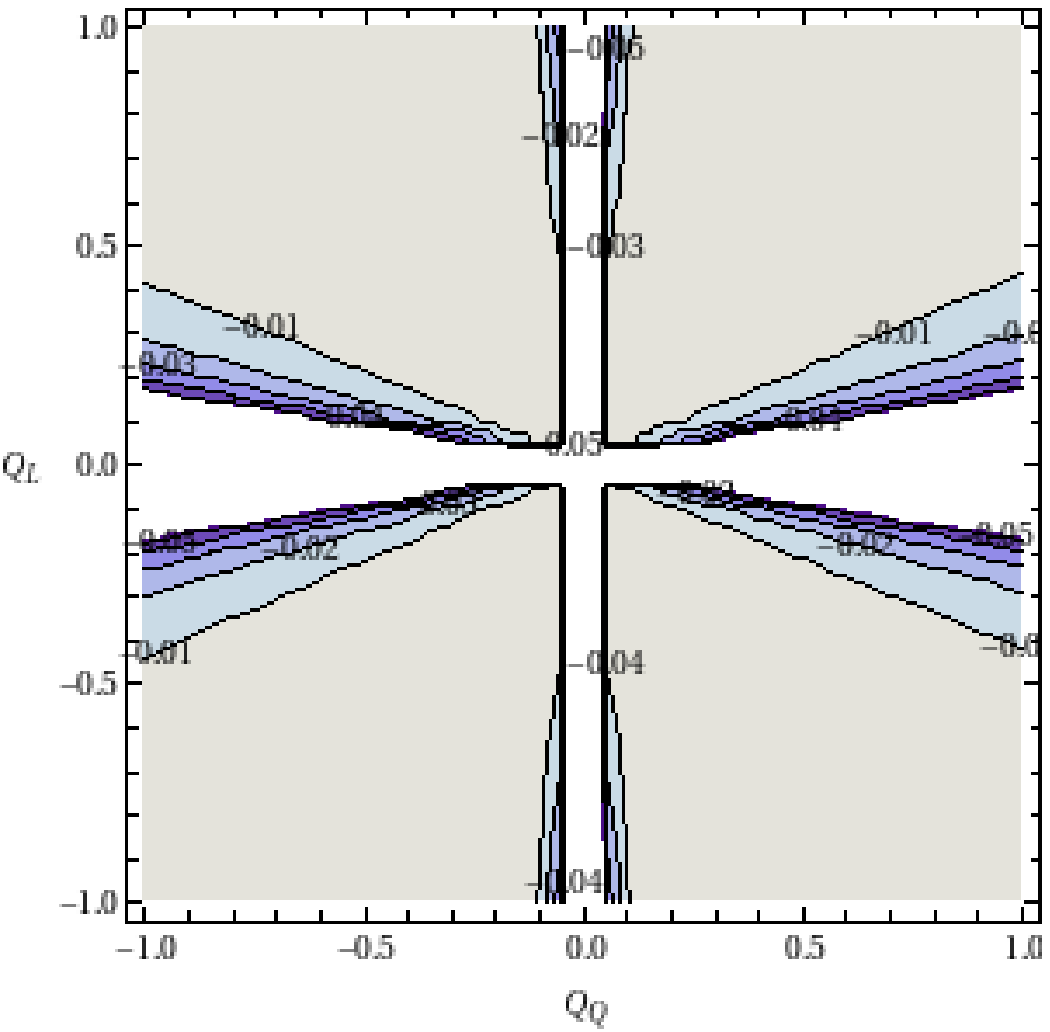}
 \includegraphics[scale=0.65]{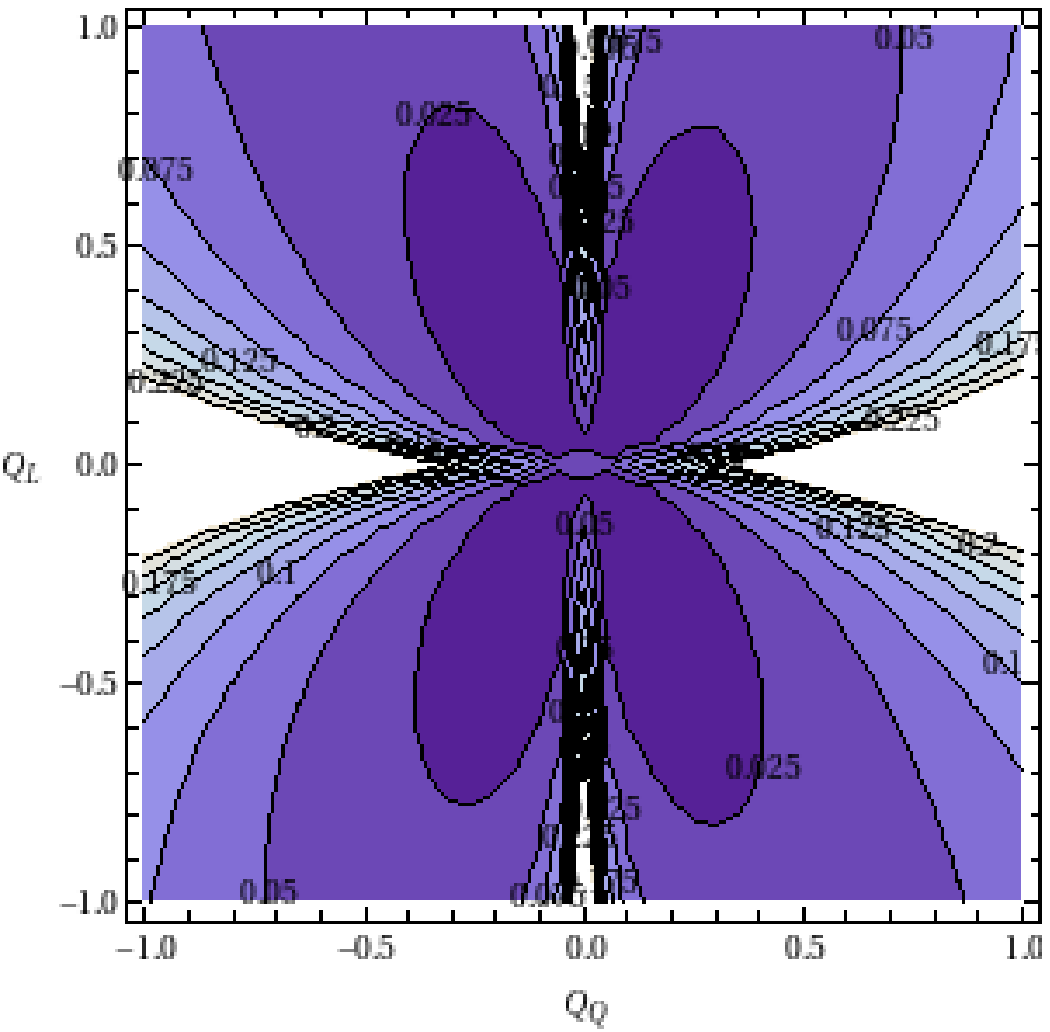}
\caption{Central asymmetry (left) and significance (right) in MiAUSSM versus $Q_Q$ and $Q_L$($Q_{H_u}=0$ and $Y_{C}=0.8$)}
\label{ac3}\end{figure} In figures \ref{ac1}, \ref{ac2} and \ref{ac3} we can see that $A_{C}$ 
is even for the exchanges $(Q_{H_u},~Q_Q,~0)\rightarrow (-Q_{H_u},~-Q_Q,~0)$,
$(Q_{H_u},~0,~Q_L)\rightarrow (-Q_{H_u},~0,~-Q_L)$, $(0,~Q_Q,~Q_L)\rightarrow (0,~-Q_Q,~Q_L)$ and 
$(0,~Q_Q,~Q_L)\rightarrow (0,~Q_Q,~-Q_L)$.

\newpage

\subsubsection{Edge asymmetry with respect to the charges}

\begin{figure}[h!]
 \includegraphics[scale=0.65]{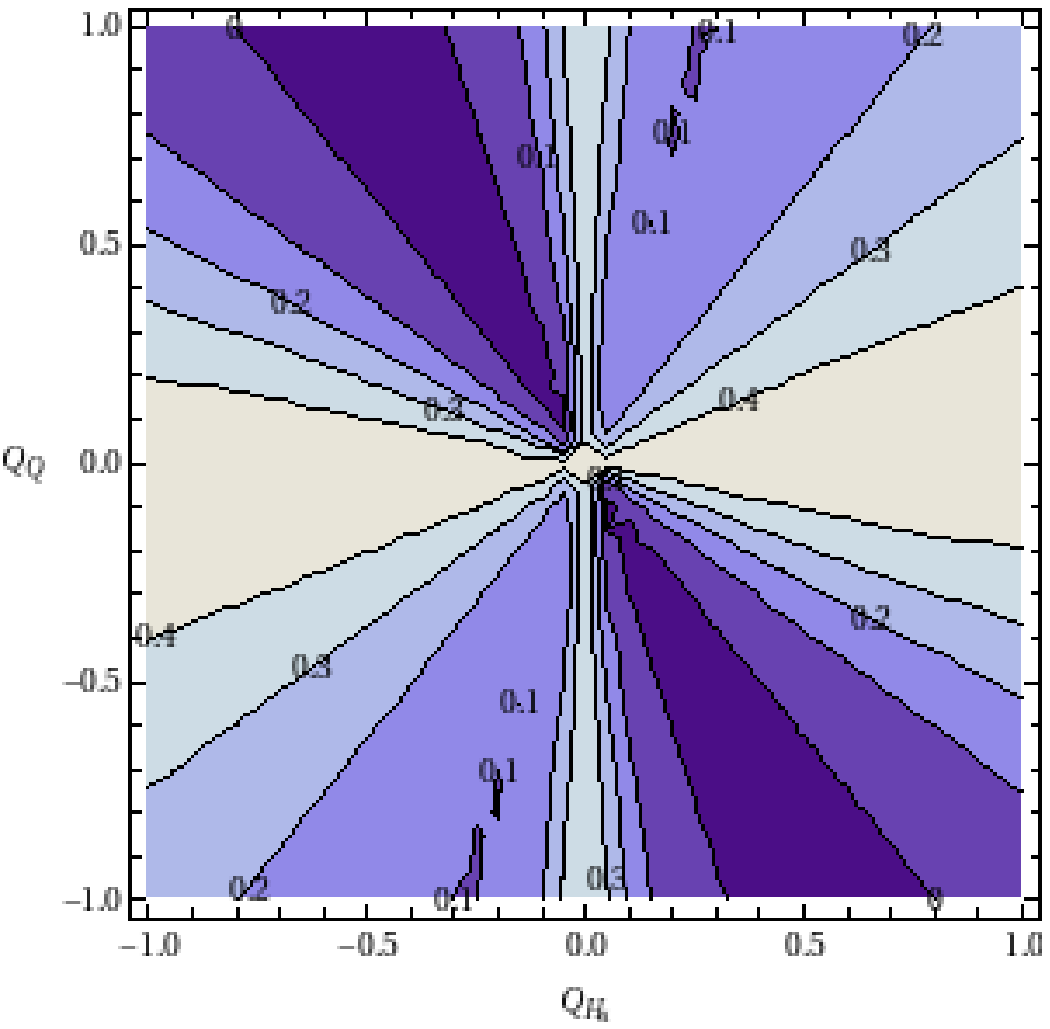}
 \includegraphics[scale=0.65]{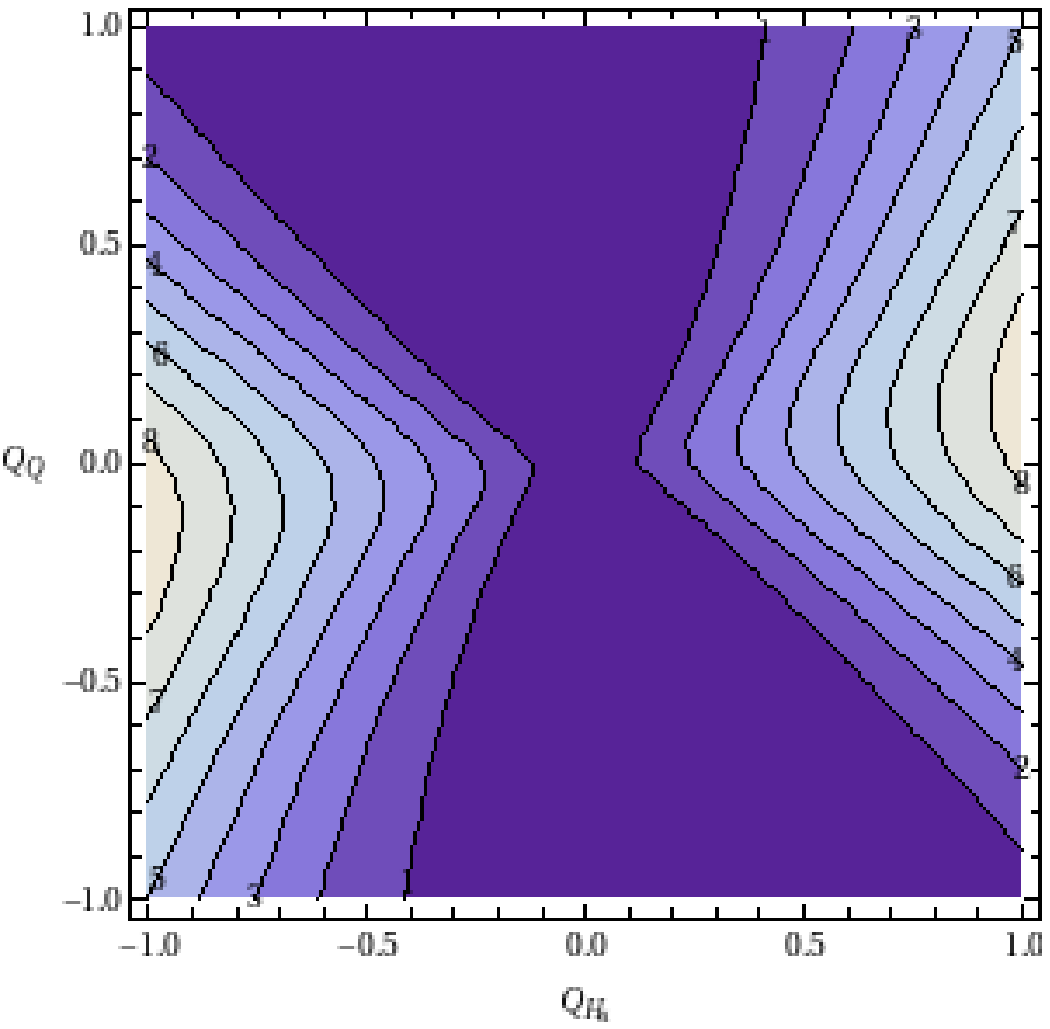}
\caption{Edge asymmetry (left) and significance (right) in MiAUSSM versus $Q_{H_u}$ and $Q_Q$($Q_L=0$ and $Y_{C}=1.4$)}
\label{ae1}\end{figure}
\begin{figure}[h!]
 \includegraphics[scale=0.65]{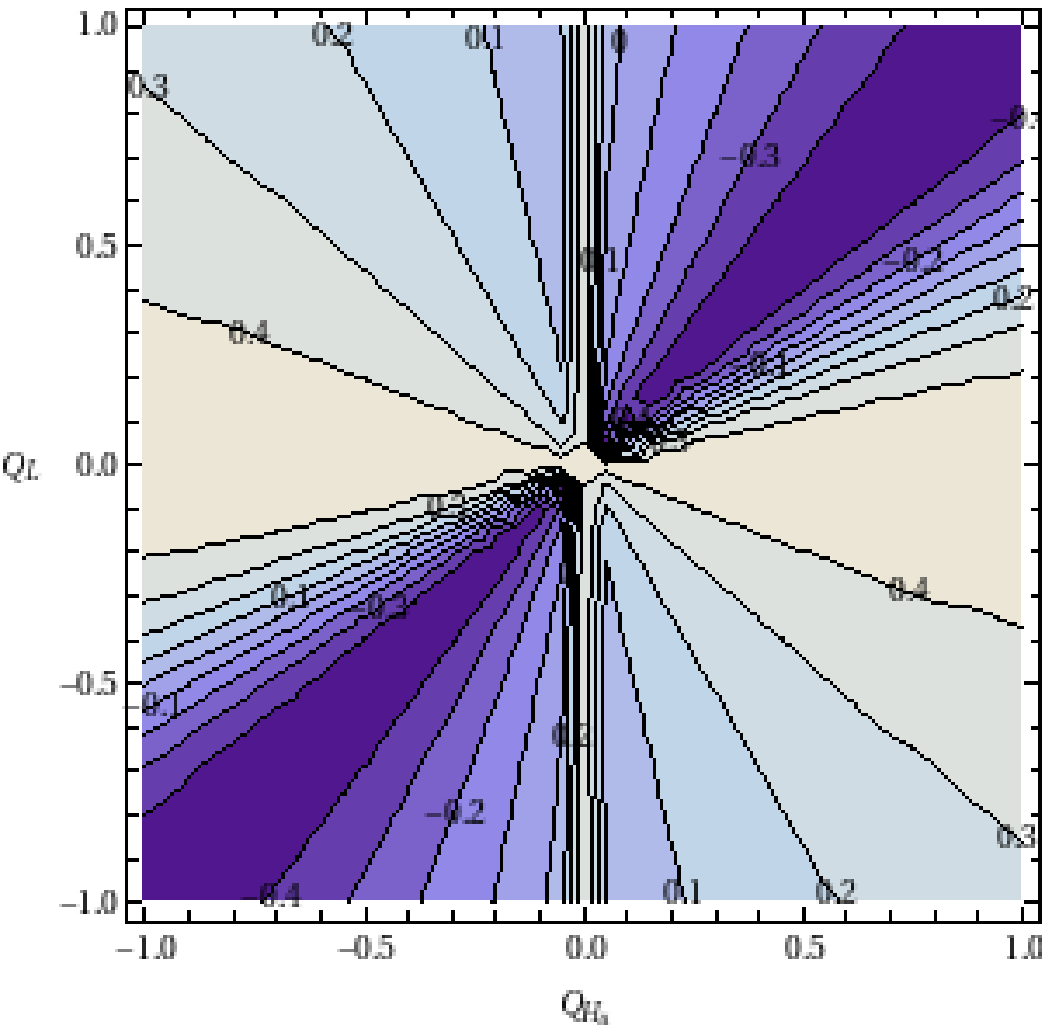}
 \includegraphics[scale=0.65]{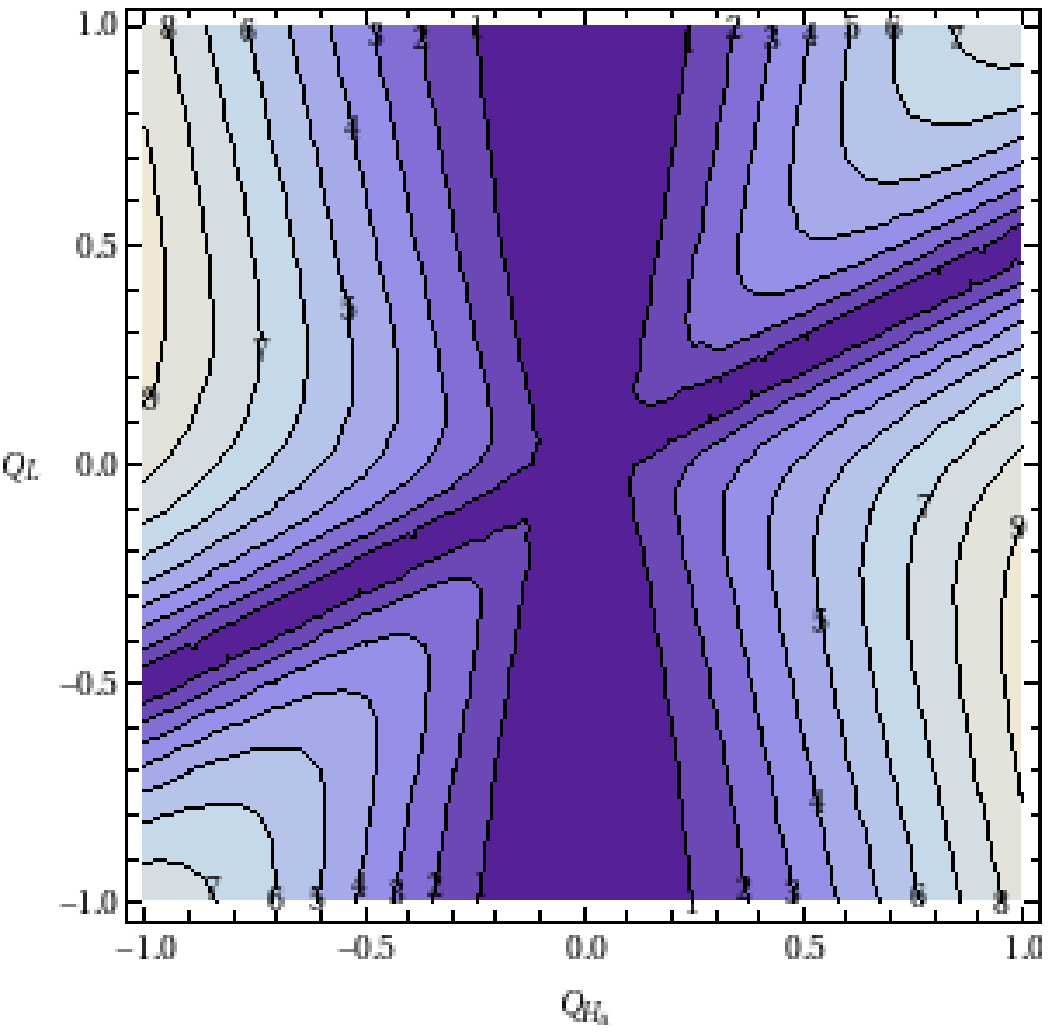}
\caption{Edge asymmetry (left) and significance (right) in MiAUSSM versus $Q_{H_u}$ and $Q_L$($Q_Q=0$ and $Y_{C}=1.4$)}
\label{ae2}\end{figure}

\newpage

\begin{figure}[h!]
 \includegraphics[scale=0.65]{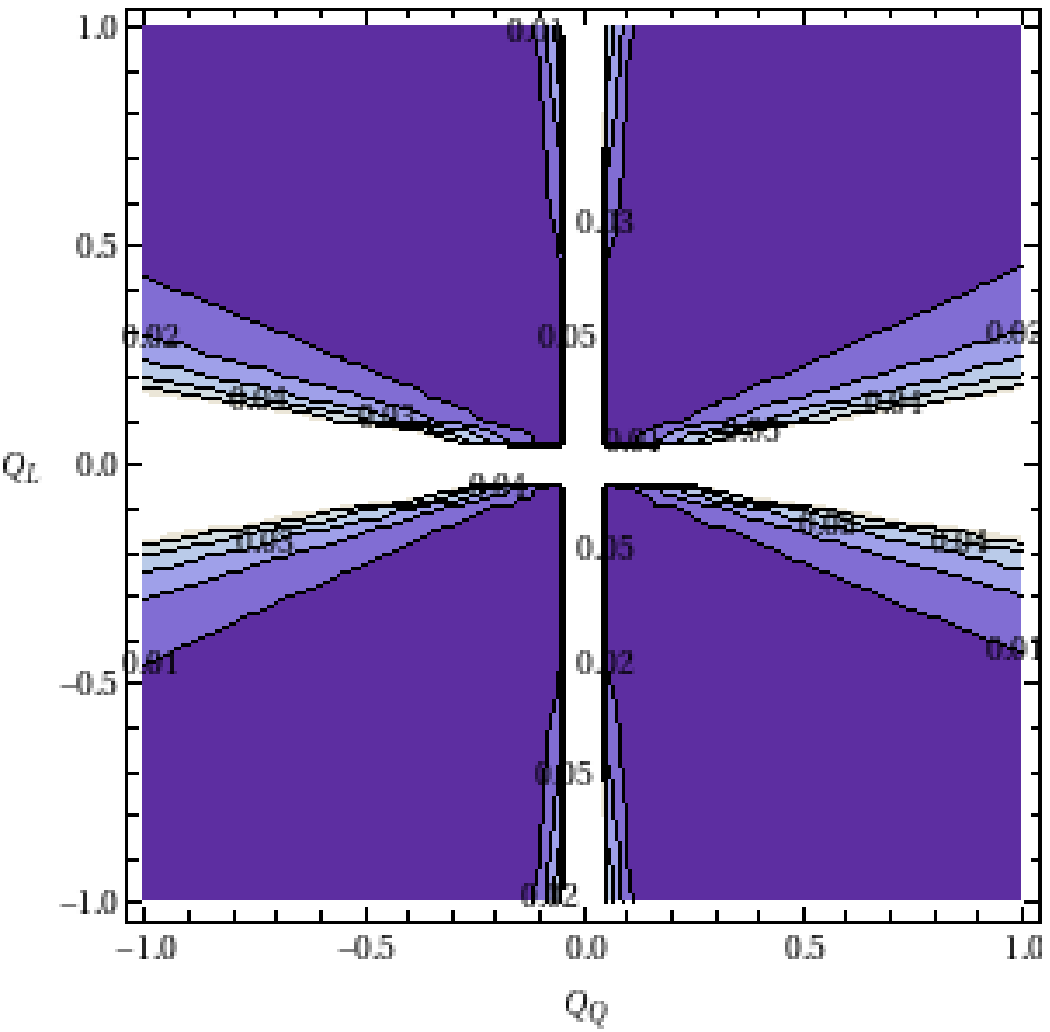}
 \includegraphics[scale=0.65]{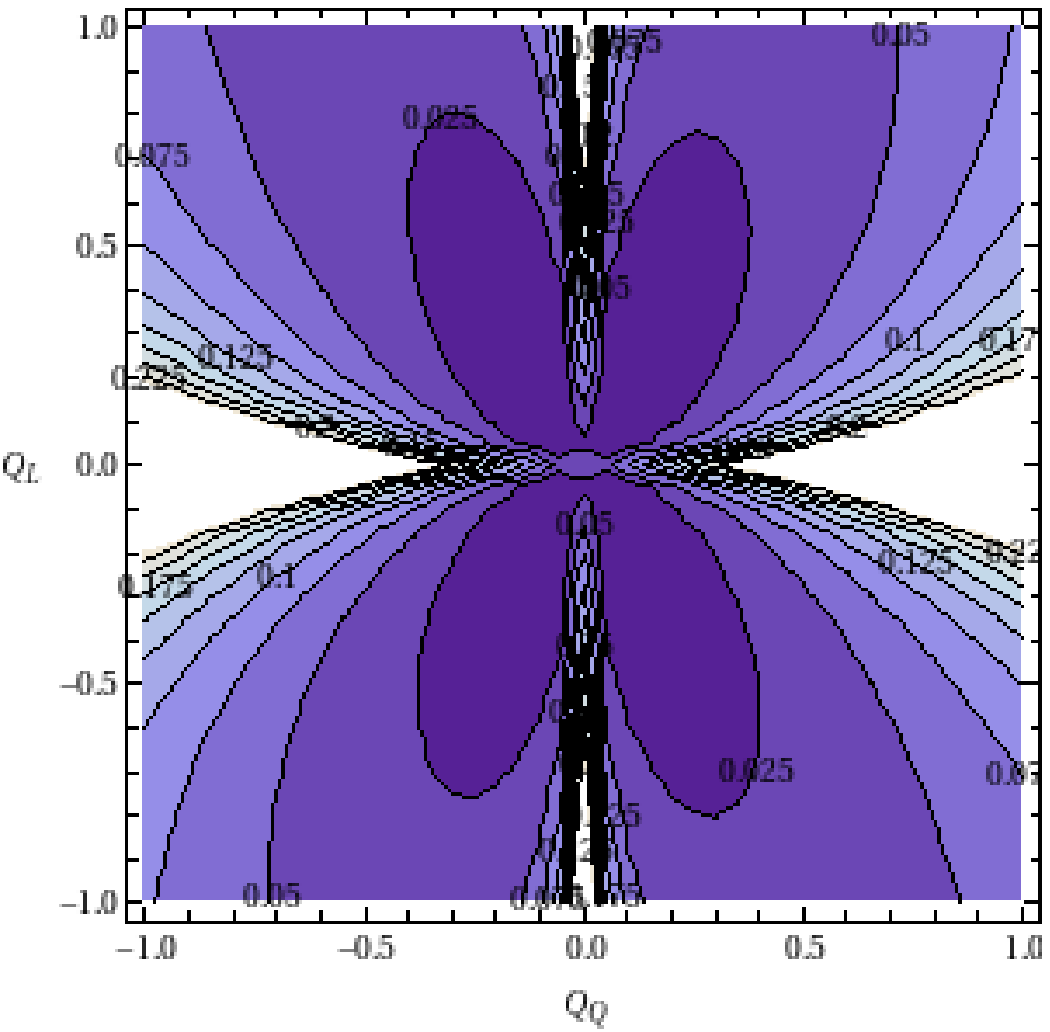}
\caption{Edge asymmetry (left) and significance (right) in MiAUSSM versus $Q_Q$ and $Q_L$($Q_{H_u}=0$ and $Y_{C}=1.4$)}
\label{ae3}\end{figure} In figures \ref{ae1}, \ref{ae2} and \ref{ae3} we can see that $A_{E}$ 
is even for the exchanges $(Q_{H_u},~Q_Q,~0)\rightarrow (-Q_{H_u},~-Q_Q,~0)$,
$(Q_{H_u},~0,~Q_L)\rightarrow (-Q_{H_u},~0,~-Q_L)$, $(0,~Q_Q,~Q_L)\rightarrow (0,~-Q_Q,~Q_L)$ and 
$(0,~Q_Q,~Q_L)\rightarrow (0,~Q_Q,~-Q_L)$.
\newpage

\subsection{Best fit for the Asymmetry in terms of the three free charges}
  
In the general case we do not impose any constraint except that the three charges can assume values between $-1$ and $1$. From
the cross section of our process, that can be found in section \ref{crsec}, we can see that the amplitude is
proportional to the fourth power of the charges. So, from the equations (\ref{ARFBchinese}), (\ref{ao}) and (\ref{ACE})
we know that the asymmetry must be a rational function in which both the numerator and
denominator are fourth grade polynomials in the charges.
  \be
 A=\frac{\sum_{i,j,k=0}^n a_{ijk}(Q_{H_u})^i(Q_Q)^j(Q_L)^k}{\sum_{i,j,k=0}^n b_{ijk}(Q_{H_u})^i(Q_Q)^j(Q_L)^k}
\label{AQi}
\ee with $i+j+k= n\le 4$.
We have written a code that numerically calculates the coefficients of this fit. The simple idea is that by using the 
experimental values for the on peak asymmetries and by 
considering only three of the four definitions  we obtain a non-linear system with three equations and three variables 
($Q_{H_u}$, $Q_Q$ and $Q_L$ ) which could be solved numerically. In this way the asymmetry is useful for 
fixing the values of the $U(1)'$ charges. Moreover, once the values of the three charges are obtained by the previous 
system, the fourth definition of asymmetry can be used as a check for the validity of the model under 
exam. In fact its hypothetical experimental value must be recovered by using (\ref{AQi}) with the charge values already 
found, within the considered error (we use the medium relative error (MRE) for each asymmetry definition).
As expected from the results of the previous section (although in that case one of the charges was fixed 
to $0$), we have found out that the odd grade polynomials have negligible coefficients.
In table \ref{tab:fits1} we have written the non zero coefficient of our polynomial for the
four asymmetries.\\

   \begin{table}[ph!]
\begin{adjustwidth}{-5em}{5em}
%       \centering
\begin{tabular}{|c||c|c|c|c|c|}
\hline
               &          $A_{RFB}$              &           $A_O$               &           $A_C$               &         $A_E$\\
\hline \hline
    $a_{000}$  & $(-0.52\pm0.02) \times 10^{-6}$ & $(-0.31\pm0.04)\times10^{-6}$ & $(1.18\pm0.04)\times10^{-6}$  & $(0.86\pm0.18)\times10^{-6}$   \\
\hline
    $a_{200}$  & $(82\pm3) \times 10^{-6} $      & $(58\pm5)\times10^{-6}$       & $(-17\pm5)\times10^{-6}$      & $(140\pm21)\times10^{-6}$ \\
\hline
    $a_{020}$  & $(9.9\pm1.5)\times 10^{-6}$     & $(9\pm2)\times10^{-6}$        & $(-27\pm3)\times10^{-6}$      & $(12\pm11)\times10^{-6}$ \\
\hline
    $a_{002} $ & $(5.2\pm1.4)\times 10^{-6}$     & $(2\pm2)\times10^{-6}$        & $(11\pm2)\times10^{-6}$       & $(61\pm10)\times10^{-6}$ \\
\hline
    $a_{110}$  & $(5\pm3)\times 10^{-6}$         & $(4\pm5)\times10^{-6}$        & $(19\pm6)\times10^{-6}$       & $(113\pm24)\times10^{-6}$ \\
\hline
    $a_{101} $ & $(-18\pm4)\times 10^{-6}$       & $(-6\pm6)\times10^{-6}$       & $(-148\pm7)\times10^{-6}$     & $(-269\pm27)\times10^{-6}$  \\
\hline
    $a_{011} $ & $(-80\pm3)\times 10^{-6}$       & $(-65\pm5)\times10^{-6}$      & $(34\pm5)\times10^{-6}$       & $(-177\pm22)\times10^{-6}$\\
\hline
    $a_{400}$  & $1 (\text{fixed})$              & $1 (\text{fixed})$            & $-1 (\text{fixed})$           & $1 (\text{fixed})$ \\
\hline
    $a_{040}$  & $0.015638\pm0.000004$           & $0.015634\pm0.000006$         & $-0.015444\pm0.000007$        & $0.01552\pm0.00003$ \\
\hline
    $a_{004}$  & $0.000969\pm0.000002$           & $0.000967\pm0.000004$         & $-0.000984\pm0.000004$        & $0.000964\pm0.000018$ \\
\hline
    $a_{310} $ & $0.88011\pm0.00005$             & $0.87460\pm0.00007$           & $-0.85003\pm0.00008$          & $0.8573\pm0.0003$ \\
 \hline
    $a_{220}$  & $0.01525\pm0.00004$             & $0.01544\pm0.00006$           & $-0.01565\pm0.00007$          & $0.0163\pm0003$ \\
\hline
    $a_{130} $ & $-0.000729\pm0.000019$          & $-0.00061\pm0.00003$          & $0.00077\pm0.00003$           & $-0.00113\pm0.00013$ \\
\hline
    $a_{301}$  & $-1.99340\pm0.00005$            & $-1.99305\pm0.00008$          & $1.99360\pm0.00009$           & $-1.9930\pm0.0004$  \\
\hline
    $a_{211}$  & $-1.75973\pm0.00010$            & $-1.74845\pm0.00016$          & $1.69973\pm0.00018$           & $-1.7141\pm0.0007$  \\
\hline
    $a_{121}$  & $-0.00388\pm0.00007$            & $-0.00403\pm0.00011$          & $0.00429\pm0.00012$           & $-0.0052\pm0.0005$\\
\hline 
    $a_{031}$  & $0.00165\pm0.00002$             & $0.00128\pm0.00003$           & $-0.00188\pm0.00004$          & $0.00244\pm0.00015$  \\
 \hline  
    $a_{202}$  & $0.00456\pm0.00010$             & $0.00438\pm0.00016$           & $-0.00463\pm0.00018$          & $0.0048\pm0.0007$ \\
\hline
    $a_{112}$  & $-0.00049\pm0.00009$            & $-0.00044\pm0.00014$          & $0.00013\pm0.00015$           & $-0.0011\pm0.0006$ \\
\hline
    $a_{022}$  & $0.00773\pm0.00003$             & $0.07740\pm0.00005$           & $-0.00778\pm0.00006$          & $0.0068\pm0.0002$ \\
\hline  
    $a_{103}$  & $-0.001244\pm0.000013$          & $-0.00120\pm0.00002$          & $0.00136\pm0.00002$           & $-0.00170\pm0.00009$ \\
\hline   
    $a_{013}$  & $0.000181\pm0.000015$           & $0.00023\pm0.00002$           & $-0.00014\pm0.00003$          & $0.00047\pm0.00011$ \\
\hline \hline
    $b_{000}$  & $(-1.19\pm0.06)\times10^{-6}$   & $(-0.70\pm0.09)\times10^{-6}$ & $(-3.19\pm0.12)\times10^{-6}$ & $(2.2\pm0.4)\times10^{-6}$   \\
\hline
    $b_{200}$  & $(121\pm7)\times10^{-6}$        & $(181\pm12)\times10^{-6}$     & $(-88\pm16)\times10^{-6}$     & $(-637\pm57)\times10^{-6}$ \\
\hline
    $b_{020}$  & $(23\pm4)\times10^{-6}$         & $(21\pm6)\times10^{-6}$       & $(74\pm7)\times10^{-6}$       & $(29\pm28)\times10^{-6}$ \\
\hline
    $b_{002} $ & $(12\pm3)\times10^{-6}$         & $(5\pm5)\times10^{-6}$        & $(-29\pm6)\times10^{-6}$      & $(154\pm25)\times10^{-6}$ \\
\hline
    $b_{110}$  & $(-58\pm26)\times10^{-6}$       & $(51\pm41)\times10^{-6}$      & $(-282\pm52)\times10^{-6}$    & $(2392\pm204)\times10^{-6}$ \\
\hline
    $b_{101} $ & $(-116\pm13)\times10^{-6}$      & $(-207\pm20)\times10^{-6}$    & $(408\pm27)\times10^{-6}$     & $(-392\pm100)\times10^{-6}$  \\
\hline
    $b_{011} $ & $(85\pm39)\times10^{-6}$        & $(-82\pm62)\times10^{-6}$     & $(-104\pm79)\times10^{-6}$    & $(-1120\pm306)\times10^{-6}$\\
\hline
    $b_{400}$  & $1.90571\pm0.00004$             & $1.90585\pm0.00006$           & $2.28541\pm0.00008$           & $2.1465\pm0.0003$ \\
\hline
    $b_{040}$  & $0.036021\pm0.000010$           & $0.036021\pm0.000016$         & $0.04190\pm0.00002$           & $0.03945\pm0.00008$ \\
\hline
    $b_{004}$  & $0.002232\pm0.000006$           & $0.002229\pm0.000009$         & $0.002670\pm0.000012$         & $0.00245\pm0.00005$ \\
\hline
    $b_{310} $ & $1.40007\pm0.00018$             & $1.3899\pm0.0003$             & $1.3347\pm0.0004$             & $1.2650\pm0.0015$ \\
 \hline
    $b_{220}$  & $3.8269\pm0.0003$               & $3.8323\pm0.0004$             & $4.5879\pm0.0006$             & $4.338\pm0.002$ \\
\hline
    $b_{130} $ & $-0.00386\pm0.00018$            & $-0.0033\pm0.0003$            & $-0.0035\pm0.0004$            & $-0.0103\pm0.0014$ \\
\hline
    $b_{301} $ & $-3.79461\pm0.00014$            & $-3.7943\pm0.0002$            & $-4.5504\pm0.0003$            & $-4.2761\pm0.0011$  \\
\hline
    $b_{211}$  & $-2.7984\pm0.0004$              & $-2.7775\pm0.0007$            & $-2.6661\pm0.0009$            & $-2.533\pm0.003$  \\
\hline
    $b_{121}$  & $-7.5882\pm0.0007$              & $-7.5998\pm0.0011$            & $-9.1030\pm0.0014$            & $-8.592\pm0.005$\\
\hline 
    $b_{031}$  & $0.0081\pm0.0002$               & $0.0061\pm0.0003$             & $0.0098\pm0.0005$             & $0.0168\pm0.0017$  \\
 \hline 
    $b_{202}$  & $3.8004\pm0.0003$               & $3.8021\pm0.0004$             & $4.5567\pm0.0006$             & $4.291\pm0.002$ \\
\hline
    $b_{112}$  & $2.8003\pm0.0005$               & $2.7871\pm0.0009$             & $2.6705\pm0.0011$             & $2.524\pm0.04$ \\
\hline
    $b_{022}$  & $7.6032\pm0.0006$               & $7.6124\pm0.0009$             & $9.1181\pm0.0012$             & $8.599\pm0.004$ \\
\hline
    $b_{103}$  & $-0.0023\pm0.0002$              & $-0.0012\pm0.0004$            & $-0.0026\pm0.0005$            & $-0.0120\pm0.0018$ \\
\hline
    $b_{013}$  &  $-0.00009\pm0.00035$           & $0.0002\pm0.006$              & $-0.0021\pm0.0007$            & $0.017\pm0.003$ \\
\hline
\end{tabular}
\end{adjustwidth} 
\caption{Coefficients of the fits for  the four definitions of asymmetry} \label{tab:fits1}
\end{table} Note that this table contains only the statistical error and not the systematic error
due to the choice of the PDFs.\\ We have calculated the goodness of the fit parameter $R^2$ (a perfect fit has $R^2=1$) 
and the medium relative error for these results, obtaining the results that are showed in 
table \ref{controlfit}, attesting the accuracy of the procedure. In particular the $R^2$ value states (as we expect)
that the errors in our fit are only due to numerical approximation in the calculation
of the integrals that, given the cross section, give back the asymmetry .
\begin{table}[h]
 \centering
\begin{tabular}[h]{|c||c|c|c|c|}
 \hline & $A_{RFB}$ & $A_{O}$ & $A_C$ & $A_E$  \\
 \hline \hline $R^2$ & $0,999$ & $0,999$ & $0,999$ & $0,999$ \\
 \hline $MRE$ & $0.008$ & $0.009$ & $0.019$ & $0.017$ \\
 \hline
\end{tabular}
\caption{$R^2$ and Medium Relative Error for the polynomial fit of the 
asymmetry with respect to the three charges}
\label{controlfit}
\end{table}

\noindent \vskip 1cm

\newpage

\chapter{Conclusion}
In this thesis we have studied many phenomenological aspect of the MiAUSSM. First of all,
we have obtained the neutralino mass matrix and we have derived the conditions to decouple
the anomalous sector from the MSSM sector. We have found that at the tree level
this condition is simply $Q_{H_u}=0$.\\
In order to calculate the cross sections we have calculated all the vertices of the model
with and without mixing between the two sectors of the model. Then we have used these results
to calculate the cross sections of the LSP-LSP annihilation and of the LSP-NLSP coannihilation
in the case that the LSP is a St\"{u}ckelino and the NLSP is a MSSM neutralino.\\
We have found that in the decoupled case the annihilation of two LSPs has a much weaker cross
section than that needed to be in agreement with the WMAP results. So the St\"{u}ckelino 
cannot explain alone the DM relic density. Differently, we have found that if the St\"{u}ckelino
coannihilates with an MSSM neutralino plus possibly other particles we can obtain the WMAP
results. In the case of coannihilation with an NLSP that is almost an MSSM bino we have obtained
that up to a mass gap of $10\%$ we can satisfy the experimental constraints; in the case of 
coannihilation with an NLSP that is almost an MSSM wino and with its related chargino we 
find an agreement with the experimental results up to a $20\%$ mass gap.\\
We have also studied the general case, in which we do not decouple the two sectors of the 
neutralinos mass matrix. In this case we cannot have a pure St\"{u}ckelino as LSP, a pure bino 
as NLSP, etc. So the cross section of the LSP coannihilation is much more complicated that in 
the decoupled case. To deal with it we have modified the DarkSUSY package.\\
Our motivation is simple: this package already contains routines to numerically calculate the relic 
density of particles in the most general cases. It is designed to work
in an MSSM background, so we have added the new particles, couplings and all the interaction vertices 
introduced by the MiAUSSM. We have checked that our modifications do not affect the results
given by the program if we shut down the extra $U(1)$.\\
We have used the modified version of DarkSUSY to extensively study the relic density of the LSP in the
MiAUSSM with respect to the general set of free parameters given by the seven of the MSSM-7 and the five
added by the anomalous extension. As in the case of decoupling of the anomalous and the MSSM sectors
in the neutralinos mass matrix, we have studied separately the case in which the NLSP is
mostly a bino and the case in which it is mostly a wino.\\
In both cases we have checked that the introduction of the extra $U(1)$ does not lead to divergences or
unphysical behaviour in the relic density. We have also found that the relic density dependence on
the MSSM parameters in our model is similar to that in the MSSM. For example we have found an example
of funnel region, a phenomenon that is well known to happen in the MSSM if $M_{A_0}\sim 2 M_{LSP}$
and it is found in our model under the same condition. We have also checked that to satisfy experimental
constraints we must obey to $-1\lesssim Q_{H_u}\lesssim 1$.\\
Our main result in this context is the study of the dependence of the relic density on the 
anomalous masses $M_S$ and $M_{A_0}$. We have found that for mass gap of the order of $5\%$
we satisfy the WMAP data in a region much wider than that in the case of mass gap around $10\%$ 
and above. This implies that for small mass gap our model can satisfy the experimental constraints 
significantly better than for bigger mass gap.\\
In the second part of the thesis we have studied the asymmetry in the MiAUSSM for the process
$pp\rightarrow e^-e^+$ that is studied at the LHC. The asymmetry is used to impose constraints on theoretical
models, to perform consistency checks and to calculate some quantities, usually the charges.\\
We have calculated the elementary cross sections that contribute to this process, adding the
contribution of the $Z'$ to those of the $\g$ and the $Z_0$, already present in the SM.
Then we have calculated the asymmetry using four different definitions that are viable at
the LHC, namely rapidity dependent forward backward asymmetry, one side asymmetry, central
asymmetry and edge asymmetry ($A_{RFB}$, $A_O$, $A_C$ and $A_E$ respectively).\\
All of these definitions contain a cut in the parameter space, necessary to obtain a result
different from $0$. We have calculated the value for the cut in each definition to have maximum
significance and then we have studied the asymmetries (now \textquotedblleft optimized \textquotedblright).\\
The asymmetries and their related significancies are functions of the three free charges of the MiAUSSM. 
We have studied their dependence on the three possible couples
of charges keeping each time one of them fixed. We have found that the asymmetries are even for exchange of sign 
of the charges as we expected.\\
We have also studied the general case, in which no charge is kept fixed. In this case we have calculated 
the value of the four definitions of asymmetry for discretized values of the charges in the interval $[-1,1]$.
We have used this result to calculate a fit of the asymmetries as rational functions of the charges.
Using these fits and using eventually founded experimental results, we have shown how can be obtained
constrained values for the charges.\\

%\addcontentsline{toc}{chapter}{Bibliography}

\end{document}